\documentclass[aps,prb,twocolumn,groupedaddress,floatfix,letterpaper,nofootinbib]{revtex4-2}

\definecolor{refkey}{rgb}{0.5,0.8,0.5}
\definecolor{labelkey}{rgb}{0.5,0.8,0.5}

\usepackage{makecell}
\usepackage{array}

\usepackage[normalem]{ulem} 

\newcommand{\id}{\mathrm{id}}
\newcommand{\w}{\mathrm{w}}
\newcommand{\RZ}{{\mathbb{R}/\mathbb{Z}}}

\newcommand{\eGau}{\eG\mathrm{au}}

\newcommand{\cQL}{\mathcal{QL}}

\newcommand{\psymm}{patch symmetry}
\newcommand{\psymms}{patch symmetries}
\newcommand{\symmTO}{symmetry TO}
\newcommand{\symmTOs}{symmetry TOs}

\newcommand{\hcatsymm}{holo-cat symmetry}

\newcommand{\catsymm}{{\sf categorical symmetry$^{\bigcirc\hskip -0.65em h \hskip 0.2em}$}}
\newcommand{\catsymms}{{\sf categorical symmetries$^{\bigcirc\hskip -0.65em h \hskip 0.2em}$}}
\newcommand{\nBFcat}{nBF category}
\newcommand{\nBFcats}{nBF categories}
\newcommand{\ord}{\textrm{ord}}

\newcommand{\xZ}{{\Z \hskip -0.7em \mathsf{X}}}

\begin{document}

\begin{titlepage}

\title{Holographic theory for continuous phase transitions \\
-- the emergence and symmetry protection of gaplessness }

\author{Arkya Chatterjee} 
\affiliation{Department of Physics, Massachusetts Institute of Technology,
Cambridge, Massachusetts 02139, USA}

\author{Xiao-Gang Wen} 
\affiliation{Department of Physics, Massachusetts Institute of Technology,
Cambridge, Massachusetts 02139, USA}

\begin{abstract} 

Two global symmetries are \emph{holo-equivalent} if their algebras of local
symmetric operators are isomorphic.  Holo-equivalent classes of global
symmetries are classified by gappable-boundary topological orders (TO) in
one higher dimension (called symmetry TO), which leads to a
symmetry/topological-order (Symm/TO) correspondence.  We establish that: (1)
For systems with a symmetry described by symmetry TO $\EuScript{M}$, their
gapped and gapless states are classified by condensable algebras $\mathcal{A}$,
formed by elementary excitations in $\EuScript{M}$ with trivial self/mutual
statistics. Such classified states (called $\mathcal{A}$-states) can describe
symmetry breaking orders, symmetry protected topological orders, symmetry
enriched topological orders, gapless critical points,  {\it etc}., in a unified
way.  (2) The local low-energy properties of an $\mathcal{A}$-state can be
calculated from its reduced symmetry TO $\EuScript{M}_{/\mathcal{A}}$, using
holographic modular bootstrap (holoMB) which takes
$\EuScript{M}_{/\mathcal{A}}$ as an input.  Here $\EuScript{M}_{/\mathcal{A}}$
is obtained from $\EuScript{M}$ by condensing excitations in $\mathcal{A}$.
Notably, an $\mathcal{A}$-state must be gapless if
$\EuScript{M}_{/\mathcal{A}}$ is nontrivial.  This provides a unified
understanding of the emergence and symmetry protection of gaplessness that
applies to symmetries that are anomalous, higher-form, and/or non-invertible.
(3) The relations between condensable algebras  constrain the structure of the
global phase diagram.  We find that, for 1+1D $\mathbb{Z}_2 \times
\mathbb{Z}_2'$ symmetry with mixed anomaly, there is a stable continuous
transition (deconfined quantum critical point) between the
$\mathbb{Z}_2$-breaking-$\mathbb{Z}_2'$-symmetric phase and the
$\mathbb{Z}_2$-symmetric-$\mathbb{Z}_2'$-breaking phase. The critical point is
the same as a $\mathbb{Z}_4$ symmetry breaking critical point.  (4) 1+1D
bosonic systems with $S_3$ symmetry have four gapped phases with unbroken
symmetries $S_3$, $\mathbb{Z}_3$, $\mathbb{Z}_2$, and $\mathbb{Z}_1$.  We find
a duality between two transitions $S_3 \leftrightarrow \mathbb{Z}_1$ and
$\mathbb{Z}_3 \leftrightarrow \mathbb{Z}_2$: they are either both first order
or both (stably) continuous, and in the latter case, they are described by the
same conformal field theory (CFT).  (5) The gapped and gapless states for 1+1D
bosonic systems with anomalous $S_3$ symmetries are obtained as well.  For
example, anomalous $S_3^{(1)}$ and $S_3^{(2)}$ symmetries can have 
symmetry protected stable chiral gapless phases.

\end{abstract}

\maketitle

\end{titlepage}

\setcounter{tocdepth}{1} 
{\small \tableofcontents }

\section{Introduction}

For a long time, Landau's symmetry breaking theory \cite{L3726,L3745} was
regarded as the standard theory for continuous phase transitions.  In
particular, it was believed that all continuous phase transitions are
spontaneous symmetry breaking transitions, where the symmetry groups for the
two phases across a transition have a special relation 
\begin{equation*}
G_\text{small} \subset G_\text{large}.
\end{equation*}
\ie the symmetry group for the phase with less symmetry is a strict subgroup of
the symmetry group for the phase with more symmetry.  

\subsection{Symmetry/Topological-Order (Symm/TO) correspondence }

However, in the last 30 years, increasingly many examples of continuous phase
transitions have been discovered in quantum systems whose description is beyond
Landau's theory.  Continuous quantum phase transitions were found between two
states with the same symmetry \cite{WW9301,CFW9349,SMF9945,RG0067,W0050,RW0620}
(but different topological orders \cite{W9039,KW9327}). Continuous quantum
phase transitions are also possible between two states with \emph{incompatible}
symmetries \cite{SVB0490}, \ie the symmetry groups of the two phases across the
transition do not have the group-subgroup relation. These ``deconfined quantum
critical points" (DQCPs) have been found to be related to mutual anomalies
between internal and lattice symmetries.\cite{WS170302426,MT170707686} Even
symmetry-breaking transitions with well-defined order parameters are sometimes
not described by Landau's symmetry breaking theory \cite{RW0620}.  In light of
these examples, it appears that many continuous quantum phase transitions are
not described by Landau theory, regardless of whether they have symmetry
breaking and order parameters or not.  There are many situations and mechanisms
that can lead to continuous quantum phase transitions that go beyond Landau
symmetry breaking theory. It is interesting to ask whether there is a unified
theory to understand these various beyond-Landau continuous quantum phase
transitions.

To systematically understand gapless critical points at continuous
transitions, it is fruitful to identify all the emergent symmetries in the gapless states.
Emergent symmetry can be very rich and may include 0-symmetry,
higher\footnote{We use ``higher symmetry'' to cover both higher-form symmetry
and higher-group symmetry.} symmetry
\cite{NOc0605316,NOc0702377,KT13094721,GW14125148}, anomalous symmetry
\cite{H8035,CGL1314,W1313,KT14030617}, anomalous higher symmetry
\cite{KT13094721,GW14125148,TK151102929,T171209542,P180201139,DT180210104,BH180309336,ZW180809394,WW181211968,WW181211967,GW181211959,WW181211955,W181202517},
beyond-anomalous symmetry \cite{CW220303596}, non-invertible symmetry
\cite{PZh0011021,CSh0107001,FSh0204148,CY180204445,TW191202817,I210315588,Q200509072},
algebraic higher symmetry \cite{KZ200308898,KZ200514178}, and/or non-invertible
gravitational anomaly
\cite{W1313,KW1458,FV14095723,M14107442,KZ150201690,KZ170200673,JW190513279}.
Recently, a symmetry/topological-order (Symm/TO) correspondence was proposed
\cite{JW191213492,KZ200514178} that can provide a unified description of all
those symmetries.  

One way to have a unified description of all these symmetries is to restrict
to the symmetric sub Hilbert space $\cV_\text{symmetric}$,
 which does not have a tensor product
decomposition
\begin{align}
\cV_\text{symmetric} \neq \bigotimes_i \cV_i.
\end{align}
Here, $\cV_i$'s are local Hilbert spaces on each lattice site.  The failure of
tensor product decomposition indicates \cite{JW190513279} a non-invertible
gravitational anomaly \cite{W1313,KW1458,FV14095723,M14107442,KZ150201690}.
This leads to the point of view that
\begin{align}
\label{symmgrav}
&\ \ \ \
\text{symmetry (restricted to } \cV_\text{symmetric})
\nonumber\\
& = \text{non-invertible
gravitational anomaly}
\end{align}
For a finite symmetry, its corresponding non-invertible gravitational anomaly
is the same as gappable-boundary topological order (TO)  \footnote{Here, the
topological order in one higher dimension is anomaly-free (\ie has UV
completion). In this paper, the term \emph{topological order} always refers to
anomaly-free topological order.  Topological order with anomaly will be
explicitly referred to as \emph{anomalous topological order}.}  in one higher
dimension.  ``Gappable-boundary topological order in one higher dimension'' is
a long name. We will refer to it as \textbf{symmetry TO}\footnote{A symmetry
TO, as a gappable-boundary topological order, always describes symmetries in
one lower dimension.} \cite{W1313,KW1458,KZ150201690}. This leads to a
holographic view of symmetry: Symm/TO correspondence
\cite{JW191213492,KZ200514178} 
\begin{align}
\label{symmTO}
&\ \ \ \
\text{symmetry (restricted to } \cV_\text{symmetric})
\nonumber\\
&= \text{symmetry TO}
\end{align}
This holographic perspective on symmetry in 1+1D was also discussed in
\Rf{CZ190312334,TW191202817,LB200304328}. 

\begin{table*}[tb] 
\caption{The first row is the classification of 2+1D topological orders (up to
$E(8)$ invertible topological order) for bosonic systems with no symmetry, up
to 10 types of anyons.  This leads to a classification of 2+1D \symmTOs, which
classify all the 1+1D global symmetries up to holo-equivalence (the second
row).  Such a classification include all finite-group symmetries with potential
anomalies (the third row).  It also includes beyond-group symmetries, such as
the Fibonacci symmetry in Fig. \ref{emFib}.  } \label{toptable} \centering
\setlength{\tabcolsep}{5pt}
\begin{tabular}{|c | c|c|c|c|c|c|c|c|c|c|}
\hline
\# of anyon types  (rank)
& 1 & 2 &  3 &  4 &  5 &  6 &  7 &  8 &  9 & 10\\
\hline
\hline
\# of 2+1D topological orders (MTC)
& 1 & 4 & 12 & 18 & 10 & 50 & 28 & 64 & 81 & 76\\
\hline
\# of \symmTOs\ (MTC in trivial Witt class) 
& 1 & 0 & 0 & 3 & 0 & 0 & 0 & 6 & 6 & 3\\
\hline
\# of finite-group symmetries (with anomaly $\om$) 
& 1 & 0 & 0 & 2$_{\Z_2^\om}$ & 0 & 0 & 0 & 6$_{S_3^\om}$ & 3$_{\Z_3^\om}$ & 0\\
\hline
\end{tabular}
\end{table*}

A second way to have a unified description of all emergent symmetries
generalizes the idea that, to describe an ordinary symmetry, we can use the
conservation law (\ie the fusion ring) of symmetry charges.  To obtain a
unified description, we use instead the fusion rings (conservation laws) of
both symmetry charges and symmetry defects at an equal footing
\cite{JW191213492}.  The resulting symmetry is called \catsymm\footnote{Here,
we use the term \catsymm\ in the original holographic sense of
\Rf{JW191213492,KZ200514178}.  However, the term ``categorical symmetry'' has
since been used by many to describe non-invertible symmetry. To avoid possible
confusions, we use \catsymm\ in Sans Serif Font with superscript
${\bigcirc\hskip -1.9ex h}\ $ to stress that we use the term in the holographic
sense. See also Appendix \ref{name} for more detailed explanations and
discussions on related concepts.} It is also necessary to include ``braiding''
properties of symmetry charges/defects \cite{JW191213492} which allow us to
describe the symmetry actions to have a full description of
symmetry.\footnote{A symmetry is described by the algebra of local symmetric
operators.  The ``braiding'' properties are features of such an algebra. See
\Rf{CW220303596} for details.  Such features become the braiding properties in
the \symmTO\ in Symm/TO correspondence, which leads to the name ``braiding''
properties.  Symmetry charges always has the trivial ``braiding'' property.
Thus, in the ordinary symmetry described by fusion ring of symmetry charge, we
do not need to introduce extra data to describe such a trivial ``braiding''
property.} Thus \catsymm\ has both fusion ring layer and ``braiding'' layer.
Just like ordinary symmetry is described by group, \catsymm\ is described by
non-degenerate braided fusion higher category in trivial Witt class
(which is referred to as \textbf{\nBFcat} in short)
\cite{JW191213492,KZ200514178}.  Here ``non-degenerate'' indicates that we have
included all the symmetry charges and the symmetry defects \cite{CW220303596},
and ``higher'' refers to the fact that the symmetry charges and defects can be
point-like, string-like, \etc.  ``In trivial Witt class'' is required if the
corresponding symmetry can be realized lattice model in the same dimension.

\begin{figure}[t]
\begin{center}
\includegraphics[scale=0.7]{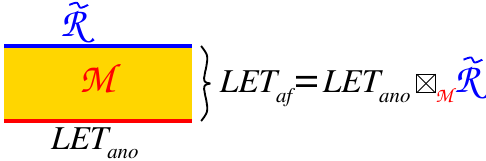}
\end{center}
\caption{ A 1+1D lattice model with emergent Fibonacci symmetry at low
energies.  The 1+1D lattice model is constructed from a slab of 2+1D lattice.
In the bulk, we have a commuting-projector Hamiltonian that realizes a
double-Fibonacci topological order \cite{LW0510} with large energy gap.  The top
boundary $\t\cR$ is a gapped boundary of the double-Fibonacci topological order
with large energy gap.  The lower boundary is described by an anomalous
low energy theory LET$_{ano}$.  The  low energy theory
LET$_{af}$ of the slab has an emergent Fibonacci symmetry below the energy gaps
of the bulk and top boundary.  } \label{emFib} 
\end{figure}

Recently, modular tensor categories with up to 10 types of anyons are
classified.  This leads to an classification of 1+1D generalized symmetries
with 10 or less symmetry charges/defects, via the classification of 2+1D
\symmTO\ up to rank 10 (see Table \ref{toptable}).  For example, for global
symmetries with 4 types of symmetry charges/defects, the three holo-equivalence
classes (which contain only one symmetry each in this case) are: (1) $\Z_2$
symmetry where the \symmTO\ is the 2+1D $\Z_2$ gauge theory; (2) anomalous
$\Z_2$ symmetry  where the \symmTO\  is the double-semion topological order;
(3) Fibonacci symmetry where the \symmTO\  is the double-Fibonacci topological
order (see Fig. \ref{emFib}).  From Table \ref{toptable}, we also see a clear
distinction between generic TO which may not allow gapped boundary and \symmTO\
which allows gapped boundary.

The above holographic view of symmetry and anomaly is motivated variously from
anomaly-inflow \cite{CH8527}, from the boundary-bulk topological holographic
relation
\cite{W9125,W9139,FT1292,KW1458,KZ150201690,KZ170200673,TW191202817,KZ200514178},
from an observation that symmetry protected topological (SPT) order
\cite{GW0931,CLW1141,CGL1314} is closely related to anomaly in one lower
dimension \cite{RML1204,W1313,TK151102929}, and from an observation that SPT
order and anomaly are closely related to braiding \cite{LG1209,W181202517}.
This holographic point of view has parallels with the AdS/CFT correspondence
\cite{Mh9711200,Wh9802150}, where a \emph{continuous} $G$-symmetry of a CFT is
associated to a $G$-gauge theory in an \emph{AdS space} in one higher
dimension. There are however some important differences between the two. In
Symm/TO correspondence, a \emph{finite} $G$-symmetry of a CFT is associated to
a $G$-gauge theory in one higher dimension with \emph{arbitrary
metric}.\footnote{The metric is arbitrary since $G$-gauge theory is topological
for finite $G$}  Moreover, in Symm/TO correspondence, the bulk theory is not
equivalent to the boundary theory. The bulk topological order (\ie the \symmTO)
just constrains the boundary dynamics.

We should note that, so far, the Symm/TO correspondence only applies to finite
symmetry. For continuous symmetry, we either need to generalize the Symm/TO
correspondence, or need to develop a new non-holographic point of view as in
\Rf{CW220303596}.  To that end, there is a third \emph{non-holographic} way to reach a unified description
of all emergent symmetries in a gapless state. Here one starts from the point of view that a
symmetry is fully described by an algebra of local symmetric operators (LSOs).  An
ordinary (global) symmetry is characterized by symmetry transformations, which
are the commutants of LSOs.\footnote{The commutants of
local symmetric operators are operators that commute with all the local
symmetric operators.} These symmetry transformations act on the whole space (or
on all the closed subspaces of codimension-$p$ for $p$-symmetry), and
correspond to the global symmetry transformations.
In this approach, we restrict to the symmetric sub
Hilbert space $\cV_\text{symmetric}$, as in the first approach.  In this case, we find that the global
symmetry transformations act trivially as identity operator.  Seemingly,
we do not see any  global symmetry after the Hilbert space restriction.  On the other hand, even after restricting to $\cV_\text{symmetric}$, symmetry clearly still
constrains the low energy dynamics and is physically meaningful.  To see the
symmetry in this case, \Rf{JW191213492,CW220303596} considered the so called
``commutant patch operators'', referred to as ``transparent patch operators''
in \Rf{CW220303596}. Commutant patch operators are operators formed by local
symmetric operators (LSOs), acting on 1-dimensional, 2-dimensional, \etc\ open subspaces
(\ie patches), and commute with all the LSOs as long as
the LSOs are far away from the boundaries of the patches and have no non-trivial linking.
Since the commutants of LSOs define global symmetry, we
say the commutant patch operators of local symmetric operators define the ``\textbf{patch symmetry}'' of the system.
\Rf{JW191213492,CW220303596} found that there are two kinds of commutant patch 
operators: the first kind are global symmetry transformations restricted on the
patches, which are called patch symmetry operators.  The boundaries of patch
symmetry operators corresponds to symmetry defects.  The second kind have empty
bulk and create neutral charge objects on their boundaries, which are called
patch charge operators.  The boundaries of patch charge operators corresponds
to symmetry charges.  We see that, in contrast to global symmetry, patch
symmetry treats symmetry charges and symmetry defects at an equal footing.  The
algebra of commutant patch operators encode the fusion ring and ``braiding''
properties of symmetry charges/defects, which is conjectured to give rise to a
\nBFcat \cite{CW220303596}.  Thus the patch symmetry is identical to \catsymm\ and they are both described by \nBFcat.

We define two symmetries to be \emph{holo-equivalent} \cite{KZ200514178} if the
algebras of their local symmetric operators are isomorphic.  We define two
patch symmetries to be the same if the algebras of their commutant patch
operators are isomorphic.  This allows us to summarize the above discussions:
\begin{align}
\label{eqsymms}
&\text{(generalized) global symmetries (restricted to } \cV_\text{symmetric})
\nonumber\\
& =
\text{\catsymms} = \text{\psymms} 
\nonumber \\
& =
\text{holo-equivalence classes of global symmetries} 
\nonumber\\
& = \text{\nBFcats}
\nonumber\\
& = \text{\symmTOs\ (for finite symmetry)}
.
\end{align}
Here, global symmetry (restricted to $ \cV_\text{symmetric}$) is viewed from
the point of view of the algebra of local symmetric operators, and ``='' means one-to-one
correspondence.  We remark that \eqref{eqsymms} is more precise than
\eqref{symmgrav} and \eqref{symmTO}.

We see that (generalized) global symmetry is different from \catsymm\ or
\psymm\ (which are two names for the same thing).  In fact, \catsymm\ (or
\psymm) only looks at a global symmetry from a local point of view, ignoring
the global features \cite{JW191213492}.  Thus  \catsymm\ (or \psymm) corresponds
to a holo-equivalence class of global symmetries.  As a result, \symmTO\ and
\nBFcat\ only describe the holo-equivalence class of (generalized) global
symmetries.  

The four terms, \catsymm, \psymm, \symmTO, and \nBFcat, describe almost the
same thing, but stress on different aspects: \catsymm\  \cite{JW191213492}
emphasizes on symmetry + dual symmetry \cite{BT170402330} (\ie treating
symmetry charges and defects at equal footing); \psymm\ emphasizes on its
difference with global symmetry; \symmTO\ emphasizes on the holographic
picture; \nBFcat\ is most accurate.  We can use any of them.  But since
``categorical symmetry'' has been used by many to mean non-invertible global
symmetry, in the rest of this paper, we will use \emph{\symmTO}.  We like to
remark that \symmTO\ can only describe finite symmetries.  For continuous
symmetries, we need to use \nBFcats\ with infinite objects/morphisms
\cite{W221204963} to describe them.  Therefore \nBFcat\ is a more accurate
term.  We use the term \symmTO\ since it is more easily associated with
symmetry and holographic picture.


Let us also point out that the \symmTO\ can be used to describe an exact UV
symmetry of a lattice model. But it can also be used to describe an emergent
symmetry that appears only at low energies (IR).  Consider a system with a
separation of energy scale, \ie some excitations have much higher energies
compared to all other excitations which may be gapped or gapless. Well below
the energy gap of the high energy excitations, the low energy properties of the
system are controlled by an emergent symmetry described by \symmTO\ $\eM$.  

If the low energy excitations are gapped, then they can be described by a
fusion $n$-category $\cC$ if the space is $n$-dimensional.  In this case, the
emergent symmetry is described by an \symmTO\ $\eM= \eZ(\cC)$, the ``Drinfeld''
center of the low energy excitations $\cC$
\cite{KW1458,KZ150201690,KZ170200673,KZ200514178}.  It was pointed out in
\Rf{PW230105261} that the emergent higher symmetry contained in $\eM= \eZ(\cC)$
is exact, while the emergent 0-symmetry contained in $\eM= \eZ(\cC)$ is
approximate.

If the low energy excitations are gapless, then the \emph{maximal} emergent
\symmTO\ \cite{CW221214432} may largely characterize the gapless state. We
know that the possible gapless states are very rich, and it is hard to believe
gapless states can be characterized by their emergent symmetries, if we only
consider emergent symmetries described by groups.  However, emergent symmetries
can be generalized symmetries that are beyond group or beyond higher group.  We
need to use \symmTOs\ to
describe those emergent generalized symmetries.  In this case, it may be
possible that emergent maximal \symmTO\ can  largely characterize gapless
states.

\subsection{Characterizing gapless liquid states using Symm/TO correspondence}

In a series of papers \cite{KZ201102859,KZ210703858,KZ220105726}, Kong and
Zheng have developed a unified mathematical theory for topological orders and
gapless quantum liquids\footnote{Here we use the notion of \emph{liquid state}
for quantum systems in the sense defined in \Rf{ZW1490,SM1403}. We will not
discuss non-liquid states, such as fractons.} in $n$-dimensional spacetime,
based on symmetric monoidal higher category, $\cQL^n$, which is called category
of quantum liquids.  $\cQL^n$ for different $n$ are related by 
delooping $\Si_* \cQL^n = \cQL^{n+1}$ (Hypothesis 5.16 in
\Rf{KZ201102859,KZ201102859}).  A quantum liquid $L$ (an object in $\cQL^n$)
contains two parts: a topological skeleton $L_\text{sk}$ and a local quantum
symmetry $L_\text{lqs}$, where the topological skeleton $L_\text{sk}$ is an
anomalous topological order.  In 1+1D, Kong and Zheng have also developed a
theory for gapless boundaries of 2+1D topological order based on categories
enriched by local quantum symmetry -- vertex operator algebra
\cite{KZ170501087,KZ190504924,KZ191201760}.

\begin{figure}[t]
\begin{center}
\includegraphics[scale=0.072]{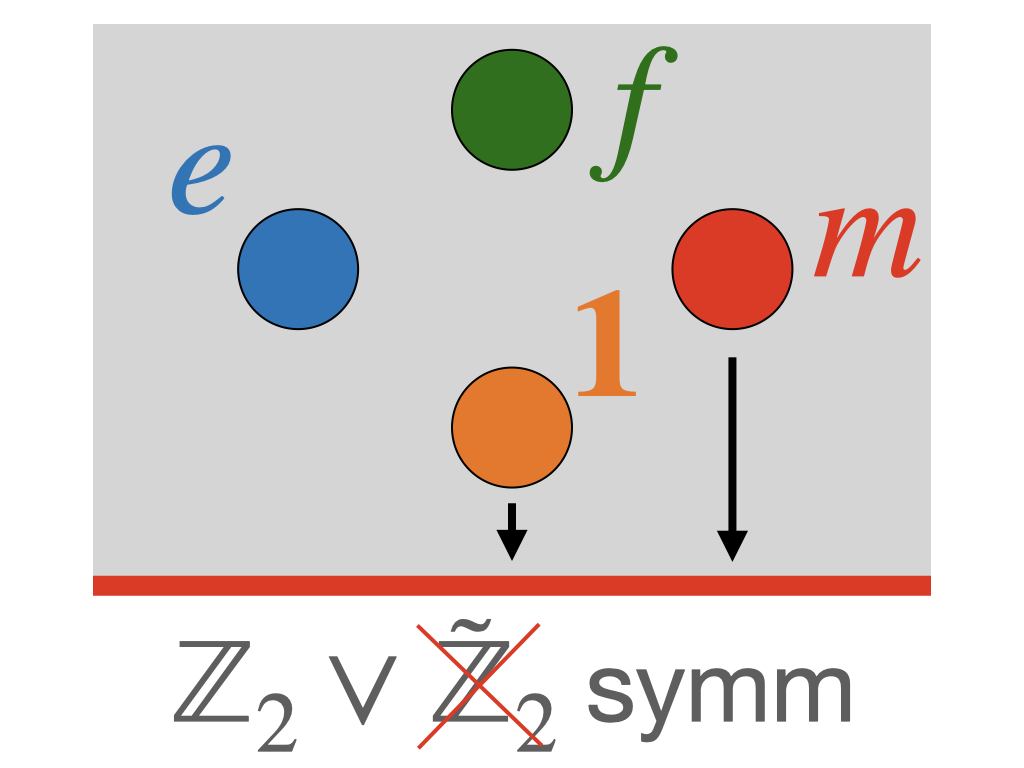}\ \ \ \
\includegraphics[scale=0.072]{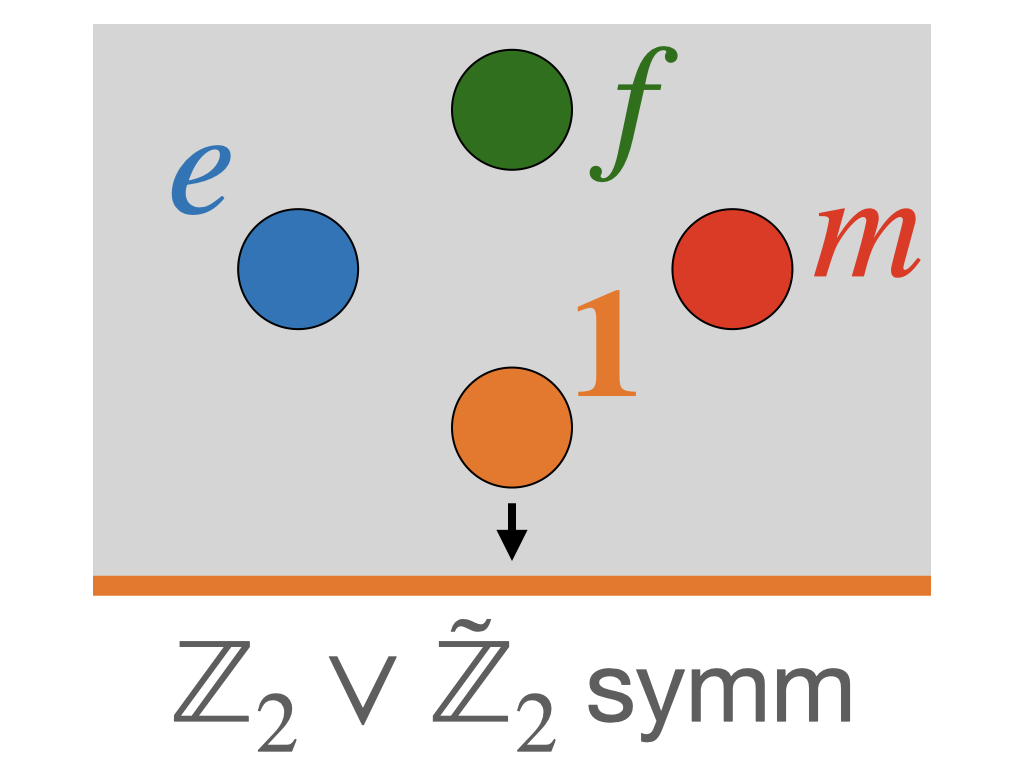}\ \ \ \
\includegraphics[scale=0.072]{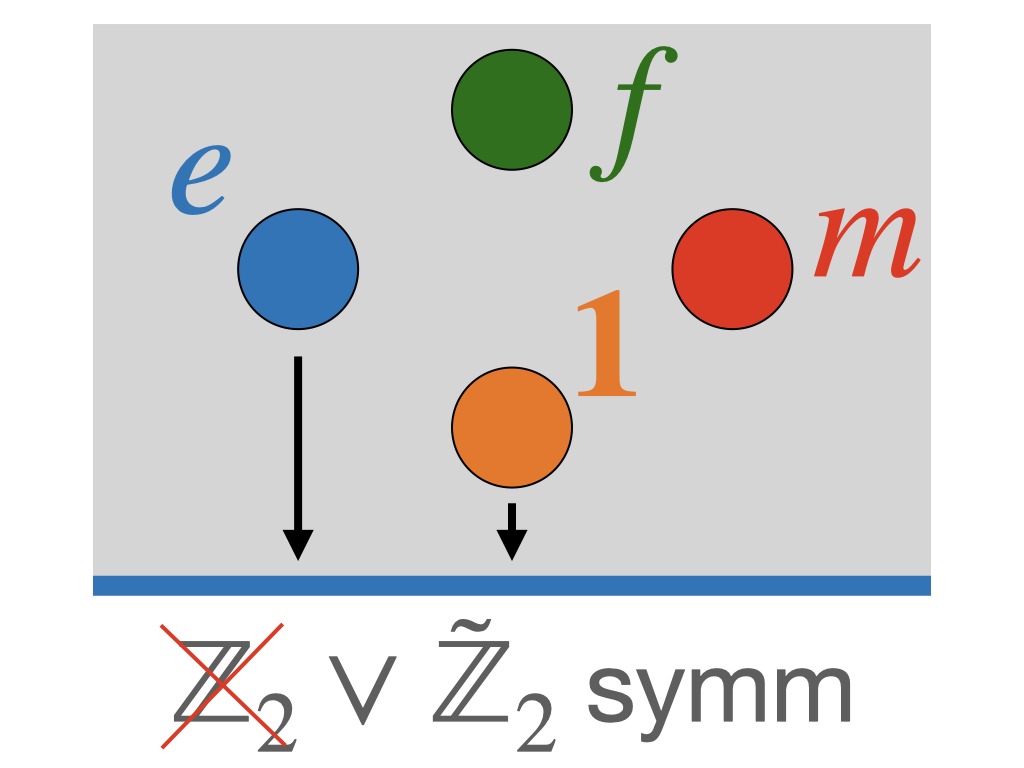}\\
(a) 
\ \ \ \ \ \ \ \ \ \
\ \ \ \ \ \ \ \ \ 
\ \ \ \
(b) 
\ \ \ \ \ \ \ \ \ \
\ \ \ \ \ \ \ \ \
\ \ \ \
(c)
\end{center}
\caption{A 1+1D $\Z_2$-symmetric system (which also has a dual $\t \Z_2$
symmetry \cite{JW191213492}) is a $\eGau_{\Z_2}$-system, \ie the system has a
\catsymm\ $\Z_2\vee \t \Z_2$, which is described by \symmTO\ $\eGau_{\Z_2}$ --
the quantum double of $\Z_2$ group.  Physically, the above statement means that
the $\Z_2$ symmetric system (when restricted to its symmetric sub-Hilbert
space) can be exactly low-energy simulated by a boundary of bulk $\Z_2$
topological order (TO), described by $\Z_2$ gauge theory.  The \symmTO\
$\eGau_{\Z_2}$ has four anyons $\one,e,m,f=e\otimes m$. The possible
condensation-induced states in $\eGau_{\Z_2}$-system are given by the
condensable algebras of the \symmTO, $\cA = \one, \one \oplus e, \one \oplus
m$.  (a) The $\one \oplus m$-state, corresponding to the $\one \oplus
m$-condensed boundary, is the $\Z_2$-symmetric state.  (c) The $\one \oplus
e$-state is the state with spontaneous $\Z_2$ symmetry breaking.  (b) The
$\one$-state is the gapless critical point at the continuous transition between
$\one \oplus m$-state and $\one \oplus e$-state.  } \label{Z2bndry} 
\end{figure}

In this paper, we are going to use Symm/TO correspondence and the
associated \symmTO\ to develop another version of the general unified theory
for topological orders and gapless quantum liquids, from a more physical point
of view.  Our theory is based on the following
proposals: 
\begin{enumerate}

\item  An $n+1$D symmetric system, when restricted to its symmetric sub-Hilbert
space, has a non-invertible gravitational anomaly \cite{JW190513279}, and can
be \emph{exactly low-energy simulated} by a boundary of a topological order in
one higher dimension \cite{JW190513279,JW191213492,KZ200514178}. Such a bulk
topological order is called a \textbf{\symmTO} and is denoted by $\eM$.  This
result allows us to say that the $n+1$D symmetry is described by a \symmTO\
$\eM$.  We will use \textbf{$\eM$-system} to refer to such a system (see Fig.
\ref{Z2bndry}).  Since the $n+2$D \symmTO\ $\eM$ is mathematically described by
a non-degenerate braided fusion $n$-category in trivial Witt class (called \textbf{\nBFcat} and also
denoted as $\eM$) \cite{KW1458,KZ200514178}, we may also say that the $n+1$D
symmetry is described by a \nBFcat\ $\eM$.  

\item The states of a $\eM$-system can be divided into classes labeled by the
condensable algebras $\cA$ in $\eM$, in the sense that a state in a class
labeled by $\cA$ (called \textbf{$\cA$-state}) is exactly low-energy simulated
by a boundary of the \symmTO\ $\eM$ induced by the condensable algebra $\cA$
\cite{BW10065479,KK11045047,K13078244}, termed an \textbf{$\cA$-condensed boundary} (see Fig.
\ref{Z2bndry}).  This way, condensable algebras can describe, in a unified way,
symmetry breaking orders, symmetry protected topological orders, symmetry
enriched topological orders, gapless critical points,  \etc. See Section
\ref{redsymmTO} for a physical description of condensable algebras.

\item A $\cA$-state can be exactly low-energy simulated by a $\one$-condensed
boundary of $\eM_{/\cA}$, where $\eM_{/\cA}$\footnote{May be read as \emph{``M
slash A"}.} is the topological order obtained from $\eM$ by condensing the
condensable algebra $\cA$ \cite{K13078244}.\footnote{Here we assume the
topological order $\eM_{/\cA}$ to have a large energy gap approaching
infinity.}  $\eM_{/\cA}$ is referred to as the \textbf{reduced \symmTO}.  
As a $\one$-condensed boundary of $\eM_{/\cA}$, the $\cA$-state in $\eM$-system
has a reduced \symmTO\ described by $\eM_{/\cA}$.  So we will also refer to
$\cA$-state in $\eM$-system as an \textbf{$\eM_{/\cA}$-state}.

We remark that it is possible that $\eM_{/\cA} =
\eM_{/\cA'}=\eM_\text{reduced}$ for two different condensible algebras $\cA$
and $\cA'$.  In this case, two different states $\cA$-state and $\cA'$-state
are both referred to as $\eM_\text{reduced}$-state.  As we will see in Section
\ref{redsymmTO}, $\cA$-state and $\cA'$-state have the same local low energy
properties.  Thus $\eM_{/\cA}$-state is a notion that is useful for gapless
states which ignores the global properties. For example, a gapped state always
has a trivial reduced \symmTO\ $\eM_{/\cA}$, and a non-tivial reduced \symmTO\
$\eM_{/\cA}$ implies gaplessness (see Section \ref{gapless}).

\item We can use \symmTO\ to constrain the possible continuous phase
transitions.  For example, if a $\cA_{12}$-state is the critical point for a
continuous phase transition between $\cA_{1}$-state and $\cA_{1}$-state, then
$\mathcal{A}_{12}$ is a sub-algebra of both $\mathcal{A}_1$ and $\mathcal{A}_2$
(see Fig. \ref{Z2bndry}).

\end{enumerate} 
In the above, we have introduced some important terms (in bold face) that we
will use in the rest of this paper.  We also used the following notion
\cite{KZ200514178}: \frmbox{ \textbf{Exactly low-energy simulate} means that
the low energy spectrum in the symmetric sub-Hilbert space is identical to the
low energy spectrum of the boundary.  It also means that there is a one-to-one
correspondence of local symmetric operators in $\eM$-system and local operators
on the boundary of the \symmTO\ $\eM$, such that the corresponding operators
have identical correlation functions (in the limit that the energy gap of the
\symmTO\ approaches infinity).  }

Very often, we can easily compute the reduced \symmTO\ $\eM_{/\cA}$, which
allows us to determine if a $\eM_{/\cA}$-state is gapless or not. If
$\eM_{/\cA}$ is trivial, then the corresponding $\eM_{/\cA}$-state can be
gapped. On the other hand, \frmbox{$\eM_{/\cA}$-state must be gapless if
$\eM_{/\cA}$ is a nontrivial.  (see Section \ref{gapless} for a proof.) }
Moreover, we can constrain its low energy properties using the $ \eM_{/\cA} $.
This is a more general version of the familiar notion of \emph{symmetry protected
gaplessness}, \ie condensation patterns in the \symmTO\ can determine whether
a state is gapless or not.

It is well known that perturbative anomalies for continuous symmetries
\cite{H8035} and perturbative gravitational anomalies \cite{AW8469,W8597} imply
gaplessness
\cite{LSM6107,CG8205,W8322,H8353,W9125,Oc9911137,Oc0002392,H0431,GM08120351,CGW1107,L13017355}.
This can be understood as \emph{perturbative-anomaly protected gaplessness}.
Even global anomalies for discrete symmetries may imply gaplessness
\cite{CLW1141,FO150307292,SV170501557,VP190506969,ET190708204,TV200806638,LZ220403131},
which can be understood as \emph{anomalous-symmetry protected gaplessness}.
Symmetry fractionalization may also imply gaplessness
\cite{W0213,L160605652,WL191206167}, which can be understood as
\emph{symmetry-fractionalization protected gaplessness}.  Our Symm/TO correspondence
provides a unified point of view to understand these different kinds of
protected gaplessness.  

We want to emphasize that a nontrivial reduced \symmTO\ $\eM_{/\cA}$ is viewed
as the reason for gaplessness in this framework.  Thus, a nontrivial
$\eM_{/\cA}$ represents the emergence of gaplessness. This suggests that the
low energy properties of the gapless state are characterized by the reduced
\symmTO\ $\eM_{/\cA}$.  In other words, we can use reduced \symmTOs\ to
systematically study, and potentially classify, gapless states and the
corresponding quantum field theories in one lower dimensions.  In particular,
we can use holographic modular bootstrap
\cite{RVh9803129,CZ190312334,JW190513279,LS230213900} to compute the low energy properties
of a 1+1D gapless state from its reduced \symmTO\ $\eM_{/\cA}$, as we will
describe in section \ref{holo-theory}.

\subsection{Organization of the paper}

The remainder of this paper is organized as follows. 
In section \ref{holo-theory}, we flesh out our Symm/TO
framework for labeling phases and their phase transitions using condensable
algebras. We discuss how to identify the patterns of condensation, \ie
the allowed condensable algebras, using a set of number theoretic constraints.
Along with the knowledge of boundary partition functions compatible with the
bulk topological order, this provides us a pathway to understanding the phase
diagram for a system with a given symmetry. In sections
\ref{Z2Z2}--\ref{anoS3}, we discuss various examples of anomalous and
anomaly-free Abelian and non-Abelian symmetries in 1+1D systems. We identify
gapped and gapless states allowed by each of these symmetries, and provide a
discussion of the gapless theories possible at the phase transitions between
these states.

The main results of this paper are summarized in the framed boxes. Gapped
and gapless states for 1+1D systems with $S_3$ symmetry (with or without
anomaly) and the corresponding condensation patterns in their \symmTO\
are summarized in the tables \ref{S3a0}, \ref{S3a1},
\ref{S3a2}.  The gapless states are potential critical points for continuous
transitions between the gapped states.

\section{Holographic theory for gapless states and for continuous phase
transitions}\label{holo-theory}

In this section, we will formulate a general holographic theory for gapless
states and for continuous phase transitions, based on Symm/TO correspondence
\cite{JW191213492,KZ200514178}.  Later, we will apply Symm/TO correspondence to
study some examples.  In fact, Symm/TO correspondence also applies to gapped
states.

\subsection{Symmetry TO reduction (analogue of symmetry breaking)}

For ordinary global symmetry, spontaneous symmetry breaking is a very important
notion, which allows us to describe gapped and gapless phases.  If we use
\symmTO\ to describe global symmetry, spontaneous symmetry breaking is replaced by
\emph{spontaneous \symmTO\ reduction}, or simply \emph{\symmTO\ reduction}.
Patterns of \symmTO\ reduction, physically induced by condensation of excitations, allow us to describe gapped and gapless phases,
as well as the critical points at continuous phase transitions. Considered thus, \symmTO\
reduction provides more information than spontaneous symmetry breaking.  In
this section, we will describe \symmTO\ reduction in details.

Symm/TO correspondence has the following meaning
\cite{JW191213492,KZ200514178}, which is the key conjecture used in this paper:
\frmbox{a system (\ie a gapped or gapless lattice Hamiltonian) with a
(generalized) global symmetry can be \emph{exactly low-energy simulated} by a
boundary (\ie a boundary Hamiltonian) of a non-invertible gappable-boundary topological order
$\eM$ in one higher dimension.  The bulk gappable-boundary topological
order is referred to as \symmTO\ $\eM$.} This is why we
can use \symmTO\ $\eM$ to describe a symmetry.  We will call such a symmetric
system as an $\eM$-system.  We remark that the above Symm/TO correspondence
works for anomalous and/or higher and/or non-invertible symmetries.  It even
works for global symmetries beyond the previous known descriptions.  Thus, it
can be viewed as a most general description of global symmetry.

We have the following mathematical result (see
\Rf{KK11045047,K13078244,KZ170501087,KZ190504924,KZ191201760} for a summary): \emph{All
(gapped or gapless) boundary states of a topological order $\eM$ are obtained
from condensing condensable algebras $\cA$ of $\eM$}.  Since different boundary
states of $\eM$ correspond to different ground states in different
$\eM$-systems, we can group all gapped or gapless ground states in
$\eM$-systems into classes labeled by the condensable algebras $\cA$, \ie
states in a class labeled by $\cA$ correspond to $\cA$-condensed boundaries of
$\eM$.  Those states are referred to as $\cA$-states.  In other words, an
$\cA$-state in an $\eM$-system has a condensation pattern $\cA$.  After
introducing those notions, we can make the following statement: \frmbox{ In
$\eM$-systems, all their gapped and gapless $\cA$-states are \emph{exactly
low-energy simulated} by the $\cA$-condensation-induced boundary states of the
\symmTO\ $\eM$.} 

\subsection{Reduced symmetry TO (analogue of unbroken symmetry)}
\label{redsymmTO}

Now let us concentrate on a topological order $\eM$ and one of its boundary
state induced by condensing a condensable algebra $\cA$. Such a boundary state
corresponds to a $\cA$-state in a $\eM$-system.  The condensable algebra $\cA$
is formed by excitations in $\eM$ that has trivial self and mutual statistics
between them.  As a result, excitations in the  condensable algebra $\cA$ can
condense together, which will change the topological order $\eM$ to another
topological order.\cite{K13078244} We will denote the resulting topological
order as $\eM_{/\cA}$, which will be called a reduced \symmTO.  

Physically a condensable algebra $\cA$ corresponds to a set of excitations in
$\eM$ that can be condensed together, \ie with trivial self/mutual statistics.
Mathematically, a condensable algebra $\cA$ is described by a \emph{composite
excitation} $\one\oplus a \oplus b \oplus \cdots$\footnote{ A \emph{composite
excitation} $\one\oplus a \oplus b \oplus \cdots$ is an excitation where the
excitations $\one$, $a$, $b$, \etc. in the composite happen to have the same
energy. For example, the bound state of two spin-1/2 excitations is a composite
excitation formed by degenerate spin-0 and spin-1 excitations: $
\text{spin-}1/2 \otimes \text{spin-}1/2 = \text{spin-}0 \oplus \text{spin-}1$.} 
which can be viewed as a ``vector space'', plus some data describing
``multiplication of vectors'' in the vector space (see \Rf{K13078244} for a
summary, and see \Rf{CZ190312334} for a detailed discussion of a simple
example).  For simplicity, in this paper, we will use the ``vector space''
$\one\oplus a \oplus b \oplus \cdots$ to denote the condensable algebra $\cA$,
and say $\one,a,b$ \etc. belong to the condensable algebra $\cA$: $\one,a,b \in
\cA$.  We will always use $\one$ to denote the trivial excitations, \ie all the
excitations that can be created by local symmetric operators.  Note that $\one$
can represent a null excitation -- the ground state itself -- an excitation
created by the identity operator.\footnote{We believe that even in higher
dimensions, the various condensation patterns associated to \symmTO\ are still classified
by condensable algebras $\cA$ in the \symmTO.  But in higher dimensions, the
notion of condensable algebras needs to be generalized beyond what is described
in this paper.}

Roughly speaking, the excitations in $\eM_{/\cA}$ all comes from the
excitations in $\eM$: the excitations in $\eM$ will become trivial excitations
in $\eM_{/\cA}$, if they are in $\cA$ (\ie if they are condensed).  The
excitations in $\eM$ will be confined in $\eM_{/\cA}$ (\ie will disappear), if
they have nontrivial mutual statistics with excitations in $\cA$ (for more
details, see Appendix \ref{MMA}).  

The condensable algebra $\cA$ is called Lagrangian if $\eM_{/\cA}$ is trivial.
In this case, the excitations in $\eM$ will be either condensed or confined.  A
Lagrangian condensable algebra is maximal in some sense, since the induced
topological order $\eM_{/\cA}$ is minimal (\ie trivial).  On the other hand, if
the condensable algebra is minimal, $\cA=\one$ (\ie nothing condenses except
trivial particle $\one$ which always condenses), the induced topological order
$\eM_{\/\one} =\eM$ is maximal.

\begin{figure}[t]
\begin{center}
\includegraphics[scale=0.175]{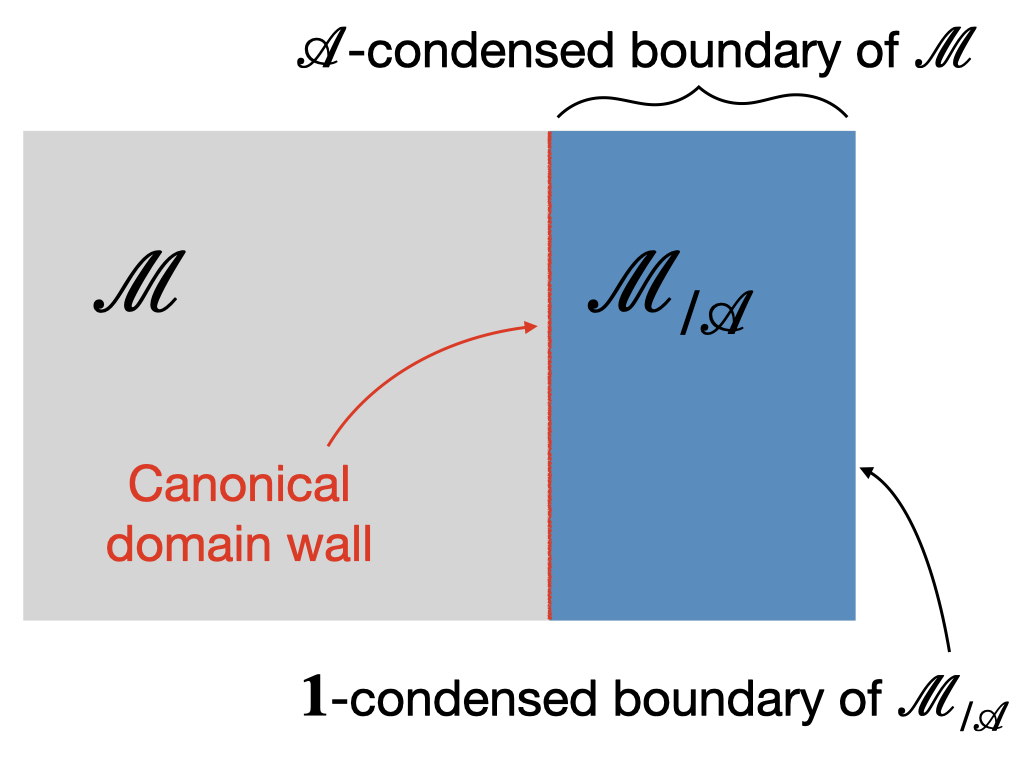}
\end{center}
\caption{ A $ \cA $-state in a $\eM$-system corresponds to a $\cA$-condensed
boundary of \symmTO\ $\eM$. Such a boundary can be obtained by attaching
$\cA$-condensation  induced topological order $\eM_{/\cA}$ with
$\one$-condensed boundary.  The $\cA$-condensation changes $\eM$ to
$\eM_{/\cA}$, causing a \emph{\symmTO\ reduction} (an analogue of spontaneous
symmetry breaking).  $\eM_{/\cA}$ is the \emph{reduced \symmTO} (an analogue
of unbroken symmetry) of the $ \cA $-state.  } \label{MAbndry} 
\end{figure}

From the $\cA$-condensation-induced topological order $\eM_{/\cA}$, we can have
an understanding of a $\cA$-condensed boundary of $\eM$, which has a special
realization as a composite boundary illustrated in Fig. \ref{MAbndry}.  There
is a canonical domain wall between $\eM$ and $\eM_{/\cA}$ which is always
gapped \cite{K13078244}.  Therefore, the local low energy properties of
$\cA$-condensed boundary of $\eM$ is same as the local low energy properties of
the $\one$-condensed boundary of $\eM_{/\cA}$.\footnote{Two systems have the
same local low energy properties if there is a correspondence of the local
symmetric operators in the two systems, such that the corresponding operators
have the same correlation function.\label{locallow}}  This result is reasonable
and expected, since the $\cA$-condensed boundary of $\eM$ also implies that
excitations outside $\cA$ do not condense.  For the composite boundary in Fig.
\ref{MAbndry}, the induced topological order $\eM_{/\cA}$ already has all the
condensations for excitations in $\cA$, and the $\one$-condensed boundary of
$\eM_{/\cA}$ implies that there is no additional condensation (for excitations
outside $\cA$).

\subsection{The emergence and the symmetry protection of gaplessness -- the
$\one$-condensed boundary of $\eM_{/\cA}$ } 
\label{gapless}

It is easy to see that the $\one$-condensed boundary of $\eM_{/\cA}$ can be
gapped if $\eM_{/\cA}$ is trivial.  The $\one$-condensed boundary of
$\eM_{/\cA}$ must be gapless if $\eM_{/\cA}$ is nontrivial.  This result is
proposed in \Rf{KZ170501087,KZ190504924,KZ191201760}, and was referred to as
topological Wick rotation.

Let us show that a $\one$-condensed boundary of a 2+1D topological orders $\eM$
must be gapless. This result is obtained in
\Rf{KZ170501087,KZ190504924,KZ191201760,L190309028} via some other methods.  As
pointed out in \Rf{LWW1414,JW190513279,LW191108470}, the partition function for
a boundary of a 2+1D topological order is a vector, whose components are
labeled by the anyon types of the bulk topological order: $\v
Z(\tau)=(Z_\one(\tau),Z_a(\tau),Z_b(\tau),\cdots)^\top$.  Here we have assumed
that the spacetime at the boundary is a torus, and $\tau$ describes the shape
of the boundary spacetime.  Under the modular transformation, vector-like
partition function transforms covariantly \cite{JW190513279}:
\begin{equation}\label{STZ}
\begin{split}
\ee^{-\ii 2\pi \frac{c_-}{24}} T^{\eM} \v Z(\tau) &= \v Z(\tau+1), \\
D^{-1} S^{\eM} \v Z(\tau) &= \v Z(-1/\tau),
\end{split} 
\end{equation}
where $S^{\eM},T^{\eM}$ are the modular data characterizing the bulk
topological order $\eM$ \cite{W9039,KW9327} (see Appendix \ref{MMA}) and $c_-$
the chiral central center of $\eM$. The physics behind the above results were
explained in \Rf{LWW1414,LW191108470,JW190513279}.

If the $\one$-condensed boundary of the bulk
topological order was gapped, the vector-like partition function would be
$\tau$ independent and would have a form
\begin{align}
\label{Z1}
Z_\one(\tau) = 1,\ \ \ 
Z_{a\neq \one}(\tau) = 0,
\end{align}
since only $\one$ condenses.  Such a partition function cannot be modular
covariant, if the bulk topological order is non-invertible.  This is because
$S^{\eM},T^{\eM}$ matrices is more than 1-dimensional for non-invertible
topological order, and the $\one$-column of the $S^{\eM}$ matrix has a form
$(d_{\one},d_{a},d_{b},\cdots)^\top$, where $d_a$ is the quantum dimension of
type-$a$ bulk anyon and $D =\sqrt{\sum_a d_a^2}$.  Since $d_\one =1$, $d_a \geq
1$, and $D>1$, \eqn{STZ} cannot be satisfied by the vector-like partition
function \eq{Z1}.  For nontrivial invertible 2+1D topological order, although
the bulk has only the trivial type-$\one$ excitations, the chiral central
charge is non-zero, and the $\one$-condensed  boundary is always gapless.  Thus
\frmbox{The $\one$-condensed boundaries of nontrivial 2+1D topological orders
are always gapless.}

A similar argument is expected to also work in higher dimensions: \frmbox{The
$\one$-condensed boundaries of nontrivial non-invertible topological orders
are always gapless.} But in higher dimensions, the $\one$-condensed boundaries
of nontrivial invertible topological orders can be gapped, such as the
$\w_2\w_3$ invertible topological order in 4+1D
\cite{WS13063238,KS14098339,K1459,T14044385,T171002545,WW170506728,WW181000844,CH211014644,FH211014654,WW211212148}.

\subsection{Canonical boundary}

To describe the gapless $\one$-condensed boundaries more precisely, we need to
introduce a notion of \emph{local-low-energy equivalence}.  Consider a low
energy theory $\cL$. We can obtain another low energy theory $\cL'$ by stacking
a gapped state to $\cL$.  If the gapped state has a nontrivial topological
order, the two low energy theories $\cL$ and $\cL'$ can have different global
properties, such as different ground state degeneracies, and different averages
of non-contractible loop operators, etc.  However, the two low energy theories
have the same local correlations for all corresponding local operators (beyond
the correlation length of the gapped state).  In this case, we say that the two
low energy theories $\cL$ and $\cL'$ are local-low-energy equivalent.  Now we
can say that the $\cA$-condensed boundary of $\eM$ is local-low-energy
equivalent to the $\one$-condensed boundary of $\eM_{/\cA}$.

The above discussion leads to the following result: \frmbox{For a $\cA$-state
in $\eM$-system, there exist a $\one$-condensed boundary state of the
$\cA$-condensation-induced topological order $\eM_{/\cA}$, such that the two
states are local-low-energy equivalent.} Let us note here that there can be
many $\cA$-states with different local low energy properties.  Similarly,
topological order $\eM_{/\cA}$ can have many different $\one$-condensed
boundary states with different local low energy properties.  What we try to say
is that there is an one-to-one correspondence between the $\cA$-states and the
$\one$-condensed boundary states of $\eM_{/\cA}$, such that the corresponding
states have identical local low energy properties.  This can be rephrased as
\frmbox{ $\cA$-states in $\eM$-system are local-low-energy equivalent to
$\one$-states in $\eM_{/\cA}$-system.  }

Some of these states are more stable if they have fewer low energy excitations.
Here, we assume that the gapless excitations all have linear dispersion
relations.  When the velocity of the gapless excitations are all the same, the
number low energy excitations can be determined by specific heat.  The states
with minimal number of  low energy excitations are most stable.  The most
stable $\one$-condensed boundaries of $\eM$ are called the \emph{canonical
boundaries} of $\eM$.  \frmbox{The local low energy properties of the most
stable $\cA$-state is same as the local low energy properties of the canonical
boundary state of $\eM_{/\cA}$.} We remark that one can also use the number of
symmetric relevant operators to define a different notion of ``most-stable''.

As we have mentioned above that the $\cA$-states are not unique.  We usually
look for the most stable states among the $\cA$-states.  Note that this aspect
is not so different from the notion of ``spontaneous $\Z_2$-symmetry breaking
state'' which does not really refer to a unique state, since we can always
stack a gapless $\Z_2$ symmetric state to it while still preserving the fact
that the $ \Z_2 $ symmetry is spontaneously broken. But the  term ``spontaneous
$\Z_2$-symmetry breaking state'' usually refers to the most stable state among
these various possible spontaneous $\Z_2$-symmetry breaking states. We use the
term $\cA$-state in an analogous fashion.

\subsection{Holographic modular bootstrap approach}

If a 2+1D topological order $\eM_{/ \cA}$ is nontrivial, there is an algebraic
number theoretical way, also called holographic modular bootstrap (holoMB)
approach \cite{RVh9803129,CZ190312334,JW190513279,LS210108343,LS230213900}, to determine
its gapless boundaries. HoloMB is a generalization of the conventional modular
bootstrap \cite{C8686},
\begin{align}
Z(\tau) =Z(\tau+1) = Z(-1/\tau),
\end{align}
in the sense that holoMB requires additional input data, \symmTO, that
describes (generalized) symmetry.  The generalization is given by \eqn{STZ},
and we want to determine the vector-valued partition function from these
conditions.

\Eqn{STZ} describes a set of algebraic equations.  In general, one
cannot determine unknown functions $\v Z(\tau)$ from  algebraic equations.
However, here a partition function for a given anyon type is the partition
function in a certain symmetry charge sector for Hamiltonian with a certain
symmetry-twist boundary condition (\ie in a certain symmetry defect sector):
\begin{align}
&\ \ \ \ \
Z_\text{symm. charge/defect}(\tau)\big|_{\text{size }L} 
\nonumber\\
&
\stackrel{\text{def}}{=} \Tr_{\text{symm. charge}} \ee^{-\Im (\tau)Lv^{-1} \hat
H_\text{symm. defect} +\ii \Re (\tau)L \hat P},
\end{align}
where $L$ is the size of 1-dimensional ring and $v$ is the velocity of our 1+1D
system.
Such a  partition function has the form
\begin{align}
Z_a(\tau) &=  q^{h_a-c/24} \bar q^{\bar h_a- \bar c/24} 
\text{Poly}^\text{non-neg-int}_{h_a,\bar h_a}( q,\bar q), 
\nonumber\\
q &= \ee^{2\pi \ii \tau} \sim \ee^{-\bt E}
\end{align}
where $h_a,\bar h_a,c,\bar c$ are rational numbers, and
$\text{Poly}^\text{non-neg-int}_{h,\bar h}$ is a polynomial of $q$ and $\bar q$
with \emph{non-negative integral coefficients}.  In fact, the non-negative
integral
coefficients are degeneracies of energy-momentum levels.  It appears that the
modular covariance conditions \eqref{STZ} can largely determine partition
functions that satisfy the ``non-negative-integer'' constraint.   We note that holomorphic
modular bootstrap was developed to solve similar problems
\cite{BCh0512011,MR220805486}.
Here, we will use a different approach.  A 1+1D gapless boundary conformal
field theory (CFT) contains right movers and left movers, described by
conformal characters $\chi_i^R(\tau)$ and $\bar \chi_j^L(\bar \tau)$.  Under
the modular transformation, the conformal characters transform as
\begin{align}
\t T_R^{ij} \chi_j^R(\tau) = \chi_i^R(\tau+1), \ \ \ \ 
\t S_R^{ij} \chi_j^R(\tau) = \chi_i^R(-1/\tau),
\nonumber\\
\t T_L^{ij} \bar\chi_j^L(\bar\tau) = \bar\chi_i^L(\bar\tau+1), \ \ \ \ 
\t S_L^{ij} \bar\chi_j^L(\bar\tau) = \bar\chi_i^L(-1/\bar\tau).
\end{align}
The multi-component partition function for the gapless boundary of $\eM_{/ \cA}$
is given by
\begin{align}
Z_a^{\eM_{/ \cA}}(\tau)=
A^{a,i,j} \chi_i^R(\tau) \bar \chi_j^L(\bar \tau),
\ \ \
A^{a,i,j} \in \N .
\end{align}
The modular covariance of $Z_a^{\eM_{/ \cA}}(\tau)$ takes a form
\begin{align}
\ee^{-\ii 2\pi \frac{c_-}{24}}T_{\eM_{/ \cA}}^{ab} Z_b^{\eM_{/ \cA}}(\tau)  &=
Z_a^{\eM_{/ \cA}}(\tau+1),
\nonumber\\
D^{-1} S_{\eM_{/ \cA}}^{ab} Z_b^{\eM_{/ \cA}}(\tau)  &= Z_a^{\eM_{/
\cA}}(-1/\tau) ,\label{STeqns}
\end{align}
where $S_{\eM_{/ \cA}},T_{\eM_{/ \cA}}$ are the $S,T$-matrices characterizing
the bulk topological order $\eM_{/ \cA}$. They constitute
the additional input, describing the \symmTO\ required in the holoMB approach.
\Eqn{STeqns} can be satisfied if non-negative integers $A^{a,i,j}$ satisfy
\begin{align}
\ee^{-\ii 2\pi \frac{c_-}{24}} T_{\eM_{/ \cA}}^{ab} \t T_R^{*ij} \t T_L^{*kl}
A^{b,j,l} &= A^{a,i,k},
\nonumber\\
D^{-1} S_{\eM_{/ \cA}}^{ab} \t S_R^{*ij} \t S_L^{*kl} A^{b,j,l} &= A^{a,i,k},
\end{align}
or more compactly
\begin{align}
\label{STAgapless}
\ee^{-\ii 2\pi \frac{c_-}{24}}  T_{\eM_{/ \cA}} \otimes \t T_R^* \otimes \t
T_L^* \v A &=\v A,
\nonumber\\
D^{-1} S_{\eM_{/ \cA}} \otimes \t S_R^* \otimes \t S_L^* \v A &=\v A,
\end{align}
where we have used the fact that the $S,T$ matrices are symmetric unitary
matrices.  Comparing \eqn{STZ} (for gapped boundary where $\Z_a$ are $\tau$
independent non-negative integers) and \eqn{STAgapless}, we see that the
mathematical method to solve for gapped and
gapless boundaries are the same.  We just need to start with different $S,T$
matrices.  In Appendix \ref{MMA}, we will describe in more details an
algebraic number theoretical method to find non-negative integer solutions of
\eqn{STZ} and \eqn{STAgapless}.  Appendix \ref{MMA} also obtains many additional
conditions on $Z_a$ and $A^{a,i,j}$ (see \eqn{Acond1}).

From the multi-component partition $Z_a^{\eM_{/ \cA}}(\tau)$ we can obtain the
scaling dimensions of operators that carry various representations of the
symmetry.  Thus the \symmTO\ in Symm/TO correspondence allows us to compute
properties of gapless state via an algebraic number theoretical method.

To summarize, using Symm/TO correspondence, the properties of gapped and
gapless states in systems with (generalized) symmetry can be studied by (1)
identifying the corresponding \symmTO\ $\eM$ that describes the symmetry, (2)
computing the condensable algebras $\cA$ of $\eM$,  which classify different
reductions  of the \symmTO\ $\eM$ (the reduced \symmTO\ is denoted as
$\eM_{/\cA}$, which is analogous to the notion of unbroken symmetry, see Section
\ref{redsymmTO}), and (3) describing the boundaries induced by condensing $\cA$ using holoMB, which correspond to different gapped
or gapless states (called $\cA$-states) for a given reduced \symmTO\
$\eM_{/\cA}$.  

\subsection{From structure of condensable algebra to structure of phase
diagram}

We have grouped the gapped and gapless states of $\eM$-systems 
into classes labeled by condensable algebras of
$\eM$.  The states in each class labeled by $\cA$ are called $\cA$-states.  If
there is a continuous transition between a $\cA_1$-state and a $\cA_2$-state,
the critical point at the transition will be described by a $\cA_{12}$-state.
The  condensable algebra $\cA_{12}$ must be a sub algebra of both  condensable
algebra $\cA_1$ and $\cA_2$:
\begin{align}
\cA_{12} \subset \cA_1,\ \ \ \ \
\cA_{12} \subset \cA_2 .
\end{align}
This is because as we approach the phase transition boundary, some anyons have
increasingly weak affinity to condense.  The condensation is absent at the
transition and the condensable algebra becomes smaller.
For more details, see Appendix \ref{phasestructure}.

To obtain more constraints on the phase diagram from condensable algebras, we
introduce a concept of \emph{competing pair}: a pair of anyons $(a,b)$ form a
competing pair if they never appear in the same condensable algebra together,
but they can appear in condensable algebras separately. In other words, the
anyons in a competing pair can condense, but cannot condense together (usually
due to the nontrivial mutual statistics between them).  Condensing one anyon
in a competing pair will uncondense the other.  We propose that continuous
phase transition is driven by condensing one anyon and uncondensing the other
anyon in a competing pair. This implies that if $\cA_1$-state and $\cA_2$-state
are connected by a continuous transition, then the union of $\cA_1$ and $\cA_2$
should contain a competing pair.  If the union of $\cA_1$ and $\cA_2$ should
contain only one competing pair, then the transition is more likely to be
stably continuous.  For more details, see Appendix \ref{phasestructure} and
Section \ref{transS3}.

\section{1+1D $\Z_2\times \Z_2'$ symmetry} \label{Z2Z2}
Let's illustrate the general discussion of the previous section with the example
of a 1+1D system with $ \Z_2\times \Z_2 $ symmetry.
Landau's symmetry-breaking framework tells us that the system can spontaneously
break
the symmetry down to various subgroups of the symmetry group, producing various
gapped states.

This symmetry breaking picture can be also be viewed through the
lens of \symmTO.  In the same way that $ \Z_2 $ gauge theory, denoted
by $ \eGau_{\Z_2} $, is the \symmTO\ of $ \Z_2 $ symmetry,
for systems with $\Z_2\times \Z_2'$ symmetry,\footnote{The prime on the second $
\Z_2 $ is used just to explicitly differentiate between the two $ \Z_2 $ groups
for purpose of identification.} the \symmTO\ is $ \eGau_{\Z_2\times \Z_2'} $,
which
refers to the 2+1D topological order described by $\Z_2\times\Z_2'$ gauge theory
with
charge and flux excitations. There are two $ e $ anyons (charges), $e_1$
and $e_2$, and two $ m $ anyons (fluxes), $m_1$ and $m_2$, that
generate all of the 16 anyons of $ \eGau_{\Z_2\times \Z_2'} $.  The \symmTO\ $
\eGau_{\Z_2\times \Z_2'} $ makes the mod 2 conservation of the flux excitations
$m_1$
and $m_2$ explicit -- this may also be described by the dual symmetry
$\t\Z_2\times \t
\Z_2'$.\cite{JW191213492} To emphasize the dual symmetry, one may denote this
\symmTO\ as $(\Z_2\times \Z_2')\vee
(\t\Z_2\times \t \Z_2')$. We will drop the discussion of dual symmetry in the
following for brevity.

Let's consider the possible gapped phases of a 1+1D system with $ \Z_2\times
\Z_2' $ symmetry from the conventional point of view first. We will then
translate that into the \symmTO\ language.

The gapped phases in 1+1D associated to symmetry group $ G $ are classified by
the unbroken subgroup $ H $, and possible SPT phases of $ H
$.\cite{CGW10083745,SPC10103732,CGW1128} For $ G=\Z_2\times \Z_2' $, the four
nontrivial symmetry-breaking gapped phases are associated to its four proper
subgroups $ \Z_1, \Z_2, \Z_2', \Z_2^d $, where $ \Z_2^d $ is the ``diagonal" $
\Z_2 $ subgroup. If we present the group $ \Z_2\times\Z_2' $ as $ \{(0,0),(0,
1),(1,0),(1,1)\} $, then these subgroups are 
\begin{align*}
\Z_1 &\simeq \{(0,0)\}\\
\Z_2 &\simeq \{(0,0),(1,0)\}\\
\Z_2' &\simeq \{(0,0),(0,1)\}\\
\Z_2^d &\simeq \{(0,0),(1,1)\}
\end{align*}
There are no nontrivial $ \Z_2 $ SPT phases in 1+1D. However, there is a
nontrivial $ \Z_2\times \Z_2' $ SPT phase, the so-called cluster state. So there
are a total of 6 gapped phases. Continuous phase transitions between the
symmetry breaking states is straightforward within Landau theory. The
transitions between the trivial $ \Z_2\times \Z_2' $ paramagnet phase and the
three $ \Z_2 $ symmetric phase are Ising transitions. The remaining symmetry in
the three $ \Z_2 $ symmetric phases can further spontaneously break via a second
Ising transition to reach the $ \Z_1 $ symmetric phase. In Landau theory, a
direct continuous transition between different $ \Z_2 $-SSB phases is not a
possibility since there is no group-subgroup relation between such pairs. The
nontrivial SPT, cluster state, also has Ising transitions to the $
\Z_2,\Z_2',\Z_2^d $ symmetric phases while a direct continuous transition to the
$ \Z_1 $ symmetric phase is not generically possible without fine tuning.
Transition from the cluster state to the trivial paramagnet proceeds via an
XY-type critical point as was shown by Kramers Wannier transformation in
\Rf{P0410416}.

\begin{figure}[t]
\begin{center}
\includegraphics[scale=0.25]{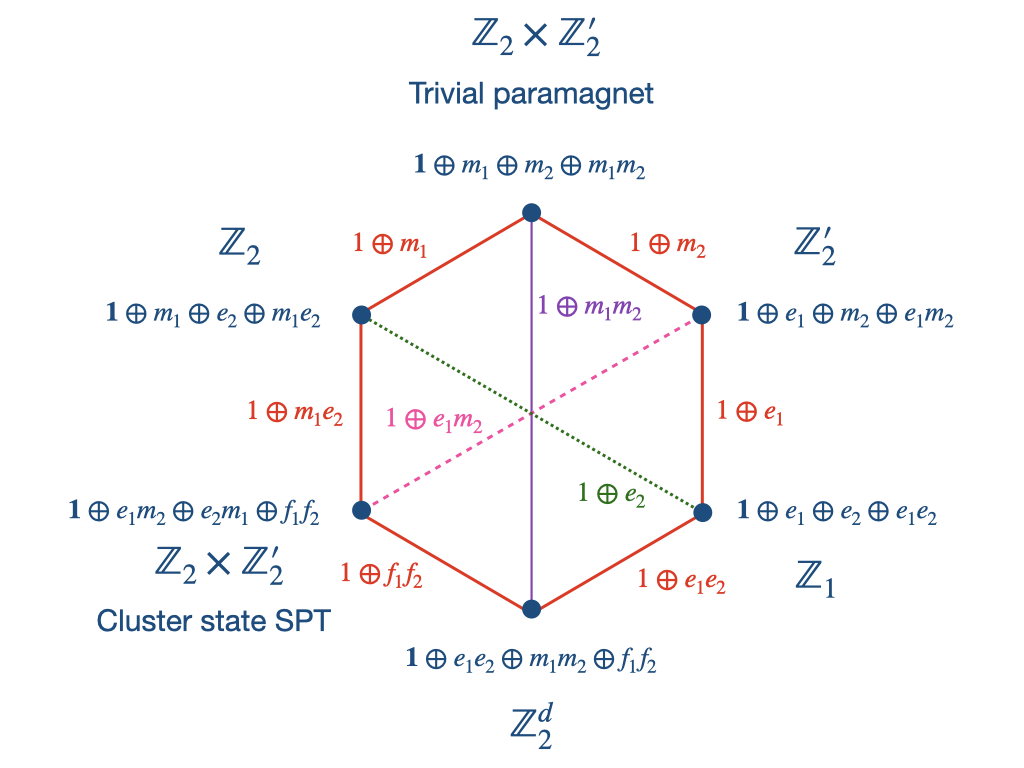}
\end{center}
\caption{A $\Z_2\times \Z_2'$ symmetric system has a symmetry
described by 2+1D $\Z_2\times \Z_2'$ topological order $ \eGau_{\Z_2\times
\Z_2'} $. Here, the 6 Lagrangian condensable algebras of $ \eGau_{\Z_2\times
\Z_2'} $ are represented by the vertices of the hexagon. The gapped phases they
correspond to are described in terms of their symmetry-breaking/SPT order. A
connecting edge between any pair represents a non-Lagrangian condensable
algebra which is strictly included in both of them, and therefore corresponds
to a phase transition between them. There is also a trivial condensable algebra
$ \one $, which is not shown in this picture; it corresponds to a
multicritical point between any two gapped phases.The inclusion relations
between condensable algebras have implication on the structure of phase
diagram.  For example, the edge labeled by $ \one\oplus m_2 $ connecting
vertices labeled by $ \one\oplus m_1\oplus m_2\oplus m_1m_2$ and $ \one\oplus
e_1\oplus m_2\oplus e_1m_2 $ suggests that the former (gapless)
$\one\oplus m_2$-state describes a critical point for the stable continuous
transition between the latter two gapped states. For this transition to be
non-fine-tuned, the gapless $\one\oplus m_2$-state must have only one symmetric
relevant operator, which is indeed the case here (see main text for details).
} \label{Z2Z2phases}
\end{figure}

Let us now phrase the above discussion in terms of \symmTO. The \symmTO\ of
the
symmetry group $ \Z_2\times\Z_2' $ is $ \eGau_{\Z_2\times \Z_2'} $, with the
following Lagrangian condensable algebras.
\begin{align}\label{z2z2Lag}
&\one\oplus e_1\oplus e_2\oplus e_1e_2,  
\one\oplus e_1\oplus m_2\oplus e_1m_2,\nonumber\\
&\one\oplus m_1\oplus e_2\oplus e_2m_1,
\one\oplus m_1\oplus m_2\oplus m_1m_2,\\
&\one\oplus m_1m_2\oplus e_1e_2\oplus f_1f_2, \ \
\one\oplus e_2m_1\oplus e_1m_2\oplus f_1f_2\nonumber
\end{align}
where $f_1=e_1\otimes m_1=e_1m_1$ and $f_2=e_2\otimes m_2=e_2m_2$.  These
correspond to gapped boundaries of  $ \eGau_{\Z_2\times \Z_2'} $ and, by our
Symm/TO correspondence, to the 6 gapped phases discussed above.  The gapped
boundary $ \one\oplus m_1\oplus m_2\oplus m_1m_2 $ condenses the two $ m $
anyons, $m_1$ and $m_2$. This phase preserves the $ \Z_2\times \Z_2' $ symmetry
since the $ \Z_2 $ and $ \Z_2' $ charges $ e_1,e_2 $ remain uncondensed (see
the top vertex of Fig. \ref{Z2Z2phases}). This is the trivial paramagnet phase.
The $ \one\oplus e_1\oplus m_2\oplus e_1m_2 $-condensed boundary corresponds to
a $ \Z_2' $ symmetric phase since the $ \Z_2' $ charge $ e_2 $ is uncondensed
while the $ \Z_2$ charge $ e_1 $ is condensed. The $ \one\oplus m_1m_2\oplus
e_1e_2\oplus f_1f_2 $ preserves $ \Z_2^d $, the diagonal $ \Z_2 $ symmetry. To
see this, note that $ e_1e_2 $ is charged under both $ \Z_2 $ and $ \Z_2' $
while it is symmetric under the action of $ \Z_2^d $. As a result, condensing $
e_1 e_2 $ must break both $ \Z_2 $ and $ \Z_2' $ but not $ \Z_2^d $. The fact
that $ m_1m_2 $ is condensed amounts to the same conclusion: we recall that,
for a single $ \Z_2 $ symmetry, condensation of $ m $ corresponds to
proliferating the disorder operator and hence preserving the $ \Z_2 $ symmetry.
Therefore, condensation of $ m_1 m_2 $ corresponds to preserving the $ \Z_2^d $
symmetry. The $ \one\oplus e_1\oplus e_2\oplus e_1e_2 $-condensed boundary
corresponds to a $ \Z_2\times \Z_2' $-SSB phase since both $ e_1 $ and $ e_2 $
are condensed. The $ \one\oplus e_2m_1\oplus e_1m_2\oplus f_1f_2 $-condensed
boundary corresponds to a 1+1D SPT phase, since none of the $ \Z_2 \times \Z_2'
$ charges condense and thus $ \Z_2 \times \Z_2' $ symmetry is not broken. This
corresponds to the SPT state . In fact, we note that this condensable algebra
actually involves a proliferation of decorated domain walls \cite{CLV13034301}
since the disorder operator of $ \Z_2 $, corresponding to $ m_1 $, is bound to
the charge of $ \Z_2' $, corresponding to $ e_2 $, and vice-versa.  In Appendix
\ref{SPTauto}, we show that the $ \one\oplus e_2m_1\oplus e_1m_2\oplus f_1f_2
$-condensed boundary is associated with an automorphism in the \symmTO, which
also indicates that $ \one\oplus e_2m_1\oplus e_1m_2\oplus f_1f_2 $-condensed
boundary gives rise to an SPT state.

Let us now discuss the possible phase transitions between these gapped phases.
Going from the $ \one\oplus m_1\oplus m_2\oplus m_1m_2 $ phase to the $
\one\oplus e_1\oplus m_2\oplus e_1m_2 $ phase, the system encounters a
continuous phase transition which, in the holographic picture, corresponds to
uncondensing $ m_1 $ and condensing $ e_1 $.  In conventional language, this is
a phase transition between a $ \Z_2\times \Z_2' $ symmetric phase to a $ \Z_2'
$-symmetric phase. At this phase transition, only $ \one\oplus m_2 $ remain
condensed. Both $ e_1,m_1 $ are uncondensed as well as inequivalent w.r.t. to
the condensed particles, which makes the corresponding boundary theory
impossible to be gapped, due to their nontrivial mutual statistics. This serves
as an argument that indeed the system becomes gapless at this phase transition,
\ie this phase transition is continuous (cf. the top right edge of Fig.
\ref{Z2Z2phases}).

In an analogous manner, one can describe the phase transition from the $
\Z_2\times \Z_2' $-symmetric phase to a $ \Z_2 $-symmetric phase as
uncondensing $ m_2 $ and condensing $ e_2 $. At the phase transition, only $
\one\oplus m_1 $ are condensed (cf.  the top left edge of Fig.
\ref{Z2Z2phases}). On the other hand, the phase transition from the $
\Z_2\times \Z_2' $ symmetric phase to the $ \Z_2^d $ symmetric phase
corresponds to uncondensing $ m_1 $ and $ m_2 $ and condensing $ e_1e_2 $. At
the phase transition, $\one \oplus m_1m_2 $ are condensed. 

At all the phase transitions discussed so far, the condensed anyons form a
non-Lagrangian condensable algebra. This is intimately connected with the
gaplessness of the critical points, as described in the previous section. Given
a \symmTO, one can in principle obtain all the gapped boundaries that it can
support by searching for Lagrangian condensable algebras. The phase transitions
between such gapped phases are described by various non-Lagrangian condensable
algebras. The example of the $ \Z_2 \times \Z_2'$ symmetry discussed here gives
us a simple example of such an analysis. The minimal condensable algebra is
just $ \one $, \ie condensation of the trivial anyon. Besides this, there are 9
other non-Lagrangian condensable algebras which correspond to the various phase
transitions between gapped phases associated to the Lagrangian condensable
algebras in \eqn{z2z2Lag}, 
\begin{align} \label{z2z2nonlag}
&\one\oplus e_1,\ \ \one\oplus e_2,\ \ \one\oplus m_1,\ \ \one\oplus m_2, \\
&
\one\oplus e_1e_2,\ \
\one\oplus m_1m_2,\ \
\one\oplus e_1m_2,\ \
\one\oplus m_1e_2,\ \
\one\oplus f_1f_2.\nonumber
\end{align}
We already discussed the phase transitions corresponding to the non-Lagrangian
condensable algebras $ \one\oplus m_1 $, $ \one\oplus m_2 $ and $ \one \oplus
m_1 m_2 $ above. The
condensable algebra $ \one\oplus e_1 $ corresponds to the transition from the $
\one\oplus e_1\oplus m_2\oplus e_1m_2 $-phase to the $ \one\oplus e_1\oplus
e_2\oplus e_1e_2 $-phase (see
the right edge of Fig. \ref{Z2Z2phases}). Similarly, $ \one\oplus e_2 $
corresponds to the transition from the $ \one\oplus
m_1\oplus e_2\oplus e_2m_1 $-phase to the $ \one\oplus e_1\oplus e_2\oplus
e_1e_2 $-phase (see the left
edge of Fig. \ref{Z2Z2phases}).  On the other hand, $ \one\oplus e_1 e_2 $
corresponds to the transition from the $
\one\oplus e_1e_2\oplus m_1m_2\oplus f_1f_2 $-phase to the $ \one\oplus
e_1\oplus e_2\oplus e_1e_2 $-phase (see
the bottom right edge of Fig. \ref{Z2Z2phases}). This is a $ \Z_2^d $ breaking
phase transition in the conventional language.

\begin{figure}[t]
\begin{center}
\includegraphics[scale=0.25]{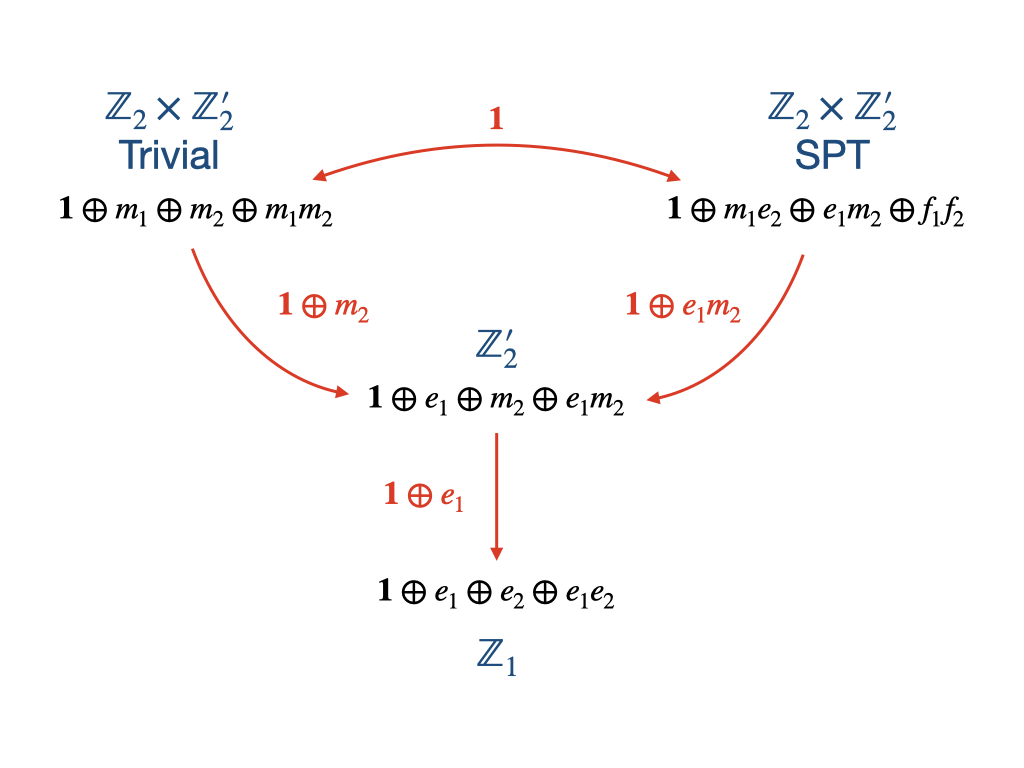}
\end{center}
\caption{The arrows denote possible continuous phase transitions. They are
labeled by the associated non-Lagrangian condensable algebras that describe the
phase transition. The three lower arrows depict possible symmetry breaking
cascades from the two $ \Z_2 \times \Z_2' $-symmetric states, the trivial
paramagnet and the cluster state SPT. By replacing $ \Z_2' $-symmetric state in
the middle of the cascade by $ \Z_2 $- and $ \Z_2^d $-symmetric state, we can
obtain two other such symmetry breaking cascades. The top arrow labeled by the
minimal condensable algebra $ \one $ represents a family of possible continuous
quantum phase transition between the trivial symmetric phase and the nontrivial
SPT phase.\cite{TTVV211007599} Note that this condensable algebra can also
describe fine-tuned continuous phase transitions between any two of the six
gapped phases of this system.
} \label{Z2Z2trans}
\end{figure}


Since the minimal condensable algebra $ \one $ is the
intersection of every pair of Lagrangian condensable algebras, it corresponds to
a direct, fine-tuned,
phase transition between any two gapped phases. In other words, it describes a
multicritical point, \eg see the top portion of Fig. \ref{Z2Z2trans}.
The gapless $\one$-state is given by the canonical boundary of
$(\eGau_{\Z_2\times
\Z_2'})_{/\one} = \eGau_{\Z_2\times \Z_2'}$, which are $(c,\bar c)=(1,1)$
$u(1)$ CFT's.  The other nine non-Lagrangian condensable algebras give rise to
gapless states that correspond to canonical boundary of $\eGau_{\Z_2}$, since
for all of these condensable algebras $ \cA $, we have
\begin{equation}\label{nLagz2z2}
(\eGau_{\Z_2\times \Z_2'})_{/\cA}=\eGau_{\Z_2}
\end{equation}
These nine gapless states are therefore all described by $(c,\bar c)=\left
(\frac12,\frac12\right )$ Ising CFT's (cf. the discussion in \Rf{JW190513279}).
However, these gapless states are distinct, despite being described by the same
CFT, since the assignment of symmetry quantum numbers to the excitations in the
CFT is different for each of them.

We see that 1+1D $\Z_2\times \Z_2'$ symmetric systems can only have two types
of ``stable'' gapless states: $(c,\bar c)=(1,1)$ $u(1)$ CFTs and $(c,\bar
c)=\left (\frac12,\frac12\right )$ Ising CFTs.  From the structure of the
condensable algebras, we see that (cf. Fig. \ref{Z2Z2phases}) \frmbox{the
trivial $\Z_2\times \Z_2'$-SPT state (the $\one\oplus m_1\oplus m_2\oplus
m_1m_2$-state) and the nontrivial $\Z_2\times \Z_2'$-SPT state  (the
$\one\oplus e_2m_1\oplus e_1m_2\oplus f_1f_2$-state) can only be connected by
the gapless $\one$-state, \ie by $(c,\bar c)=(1,1)$ $u(1)$ CFTs,} since the
overlap of the two condensable algebras $\one\oplus m_1\oplus m_2\oplus m_1m_2$
and $\one\oplus e_2m_1\oplus e_1m_2\oplus f_1f_2$ is given by $\one$. This is
consistent with the conclusions of \Rf{P0410416}.  On the other hand,
\frmbox{the nontrivial $\Z_2\times \Z_2'$-SPT state, the $\one \oplus e_2m_1
\oplus e_1m_2 \oplus f_1f_2$-state, can be connected by gapless states of
$(c,\bar c)=\left( \frac12,\frac12 \right)$ Ising CFT to each of the symmetry
breaking states: $\one\oplus m_1m_2\oplus e_1e_2\oplus f_1f_2$, $\one\oplus
m_1\oplus e_2\oplus m_1e_2$, and $\one\oplus e_1\oplus m_2\oplus e_1m_2$. The
condensable algebras for the corresponding gapless states are $\one\oplus
f_1f_2$, $\one\oplus m_1e_2$, and $\one\oplus e_1m_2$ respectively. } See also
\Rf{TL190401544,TL150306794} for a different holographic theory for the phase
transitions between SPT phases.

To summarize, in Fig. \ref{Z2Z2phases}, the six Lagrangian condensable algebras
(and corresponding gapped phases) are shown along with the nine nontrivial
non-Lagrangian condensable algebras. The vertices correspond to the various
gapped phases, while the edges describe gapless states of the 1+1D theory. An
edge that connects to a pair of vertices is understood to be describing the
gapless critical theory that mediates a phase transition between the two gapped
phases. The trivial condensable algebra $ \one $ can always mediate a
multicritical phase transition between any pair of gapped phases, as noted
above. Hence it is not shown in the figure.

In the next section, we contrast this discussion with a $ \Z_2 \times \Z_2' $
symmetry that has a mixed anomaly. The \symmTO\ of such a symmetry is distinct
from that of the anomaly-free $ \Z_2 \times \Z_2' $ symmetry. As was discussed
in \Rf{CW220303596}, the \symmTO\ of $ \Z_2 \times \Z_2' $ symmetry with mixed
anomaly is $ \eGau_{\Z_4} $. As a result the entire discussion of gapped
boundaries and condensable algebras will be completely different from the
anomaly-free case.

\section{1+1D $\Z_2 \times \Z_2'$ symmetry with mixed anomaly}

In our third example, we consider anomalous $\Z_2 \times \Z_2'$ symmetry in
1+1D. Such an anomaly is characterized by a cocycle $\om$ in $ H^3(\Z_2 \times
\Z_2'; \RZ)=\Z_2 \times \Z_2 \times \Z_2$.  The middle $\Z_2$ describes the
mixed anomaly between the $\Z_2$ and $\Z_2'$ groups.  The first and the last
$\Z_2$ describe the self anomaly of the $\Z_2$ and the $\Z_2'$ groups,
respectively. Thus we can use $(m_1,m_{12},m_2)$ to label different cocycles
$\om$. We denote an anomalous $\Z_2 \times \Z_2'$ symmetry as $(\Z_2 \times
\Z_2')^\om = (\Z_2 \times \Z_2')^{(m_1 m_{12} m_2)}$.  

1+1D systems with $(\Z_2 \times \Z_2')^{(010)}$ symmetry have a gapped state
with only the $\Z_2$-symmetry, a gapped state with only the $\Z_2'$-symmetry.,
and a third gapped state that breaks both the $\Z_2$ and the $\Z_2'$ symmetry.
However, there is no gapped state with both the $\Z_2$ and the $\Z_2'$ symmetry
due to the anomaly \cite{CLW1141}.  A state that has the full $\Z_2 \times
\Z_2'$ symmetry unbroken must be gapless.  Such a gapless state happens to be
the critical point for the continuous transition between the two gapped states
with unbroken $\Z_2$ and unbroken $\Z_2'$ symmetry respectively. Noting that
$\Z_2$ and $\Z_2'$ are not related by a group-subgroup relation, we see that
this is an example of a continuous phase transition that is beyond the
conventional Landau theory of phase transitions.

The critical point with the $\Z_2 \times \Z_2'$ symmetry also has other
symmetries.  The full symmetry of the critical point is described by the
\symmTO\ $\eGau_{\Z_2\times \Z_2}^{(010)}$, which refers to a $\Z_2 \times
\Z_2'$ twisted quantum double or a $ \Z_2 \times \Z_2' $ Dijkgraaf-Witten (DW)
gauge theory \cite{DW9093}.  The $\Z_2$ symmetry corresponds to the
$\Z_2$-gauge charge conservation in the DW theory, while the $\Z_2'$ symmetry
corresponds to the $\Z_2'$-gauge charge conservation in the DW theory.  The
$\eGau_{\Z_2\times \Z_2}^{(010)}$ DW theory also have $\Z_2$ and $\Z_2'$ gauge
flux, whose conservation give rise to additional symmetries at the critical
point.

In order to discuss the phase diagram and phase transitions of a system with
such an anomalous $ \Z_2\times \Z_2' $ symmetry, we will use the fact derived
in \Rf{EW211211394,CW220303596} that the 2+1D topological order
$\eGau_{\Z_2\times \Z_2}^{(010)}$ is the same as the 2+1D topological order $
\eGau_{\Z_4} $ (\ie the $\Z_4$ gauge theory with charge excitations).  In order
words, in 2+1D, the anomalous $ \Z_2\times \Z_2' $ symmetry and the $\Z_4$
symmetry are equivalent, since they are described by the same \symmTO.  The
anyons of $\eGau_{\Z_2\times \Z_2'}^{(010)} $ topological order can be mapped
to those of $\eGau_{\Z_4}$ topological order. This mapping is given as follows:
\begin{equation}\label{z2z2toz4map}
e_1 \to e^2, \ \ 
e_2 \to m^2, \ \
m_1 \to m, \ \
m_2 \to e
\end{equation}
where $ e $ and $ m $ are the generators of the gauge charge and the gauge flux
excitations of $ \eGau_{\Z_4} $. We argued for this mapping of anyons by
studying the patch operators and their associated braided fusion category in
\Rf{CW220303596}. We found that presence of the mixed anomaly changes the
anyon statistics from that described by $ \eGau_{\Z_2\times \Z_2'} $ to that
described by $ \eGau_{\Z_2\times \Z_2'}^{(010)} = \eGau_{\Z_4} $. Supported by
this result, we will use the language of $ \eGau_{\Z_4} $ to describe the phase
transitions of a system with $(\Z_2 \times \Z_2')^{(010)}$ symmetry.

\begin{figure}[t]
\begin{center}
\includegraphics[scale=0.6]{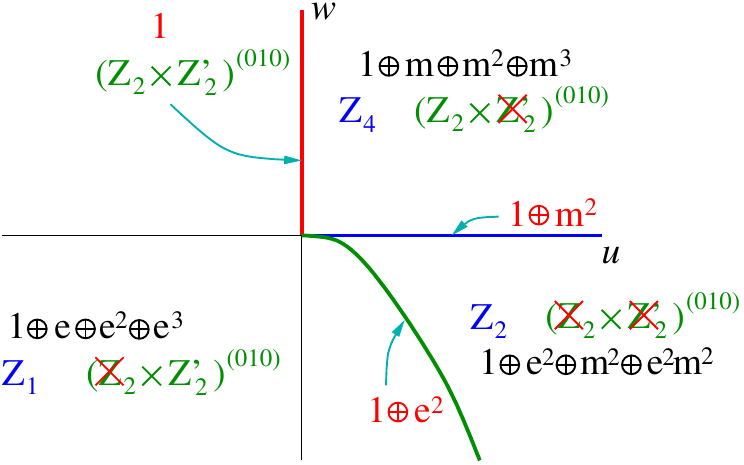}
\end{center}
\caption{ Mean-field phase diagram for systems with $\Z_4$ symmetry.  It has
three gapped phases with unbroken $\Z_4$, $\Z_2$, and $\Z_1$ symmetries.  The
two phase transitions $\Z_4 \to \Z_2$ (marked by $\one \oplus m^2$) and $ \Z_2
\to \Z_1$ (marked by $\one \oplus e^2$) are critical points described by the
same Ising CFT.  The direct phase transition $\Z_4\to \Z_1 $ (marked by $\one$)
corresponds to a critical point that does not break the anomalous symmetry
$(\Z_2\times\Z_2')^{(010)}$ and has the full  \symmTO\ $\eGau_{\Z_4}$ .  It is
a critical line that includes the $\Z_4$ parafermion CFT.  } \label{Z4phasesGL} 
\end{figure}

Let us first recall what are the different gapped phases that a system with $
\Z_4 $ symmetry may have from the Ginzburg-Landau mean field theory.  Let us
introduce two order parameter fields, a complex bosonic field $\Phi$ and a real
field $\phi$. Under the generating transformation of $\Z_4$, they transform as
\begin{align}
U_{\Z_4} \Phi = \ee^{\ii\pi/2} \Phi,\ \ \ \
U_{\Z_4} \phi = - \phi.
\end{align}
Consider the following Ginzburg-Landau functional
\begin{align}
F = \int \dd x\; &
|\prt \Phi|^2 + u |\Phi|^2 + \frac12 (\Phi^4 + h.c.) + 2 |\Phi|^4
\\
&
+ |\prt \phi|^2
+ w \phi^2 + 2 \phi^4 + \frac12 \phi (\Phi^2+h.c.)
\nonumber 
\end{align}
Since the only subgroups of $ \Z_4 $ are the trivial group and $ \Z_2 $, we can
have three different gapped phases in total: one with the full $ \Z_4 $
symmetry ($\Phi=\phi=0$ when $u,w>0$), one with unbroken $ \Z_2 $ symmetry
($\Phi=0$, $\phi\neq 0$ when $u>0$, $w<0$), and one with $ \Z_1 $ symmetry
($\Phi\neq 0$, $\phi\neq 0$ when $u<0$) (see Fig. \ref{Z4phasesGL}). 

This shape of the phase boundaries can be understood as follows. When we turn $
w $ from positive to negative in the presence of a positive $ u $, we find a
minima with non-zero $ \phi $ but still $ \Phi=0 $. This is a phase which has
an unbroken $ \Z_2 $ symmetry. The minimum is now at $ \phi_* \sim \pm
\sqrt{-w}$.  This non-zero mean-field value of $ \phi $ turns on the $ \phi
(\Phi^2+h.c.) $ term then effectively introduces a modification to the
quadratic terms for $ \Phi $ which is of the form $ \sqrt{-w}(\Phi^2+h.c.) $.
Thus we see that for $ u< O(\sqrt{-w})$, we transition into a phase with
non-zero $ \Phi $ as well as $ \phi $. This is the $ \Z_1 $ symmetric phase.
The corresponding phase transition is indicated in green in the bottom right
quadrant of Fig.  \ref{Z4phasesGL}. If we are close enough to the phase
transition regions, $ w $ is small so it is a very good approximation to drop
the corresponding higher order terms and only concentrate on the quadratic
terms and the $ \phi \Phi^2 $ term for the mean field phase boundary analysis.

Although the two phase transitions $ \Z_4 \to \Z_2 $ and $ \Z_2 \to \Z_1 $
correspond to different symmetry breaking pattern, their critical points happen
to be described by the same Ising CFT with central charge $(c,\bar
c)=\left( \frac12,\frac12 \right)$.  The third symmetry breaking pattern
$\Z_4\to \Z_1$ will
have a different critical theory.  In fact the transition is described by a
critical line  with central charge $(c,\bar c)=(1,1)$, that includes the $\Z_4$
parafermion CFT \cite{G1988,G8853}.

\begin{figure}[t]
\begin{center}
\includegraphics[scale=0.6]{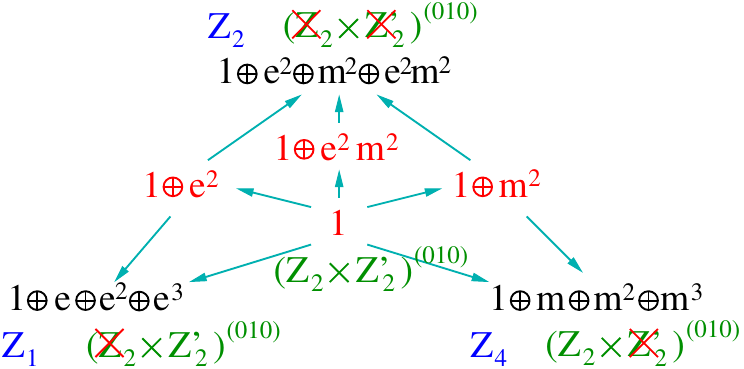}
\end{center}
\caption{ $\Z_4$-symmetric systems has a symmetry described by 2+1D
$\Z_4$ topological order $\eGau_{\Z_4}$ (\ie a $\Z_4$ gauge theory).  The
$\eGau_{\Z_4}$ topological order has seven condensable algebras, and 3 of them
are Lagrangian (the back ones above). They give rise to gapped boundaries.
Four of seven are not Lagrangian, and give rise to gapless boundaries.  The
gapless boundaries may be critical points for the transitions between gapped or
gapless boundaries, as indicated above.  The arrows indicate the directions of
more condensation.  } \label{Z4phases} 
\end{figure}

Now, from the \symmTO\ point of view, the gapped phases of
this system are the allowed gapped boundaries of $ \eGau_{\Z_4} $. Such gapped
boundaries are described by the Lagrangian condensable algebras of the
\symmTO\ $ \eGau_{\Z_4} $:
\begin{align}
& \one\oplus e\oplus e^2\oplus e^3, 
\ \ \ \ \ 
\one\oplus m\oplus m^2\oplus m^3, 
\nonumber\\
& \one\oplus e^2\oplus m^2\oplus e^2m^2  . 
\end{align}
The Lagrangian  condensable algebras match the gapped symmetric and
symmetry-breaking phases very well.  The first of these condensable algebras, $
\one\oplus  e\oplus e^2\oplus e^3 $ represents the $\Z_1$-phase that
spontaneously breaks the $ \Z_4 $ symmetry completely.  The second, $
\one\oplus m\oplus m^2\oplus m^3 $ represents the $\Z_4$-gapped phase that is
fully $ \Z_4 $ symmetric. The last one, $ \one\oplus  e^2\oplus m^2\oplus e^2
m^2 $ represents the $\Z_2$ gapped phase that breaks $ \Z_4 $ down to $ \Z_2 $. 

Let us consider these gapped phases in the dual picture with the $ (\Z_2 \times
\Z_2')^{(010)} $ symmetry. The three gapped phases obtained by breaking $ \Z_2
$, $ \Z_2' $, or $ \Z_2\times\Z_2' $ are denoted as  $(\xZ_2 \times
\Z_2')^{(010)} $-phase, $(\Z_2 \times \xZ_2')^{(010)} $-phase, $(\xZ_2 \times
\xZ_2')^{(010)} $-phase.  They have a one-to-one correspondence with the three
gapped phases discussed above.  How do we identify them? First of all, the
condensable algebra $ \one\oplus  e\oplus e^2\oplus e^3 $ may be written in
terms of $ \Z_2\times\Z_2' $ charges and fluxes -- see \eqn{z2z2toz4map} -- as $
\one\oplus m_2\oplus
e_1\oplus e_1m_2 $ which indicates a phase that has a broken $ \Z_2 $ but
unbroken $ \Z_2' $. Thus the $\Z_1$-phase (in the $ \Z_4 $ symmetry language)
corresponds to the $(\xZ_2 \times
\Z_2')^{(010)} $-phase.  (see Fig. \ref{Z4phasesGL} and \ref{Z4phases}).
Similarly, the condensable algebra $ \one\oplus m\oplus m^2\oplus m^3 $ maps on
to $ \one\oplus  m_1\oplus e_2\oplus m_1e_2 $, which corresponds to  the $(\Z_2
\times \xZ_2')^{(010)} $-phase.  This phase is mapped to the $\Z_4$-phase.
Lastly, the condensable algebra $ \one\oplus  e^2\oplus m^2\oplus e^2 m^2$ maps
on to $ \one\oplus  e_1\oplus e_2\oplus e_1e_2 $, which corresponds to the
$(\xZ_2 \times \xZ_2')^{(010)} $-phase and maps to the $\Z_2$-phase.

Next let us discuss the gapless states that describe the phase transitions of
this system.  The gapless states are given by condensation patterns
described by non-Lagrangian condensable algebras.  There are four non-Lagrangian
condensable
algebras in the $\eGau_{\Z_4}$ topological order,
\begin{align}
\one, \ \ \ \one\oplus e^2 , \ \ \ \one\oplus m^2, \ \ \ \one\oplus e^2 m^2.
\end{align}
They map to four non-Lagrangian condensable algebras in
the equivalent $\eGau_{\Z_2\times\Z_2'}^{(010)}$ topological order
\begin{align}
\one, \ \ \ \one\oplus e_1 , \ \ \ \one\oplus e_2, \ \ \ \one\oplus e_1 e_2.
\end{align}
Thus there are four condensation patterns in the \symmTO\ of this system that can give rise to gapless
states. We refer to these gapless states as $\one$-state,  $\one\oplus e^2$-state,
$\one\oplus m^2$-state, and $\one\oplus e^2m^2$-state.  

The 1+1D gapless $\one$-states are given by the
canonical boundaries of the $\one$-condensation-induced topological order,
which is nothing but the original \symmTO\ $\eGau_{\Z_4}$.  Similarly, 1+1D
gapless $ \one\oplus e^2$-state and $\one\oplus m^2$-state are given by the
canonical boundaries of $(\eGau_{\Z_4})_{/\one\oplus e^2}= \eGau_{\Z_2}$ and
$(\eGau_{\Z_4})_{/\one\oplus m^2}= \eGau_{\Z_2}$.  Last, the 1+1D gapless
$\one\oplus e^2m^2$-state is given by the canonical boundary of
$(\eGau_{\Z_4})_{/\one\oplus e^2m^2}= \eM_\text{DS}$, where $\eM_\text{DS}$ is
the double-semion topological order. To see why $(\eGau_{\Z_4})_{/\one\oplus
e^2m^2}= \eM_\text{DS}$, we can ask the question: which anyons of $ \eGau_{\Z_4}
$ have trivial mutual statistics with $ e^2 m^2 $. Out of the 16 anyons of $
\eGau_{\Z_4} $, there are 8 that satisfy this condition: 
\begin{equation*}
\one, e^2, m^2 , e^2m^2, em, e^3m^3,em^3,e^3m
\end{equation*}
Now since $ e^2m^2 $ is condensed in $ (\eGau_{\Z_4})_{/\one\oplus e^2m^2} $, we
should consider the anyons that are related by fusion with $ e^2m^2 $ as
equivalent,
\begin{align*}
&m^2 = e^2\cdot e^2m^2, && e^2m^2 = \one \cdot e^2m^2\\
&e^3m^3=em\cdot e^2m^2, && e^3 m = em^3\cdot e^2m^2
\end{align*}
Then we find that the remaining inequivalent anyons are $ \one, e^2 , em,em^3
$. Computing the self and mutual statistics of these anyons indicates that they
correspond to $\eM_\text{DS}$.

What is the canonical boundary of $ \eGau_{\Z_4}$?  We note that $\one$ is
the only condensable algebra in the overlap of two Lagrangian condensable
algebras
$ \one\oplus  e\oplus e^2\oplus e^3 $ (the $\Z_1$-phase) and $ \one\oplus
m\oplus m^2\oplus m^3 $ (the $\Z_4$-phase).  This allows us to conclude that
the canonical boundary of $\eGau_{\Z_4}$ should describe $\Z_4\to\Z_1$
symmetry breaking transition. Such a transition is described a $(c,\bar
c)=(1,1)$ critical line with only one relevant symmetric operator.  Thus the
$\Z_4\to\Z_1$ symmetry breaking transition is a stable transition, and the
canonical boundaries of $\eGau_{\Z_4}$ are described by $(c,\bar c)=(1,1)$
$u(1)$ CFT.  Similarly, we can show that canonical boundaries of $\eGau_{\Z_2}$
are described by $(c,\bar c)=\left( \frac12,\frac12 \right)$ Ising CFT with only
one
relevant symmetric operator -- the critical point of $\Z_2\to \Z_1$ symmetry
breaking transition.  The canonical boundary of double-semion topological order
$\eM_\text{DS}$ is given by the chiral boson theory \eqn{Kphi} with $K$-matrix
given by \eqn{KDS}.  So the gapless $\one\oplus e^2m^2$-state, just like the
gapless $\one$-state, is also described by $(c,\bar c)=(1,1)$ $u(1)$ CFT.

After determining the nature of gapless states $\one$, $\one\oplus e^2$,
$\one\oplus m^2$, and
$\one\oplus e^2 m^2$, we consider the harder question: how do these gapless
states get
connected by RG flow, and what is the structure of the full phase diagram?

\begin{figure}[t]
\begin{center}
\includegraphics[scale=0.6]{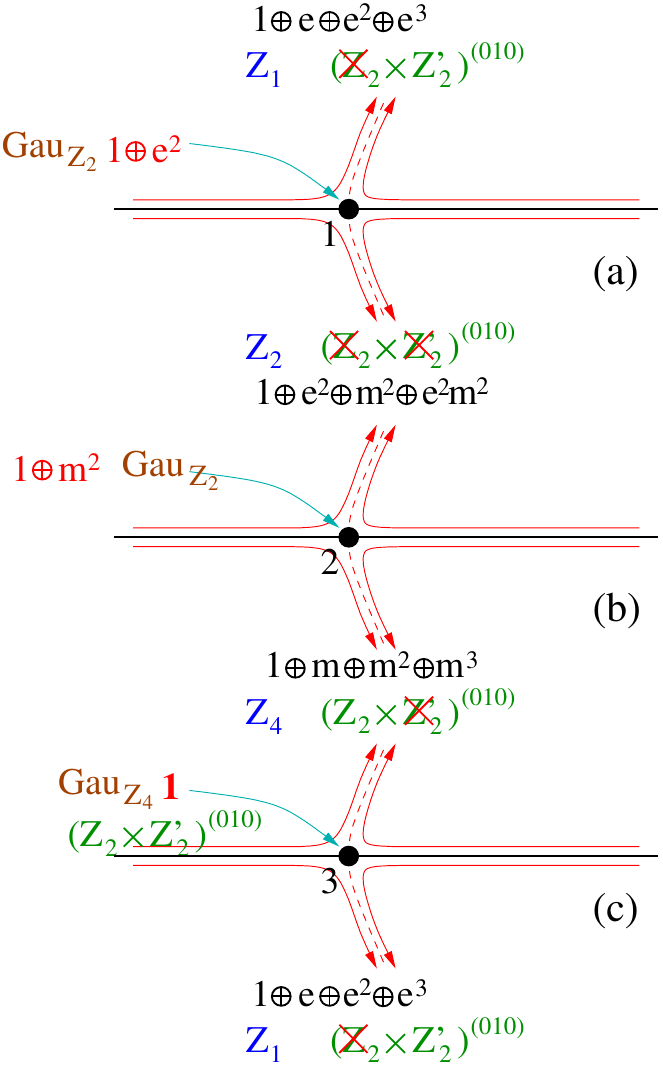}
\end{center}
\caption{ Possible local phase diagram for systems with $\Z_4$ symmetry, which
contains three gapped phases with unbroken symmetries, $\Z_4$, $\Z_2$, and
$\Z_1$. The curves with arrows represent the RG flow, and the dots are the RG
fixed points that correspond to the critical points of phase transitions.  The
plane is a space of Hamiltonians with \symmTO\ $\eGau_{\Z_4}$  (see Appendix
\ref{phasestructure} for detailed discussions).  The horizontal line in (a) is
a space of Hamiltonians whose ground states have the condensation
$\cA=\one\oplus e^2$, which is the basin of attraction of the RG fixed point 1.
The horizontal line in (b) is a space of Hamiltonians whose ground states have
the condensation $\cA=\one\oplus m^2$, the basin of attraction of the RG fixed
point 2. The horizontal line in (c) is a space of of Hamiltonians whose ground
states have the condensation $\cA=\one$, the basin of attraction of the RG
fixed point 3.  The critical point 3 is actually part of a critical line of
$(c,\bar c)=(1,1)$ $u(1)$ CFT (the canonical boundary of $\eGau_{\Z_4}$
topological order).  The critical point 1 and 2 are the $(c,\bar c)=\left(
\frac12,\frac12 \right)$ Ising CFT (the canonical boundary of $\eGau_{\Z_2}$
topological order).  We also list the corresponding condensable algebras, for each gapped phase and gapless
critical point. } \label{Z4phasetrans} \end{figure}

\begin{figure}[t]
\begin{center}
\includegraphics[scale=0.6]{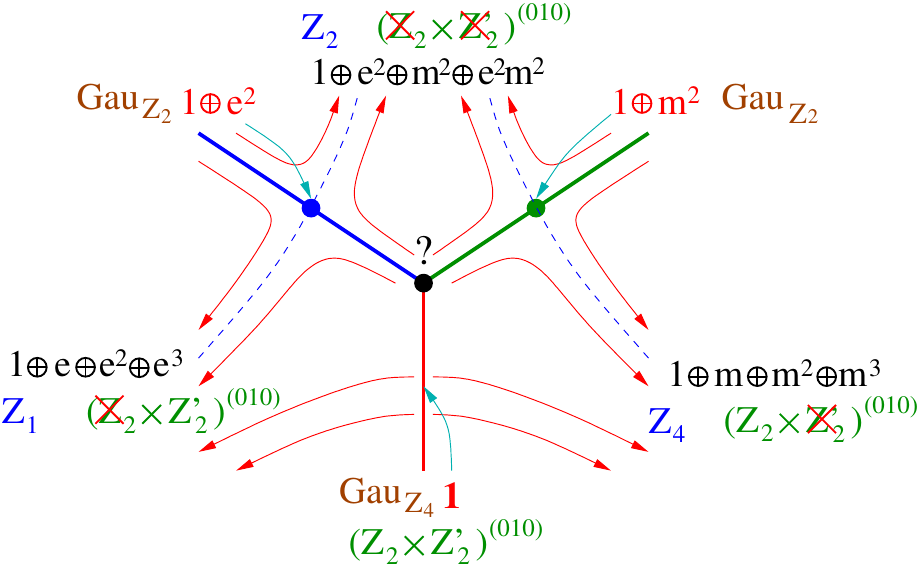}
\end{center}
\caption{ A possible global phase diagram for systems with $\Z_4$ symmetry or
$(\Z_2\times\Z_2')^{(010)}$ symmetry, which has a similar topology with the
mean-field phase diagram Fig. \ref{Z4phasesGL}. We are not sure about the phase
structure near the center of the phase diagram.} \label{Z4phasetransY}
\end{figure}

The condensable algebra $\one\oplus e^2$ differs from Lagrangian condensable
algebras by condensing one excitations.  In fact, condensing $e$ changes
$\one\oplus e^2$ to $\one\oplus e \oplus e^2 \oplus e^3$, and condensing $m^2$
changes $\one\oplus e^2$ to $\one\oplus e^2 \oplus m^2 \oplus e^2m^2$.  Here
$(e,m^2)$, having a nontrivial mutual statistics, form a competing pair.  We
either have an $e$-condensation that gives rise to the condensable algebra  $\one\oplus e \oplus e^2 \oplus e^3$, or we have an $m^2$-condensation
that gives rise to the condensable algebra $\one\oplus e^2 \oplus m^2
\oplus e^2m^2$. However $e$ and $m^2$ cannot both condense.  If we fine tune, we
can
ensure neither of them condense; that gives rise to the condensable algebra $\one \oplus e^2 $.  The gapless $\one\oplus e^2$-state is described
by $(c,\bar c)=\left( \frac12,\frac12 \right)$ Ising CFT with only one relevant
symmetric
operator.  The condensable algebra $\one\oplus e^2$ only allows one competing
pair $(e,m^2)$.  Thus the RG flow in the relevant direction will cause the
condensation of the competing pair.  This gives rise to the phase diagram Fig.
\ref{Z4phasetrans}(a) near the gapless $\one\oplus e^2$-state.  Similarly, the
phase diagram near the gapless $\one\oplus m^2$-state is given by Fig.
\ref{Z4phasetrans}(b).  The condensable algebra $\one$ differs from the
Lagrangian
condensable algebras $\one\oplus e \oplus e^2 \oplus e^3$ and $\one\oplus e^2
\oplus m^2
\oplus e^2m^2$ by condensing one excitation.  The competing pair
involved is $(e,m)$.  Since the gapless $\one$-state has only one relevant
operator, if that corresponds to this competing pair, the phase diagram near the
gapless $\one$-state is given by Fig.
\ref{Z4phasetrans}(c).  Putting the three local phase diagram together, we
obtain a possible global phase diagram Fig.  \ref{Z4phasetransY}.

\begin{figure}[t]
\begin{center}
\includegraphics[scale=0.6]{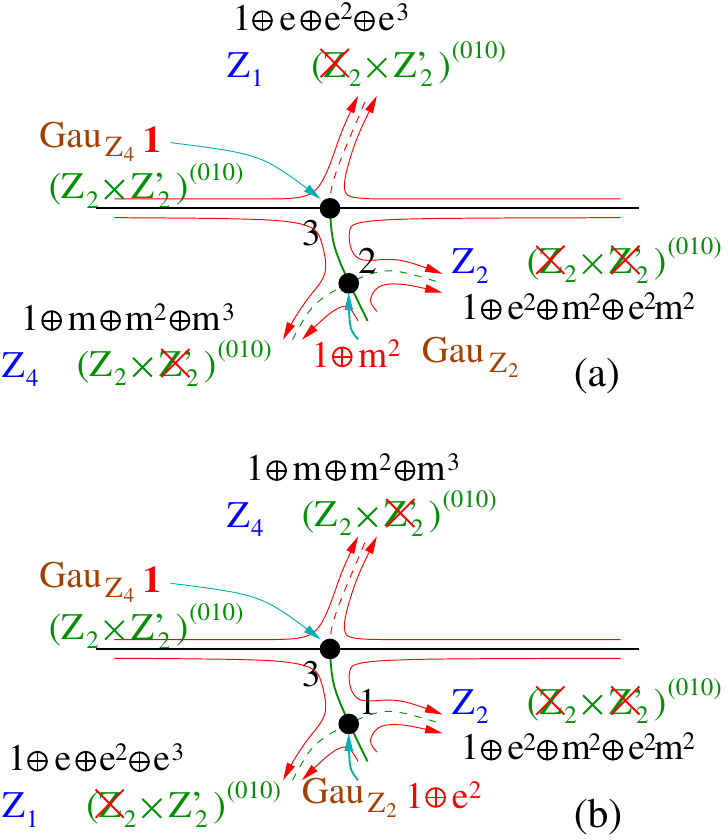}
\end{center}
\caption{Other possible local phase diagrams near the gapless $\one$-state, if
the relevant direction corresponds to the condensation of (a) competing pair
$(e, m^2)$ or (b) competing pair $(m, e^2)$.  This requires the relevant
direction of certain $(c,\bar c)=(1,1)$ $u(1)$ CFT's to flow to the $(c,\bar
c)=\left( \frac12,\frac12 \right)$ Ising CFT.  } \label{Z4phasetransT}
\end{figure}

However, the gapless $\one$-state also allows competing pairs $(e, m^2)$ and
$(m, e^2)$,
in addition to $(e,m)$.  If instead, it is the competing pair $(e, m^2)$ that
corresponds
to the relevant direction, the phase diagram will be Fig.
\ref{Z4phasetransT}(a), which implies a stable $\Z_2\to\Z_1$ symmetry breaking
transition described by the critical line of $(c,\bar c)=(1,1)$ $u(1)$ CFT that
includes the Ising critical point. On the other hand, if the competing pair $(m,
e^2)$
corresponds to the relevant direction, the phase diagram will be Fig.
\ref{Z4phasetransT}(b), which implies a stable $\Z_4\to\Z_2$ symmetry breaking
transition described by the critical line of $(c,\bar c)=(1,1)$ $u(1)$ CFT that
involves the Ising critical point.  The direct phase transition $\Z_4\to\Z_1$
has been observed which is not a critical line and does not involve Ising
critical point.  This implies that the competing pair $(e, m)$ corresponds to
the
relevant direction, and  phase diagram Fig.  \ref{Z4phasetrans}(c) is realized.
However, since $(c,\bar c)=(1,1)$ $u(1)$ CFT is a critical line with a marginal
direction, it is not
clear if either of the phase diagrams in Fig.  \ref{Z4phasetransT} can be
realized in some
parts of the critical line.

To obtain a concrete global phase diagram for $\Z_4$ symmetric systems, we
consider a $\Z_4$ symmetric statistical model on square lattice, which has
degree of freedoms $(\th_{\v i}, \phi_{\v i})$, $\th_{\v i}=0,1,2,3$, $\phi_{\v
i}=0,1,2,3$, on site $\v i$.  The energy is given by
\begin{align}
\label{Z4model}
E &= - \sum_{\v i} J_1( 
\del_{\th_{\v i},\th_{\v i+\v x}}+ \del_{\th_{\v i},\th_{\v i+\v y}} 
)
\nonumber\\
&\ \ \ \
- \sum_{\v i} J_2( 
\del_{\text{mod}(\phi_{\v i}-\phi_{\v i+\v x},2)}+ \del_{\text{mod}(\phi_{\v
i}-\phi_{\v i+\v y},2)} 
) 
\nonumber\\
&\ \ \ \
-  \sum_{\v i}  J  \del_{\th_{\v i},\phi_{\v i}}
+J_c \sin\left (\frac{\pi(\th_{\v i}-\phi_{\v i})}{2}\right ) 
\end{align}
The $J_c$ term breaks the $\th_{\v i} \to \text{mod}(-\th_{\v i},4)$, $\phi_{\v
i} \to \text{mod}(-\phi_{\v i},4)$ symmetry, so the full internal symmetry of
the model is $\Z_4$.  The $J_2$ term helps to realize the $\Z_2$-phase.

\begin{figure}[t]
\begin{center}
\includegraphics[scale=0.45]{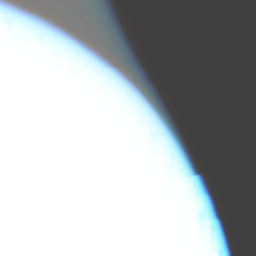}\ \ \ \
\includegraphics[scale=0.45]{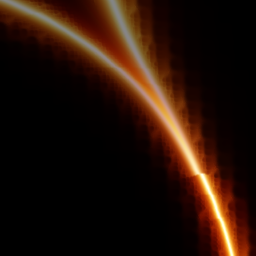}
\end{center}
\begin{center}
(a) 
\ \ \ \ \ \ \ \ \ \ \ \ \ \ \ \ \ \
\ \ \ \ \ \ \ \ \ \ \ \ \ \ \ \ \ \
(b)
\end{center}
\caption{ A phase diagram for the model \eq{Z4model} with $0.993 < J_1\bt <
1.083$ (horizontal axis), $0.533 < J_2\bt < 0.773$ (vertical axis), and $J\bt
=J_c\bt =1.0$.  (a) A plot of $1/GSD$, where $GSD$ is obtained from the
partition function: $GSD\equiv Z^2(L, L)/Z(L, 2L)$ where $Z(L_1, L_2)$ is the
partition function for system of size $L_1\times L_2$.  For gapped quantum
systems, $GSD$ happen to be the ground state degeneracy.  (b) A plot of central
charge $c$.  The central charge $c$ is also obtained from the partition
function $Z(L, L_\infty)$, which has a form $\ee^{- L_\infty[ \eps L
-\frac{2\pi c }{24 L} +o(L^{-1})]}$ when $L_\infty \gg L$, where $c$ is the
central charge.  This way,  the central charge $c$ is defined even for
non-critical states.  The red-channel of the colored image is for system of
size $64\times 64$, green-channel for $128\times 128$, and blue-channel for
$256\times 256$.  } \label{Z4_10-10_10-11_5-8} 
\end{figure}

We use the spacetime tensor network renormalization approach \cite{LN0701} to
study the above statistical model. In fact, we use a particular version of the
tensor network approach which is described in detail in \Rf{GW0931}.  We obtain
the phase diagram Fig.  \ref{Z4_10-10_10-11_5-8}.  The lower left of Fig.
\ref{Z4_10-10_10-11_5-8} is the $\Z_4$-phase.  The upper left is the
$\Z_2$-phase, and the right is the $\Z_1$-phase.  The numerical phase diagram
qualitatively agrees with Fig. \ref{Z4phasetransY}.

We have described the gapless states and continuous transitions from the $\Z_4$
symmetry point of view.  We can repeat the above discussion using  the
$(\Z_2\times\Z_2')^{(010)}$ symmetry point of view, and obtain analogous
results.

It is interesting to compare the $\Z_2\times\Z_2'$ symmetry and
$(\Z_2\times\Z_2')^{(010)}$ symmetry.  The $\xZ_2\times\Z_2'\to
\Z_2\times\xZ_2' $ DQCP-type transition can be a continuous transition
described by a gapless $\one$-state. Such a gapless
state is given by the canonical boundary of the \symmTO\ $\eGau_{\Z_2\times
\Z_2'}$, which is the $(c,\bar c)=(1,1)$ Ising$\times$Ising CFT.  Such a
gapless state has two $\Z_2\times\Z_2'$ symmetric relevant operators.  So the
$\xZ_2\times\Z_2'\to \Z_2\times\xZ_2' $ symmetry breaking transition can be a
direct continuous transition, but it must be a multicritical point.  In
contrast, the $(\xZ_2\times\Z_2')^{(010)}\to (\Z_2\times\xZ_2')^{(010)} $
symmetry breaking transition is described by the canonical boundary of the
\symmTO\ $\eGau_{\Z_2\times \Z_2'}^{(010)}$, which is a $(c,\bar c)=(1,1)$
$u(1)$ CFT that has only one $(\Z_2\times\Z_2')^{(010)}$ symmetric relevant
operator.

\section{1+1D anomaly-free $S_3$ symmetry}

\subsection{A Ginzburg-Landau approach for phases and phase transitions}

The description of $ S_3 $ symmetry-breaking phases in Ginzburg-Landau theory
is based on order parameters that transform nontrivially under the relevant
broken symmetries. The group $ S_3 $ has two inequivalent nontrivial
subgroups, $ \Z_2 $ and $ \Z_3 $. In terms of permutations, $ S_3 $ is
represented as
\begin{equation}\label{S3}
S_3=\{\mathbb{1},(1,2),(2,3),(1,3),(1,2,3),(1,3,2)\} 
\end{equation}
with subgroups
\begin{align}\label{S3sub}
\Z_2 &\simeq \{\mathbb{1},(1,2)\},\{\mathbb{1},(1,3)\},\{\mathbb{1},(2,3)\},\\
\Z_3 &\simeq \{\mathbb{1},(1,2,3),(1,3,2)\}
\end{align}
The two elements $ (1,2) $ and $ (1,2,3) $ generate the group $ S_3 $. There
are two nontrivial representations of this group, a one-dimensional
representation and a two-dimensional one. The first one, which we call $ a_1 $,
may be realized by a real-valued Ising-like field, $ \phi^{a_1} $ that
transforms under the generators as
\begin{align}
(1,2) \circ \phi^{a_1} = -\phi^{a_1},\ \ \
(1,2,3) \circ \phi^{a_1} = \phi^{a_1}.
\end{align}
The condensation of $\phi^{a_1}$, that gives it a non-zero vacuum expectation
value, breaks $S_3$ symmetry down to $S_3/\Z_2=\Z_3$ symmetry. The second
nontrivial representation, which we call $ a_2 $, may be realized by a complex
two-component bosonic field $\Phi^{a_2}_\al$, $\al=1,2$. It transforms under
the $ S_3 $ generators as
\begin{align}
(1,2) \circ \Phi^{a_2} &= 
\begin{pmatrix}
0 & 1\\
1 & 0\\
\end{pmatrix}
\Phi^{a_2},
\nonumber\\
(1,2,3) \circ \Phi^{a_2} &= 
\begin{pmatrix}
\ee^{\ii 2\pi/3} &0 \\
0 & \ee^{-\ii 2\pi/3}\\
\end{pmatrix}
\Phi^{a_2}.
\end{align}
This representation is fully faithful in its $ S_3 $ action. The condensation
of $\Phi^{a_2}$ satisfying $\Phi^{a_2}_1 = \Phi^{a_2}_2$ breaks $S_3$ symmetry
down to $\Z_2$. The condensation of $\Phi^{a_2}$ that does not satisfy
$ \Phi^{a_2}_1 = \Phi^{a_2}_2$ breaks $S_3$ symmetry completely down to $\Z_1$
symmetry (\ie the trivial group). To study the different phases allowed by the
symmetry breaking structure, we can work with the following Ginzburg-Landau
functional (we have dropped the superscripts $ a_1,a_2 $ for readability):
\begin{align}
\label{LG}
& F[\phi,\Phi_\al] = 
u \phi^2 +\phi^4 
+v (|\Phi_1|^2+|\Phi_2|^2) 
\nonumber\\
&\ \ \ \
+ \al (\Phi_1^3+\Phi_2^3) 
+ \bt (\Phi_1^3-\Phi_2^3) \phi
+ \ga \phi^2(|\Phi_1|^2+|\Phi_2|^2) 
\nonumber\\
&\ \ \ \
+ |\Phi_1-\Phi_2^*|^2 +(|\Phi_1|^2+|\Phi_2|^2)^2 
+c.c.  
\end{align}
It is straightforward to check that $ V[\phi,\Phi_\al] $ is symmetric under the
action of the two generators of $ S_3 $ and hence fully symmetric under $ S_3 $
transformations. The mean-field solution is obtained by minimizing this
functional with the assumption that the fields are independent of the spatial
coordinates. 
The mean-field phase diagrams are plotted using  
\begin{align}
\Z_2 \text{ order parameter: }& |\phi|,
\nonumber\\
S_3 \text{ order parameter: }& \sqrt{|\Phi_1|^2+|\Phi_2|^2}.
\end{align}

\begin{figure}[t]
\centering
\includegraphics[height=1.5in]{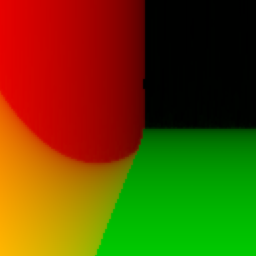} \hfill
\includegraphics[height=1.1in]{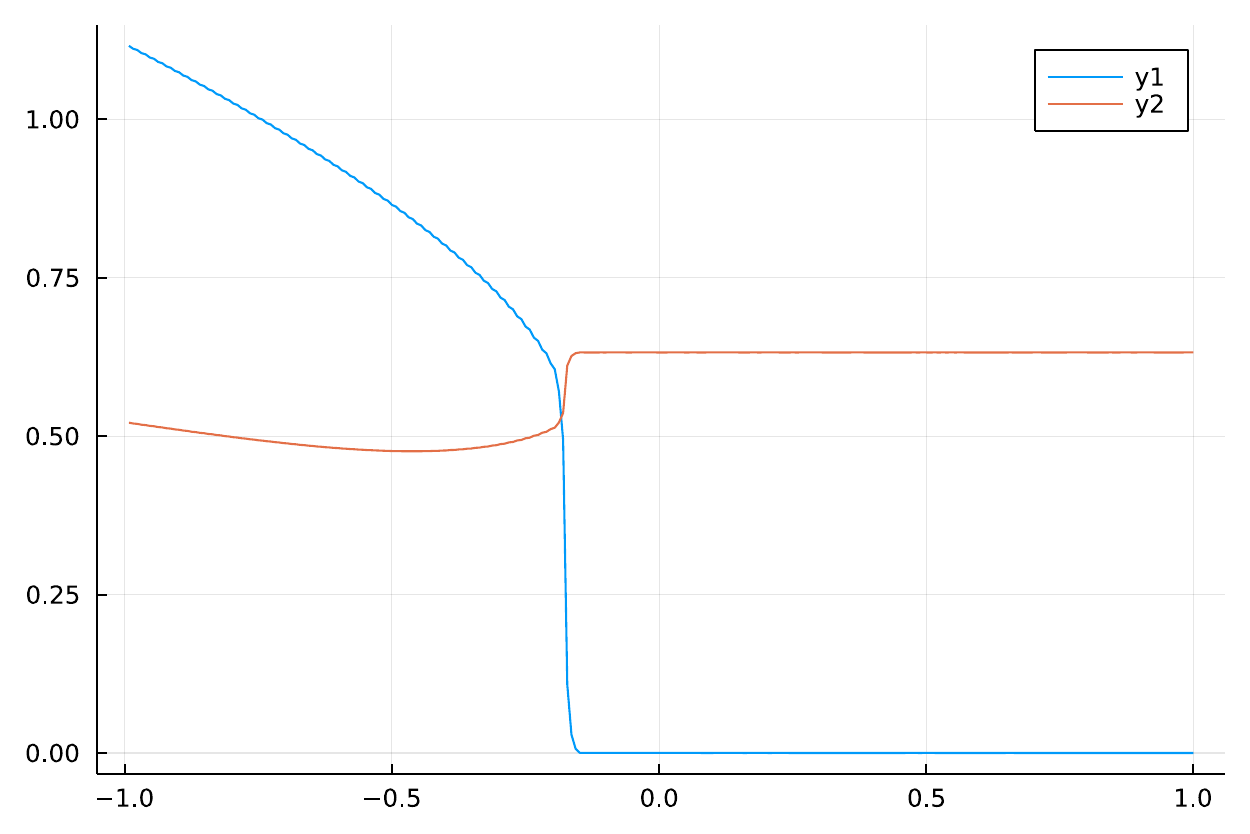} \\
\ \ \ \ \ \ \ (a) \hskip 1.3in $\Z_1$ \ \ \ \ \ (b) \ \ \ \ \ $\Z_3$ \\
\includegraphics[height=1.1in]{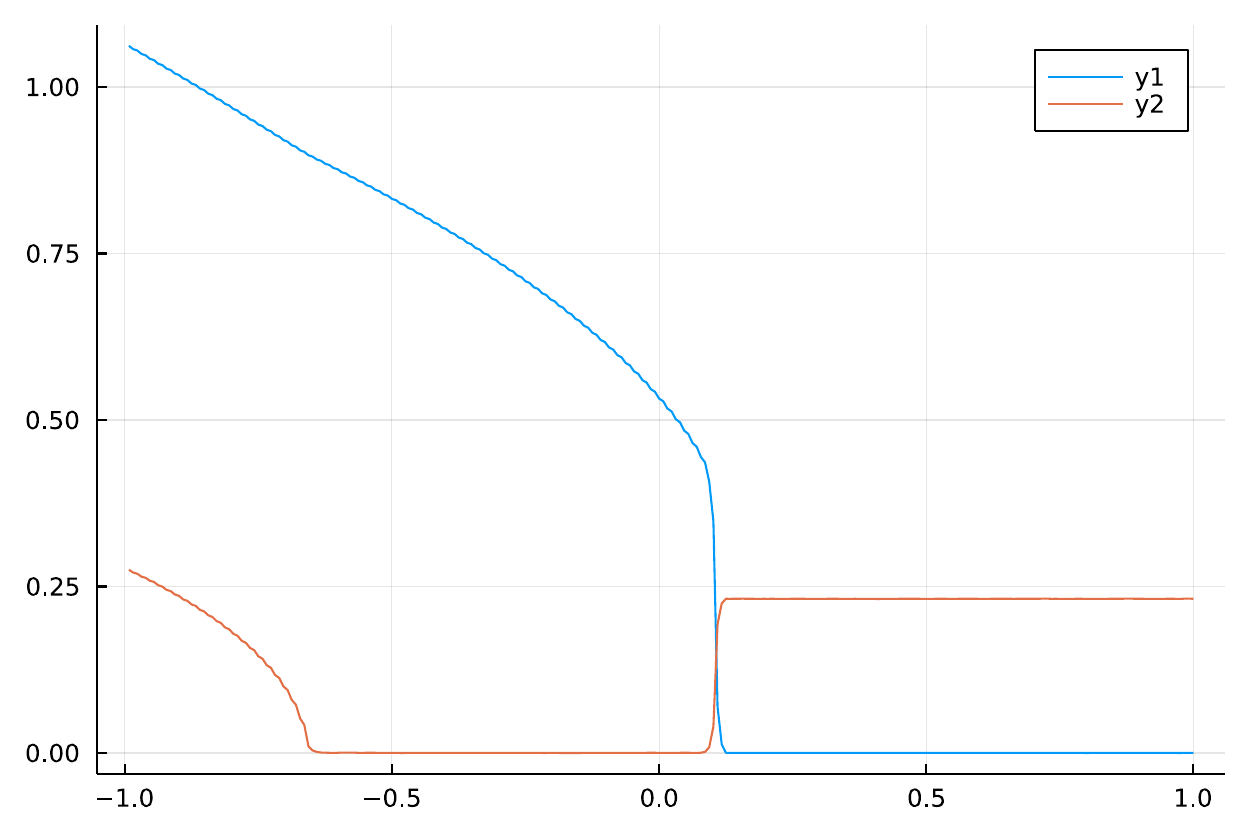} \hfill
\includegraphics[height=1.1in]{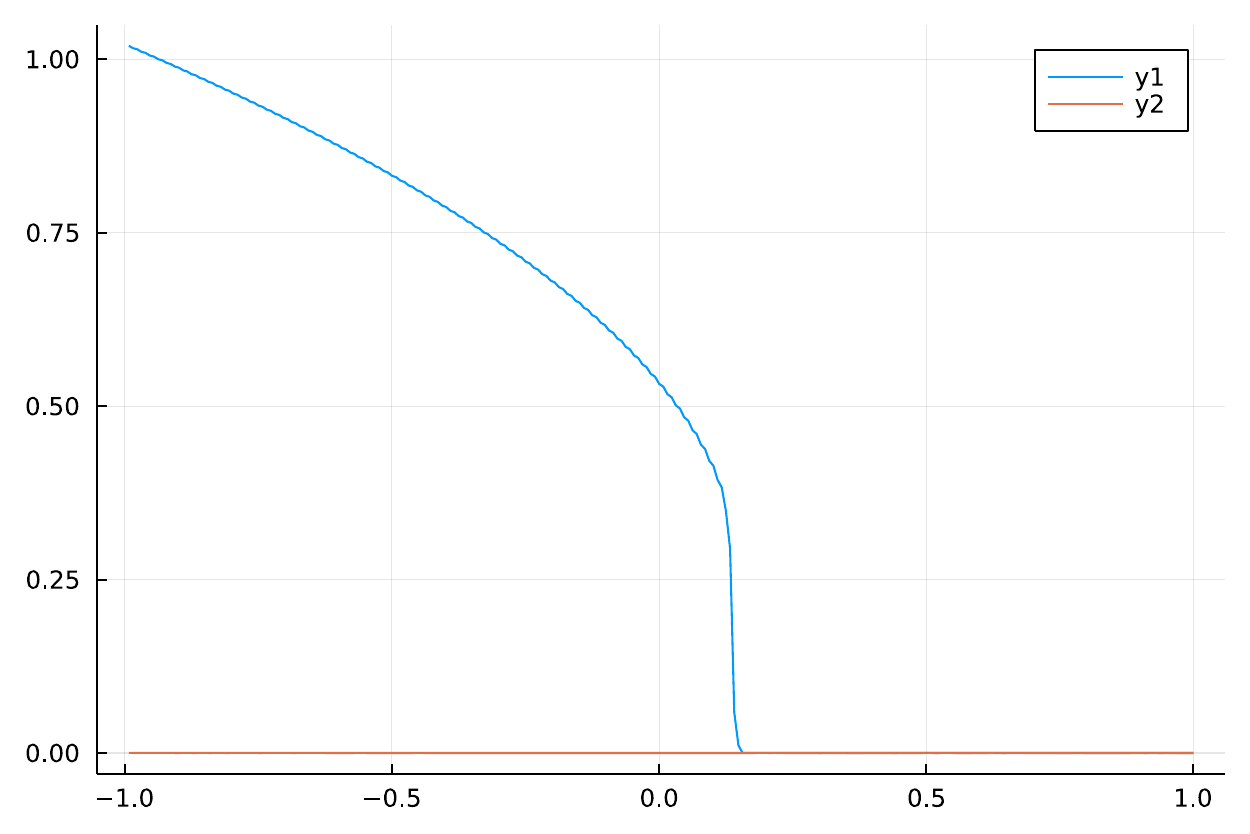} \\
$\Z_1$ \ \ \ \ $\Z_2$ \ \ \ \ \  (c)\ \ \ \ \  $\Z_3$ \hskip 0.8in $\Z_2$ \ \ \
(d) \ \ \ \ \   $S_3$ \\
\caption{Mean-field phase diagram of \eqref{LG} in $ (u,v) $ space, for
$ \al=\bt=\ga=1$. (a): the horizontal axis is $u\in [-1,1]$  and  the vertical
axis is $v\in [-1,1]$.  The red channel corresponds to the value of the $ S_3 $
order parameter $\sqrt{|\Phi_1|^2 + |\Phi_2|^2}$.  The green channel
corresponds to the value of the $ \Z_2 $ order parameter $\sqrt{|\phi|}$.  The
black region corresponds to a $ S_3 $ symmetric phase, the red region to a $
\Z_2 $ symmetric phase, the green region to a $ \Z_3 $ symmetric phase,  and
the yellow region to a $ \Z_1 $ phase.  (b),(c),(d): Plots of $\sqrt{|\Phi_1|^2
+
|\Phi_2|^2}$ (y1) and $|\phi|$ (y2), for  $u\in [-1,1]$, and (b) $v=-0.8$, (c)
$v=-0.1$, (d) $v=0.5$.  } \label{phaseMF_10_10_10_-10_10_-10_10ord}
\end{figure}

From the mean-field phase diagrams, we see that all four symmetry breaking
phases, $S_3$-phase, $\Z_3$-phase, $\Z_1$-phase, and $\Z_1$-phase are realized.
Let us consider the possible phase transitions between the various phases.
Landau theory tells us that we should expect continuous phase transitions
between pairs of groups that have a group-subgroup relation. The proper
subgroups of $ S_3 $ are three $ \Z_2 $ subgroups, a $ \Z_3 $ subgroup and the
trivial subgroup $ \Z_1 $. There are five distinct group-subgroup pairs that
one can find among these groups: \[ \Z_1\subset S_3, \ \ \Z_1\subset \Z_2,\ \
\Z_1\subset \Z_3,\ \ \Z_2\subset S_3,\ \ \Z_3\subset S_3 \] There are two
questions one can immediately ask: 
\begin{enumerate}
\item Are these transitions all distinct? 
\item Are these transitions all stably continuous?
\end{enumerate}

\begin{figure}[t]
\centering
\includegraphics[height=1.5in]{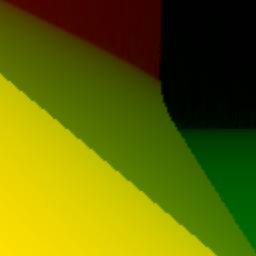} \hfill
\includegraphics[height=1.1in]{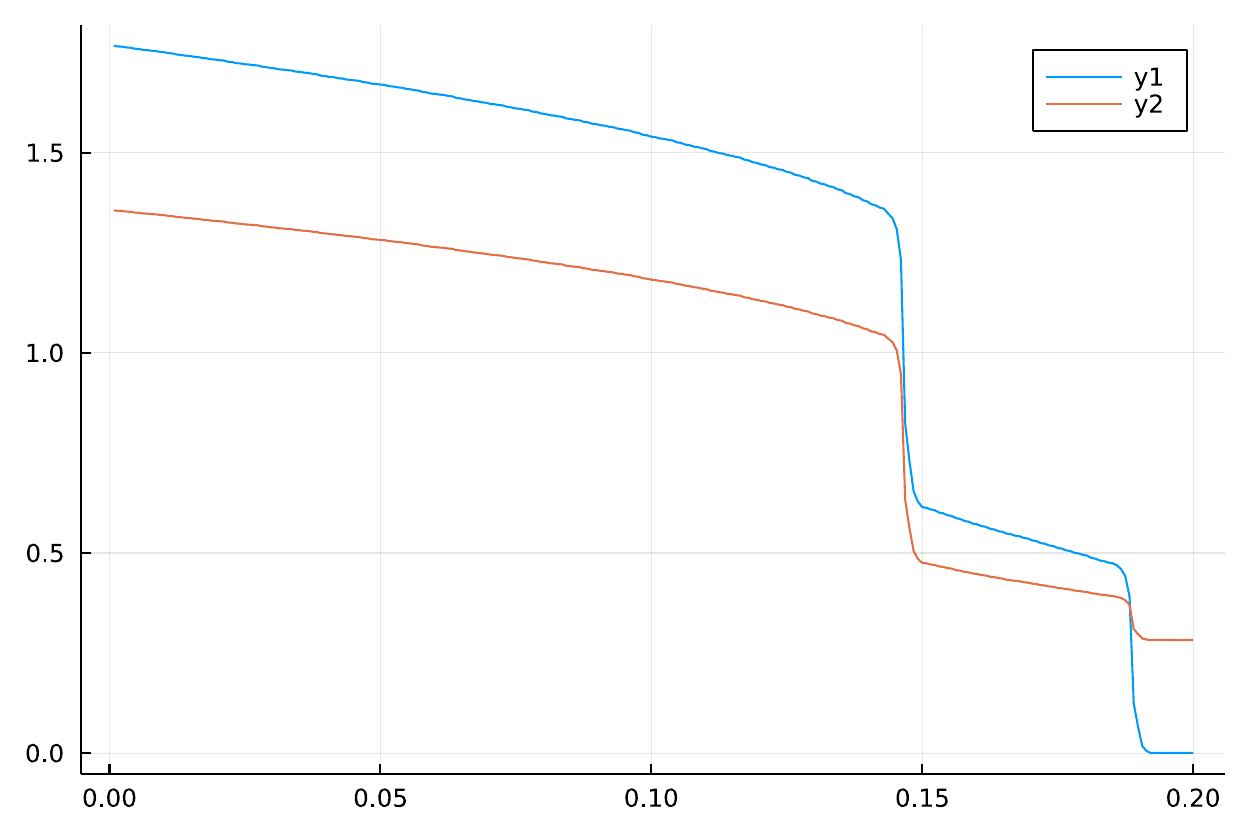} \\
\ \ \ \ \ \ \ \ \ \ \ \ \ \ \ \ \
(a) \hskip 1.4in  (b) \ \ \ \ \ \ \ \ \ $\Z_1$ \ \ \ $\Z_3$\\
\includegraphics[height=1.1in]{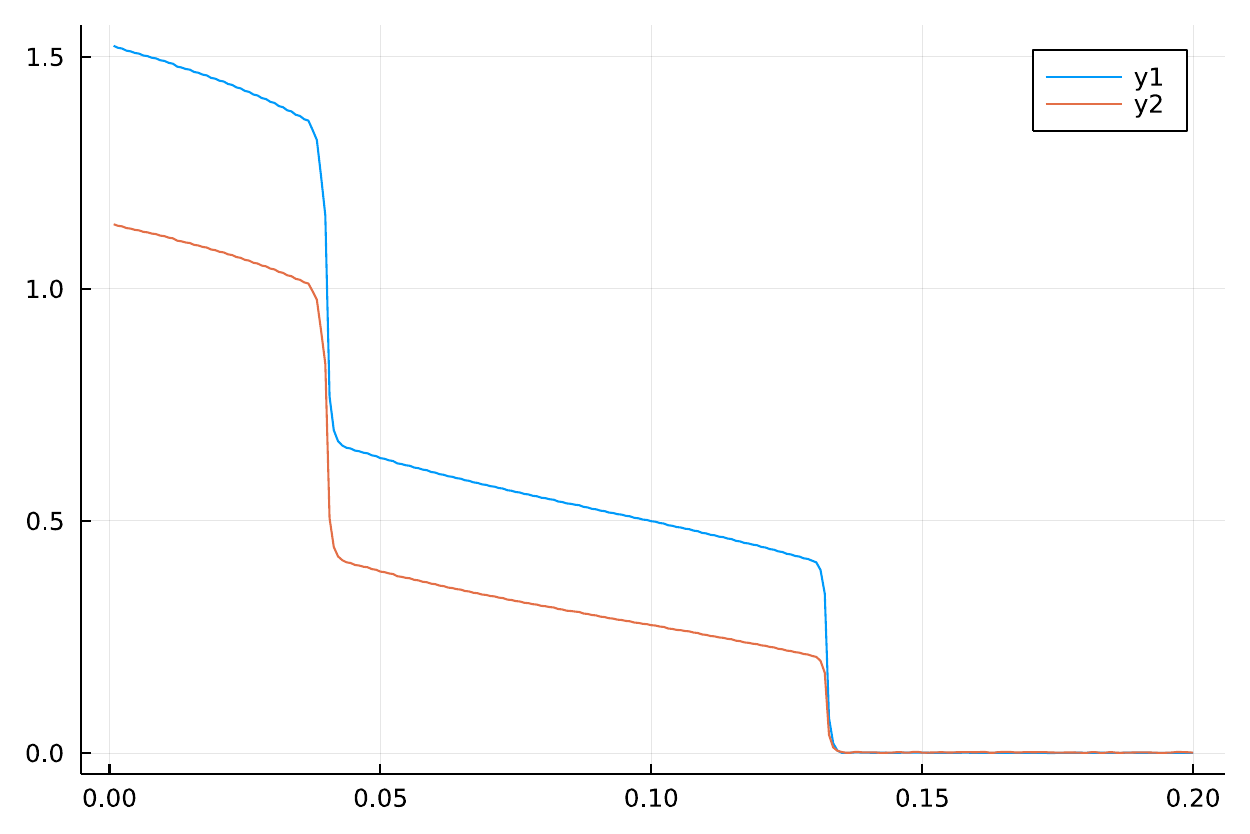} \hfill
\includegraphics[height=1.1in]{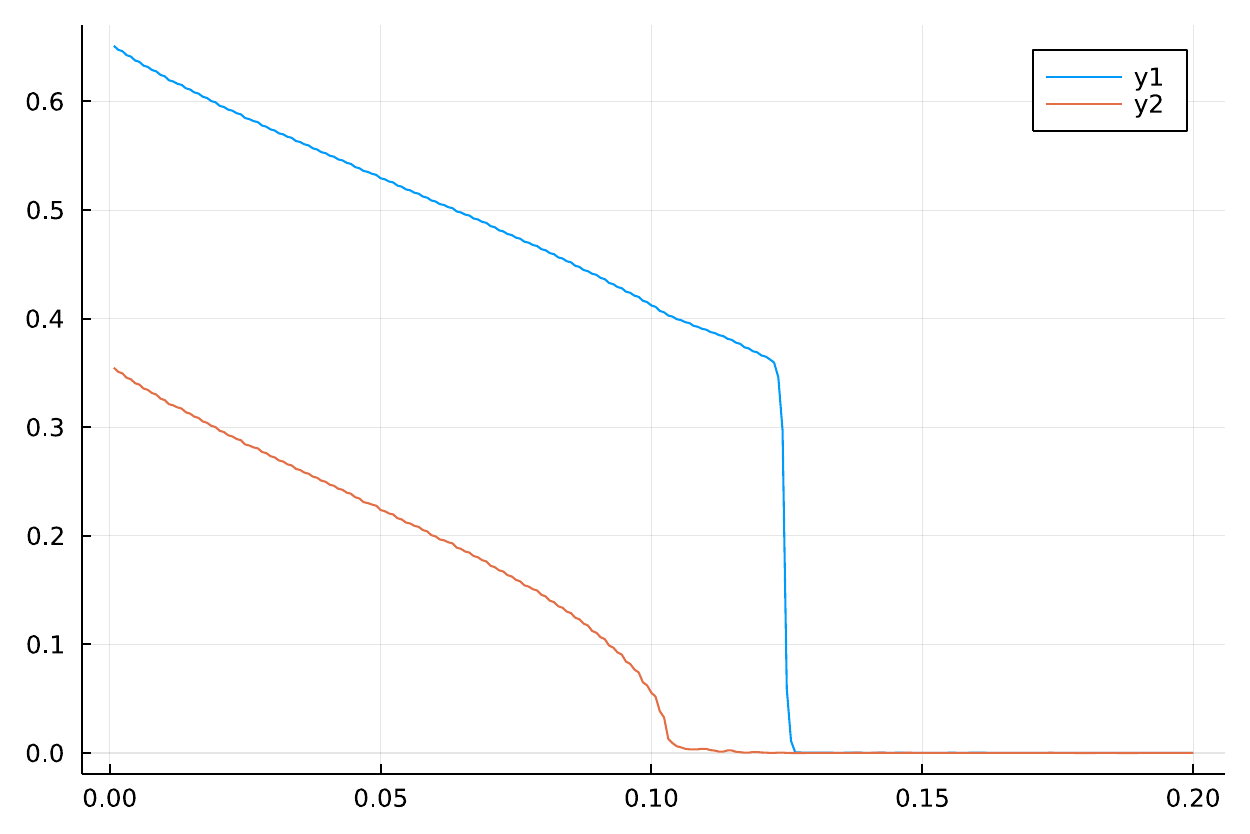} \\
\ \ $\Z_1$\ \ \ \ \ \ \ (c)\ \ \ \ \ \ \ \ \  $S_3$\hskip 0.6in  $\Z_1$\ \ \ \ \
\ (d) $\Z_2$\ \ \ \ \ $S_3$\\
\caption{Mean-field phase diagram of \eqref{LG} in $ (u,v) $ space, for
$ \al=\bt=1, \ga=-0.5$. (a): the horizontal axis is $u\in [0,0.2]$  and  the
vertical axis is $v\in [-0.2,0.2]$.  The red channel corresponds to the value
of the $ S_3 $ order parameter $\sqrt{|\Phi_1|^2 + |\Phi_2|^2}$.  The green
channel corresponds to the value of the $ \Z_2 $ order parameter
$\sqrt{|\phi|}$.  The black region corresponds to a $ S_3 $ symmetric phase,
the red region to a $ \Z_2 $ symmetric phase, the green region to a $ \Z_3 $
symmetric phase'  and the yellow region to a $ \Z_1 $ phase.  (b),(c),(d): Plots
of $\sqrt{|\Phi_1|^2 + |\Phi_2|^2}$ (y1) and $|\phi|$ (y2), for  $u\in
[0,0.2]$, and (b) $v=-0.16$, (c) $v=0.02$, (d) $v=0.1$.  }
\label{phaseMF_10_10_-5_0_2_-2_2ord}
\end{figure}

From Landau's theory of phase transitions, one expects that a symmetry-breaking
phase transition should depend only on the pair of symmetry groups across this
transition. Moreover, we also expect from this point of view that any pair of
groups related by a group-subgroup relation should have a corresponding
continuous transition between gapped states that have those symmetries. By the
same token, a pair of gapped phases that are symmetric under groups not related
by a group-subgroup relation are generically expected to have a first-order
discontinuous transition; sometimes a continuous transition that is
multicritical (\ie fine-tuned continuous) may also be allowed.

Thus we expect the two transitions $S_3\leftrightarrow \Z_3$ and
$\Z_2\leftrightarrow \Z_1$ to be stably continuous and identical, since both
transitions break a $\Z_2$ symmetry and is controlled by the change of a $\Z_2$
order parameter.  However, although the two Ginzburg-Landau theories describing
the two transitions are controlled by the same $\Z_2$ order parameter, the
Ginzburg-Landau theory for the transition $S_3\leftrightarrow \Z_3$ has a
$\Z_3$ symmetry, while the Ginzburg-Landau theory for the transition
$\Z_2\leftrightarrow \Z_1$ does not have any additional symmetry.  Since the
$\Z_3$ symmetry has trivial actions on the $\Z_2$ order parameter, the two
Ginzburg-Landau theories are actually identical.  Therefore Ginzburg-Landau
theory predicts that the two transitions $S_3\leftrightarrow \Z_3$ and
$\Z_2\leftrightarrow \Z_1$ are indeed described by the same CFT.  This result
is confirmed by numerical calculations and the \symmTO\ approach, presented in
the next few subsections.

We might also expect the two transitions $S_3\leftrightarrow \Z_3$ and
$\Z_2\leftrightarrow \Z_1$ to be stably continuous and identical, since both
transitions break a $\Z_3$ symmetry and is controlled by the change of a $\Z_3$
order parameter.  However, the Ginzburg-Landau theory for the transition
$S_3\leftrightarrow \Z_2$ has a $\Z_2$ symmetry, while the Ginzburg-Landau
theory for the transition $\Z_3\leftrightarrow \Z_1$ does not have this
symmetry.  Also the $\Z_2$ symmetry acts nontrivially on the $\Z_3$ order
parameter.  Thus the Ginzburg-Landau theories for the two transitions are not
really the same.  Also, due to the cubic term, the two transitions  must be
first order at mean-field level.  Later, we will see that the fluctuations turn
the  two first order transitions into stable continuous transitions.  The CFT's
for the two transitions are different.  However, the two CFT's are both
constructed from the $(6,5)$ minimal model.

Due to the group-subgroup relation between $S_3$ and $\Z_1$, we might expect
the transition $S_3\leftrightarrow \Z_1$ to be stably continuous, which is
described by a CFT with one relevant direction.  But using the \symmTO\
approach, we only find three gapless states with one relevant direction and
with small central charge less than $(c,\bar c)=(1,1)$, for $S_3$ symmetric
systems.  The first gapless state describes the transitions $S_3\leftrightarrow
\Z_3$ and $\Z_2\leftrightarrow \Z_1$. The second gapless state describes the
transition $\Z_3\leftrightarrow \Z_1$ and the third one describes the
transition $S_3\leftrightarrow \Z_2$.  So which gapless state describes the
transition $S_3\leftrightarrow \Z_1$?  May be the stable continuous transition
$S_3\leftrightarrow \Z_1$ is described by a CFT with central charge larger than
$(c,\bar c)=(1,1)$, or may be the stable continuous transition
$S_3\leftrightarrow \Z_1$ does not exist.  In the next section, we perform some
numerical calculations to study this issue.

Lastly, we expect the stable transition $\Z_3\leftrightarrow \Z_2$ to be first
order.  We know that the transitions $\Z_3\leftrightarrow \Z_1$ and
$\Z_1\leftrightarrow \Z_2$ can be stably continuous.  Can we fine tune a
parameter to make the two transitions to coincide and obtain a direct
continuous transition $\Z_3\leftrightarrow \Z_2$?  If $\Z_2$ and $\Z_3$ were
independent (\ie if the total symmetry were $\Z_2\times \Z_3$), then the answer
is yes.  But for total symmetry $S_3$, $\Z_2$ and $\Z_3$ are not independent
since $ S_3= \Z_3 \rtimes \Z_2$, so we are not sure.  In next subsection, we
find that the transition $\Z_3\leftrightarrow \Z_2$ can indeed be continuous
and multicritical.  In fact, the same multicritical point describes both the
transitions $S_3\leftrightarrow \Z_1$ and $\Z_3\leftrightarrow \Z_2$. 

\subsection{Numerical result from tensor network calculations}

The 3-state Potts model is a well studied statistical model with $S_3$ symmetry.
However,
this model only realizes two phases, the $S_3$-symmetric phase and the
$\Z_2$-symmetric phase.  Here we construct an $S_3$-symmetric statistical
model on a square lattice that can realize all four phases: $S_3$-phase,
$\Z_3$-phase, $\Z_2$-phase, and $\Z_1$-phase.

The first model has degrees of freedom $(\th_{\v i}, s_{\v i})$, $\th_{\v
i}=0,1,2$, $s_{\v i}=0,1$, on site $\v i$.  The energy is given by
\begin{align}
\label{Pottsmodel}
E &= - \sum_{\v i} J_1( 
\del_{\th_{\v i},\th_{\v i+\v x}}+ \del_{\th_{\v i},\th_{\v i+\v y}} 
)
+ J_2( 
\del_{s_{\v i},s_{\v i+\v x}}+ \del_{s_{\v i},s_{\v i+\v y}} 
) 
\nonumber\\
&\ \ \ \
+ \frac13 J_c \sum_{\v i = \text{even}} 
\Big[
\text{sgn}(\th_{\v i}-\th_{\v i+\v x})
+\text{sgn}(\th_{\v i+\v x}-\th_{\v i+\v x+\v y})
\nonumber\\
&\ \ \ \ \ \ \ \
+\text{sgn}(\th_{\v i+\v x+\v y}-\th_{\v i+\v y})
+\text{sgn}(\th_{\v i+\v y}-\th_{\v i})
\Big] 
\nonumber\\
&\ \ \ \ \ \ \ \
(
s_{\v i}
+s_{\v i+\v x}
+s_{\v i+\v x +\v y}
+s_{\v i +\v y}
-2)
\end{align}
where $\text{sgn}(\th)\equiv \text{mod}(\th+1,3)-1$.  The $J_1$ and $J_2$ terms
give rise to $q=3$ Potts model and Ising model. If we view $\th_{\v i}$ as a
planer vector that can points to three directions separated by 120$^\circ$
degree, then the term $ \text{sgn}(\th_{\v i}-\th_{\v i+\v x})
+\text{sgn}(\th_{\v i+\v x}-\th_{\v i+\v x+\v y}) +\text{sgn}(\th_{\v i+\v x+\v
y}-\th_{\v i+\v y}) +\text{sgn}(\th_{\v i+\v y}-\th_{\v i}) $ has a meaning
chirality: it measures whether the vectors turn clock-wise or anti-clock-wise as
we go round a square.  The coupling of the chirality with the Ising order
parameter $s_{\v i}$ breaks the $S_3\times \Z_2$ symmetry to $S_3$ symmetry.

\begin{figure}[t]
\begin{center}
\includegraphics[scale=0.45]{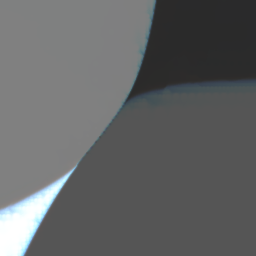}\ \ \ \
\includegraphics[scale=0.45]{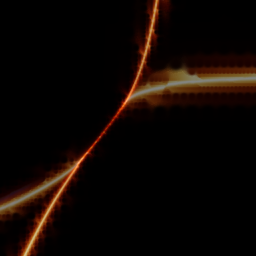}
\\[-1.4in]
\hskip 1.8in \white{$\Z_3$} \hskip 0.8in \white{$\Z_1$}\\[1.0in]
\hskip 1.35in \white{$S_3$} \hskip 0.8in \white{$\Z_2$}
\end{center} 
\vskip -2mm
\begin{center}
(a) 
\ \ \ \ \ \ \ \ \ \ \ \ \ \ \ \ \ \
\ \ \ \ \ \ \ \ \ \ \ \ \ \ \ \ \ \
(b)
\end{center}
\caption{ A phase diagram for the first model \eq{Pottsmodel} with $1.2 <
J_1\bt < 1.4$ (horizontal axis), $0.6 < J_2\bt < 1.0$ (vertical axis), and
$J_c\bt =1.5$.  (a) A plot of $1/GSD$, where $GSD\equiv Z^2(L, L)/Z(L, 2L)$ and
$Z(L_1, L_2)$ is the partition function for system of size $L_1\times L_2$.
(b) A plot of central charge $c$.  $Z(L, L_\infty)$ has a form $\ee^{-
L_\infty[ \eps L -\frac{2\pi c }{24 L} +o(L^{-1})]}$ when $L_\infty \gg L$,
where $c$ is the central charge.  The red-channel of the colored image is for
system of size $64\times 64$, green-channel for $128\times 128$, and
blue-channel for $256\times 256$.  } \label{Potts15_12-14_06-10} 
\end{figure}

\begin{figure}[t]
\begin{center}
\includegraphics[scale=0.225]{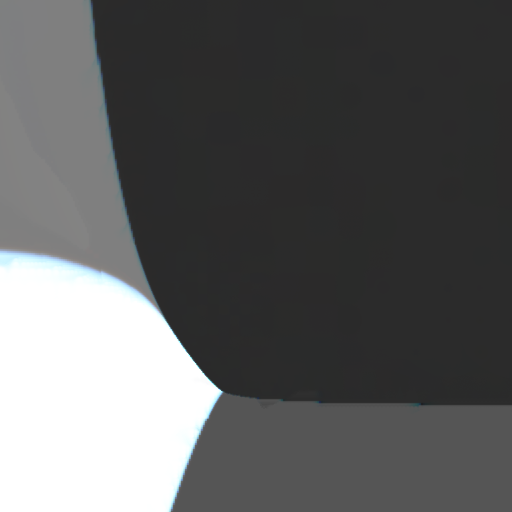}\ \ \ \
\includegraphics[scale=0.225]{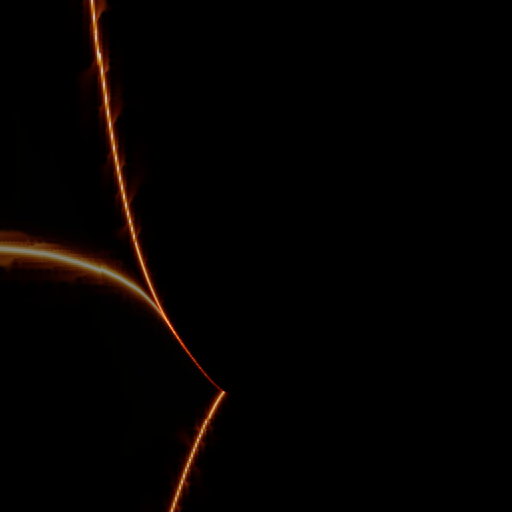}
\\[-1.4in]
\hskip 1.5in \white{$\Z_3$} \hskip 0.8in \white{$\Z_1$}\\[1.0in]
\hskip 1.5in \white{$S_3$} \hskip 0.8in \white{$\Z_2$}
\end{center}
\begin{center}
(a) 
\ \ \ \ \ \ \ \ \ \ \ \ \ \ \ \ \ \
\ \ \ \ \ \ \ \ \ \ \ \ \ \ \ \ \ \
(b)
\end{center}
\caption{ A phase diagram for the second model \eq{S3model} with $0.57 < J_1\bt
< 1.29$ (horizontal axis), $0.4775 < J_2\bt < 1.2275$ (vertical axis), and
$J\bt =1-J_2\bt $, $J_c\bt =1$.  (a) A plot of $1/GSD$.  (b) A plot of central
charge $c$.  The red-channel of the colored image is for system of size
$64\times 64$, green-channel for $128\times 128$, and blue-channel for
$256\times 256$.  } \label{S3_10-10_6-13_5-12} 
\end{figure}

\begin{figure}[t]
\begin{center}
\includegraphics[scale=0.45]{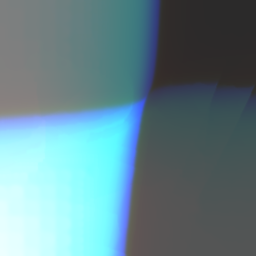}\ \ \ \
\includegraphics[scale=0.45]{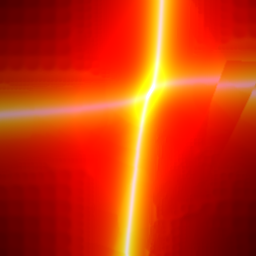}
\\[-1.5in]
\hskip 1.8in \white{$\Z_3$} \hskip 0.8in \white{$\Z_1$}\\[0.9in]
\hskip 1.8in \white{$S_3$} \hskip 0.8in \white{$\Z_2$}\\[0.2in]
\end{center}
\begin{center}
(a) 
\ \ \ \ \ \ \ \ \ \ \ \ \ \ \ \ \ \
\ \ \ \ \ \ \ \ \ \ \ \ \ \ \ \ \ \
(b)
\end{center}
\caption{ A phase diagram for the first model \eq{Pottsmodel} with $1.02125 <
J_1\bt < 1.04125$ (horizontal axis), $0.851875 < J_2\bt < 0.891875$ (vertical
axis), and $J_c\bt =0.5$.  (a) A plot of $1/GSD$.  (b) A plot of central charge
$c$.  The red-channel of the colored image is for system of size $64\times
64$, green-channel for $256\times 256$, and blue-channel for $1024\times 1024$.
} \label{Potts_5_10-10_9-9} 
\end{figure}

The second $S_3$ symmetric statistical model has degree of freedoms $(\th_{\v
i}, \phi_{\v i}, s_{\v i})$, $\th_{\v i}=0,1,2$, $\phi_{\v i}=0,1,2$, $s_{\v
i}=0,1$, on site $\v i$.  The energy is given by
\begin{align}
\label{S3model}
E &= - \sum_{\v i} J_1( 
\del_{\th_{\v i},\th_{\v i+\v x}}+ \del_{\th_{\v i},\th_{\v i+\v y}} 
+\del_{\phi_{\v i},\phi_{\v i+\v x}}+ \del_{\phi_{\v i},\phi_{\v i+\v y}} 
)
\nonumber\\
&\ \ \ \
- \sum_{\v i} J_2( 
\del_{s_{\v i},s_{\v i+\v x}}+ \del_{s_{\v i},s_{\v i+\v y}} 
) + J  \del_{\th_{\v i},\phi_{\v i}}
\nonumber\\
&\ \ \ \
+ J_c \sum_{\v i}  \text{sgn}(\th_{\v i}-\phi_{\v i}) (s_{\v i}-\frac12)
\end{align}
The $J_c$ term is also a coupling of the chirality with the Ising order
parameter $s_{\v i}$.

Again, we use a particular version of the tensor network approach in
\Rf{GW0931} to study the above two statistical models.  We obtain the phase
diagrams Fig.  \ref{Potts15_12-14_06-10}, Fig.  \ref{S3_10-10_6-13_5-12}, and
Fig.  \ref{Potts_5_10-10_9-9}.  All three phase diagrams contain all the four
phases: $S_3$-phase in lower left, $\Z_3$-phase in upper left, $\Z_2$-phase in
lower right, and $\Z_1$-phase in upper right, 

In phase diagram Fig.  \ref{Potts15_12-14_06-10}, we see five stable direct
transitions $S_3 \leftrightarrow \Z_2$, $S_3 \leftrightarrow \Z_3$, $\Z_3
\leftrightarrow \Z_2$, $\Z_3 \leftrightarrow \Z_1$, and $\Z_2 \leftrightarrow
\Z_1$. We also computed the central charge $\frac{c+\bar c}{2}$, along the
transition lines.  The non-zero central charges suggest that all the five
transitions 
are  stable continuous transitions.  

In phase diagram Fig.  \ref{S3_10-10_6-13_5-12}, we see five stable direct
transitions $S_3 \leftrightarrow \Z_2$, $S_3 \leftrightarrow \Z_3$, $S_3
\leftrightarrow \Z_1$, $\Z_3 \leftrightarrow \Z_1$, $\Z_2 \leftrightarrow
\Z_1$ .  From the computed central charge, we find that the transitions $S_3
\leftrightarrow \Z_2$, $\Z_3 \leftrightarrow \Z_1$, and $S_3 \leftrightarrow
\Z_1$ are stably continuous.  The transitions $\Z_2 \leftrightarrow \Z_1$ is
first order since the central charge $c=\bar c =0$ along the transition line.

In phase diagram Fig.  \ref{Potts_5_10-10_9-9}, we reduce $J_c$ from $J_c=1.5$
in Fig. \ref{Potts15_12-14_06-10} to  $J_c=0.5$.  We see an evidence of a
multicritical point connecting the four phases $S_3$, $\Z_3$, $\Z_2$, and
$\Z_1$. More systematic and detailed studies are needed.

Among these stable continuous transitions, the  $\Z_2\leftrightarrow \Z_1$ and
$S_3\leftrightarrow \Z_3$ critical points are the well known Ising critical
point, which is described by a conformal field theory (CFT) constructed from
(4,3) minimal model.  $S_3\leftrightarrow \Z_2$ critical point is the well
known critical point of $q=3$ Potts model, which is described by a CFT
constructed from (6,5) minimal model.  But what are the  $\Z_3 \leftrightarrow
\Z_2$, $S_3\leftrightarrow \Z_2$, and  $\Z_3\leftrightarrow \Z_1$ critical
points?  What is the
$(S_3, \Z_2, \Z_1,\Z_3)$ multicritical point?
In next section, we will use \symmTO\ to understand the above global phase
diagrams of $S_3$ symmetric systems and the critical points.  In particular, we
will show a duality relation between $S_3\leftrightarrow \Z_1$ transition and
$\Z_3\leftrightarrow \Z_2$ transition.  For example, the existence of a stable
continuous  transition $S_3\leftrightarrow \Z_1$ implies the existence of
stable continuous  transition $\Z_3\leftrightarrow \Z_2$. The two stable
continuous  transitions, if they exist, are described by the same CFT.

\subsection{A \symmTO\ approach for gapped and gapless phases}

\begin{table*}[tb]
\caption{The point-like excitations and their fusion rules in 2+1D $\eGau_{S_3}$
topological order (\ie $S_3$ gauge theory with charge excitations).  The $S_3$
group are generated by $(1,2)$ and $(1,2,3)$.  Here $\bm 1$ is the trivial
excitation.  $a_1$ and $a_2$ are pure $S_3$ charge excitations, where $a_1$
corresponds to the 1-dimensional representation, and $a_2$ the 2-dimensional
representation of $S_3$.  $b$ and $c$ are pure $S_3$ flux excitations, where
$b$ corresponds to the conjugacy class $\{(1,2,3),(1,3,2)\}$, and $c$ conjugacy
class $\{(1,2),(2,3),(1,3)\}$.  $b_1$, $b_2$, and $c_1$ are charge-flux bound
states.  $d,s$ are the quantum dimension and the topological spin of an
excitation.  } \label{S3FusionRules} 
\setlength\extrarowheight{4pt}
\setlength{\tabcolsep}{6pt}
\centering
\begin{tabular}{c | c|c|c|c|c|c|c|c}
\Xhline{4\arrayrulewidth}
$d,s$ & $1,0$ & $1,0$ & $2,0$ & $2,0$ & $2,\frac13$ & $2,-\frac13$ & $3,0$ &
$3,\frac12$\\
\hline
$\otimes$ & $\bm 1$ &  $a_1$  & $a_2$ &  $b$  &  $b_1$  &  $b_2$  & $c$  & $c_1$
\\
\Xhline{2.5\arrayrulewidth}
$ \bm 1 $ & $\bm 1$ & $a_1$ & $a_2$   & $b$  & $b_1$  &  $b_2$          & $c$  &
$c_1$    \\
$a_1$ & $a_1$ & $\bm 1$ & $a_2$  & $b$  &$b_1$  & $b_2$      &  $c_1$  & $c$ 
\\
$a_2$ & $a_2$ & $a_2$   & $\bm 1\oplus a_1\oplus a_2$      & $b_1\oplus b_2$   
& $b\oplus b_2$ & $b\oplus b_1$      & $c\oplus c_1$  & $c\oplus c_1$\\
$b$  & $b$  & $b$ & $b_1\oplus b_2$     & $\bm 1\oplus a_1\oplus b$ & $b_2\oplus
a_2$  & $b_1\oplus a_2$     & $c\oplus c_1$   & $c\oplus c_1$  \\
$b_1$  & $b_1$   & $b_1$ & $b\oplus b_2$      & $b_2\oplus a_2$ & $\bm 1\oplus
a_1\oplus b_1$  & $b\oplus a_2$      & $c\oplus c_1$ & $c\oplus c_1$ \\
$b_2$  & $b_2$  & $b_2$  & $b\oplus b_1$    & $b_1\oplus a_2$ & $b\oplus a_2$  &
$\bm 1\oplus a_1\oplus b_2$      & $c\oplus c_1$  & $c\oplus c_1$ \\
$c$ & $c$ & $c_1$ & $c\oplus c_1$     & $c\oplus c_1$   & $c\oplus c_1$  &
$c\oplus c_1$     & $\bm 1\oplus a_2\oplus b\oplus b_1\oplus b_2$ & $a_1 \oplus
a_2\oplus b\oplus b_1\oplus b_2$  \\
$c_1$ & $c_1$ & $c$  & $c\oplus c_1$     & $c\oplus c_1$ & $c\oplus c_1$  &
$c\oplus c_1$     & $a_1 \oplus a_2\oplus b\oplus b_1\oplus b_2$  & $\bm 1\oplus
a_2\oplus b\oplus b_1\oplus b_2$ \\
\Xhline{4\arrayrulewidth}
\end{tabular}
\end{table*}

\begin{figure}[t]
\begin{center}
\includegraphics[scale=0.6]{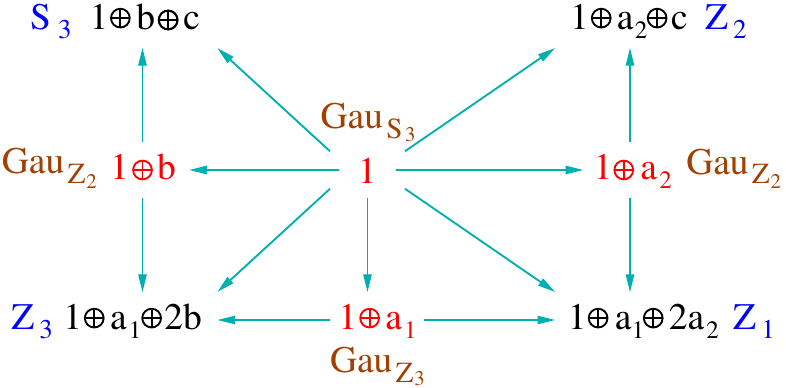}
\end{center}
\caption{1+1D $S_3$ symmetry is described by  \symmTO\ 
$\eGau_{S_3} $ ($S_3$ gauge theory with
charge excitations). The $\eGau_{S_3}$ topological order has eight condensable
algebras, and four of them are Lagrangian (the black ones above). They give rise
to
gapped phases as the corresponding gapped boundary states. We also indicate the
unbroken symmetry groups of these phases.  Four of eight are not Lagrangian, and
give rise to gapless states.  We also indicate the condensation-induced
topological orders whose canonical boundaries give rise to these gapless states.
The gapless states are critical points for the transitions between gapped or
gapless states, as indicated above.  The arrows indicate the embedding maps of
condensable algebra and the directions of more condensation.  All critical
points has only one symmetric local relevant operator and are stable critical
points.} \label{S3phases} 
\end{figure}

1+1D $S_3$ symmetry is described a \symmTO\ (\ie a 2+1D topological
order) whose topological excitations are described by $S_3$ quantum double
$\eGau_{S_3}$ (\ie $S_3$ gauge theory with both charge and flux excitations, as
described in Table \ref{S3FusionRules}).  From Appendix \ref{MMA}, we find that
a condensable algebra $\cA=\bigoplus_{a\in \eM} A^a a$ in $\eM$ must satisfies
\begin{align}
\label{Acond}
A^\one &= 1,\ \ \ A^a \in \N,\ \ \ A^a = A^{\bar a},
\nonumber\\
s_a &= 0 \ \text{ for } a \in \cA, 
\nonumber\\
\frac{\sum_{b\in \eM} S_{\eM}^{ab} A^b}{ \sum_{b\in \eM} 
S_{\eM}^{\one b}  A^b}
&= 
\text{cyclotomic integer for all } a\in \eM  
\nonumber\\
A^a &\leq d_a -\del(d_a),
\nonumber\\
A^a A^b & \leq \sum_c N^{ab}_{\eM,c} A^c -\del_{a,\bar b} \del(d_a),
\nonumber\\
A^a
&=
\sum_b S_{\eM}^{ab} A^b \ \text{ if } \cA  \text{ is Lagrangian}.
\end{align}
Solving the above conditions for $\eM=\eGau_{S_3}$, we find the following
potential condensable algebras
\begin{align}
& \one\oplus  b\oplus c, &
& \one\oplus  a_2\oplus c,
&
& \one\oplus  a_1\oplus  2b, &
& \one\oplus  a_1\oplus  2a_2 .
\nonumber\\
& \one\oplus  b, &
& \one\oplus  a_2, 
&
& \one\oplus  a_1, &
& \one. 
\end{align}
Since \eqn{Acond} are only necessary conditions, some of the above $\cA$'s may
not be valid. For $\eGau_{S_3}$, using the following physical considerations,
we argue that the above $\cA$'s are all valid and describe the actual condensation patterns in physical systems (see Fig. \ref{S3phases}).

We know that 1+1D $S_3$ symmetric systems can have four gapped phases with
unbroken symmetry group $S_3$, $\mathbb{Z}_3$, $\mathbb{Z}_2$, and
$\mathbb{Z}_1$, and they correspond to the four Lagrangian condensable algebras
that we find above:
\begin{align}
\label{cAphase}
&\one\oplus  b\oplus c \to S_3\text{-phase},\ \ \ \ \  \one\oplus  a_2\oplus c
\to \Z_2\text{-phase},
\\
& \one\oplus  a_1\oplus  2b \to \Z_3\text{-phase}, \ \ \one\oplus  a_1\oplus 
2a_2 \to \Z_1\text{-phase}.
\nonumber 
\end{align}
To understand the above result, we note that 
a condensation of real field $\phi^{a_1}$ carrying the 1-dimensional
representation of $S_3$ breaks $S_3$ symmetry down to $\Z_3$ symmetry.  Thus
the condensation of the corresponding Lagrangian condensable algebra $\one\oplus
a_1\oplus  2b $ induces a gapped symmetry breaking phase where the unbroken
symmetry is $\Z_3 \subset S_3$.  A condensation of complex two-component
bosonic field $\Phi^{a_2}_\al$, $\al=1,2$, carrying the 2-dimensional
representation of $S_3$ and satisfying $\Phi^{a_2}_1 = \Phi^{a_2}_2$ breaks
$S_3$ symmetry down to $\Z_2$ symmetry.  Thus the condensation of Lagrangian
condensable algebra $\one\oplus a_c\oplus  c $ induces a gapped symmetry
breaking phase where the unbroken symmetry is $\Z_2 \subset S_3$.  Similarly,
the condensation of $\phi^{a_1}$ and $\Phi^{a_2}$ breaks $S_3$ symmetry down to
$\Z_1$ symmetry.  So the condensation of Lagrangian condensable algebra
$\one\oplus  a_1\oplus 2a_2 $ induces a gapped symmetry breaking phase where
the unbroken symmetry is $\Z_1 \subset S_3$. 

Since $\one\oplus  b\oplus c$, $\one\oplus  a_2\oplus c$, $\one\oplus
a_1\oplus  2b$, $\one\oplus  a_1\oplus  2a_2$ are Lagrangian \ie $ D_{\eM}/
d_\cA$
= $D_{\eM_{/\cA}}$ = 1, the corresponding condensation-induced topological
orders, $(\eGau_{S_3})_{/\one\oplus  b\oplus c}$, $(\eGau_{S_3})_{/\one\oplus
a_2\oplus c}$, $(\eGau_{S_3})_{/\one\oplus  a_1\oplus  2b}$,
$(\eGau_{S_3})_{/\one\oplus  a_1\oplus  2a_2}$, are all trivial.  On the other
hand, $\one\oplus  a_2$, $\one\oplus  b$, $\one\oplus  a_1$, $\one$ are not
Lagrangian, and their condensation-induced topological orders are not trivial. 
We
have
\begin{align}
\label{S3toZs}
(\eGau_{S_3})_{/\one\oplus  a_2} &= \eGau_{\Z_2}, &
(\eGau_{S_3})_{/\one\oplus  b} &= \eGau_{\Z_2},
\nonumber\\
(\eGau_{S_3})_{/\one\oplus  a_1} &= \eGau_{\Z_3}, &
(\eGau_{S_3})_{/\one} &= \eGau_{S_3} .
\end{align}

The above results can be understood using the usual Anderson-Higgs condensation
in $S_3$ gauge theory.  The $\one\oplus  a_2$-condensation correspond to the
condensation of the $\Phi^{a_2}$ field, which change the topological order
$\eGau_{S_3}$ described by $S_3$ gauge theory to a topological order
$\eGau_{\Z_2}$ described by $\Z_2$ gauge theory.  Similarly,  the $\one\oplus
a_1$-condensation correspond to the condensation of the $\phi^{a_1}$ field,
which change the topological order $\eGau_{S_3}$ described by $S_3$ gauge
theory to a topological order $\eGau_{\Z_3}$ described by $\Z_3$ gauge theory.
$(\eGau_{S_3})_{/\one} = \eGau_{S_3}$ is obvious since $\one$ is trivial.  To
understand $ (\eGau_{S_3})_{/\one\oplus  b} = \eGau_{\Z_2}$, we note that
$\eGau_{S_3}$ is invariant under exchanging $a_2$ and $b$ \cite{BW10065479}. Such an automorphism
exchanges $\one\oplus  a_2$ and $\one\oplus  b$.  Thus $
(\eGau_{S_3})_{/\one\oplus  b} =  (\eGau_{S_3})_{/\one\oplus  a_2} =
\eGau_{\Z_2}$.  

We can also derive \eqn{S3toZs} using the boundary theory of topological order
summarized in Appendix \ref{MMA}.  For example, $(\eGau_{S_3})_{/\one\oplus a_2}
= \eGau_{\Z_2}$ implies that there is a canonical domain wall between
$\eGau_{S_3}$ and $\eGau_{\Z_2}$, that is described by a non-negative integer
matrix $(A^{ai})$.  $A^{ai}$ satisfy \eqn{STA}, \eqn{Aainzero}, \eqn{AMA1}, and
\eqn{AMA2}, as well as other properties listed in  Appendix \ref{MMA}. We find
that such $(A^{ai})$ does exist
\begin{align}
\left (A^{ai}_{\eGau_{S_3}\mid \eGau_{\Z_2}}\right )=
\begin{pmatrix}
1&  0&  0&  0 &\one\\
0&  0&  1&  0 & a_1\\
1&  0&  1&  0 & a_2\\
0&  0&  0&  0 & b\\
0&  0&  0&  0 & b_1\\
0&  0&  0&  0 & b_2\\
0&  1&  0&  0 & c\\
0&  0&  0&  1 & c_1\\
\one&  e&  m&  f & \\
\end{pmatrix}.
\end{align}
In the above $A^{ai}_{\eGau_{S_3}\mid \eGau_{\Z_2}}$, the non-zero entries in
the first row of $(A^{ai})$ indicate the condensation of corresponding anyons
in $\eGau_{\Z_2}$ (\ie these anyons become $\one$ at the domain wall).  Thus
the first row of $(A^{ai})$ gives rise to a condensable algebra in
$\eGau_{\Z_2}$: $\bigoplus_i A^{\one i}_{\eGau_{S_3}\mid \eGau_{\Z_2}} i=\one$,
which indicates that the domain wall is a $\one$-condensed boundary of
$\eGau_{\Z_2}$. This in turn indicates that $\eGau_{\Z_2}$ comes from
$\eGau_{S_3}$ via a condensation.

Similarly, the first column  of $(A^{ai})$ gives rise to a condensable algebra
in $\eGau_{S_3}$: $\bigoplus_a A^{a \one }_{\eGau_{S_3}\mid \eGau_{\Z_2}} a =
\one\oplus a_2$, which tells us that the domain wall is a $\one\oplus
a_2$-condensed boundary of $\eGau_{S_3}$, and the  $\one\oplus  a_2$
condensation changes $\eGau_{S_3}$ to $\eGau_{\Z_2}$.  This confirms our above
result from physical considerations.

$\eGau_{S_3}$ and $\eGau_{\Z_2}$ has another
canonical boundary described by
\begin{align}
\left (A^{ai}_{\eGau_{S_3}\mid \eGau_{\Z_2}}\right )=
\begin{pmatrix}
1&  0&  0&  0 &\one\\
0&  0&  1&  0 & a_1\\
0&  0&  0&  0 & a_2\\
1&  0&  1&  0 & b\\
0&  0&  0&  0 & b_1\\
0&  0&  0&  0 & b_2\\
0&  1&  0&  0 & c\\
0&  0&  0&  1 & c_1\\
\one&  e&  m&  f & \\
\end{pmatrix}.
\end{align}
which describes the condensation of $ \one\oplus  b$. Such a condensation
also changes $\eGau_{S_3}$ to $\eGau_{\Z_2}$.

Similarly, the canonical domain wall between $\eGau_{S_3}$ and $\eGau_{\Z_3}$
is given by
\begin{align}
\left (A^{ai}_{\eGau_{S_3}\mid \eGau_{\Z_3}}\right )=
\begin{pmatrix}
1&  0&  0&  0&  0&  0&  0&  0&  0 &\one\\
1&  0&  0&  0&  0&  0&  0&  0&  0 & a_1\\
0&  0&  0&  1&  1&  0&  0&  0&  0 & a_2\\
0&  1&  1&  0&  0&  0&  0&  0&  0 & b\\
0&  0&  0&  0&  0&  0&  0&  1&  1 & b_1\\
0&  0&  0&  0&  0&  1&  1&  0&  0 & b_2\\
0&  0&  0&  0&  0&  0&  0&  0&  0 & c\\
0&  0&  0&  0&  0&  0&  0&  0&  0 & c_1\\
\end{pmatrix}
\end{align}
From the first row of $ (A^{a i}_{\eGau_{S_3}\mid \eGau_{\Z_3}})$, we obtain
the corresponding condensable algebra in $\eGau_{\Z_3}$: $\bigoplus_i A^{\one
i}_{\eGau_{S_3}\mid \eGau_{\Z_3}} i=\one$, which suggests that $\eGau_{\Z_3}$
comes from $\eGau_{S_3}$ by a condensation.  From the first column of $ (A^{a
i}_{\eGau_{S_3}\mid \eGau_{\Z_3}})$, we obtain the corresponding condensable
algebra in $\eGau_{S_3}$: $\bigoplus_a A^{a \one }_{\eGau_{S_3}\mid
\eGau_{\Z_3}} a =\one\oplus a_1 $, which tells us that the
condensation is given by $\one\oplus  a_1$.  This again confirms our above
result.

Therefore, although we have four non-Lagrangian condensable algebras, we only
have three different condensation-induced topological orders, $\eGau_{\Z_2}$,
$\eGau_{\Z_3}$, $\eGau_{S_3}$, which correspond to three reduced \symmTOs\
and give rise to three types of gapless states.  But what are these gapless
states?

First, a reduced \symmTO\ allows many different gapless states. Here we want to
know which is the most stable gapless state with minimal number of relevant
operators.  If the reduced \symmTO\ is trivial,  the most stable state
with the trivial reduced \symmTO\ is gapped.  If the reduced \symmTO\ is
nontrivial, the state with minimal low energy excitations is still gapless.
What is this minimal gapless state?  How to calculate the  canonical boundary
of the topological order $\eM_{/\cA}$, which corresponds to the minimal gapless
state with reduced \symmTO\ $\eM_{/\cA}$?

To calculate the canonical boundary, we will use holoMB
\cite{LS190404833,JW190513279,JW191209391,JW210602069}  which is a
generalization of modular
bootstrap.\cite{H09022790,KO12094649,FK13076562,CY160806241,CY170505865,CY180204445,KP210713557,LS210108343}
Modular bootstrap looks for single-component modular invariant partition
functions.  HoloMB looks for multi-component boundary partition functions for a
boundary spacetime that has a form of torus.  The shape of the boundary
spacetime torus is described by a complex number $\tau$. A multi-component
boundary partition function $Z_\al(\tau)$ transform covariantly under the
modular transformation, according to the $S,T$ matrices that characterize the
bulk topological order (see \eqn{STZ}). The physical reason for such a
bulk-boundary connection is discussed in \Rf{JW190513279,LW191108470}.  Thus,
in contrast to the modular bootstrap, holoMB requires additional input data, the
$S,T$-matrices, to describe the \symmTO.

If $Z_\al(\tau)$ is independent of $\tau$, the corresponding boundary is
gapped. If $Z_\al(\tau)$ depend on $\tau$, the corresponding boundary is
gapless.  For gapless boundary, the multi-component partition function
$Z_\al(\tau)$ is formed by conformal characters of certain CFT.  So we look
for a CFT, whose conformal characters, after a suitable combination, can form
multi-component boundary partition function that transform under modular
transformation according to the bulk $S,T$ matrices.  The method of computing a
suitable combination is the same as computing gapped boundary (see Appendix
\ref{MMA}) of some properly constructed topological order.  For details see
\Rf{JW190513279}.

\begin{table*}[tb]
\caption{ Possible gapped and gapless states for systems with $S_3$ symmetry.
The most stable gapped or gapless state with reduced \symmTO\ $\eM_{/\cA}$
is given by the canonical boundary of $\eM_{/\cA}$.  } \label{S3a0}
\setlength\extrarowheight{4pt} \setlength{\tabcolsep}{6pt} \centering
\begin{tabular}{c|c|c}
\hline
condensable algebra $\cA$ & reduced \symmTO\ $\eM_{/\cA}$ & most stable low
energy state\\
\hline
$\one\oplus a_1\oplus 2a_2$
& $(\eGau_{S_3})_{/\one\oplus a_1\oplus 2a_2} = $ trivial 
& gapped $\Z_1$-state  ($S_3$ completely broken) \\
\hline
$\one\oplus a_1\oplus 2b$
& $(\eGau_{S_3})_{/\one\oplus a_1\oplus 2b} = $ trivial 
& gapped $\Z_3$-state ($S_3$ broken to $\Z_3$) \\
\hline
$\one\oplus a_2\oplus c$ 
& $(\eGau_{S_3})_{/\one\oplus a_2\oplus c} = $ trivial 
& gapped $\Z_2$-state ($S_3$ broken to $\Z_2$) \\
\hline
$\one\oplus b\oplus c$
& $(\eGau_{S_3})_{/\one\oplus b\oplus c} = $ trivial 
& gapped $S_3$-symmetric state \\
\hline
$\one\oplus a_1$ 
& $(\eGau_{S_3})_{/\one\oplus a_1} =\eGau_{\Z_3} $ 
& the $(c,\bar c)= \left( \frac45,\frac45 \right)$ CFT \eq{Z3bdy} \\
\hline
$\one\oplus a_2$ 
& $(\eGau_{S_3})_{/\one\oplus a_2} =\eGau_{\Z_2} $ 
& the $(c,\bar c)= \left( \frac12,\frac12 \right)$ Ising CFT \eq{Z2bdy} \\
\hline
$\one\oplus b$ 
& $(\eGau_{S_3})_{/\one\oplus b} =\eGau_{\Z_2} $ 
& the $(c,\bar c)= \left( \frac12,\frac12 \right)$ Ising CFT  \eq{Z2bdy}\\
\hline
$\one$ & $(\eGau_{S_3})_{/\one} = \eGau_{S_3}$ 
&
the $(c,\bar c)= \left( \frac45,\frac45 \right)$ CFT \eq{S3bdy}
\\
\hline
\end{tabular}
\end{table*}

Using such a method, we can obtain the properties of the gapless $\one\oplus
a_1$-state with reduced \symmTO\ $(\eGau_{S_3})_{/\one\oplus
a_1}=\eGau_{\Z_3}$.  First, we find that one of gapless boundaries of
$\eGau_{\Z_3}$ is given by the following multi-component partition function.
Note that the $\eGau_{\Z_3}$ topological order has nine anyons $\one,e,e^2, m,
em, e^2m, m^2, em^2, e^2m^2 $.  The  multi-component boundary partition
function for $\eGau_{\Z_3}$ are labeled by these nine anyons:
\begingroup
\allowdisplaybreaks
\begin{align}
\label{Z3bdy}
Z^{\eGau_{\Z_3}}_{\bf 1} &=  |\chi^{m6}_0 + \chi^{m6}_3|^2 + 
|\chi^{m6}_\frac{2}{5} +  \chi^{m6}_\frac{7}{5}|^2,  
\nonumber \\ 
Z^{\eGau_{\Z_3}}_e &=  |\chi^{m6}_\frac{2}{3}|^2 +  |\chi^{m6}_\frac{1}{15}|^2, 
\\ 
Z^{\eGau_{\Z_3}}_{e^2} &=  |\chi^{m6}_\frac{2}{3}|^2 + 
|\chi^{m6}_\frac{1}{15}|^2,  
\nonumber \\ 
Z^{\eGau_{\Z_3}}_m &=  |\chi^{m6}_\frac{2}{3}|^2 +  |\chi^{m6}_\frac{1}{15}|^2, 
\nonumber \\ 
Z^{\eGau_{\Z_3}}_{me} &=  \chi^{m6}_0 \bar\chi^{m6}_\frac{2}{3} +  \chi^{m6}_3
\bar\chi^{m6}_\frac{2}{3} +  \chi^{m6}_\frac{2}{5} \bar\chi^{m6}_\frac{1}{15} + 
\chi^{m6}_\frac{7}{5} \bar\chi^{m6}_\frac{1}{15} , 
\nonumber \\ 
Z^{\eGau_{\Z_3}}_{me^2} &=  \chi^{m6}_\frac{2}{3} \bar\chi^{m6}_0 + 
\chi^{m6}_\frac{2}{3} \bar\chi^{m6}_3 +  \chi^{m6}_\frac{1}{15}
\bar\chi^{m6}_\frac{2}{5} +  \chi^{m6}_\frac{1}{15} \bar\chi^{m6}_\frac{7}{5},  
\nonumber \\ 
Z^{\eGau_{\Z_3}}_{m^2} &=  |\chi^{m6}_\frac{2}{3}|^2 + 
|\chi^{m6}_\frac{1}{15}|^2, 
\nonumber \\ 
Z^{\eGau_{\Z_3}}_{m^2e} &=  \chi^{m6}_\frac{2}{3} \bar\chi^{m6}_0 + 
\chi^{m6}_\frac{2}{3} \bar\chi^{m6}_3 +  \chi^{m6}_\frac{1}{15}
\bar\chi^{m6}_\frac{2}{5} +  \chi^{m6}_\frac{1}{15} \bar\chi^{m6}_\frac{7}{5},  
\nonumber \\ 
Z^{\eGau_{\Z_3}}_{m^2e^2} &=  \chi^{m6}_0 \bar\chi^{m6}_\frac{2}{3} + 
\chi^{m6}_3 \bar\chi^{m6}_\frac{2}{3} +  \chi^{m6}_\frac{2}{5}
\bar\chi^{m6}_\frac{1}{15} +  \chi^{m6}_\frac{7}{5} \bar\chi^{m6}_\frac{1}{15} .
\nonumber 
\end{align}
\endgroup
where $\chi^{m6}_{h}=\chi^{m6}_{h}(\tau)$ are conformal characters with
conformal dimension $h$, for $(6,5)$ minimal model.  The above result used the
expression of $S$-matrix of $(p,q)$ minimal model in \Rf{R160909523}.  Such CFT
has a chiral central charge $c=\frac 45$.  

The above boundary is a $\one$-condensed boundary of the topological order
$\eGau_{\Z_3}$.  This is because a condensation of an anyon $a$, will cause
the correspond partition function $Z_a(\tau)$ to contain the $|\chi^{m6}_0|^2$
term (see \eqn{S3Ba1}).  In the above partition, the term $|\chi^{m6}_0|^2$
appears only in $Z_\one(\tau)$, and thus the boundary is a $\one$-condensed
boundary.  Also if $a$ condense, there must be a nontrivial anyon $b$ that has
a nontrivial mutual statistics with $a$.  (This is due to the
remote-detectability principle of anomaly-free topological order.) The
condensation of $a$ will confine the anyon $b$ and cause $Z_b$ to vanish (see
\eqn{S3Ba1}).  This does not happen for the above partition function. Thus
there is no condensation of nontrivial anyons.

We have checked other CFT's with smaller central charges. We find that although
these CFT's can be gapless boundaries of $\eGau_{\Z_3}$, but they cannot be
$\one$-condensed boundaries of $\eGau_{\Z_3}$.  This implies that the above
boundary is also a canonical boundary (\ie a minimal $\one$-condensed boundary)
of the topological order $\eGau_{\Z_3}$.

In the above, we obtained the $\one\oplus a_1$-condensed boundary of
$\eGau_{S_3}$ via the $\one$-condensed boundary of $\eGau_{\Z_3}$.  This works
since $\eGau_{\Z_3}$ is the $\one\oplus a_1$-condensation-induced topological
order from $\eGau_{S_3}$: $(\eGau_{S_3})_{/\one\oplus a_1} = \eGau_{\Z_3}$.  In
fact, we can directly obtain $\one\oplus a_1$-condensed boundary of
$\eGau_{S_3}$. The topological order $\eGau_{S_3}$ has many gapless
boundaries, described by various multi component partition functions, labeled
by the anyons in $\eGau_{S_3}$.  One such multi component partition function is
given by
\begingroup
\allowdisplaybreaks
\begin{align}
\label{S3Ba1}
Z^{\eGau_{S_3}}_{\one} &=  |\chi^{m6}_{0}+\chi^{m6}_{3}|^2 + 
|\chi^{m6}_{\frac{2}{5}}+\chi^{m6}_{\frac{7}{5}}|^2 , 
\nonumber \\ 
Z^{\eGau_{S_3}}_{a_1} &= |\chi^{m6}_{0}+\chi^{m6}_{3}|^2 + 
|\chi^{m6}_{\frac{2}{5}}+\chi^{m6}_{\frac{7}{5}}|^2,
\nonumber \\ 
Z^{\eGau_{S_3}}_{a_2} &=  |\chi^{m6}_{\frac{2}{3}}|^2 + 
|\chi^{m6}_{\frac{1}{15}}|^2 , 
\nonumber \\ 
Z^{\eGau_{S_3}}_{b} &=  |\chi^{m6}_{\frac{2}{3}}|^2 + 
|\chi^{m6}_{\frac{1}{15}}|^2 , 
\nonumber \\ 
Z^{\eGau_{S_3}}_{ b_1} &=  \chi^{m6}_{0} \bar\chi^{m6}_{\frac{2}{3}} + 
\chi^{m6}_{3} \bar\chi^{m6}_{\frac{2}{3}} +  \chi^{m6}_{\frac{2}{5}}
\bar\chi^{m6}_{\frac{1}{15}} +  \chi^{m6}_{\frac{7}{5}}
\bar\chi^{m6}_{\frac{1}{15}} ,
\nonumber \\ 
Z^{\eGau_{S_3}}_{ b_2} &=  \chi^{m6}_{\frac{2}{3}} \bar\chi^{m6}_{0} + 
\chi^{m6}_{\frac{2}{3}} \bar\chi^{m6}_{3} +  \chi^{m6}_{\frac{1}{15}}
\bar\chi^{m6}_{\frac{2}{5}} +  
\chi^{m6}_{\frac{1}{15}} \bar\chi^{m6}_{\frac{7}{5}} ,
\nonumber \\ 
Z^{\eGau_{S_3}}_{c} &= 0,
\nonumber \\ 
Z^{\eGau_{S_3}}_{ c_1} &= 0.
\end{align}
\endgroup
We see that the term $|\chi^{m6}_{0}|^2$ appear in and only in
$Z^{\eGau_{S_3}}_{\one}(\tau) $ and $ Z^{\eGau_{S_3}}_{a_1}(\tau)$.  Thus
$\one$ and $a_1$ condense, or more precisely, the condensable algebra is
$\one\oplus a_1$.  We also see that
$Z^{\eGau_{S_3}}_{c}=Z^{\eGau_{S_3}}_{c_1}=0$.  So $c$ and $c_1$ remain gapped
on the boundary.  These properties suggest that $\one\oplus a_1$ is a
condensable algebra and the above boundary is produced by condensing such a
condensable algebra. Such a gapless boundary of $\eGau_{S_3}$ is the canonical
boundary of $(\eGau_{S_3})_{/\one\oplus a_1} = \eGau_{\Z_3}$ given in
\eqn{Z3bdy}.

Similarly, the gapless $\one\oplus a_2$-state, with reduced \symmTO\
$(\eGau_{S_3})_{/\one\oplus a_2}=\eGau_{\Z_2}$, is given by the canonical
boundary of $\eGau_{\Z_2}$, which is described by a $(c,\bar
c)=\left( \frac12,\frac12 \right)$ Ising CFT.  The corresponding multi-component
boundary
partition function is labeled by four anyons $\one,e,m,f$
of $\eGau_{\Z_2}$:
\begin{align}
\label{Z2bdy}
Z_{{\bf 1}}^{\eGau_{\Z_2}} &=  |\chi^\text{Ising}_{0}|^2 + 
|\chi^\text{Ising}_{\frac{1}{2}}|^2 
\nonumber \\ 
Z_{e}^{\eGau_{\Z_2}} &=  |\chi^\text{Ising}_{\frac{1}{16}}|^2 
\nonumber \\ 
Z_{m}^{\eGau_{\Z_2}} &=  |\chi^\text{Ising}_{\frac{1}{16}}|^2 
\nonumber \\ 
Z_{f}^{\eGau_{\Z_2}} &=  \chi^\text{Ising}_{0}
\bar\chi^\text{Ising}_{\frac{1}{2}} +  \chi^\text{Ising}_{\frac{1}{2}}
\bar\chi^\text{Ising}_{0} 
\end{align}
where $\chi^\text{Ising}_{h}=\chi^\text{Ising}_{h}(\tau)$ are conformal
characters with conformal dimension $h$, for $(4,3)$ minimal model (the Ising
CFT).  The gapless $\one\oplus b$-state is also described by a $(c,\bar
c)=\left( \frac12,\frac12 \right)$ Ising CFT, since its reduced \symmTO\ is also
given by
$\eGau_{\Z_2}$.

Again, the gapless $\one\oplus a_2$-state, with reduced \symmTO\
$(\eGau_{S_3})_{/\one\oplus a_2}=\eGau_{\Z_2}$, can also be given directly by
the $\one\oplus a_2$-condensed boundary of $\eGau_{S_3}$.  Indeed, one of the
gapless boundary of $\eGau_{S_3}$ is given by the following multi component
partition function 
\begingroup
\allowdisplaybreaks
\begin{align}
Z_\one^{\eGau_{S_3}} &=  |\chi^{m4}_{0}|^2 +  |\chi^{m4}_{\frac{1}{2}}|^2 
\nonumber \\ 
Z_{a_1}^{\eGau_{S_3}} &=  |\chi^{m4}_{\frac{1}{16}}|^2 
\nonumber \\ 
Z_{a_2}^{\eGau_{S_3}} &=  |\chi^{m4}_{0}|^2 +  |\chi^{m4}_{\frac{1}{16}}|^2 + 
|\chi^{m4}_{\frac{1}{2}}|^2 
\nonumber \\ 
Z_{b}^{\eGau_{S_3}} &= 0
\nonumber \\ 
Z_{b_1}^{\eGau_{S_3}} &=  0
\nonumber \\ 
Z_{b_2}^{\eGau_{S_3}} &=  0
\nonumber \\ 
Z_{c}^{\eGau_{S_3}} &=  |\chi^{m4}_{\frac{1}{16}}|^2 
\nonumber \\ 
Z_{c_1}^{\eGau_{S_3}} &=  \chi^{m4}_{0} \bar\chi^{m4}_{\frac{1}{2}} + 
\chi^{m4}_{\frac{1}{2}} \bar\chi^{m4}_{0} 
\end{align}
\endgroup
The above is a $\one\oplus a_2$-condensed boundary since only
$Z_{\one}^{\eGau_{S_3}}$ and $Z_{a_2}^{\eGau_{S_3}}$ contain
$|\chi^{m4}_{0}|^2$ term.

To obtain the minimal gapless $\one$-state with the full \symmTO\
$\eGau_{S_3}$, we find that, in addition to the gapless boundary described by
\eqn{S3Ba1}, $\eGau_{S_3}$ has another gapless boundary described by the
following multi component partition function, labeled by the anyons in
$\eGau_{S_3}$:
\begingroup
\allowdisplaybreaks
\begin{align}
\label{S3bdy}
Z^{\eGau_{S_3}}_{\one} &=  |\chi^{m6}_{0}|^2 +  |\chi^{m6}_{3}|^2 + 
|\chi^{m6}_{\frac{2}{5}}|^2 +  |\chi^{m6}_{\frac{7}{5}}|^2 
\nonumber \\ 
Z^{\eGau_{S_3}}_{a_1} &=  \chi^{m6}_{0} \bar\chi^{m6}_{3} +  \chi^{m6}_{3}
\bar\chi^{m6}_{0} +  \chi^{m6}_{\frac{2}{5}} \bar\chi^{m6}_{\frac{7}{5}} + 
\chi^{m6}_{\frac{7}{5}} \bar\chi^{m6}_{\frac{2}{5}} 
\nonumber \\ 
Z^{\eGau_{S_3}}_{a_2} &=  |\chi^{m6}_{\frac{2}{3}}|^2 + 
|\chi^{m6}_{\frac{1}{15}}|^2 
\nonumber \\ 
Z^{\eGau_{S_3}}_{b} &=  |\chi^{m6}_{\frac{2}{3}}|^2 + 
|\chi^{m6}_{\frac{1}{15}}|^2 
\\ 
Z^{\eGau_{S_3}}_{b_1} &=  \chi^{m6}_{0} \bar\chi^{m6}_{\frac{2}{3}} + 
\chi^{m6}_{3} \bar\chi^{m6}_{\frac{2}{3}} +  \chi^{m6}_{\frac{2}{5}}
\bar\chi^{m6}_{\frac{1}{15}} +  \chi^{m6}_{\frac{7}{5}}
\bar\chi^{m6}_{\frac{1}{15}} 
\nonumber \\ 
Z^{\eGau_{S_3}}_{b_2} &=  \chi^{m6}_{\frac{2}{3}} \bar\chi^{m6}_{0} + 
\chi^{m6}_{\frac{2}{3}} \bar\chi^{m6}_{3} +  \chi^{m6}_{\frac{1}{15}}
\bar\chi^{m6}_{\frac{2}{5}} +  \chi^{m6}_{\frac{1}{15}}
\bar\chi^{m6}_{\frac{7}{5}} 
\nonumber \\ 
Z^{\eGau_{S_3}}_{c} &=  |\chi^{m6}_{\frac{1}{8}}|^2 + 
|\chi^{m6}_{\frac{13}{8}}|^2 +  |\chi^{m6}_{\frac{1}{40}}|^2 + 
|\chi^{m6}_{\frac{21}{40}}|^2 
\nonumber \\ 
Z^{\eGau_{S_3}}_{c_1} &=  \chi^{m6}_{\frac{1}{8}} \bar\chi^{m6}_{\frac{13}{8}} +
\chi^{m6}_{\frac{13}{8}} \bar\chi^{m6}_{\frac{1}{8}} + 
\chi^{m6}_{\frac{1}{40}} \bar\chi^{m6}_{\frac{21}{40}} + 
\chi^{m6}_{\frac{21}{40}} \bar\chi^{m6}_{\frac{1}{40}} .
\nonumber 
\end{align}
\endgroup
which is a $(c,\bar c)=\left( \frac45,\frac45 \right)$ CFT.  In contrast to the
boundary
\eq{S3Ba1}, the above boundary is $\one$-condensed because only $Z_\one$
contains the term $|\chi^{m6}_{0}|^2$.  The boundary is $\one$-condensed also
because no components of the partition function is zero, \ie no other
topological excitations in $\eGau_{S_3}$ remain gapped on the boundary.  These
gapped topological excitations on the boundary become confined in
the condensation-induced topological order $(\eGau_{S_3})_{/\cA}$.  If there are
no confined topological excitations, then $\cA$ must be trivial $\cA=\one$.  

We have checked other CFT's with smaller central charges, and find that those
CFT's cannot be $\one$-condensed boundaries of the topological order
$\eGau_{S_3}$ (see Appendix \ref{S3boundaries}).  Among $(c,\bar
c)=\left( \frac45,\frac45 \right)$ CFT's, the above boundary is the only
$\one$-condensed
boundary.  This implies that the above boundary is also a canonical boundary
(\ie a minimal $\one$-condensed boundary) of the topological order
$\eGau_{S_3}$, and is the only canonical boundary.  

We remark that although the gapless $\one\oplus a_1$-state in \eqn{S3Ba1} and
the gapless $\one$-state in \eqn{S3bdy} are both related to (6,5) minimal model
with the same central charge $(c,\bar c) =\left( \frac45,\frac45 \right)$, they
are
described by different CFT's.  For example from the partition function
$Z_{a_1}^{\eGau_{S_3}}$, we find that the operator carrying the 1-dimensional
representation $a_1$ of $S_3$ has a minimal scaling dimension of $0$ for the
$\one\oplus a_1$-state (since $a_1$ is condensed) and has a minimal scaling
dimension of $\frac25+\frac75 =\frac95$ for the $\one$-state.

The above three types of gapless states \eqn{Z3bdy}, \eqn{Z2bdy}, and
\eqn{S3bdy}, together with four types pf gapped states, are the gapped and
gapless phases for systems with $S_3$ symmetry. They are summarized in Table
\ref{S3a0}.

\subsection{An automorphism in \symmTO\ $\eGau_{S_3}$}
\label{dual}

The excitations in the 2+1D topological order $\eGau_{S_3}$ (\ie the \symmTO)
are listed in Table \ref{S3FusionRules}, together with their quantum
dimensions, topological spins, and fusion rules.  From the table, we see that
the \symmTO\ $\eGau_{S_3}$ has a automorphism that exchange $a_2
\leftrightarrow b$ \cite{BW10065479}.  In fact, the $S^{S_3},T^{S_3}$-matrices
for the  $\eGau_{S_3}$ topological order are invariant under the exchange $a_2
\leftrightarrow b$.  From the correspondence between the condensable algebras
and phases of matter \eq{cAphase}, we see that the  automorphism exchanges
$S_3$-phase with $\Z_2$-phase, and $\Z_3$-phase with $\Z_1$-phase.  In other
words, the  automorphism flips the phase diagrams, Fig.
\ref{Potts15_12-14_06-10}, \ref{S3_10-10_6-13_5-12}, \ref{Potts_5_10-10_9-9},
as well as Fig. \ref{S3phases}, horizontally.

As a result, the transitions $S_3 \leftrightarrow \Z_1$ and $\Z_3
\leftrightarrow \Z_2$ are related, \ie they are either both first order, both
stably continuous, or both unstably continuous.  More precisely, \frmbox{if the
transition $S_3 \leftrightarrow \Z_1$ is continuous in a $S_3$ symmetric
system, then there is exist another $S_3$ symmetric system where the transition
$\Z_3 \leftrightarrow \Z_2$ is also continuous, and the two continuous
transitions are described by the same CFT.  }

\subsection{A \symmTO\ approach for phase transitions}
\label{transS3}

\begin{figure}[t]
\begin{center}
\includegraphics[scale=0.55]{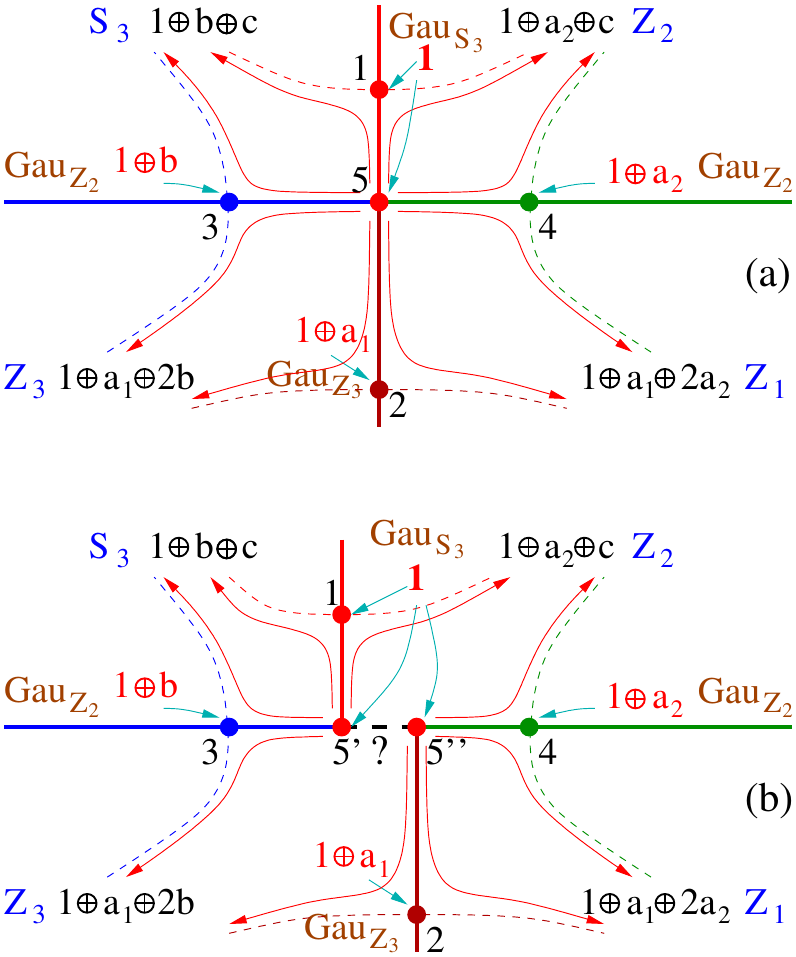}
\end{center}
\caption{Two possible global phase diagrams for systems with $S_3$ symmetry,
which contains four gapped phases with unbroken symmetries, $S_3$, $\Z_3$,
$\Z_2$, and $\Z_1$. The curves with arrow represent the RG flow, and the dots
are the RG fixed points that correspond to the critical points of phase
transitions.  The right horizontal line is the space of Hamiltonians whose
ground states have a condensation $\cA=\one\oplus a_2$ (see Appendix
\ref{phasestructure}), which is the basin of attraction of the RG fixed point
4. The left horizontal line is the space of Hamiltonians whose ground states
have a condensation $\cA=\one\oplus b$, the basin of attraction of the RG fixed
point 3.  The upper vertical line is the space of Hamiltonians whose ground
states have a condensation $\cA=\one$, the basin of attraction of the RG fixed
point 1.  The lower vertical line is the space of Hamiltonians whose ground
states have a condensation $\cA=\one\oplus a_1$, the basin of attraction of the
RG fixed point 2.  The critical point 3 and 4 are the $(c,\bar c)=\left(
\frac12,\frac12 \right)$ Ising CFT (the canonical boundary of $\eGau_{\Z_2}$
topological order).  The critical point 1 is a $(c,\bar c)=\left(
\frac45,\frac45 \right)$ CFT \eq{S3bdy} from $(6,5)$ minimal model (the
canonical boundary of $\eGau_{S_3}$ topological order).  The critical point 2
is another $(c,\bar c)=\left( \frac45,\frac45 \right)$ CFT \eq{Z3bdy} also from
$(6,5)$ minimal model (the canonical boundary of $\eGau_{\Z_3}$ topological
order).  The tricritical points $5$, $5'$, and $5''$ are described by gapless
$\one$-condensed states with two $S_3$ symmetric relevant operators.  The
$(c,\bar c)=\left( \frac56,\frac56 \right)$ CFT \eq{S3m76} from $(7,6)$ minimal
model is one such gapless $\one$-condensed state.  We also list the corresponding condensable algebras for these gapped
phases and gapless critical points.  } \label{S3phasetrans}
\end{figure}

We have used \symmTO\ approach to study possible gapped and gapless states in
$S_3$ symmetric systems.  Now let us discuss a more difficult problem: how are
these gapped and gapless states connected by continuous phase transitions and
what are the critical points at the transitions?  The gapless states
\eqn{Z3bdy}, \eqn{Z2bdy}, \eqn{S3bdy}, and others constructed from $(5,4)$ and
$(7,6)$ minimal models and $(c,\bar c) \geq (1,1)$ CFTs should describe the
(multi-)critical points for the transitions between the four gapped phases,
$S_3$-phase, $\Z_3$-phase, $\Z_2$-phase, $\Z_1$-phase (see Fig.
\ref{S3phasetrans}).  But which pair of gapped states are connected by which
gapless state, as the critical point of the continuous transition?

To address this issue, we first consider the gapless $\one \oplus a_1$-state,
which is described by the canonical boundary of $(\eGau_{S_3})_{/\one\oplus
a_1}= \eGau_{\Z_3}$ topological order (\ie 2+1D $\Z_3$ gauge theory).  Its
multi-component boundary partition function is given by \eqn{Z3bdy}.  From the
$|\chi^{m6}_\frac{2}{5} + \chi^{m6}_\frac{7}{5}|^2$ term in $Z_\one$ in
\eqn{Z3bdy}, we see that there is only one $\Z_3$ symmetric relevant operator,
which has a scaling dimension $(h,\bar h)=(\frac 25,\frac 25)$.  So the gapless
$\one\oplus a_1$-state has only one relevant direction.  To see what kind of
phase transition the relevant operator induces, we note that the condensable
algebra $\one\oplus a_1$ only allows one competing pair $(a_2,b)$.  So the
single relevant direction must correspond to the switching between the two
condensations of the competing pair $(a_2,b)$.  This induces a stable
continuous phase transition between the $\one\oplus  a_1\oplus
a_2\oplus\cdots = \one\oplus a_1\oplus 2a_2 $-state (the $\Z_1$-state) and
$\one\oplus  a_1\oplus b\oplus\cdots = \one\oplus a_1\oplus 2b $ (the
$\Z_3$-state). The local phase diagram for such transition is given by Fig.
\ref{phasetrans}(a). Thus the $\Z_3\to \Z_1$ symmetry breaking critical point
is described by a $(c,\bar c)=\left( \frac45,\frac45 \right)$ CFT constructed
from $(6,5)$
minimal model.  This example demonstrates how to use \symmTO\ to study
continuous phase transitions and their critical points.

Next, we consider the gapless $\one\oplus a_2$-state which is the canonical
boundary of $(\eGau_{S_3})_{/\one\oplus a_2}= \eGau_{\Z_2}$ (\ie 2+1D $\Z_2$
gauge theory).  Its multi-component boundary partition function is given by
\eqn{Z2bdy}.  From the $|\chi^\text{Ising}_\frac{1}{2}|^2$ term in $Z_\one$, we
see that there is only one $\Z_2$ symmetric relevant operator, with a scaling
dimension $(h,\bar h)=(\frac 12,\frac 12)$.  So the gapless $\one\oplus
a_2$-state has only one relevant direction.  To see which phase transition is
induced
by the relevant operator, we note that the condensable algebra $\one\oplus a_2$
allows only one competing pair $(a_1,c)$.  Thus, the gapless $\one\oplus
a_2$-state is the critical point for a stable continuous phase transition
between
$\one\oplus a_2\oplus a_1\oplus \cdots =\one\oplus  a_1\oplus  2a_2$-state (the
$\Z_1$-state) and $\one\oplus  a_2\oplus c\oplus \cdots =\one\oplus a_2
\oplus c $-state (the $\Z_2$-state), whose local phase diagram is given in Fig.
\ref{phasetrans}(a).

The gapless $\one\oplus b$-state is similar to the gapless $\one\oplus
a_2$-state, due to the automorphism of $\eGau_{S_3}$ that exchange $a_2$ and
$b$.
Both are described by $(c,\bar c)=\left( \frac12,\frac12 \right)$ Ising CFT
\eq{Z2bdy}.  The
gapless $\one\oplus b$-state also allows only one competing pair $(a_1,c)$, and
describes a stable continuous phase transition between $\one\oplus
b\oplus a_1\oplus \cdots =\one\oplus  a_1\oplus  2b$-state (the $\Z_3$-state)
and
$\one\oplus  b\oplus c\oplus \cdots =\one\oplus b\oplus c$-state (the
$S_3$-state).  The $S_3\to \Z_3$ symmetry breaking transition looks different
from the $\Z_2\to \Z_1$ symmetry breaking transition.  However, the above
discussion suggests that the two transitions are described by the same critical
theory.  This result is supported by the standard Ginzburg-Landau theory.

Last, let us consider the gapless $\one$-state and its neighborhood.  The state
is given by a canonical boundary of $\eGau_{S_3}$, whose multi-component
boundary partition function is given by \eqn{S3bdy}.  The gapless  $\one$-state
has only one $S_3$ symmetric relevant operator with dimension $(h,\bar h) =
(\frac 25, \frac 25)$, as one can see from the $|\chi^{m6}_\frac{2}{5} |^2$
term in $Z_\one$.  However, the condensable algebra $\one$ allows two competing
pairs: $(a_2,b)$ and $(a_1,c)$.  Which competing pair corresponds to the
relevant direction? 

\begin{figure}[t]
\begin{center}
\includegraphics[scale=0.6]{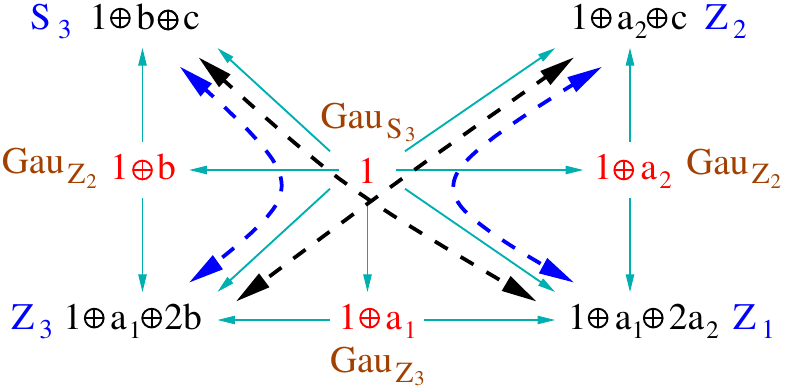}
\end{center}
\caption{The gapless $\one$-state described by \eqn{S3bdy} has one relevant
direction. If this relevant direction corresponds to the competing pair
$(a_1,c)$, the gapless $\one$-state may be the critical point of the potential
continuous transitions represented by the four curved double-arrows.  These
potential continuous transitions are the ones that we cannot rule out at the
moment.  } \label{S3trans1} 
\end{figure}

If we assume the competing pair $(a_1,c)$ corresponds to the relevant
direction, then on one side of transition, the condensable algebra $\one$ is
enlarged to include $a_1$ condensation: $\one \to \one\oplus a_1\oplus \cdots$.
Here $\cdots$ represents the additional condensation, after $a_1$ condense.
Such additional condensations must be compatible with $a_1$-condensation.
We have three possible additional condensations: (1) we may get a gapless
$\one\oplus a_1\oplus \cdots = \one\oplus a_1$-state (\ie no additional
condensation).  (2) we may get a gapped $\one\oplus a_1\oplus \cdots =
\one\oplus a_1\oplus 2 a_2$-state (\ie with additional $a_2$-condensation).  (3)
we
may get a gapped $\one\oplus a_1\oplus \cdots = \one\oplus a_1\oplus 2 b$-state
(\ie with additional $b$-condensation).  On the other side of transition where
$c$ condenses, we have two possible additional condensations: (1') we may get a
gapped $\one\oplus c\oplus \cdots = \one\oplus a_2\oplus c$-state (\ie with
additional $a_2$-condensation).  (2') we may get a gapped $\one\oplus
c\oplus \cdots = \one\oplus a_2\oplus c$-state (\ie with additional
$a_2$-condensation).  Combining the above possibilities, we obtain the following
scenarios (see Fig. \ref{S3trans1}):
\begin{itemize}

\item[(11')] stable continuous transition between gapless $\one\oplus
a_1$-state and $\one\oplus a_2\oplus c$-state, described by the $(c,\bar
c)=\left( \frac45,\frac45 \right)$ CFT \eq{S3bdy}.  Further instability from
dangerously
irrelevant operators may change the gapless $\one\oplus a_1$-state to
$\Z_3$-state or $\Z_1$-state.  (Not likely. This scenario assumes that the
condensation of $c$ also induce the condensation of $a_2$. As we switch the
condensation of $c$ to the condensation of $a_1$, the condensation of $a_1$ is
compatible with the condensation of $a_2$ and does not suppress the
condensation of $a_2$. The condensation of $a_2$ will destabilize the gapless
$\one\oplus a_1$-state and change it to the gapped $\one\oplus a_1\oplus
a_2$-state. This turns the scenario (11') to scenario (21')).

\item[(12')] stable continuous transition between gapless $\one\oplus
a_1$-state and $\one\oplus b\oplus c$-state, described by the $(c,\bar
c)=\left( \frac45,\frac45 \right)$ CFT \eq{S3bdy}.  Further instability from
dangerous
irrelevant operators may change the gapless $\one\oplus a_1$-state to
$\Z_3$-state or $\Z_1$-state.
(Not likely. This scenario assumes that the
condensation of $c$ also induce the condensation of $b$. As we switch the
condensation of $c$ to the condensation of $a_1$, the condensation of $a_1$ is
compatible with the condensation of $b$ and does not suppress the
condensation of $b$. The condensation of $b$ will destabilize the gapless
$\one\oplus a_1$-state and change it to the gapped $\one\oplus a_1\oplus
b$-state. This turns the scenario (12') to scenario (32')).

\item[(21')] stable continuous transition between $\one\oplus
a_1\oplus 2 a_2$-state and $\one\oplus a_2\oplus c$-state, described the
$(c,\bar c)=\left( \frac45,\frac45 \right)$ CFT \eq{S3bdy}.  

\item[(22')] stable continuous transition between $\one\oplus
a_1\oplus 2 a_2$-state and $\one\oplus b\oplus c$-state, described by the
$(c,\bar
c)=\left( \frac45,\frac45 \right)$ CFT \eq{S3bdy}.

\item[(31')] stable continuous transition between $\one\oplus
a_1\oplus 2 b$-state and $\one\oplus a_2\oplus c$-state, described by the
$(c,\bar c)=\left( \frac45,\frac45 \right)$ CFT \eq{S3bdy}.

\item[(32')] stable continuous transition between $\one\oplus
a_1\oplus 2 b$-state and $\one\oplus b\oplus c$-state, described by the $(c,\bar
c)=\left( \frac45,\frac45 \right)$ CFT \eq{S3bdy}.

\end{itemize}
Some scenarios, (11') and (12'), are not likely. We remark that the above
scenarios may not be mutually exclusive.  Different scenarios may be realized
at different parts of the phase diagram.  We also remark that the scenario
(21'), if realized, will represent a non-Ising critical point for the
transition between the $\Z_2$-state and $\Z_1$-state.  This  scenario
represents a mechanism that two phases may be connected by different continuous
transitions described by different CFTs.

\begin{figure}[t]
\begin{center}
\includegraphics[scale=0.6]{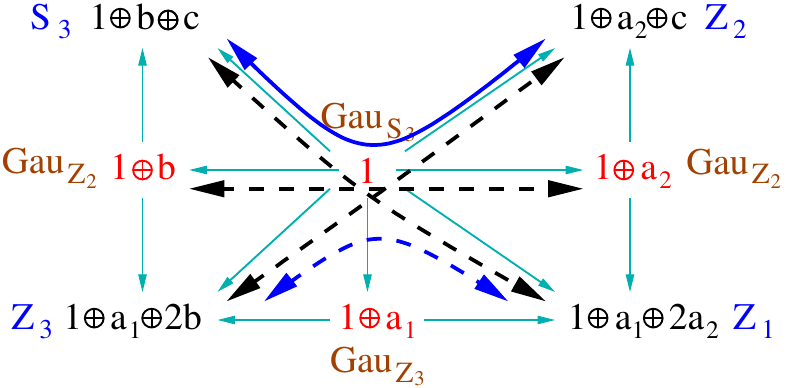}
\end{center}
\caption{The gapless $\one$-state described by \eqn{S3bdy} has one relevant
direction. If this relevant direction corresponds to the competing pair
$(a_2,b)$, the gapless $\one$-state may be the critical point of the potential
continuous transitions represented by the eight curved double-arrows. The
dashed curves are the potential continuous transitions that we cannot
rule out at the moment.  The solid curve is the continuous transition that is
known to be realized by 3-state Potts model.  } \label{S3trans2} 
\end{figure}

Next, we assume the relevant direction corresponds to the competing pair
$(a_2,b)$.  On one side of transition, $a_2$ condenses, which may give rise to
the following possible states: (a) the gapless $\one\oplus a_2\oplus
\cdots = \one\oplus a_2$-state; (b) the gapped $\one\oplus
a_2\oplus\cdots=\one\oplus a_1\oplus 2 a_2$-state; (c) the gapped $\one\oplus
a_2\oplus \cdots = \one\oplus a_2 \oplus c$-state.  On one side of transition,
$b$
condenses, which may give rise to the following possible states: (a') the
gapless $\one\oplus b\oplus \cdots = \one\oplus b$-state; (b') the gapped
$\one\oplus b\oplus\cdots=\one\oplus a_1\oplus 2 b$-state; (c') the gapped
$\one\oplus b\oplus \cdots = \one\oplus b \oplus c$-state.  Combining the above
possibilities, we obtain the following scenarios (see Fig. \ref{S3trans2}):
\begin{itemize}

\item[(aa')] stable continuous transition between gapless $\one\oplus
a_2$-state and gapless $\one\oplus b$-state, described by the $(c,\bar
c)=\left( \frac45,\frac45 \right)$ CFT \eq{S3bdy}.  Further instability from
dangerous
irrelevant operators change the gapless $\one\oplus a_2$-state to
$\Z_2$-state or $\Z_1$-state, and change the gapless $\one\oplus b$-state to
$S_3$-state or $\Z_3$-state.

\item[(ab')] stable continuous transition between gapless $\one\oplus
a_2$-state and gapped $\one\oplus a_1\oplus 2b$-state, described by the
$(c,\bar c)=\left( \frac45,\frac45 \right)$ CFT \eq{S3bdy}.  Further instability
from
dangerously irrelevant operators change the gapless $\one\oplus a_2$-state to
$\Z_2$-state or $\Z_1$-state. (Not likely. See discussion in scenario (11').)

\item[(ac')] stable continuous transition between gapless $\one\oplus
a_2$-state and gapped $S_3$-$\one\oplus b\oplus c$-state, described by the
$(c,\bar
c)=\left( \frac45,\frac45 \right)$ CFT \eq{S3bdy}.  Further instability from
dangerous
irrelevant operators change the gapless $\one\oplus a_2$-state to
$\Z_2$-state or $\Z_1$-state. (Not likely. See discussion in scenario (11').)

\item[(ba')] stable continuous transition between gapped $\one\oplus a_1\oplus 2
a_2$-state and gapless $\one\oplus b$-state, described by the
$(c,\bar c)=\left( \frac45,\frac45 \right)$ CFT \eq{S3bdy}.  Further instability
from
dangerously irrelevant operators change the gapless $\one\oplus b$-state to
$S_3$-state or $\Z_3$-state. (Not likely. See discussion in scenario (11').)

\item[(bb')] stable continuous transition between gapped $\Z_1$-$\one\oplus
a_1\oplus 2 a_2$-state and gapped $\Z_3$-$\one\oplus a_1\oplus 2b$-state,
described by the $(c,\bar c)=\left( \frac45,\frac45 \right)$ CFT \eq{S3bdy}. 
(Not valid.
Such a transition should be described by a different  $(c,\bar
c)=\left( \frac45,\frac45 \right)$ CFT \eq{Z3bdy}.)

\item[(bc')] stable continuous transition between gapped $\Z_1$-$\one\oplus
a_1\oplus 2 a_2$-state and gapped $S_3$-$\one\oplus b\oplus c$-state,
described by the $(c,\bar c)=\left( \frac45,\frac45 \right)$ CFT \eq{S3bdy}.  

\item[(ca')] stable continuous transition between gapped $\Z_2$-$\one\oplus
a_2\oplus c$-state and gapless $\one\oplus b$-state, described by the
$(c,\bar c)=\left( \frac45,\frac45 \right)$ CFT \eq{S3bdy}.  Further instability
from
dangerously irrelevant operators change the gapless $\one\oplus b$-state to
$S_3$-state or $\Z_3$-state. (Not likely. See discussion in scenario (11').)

\item[(cb')] stable continuous transition between gapped $\Z_2$-$\one\oplus
a_2\oplus c$-state and gapped $\Z_3$-$\one\oplus a_1\oplus 2b$-state,
described by the $(c,\bar c)=\left( \frac45,\frac45 \right)$ CFT \eq{S3bdy}.  

\item[(cc')] stable continuous transition between gapped $\Z_2$-$\one\oplus
a_2\oplus c$-state and gapped $S_3$-$\one\oplus b\oplus c$-state,
described by the $(c,\bar c)=\left( \frac45,\frac45 \right)$ CFT \eq{S3bdy}.  

\end{itemize}

We believe that scenario (cc') is realized in the 3-state Potts model, which
has a $S_3\leftrightarrow \Z_2$ transition described a $(c,\bar
c)=\left( \frac45,\frac45 \right)$ CFT.  We believe such a CFT to be the one
given in
\eq{S3bdy}, rather than the $(c,\bar c)=\left( \frac45,\frac45 \right)$ CFT 
given in
\eq{S3Ba1}.

Note that the stable continuous $S_3\leftrightarrow \Z_1$ transition should be
described by a $\one$-condensed boundary of $\eGau_{S_3}$ with one and only one
$S_3$ symmetric operator.  The $(c,\bar c)=\left( \frac45,\frac45 \right)$ CFT
\eq{S3bdy} is
one such $\one$-condensed boundary. But such a CFT is already used to described
the stable continuous $S_3\leftrightarrow \Z_2$ transition.  We need to find
another $\one$-condensed boundary of $\eGau_{S_3}$ to describe the stable
continuous $S_3\leftrightarrow \Z_1$ transition.

Summarizing the above result and assuming the scenario (cc'), we obtain several
possible global phase diagrams.  One of them is Fig. \ref{S3phasetrans}(a) and
another is Fig. \ref{S3phasetrans}(b).  Both possibilities are realized by
numerical calculations in Fig. \ref{Potts_5_10-10_9-9} and Fig.
\ref{S3_10-10_6-13_5-12}.  The above two global phase diagrams suggest three
tricritical points $5$, $5'$, and $5''$.  From the phase diagram, we see that
tricritical points $5$ and $5'$ are connected to the $S_3$-phase.  Thus they
are $\one$-condensed boundaries of topological order $\eGau_{S_3}$.  From the
phase diagram, we also see that tricritical point $5''$ connects to both
$\Z_3$- and $\Z_2$-phases, and thus has both $\Z_3$ and $\Z_2$ symmetries.
Therefore, tricritical point $5''$ has the full $S_3$ symmetry and is also a
$\one$-condensed boundary of topological order $\eGau_{S_3}$.

The three tricritical points $5$, $5'$, and $5''$ are not the canonical
boundaries of $\eGau_{S_3}$, since they have two symmetric relevant operators
and are more unstable.  By examining other $\one$-condensed boundaries of
$\eGau_{S_3}$, we find the following multi-component partition function:
\begingroup
\allowdisplaybreaks
\begin{align}
\label{S3m76}
Z^{\eGau_{S_3}}_{\one} &=  |\chi^{m7}_{0}|^2 +  |\chi^{m7}_{\frac{1}{7}}|^2 + 
|\chi^{m7}_{\frac{5}{7}}|^2 +  |\chi^{m7}_{\frac{12}{7}}|^2 
\nonumber\\
&\ \ \ \
+  |\chi^{m7}_{\frac{22}{7}}|^2 +  |\chi^{m7}_{5}|^2 
\nonumber \\ 
Z^{\eGau_{S_3}}_{a_1} &=  \chi^{m7}_{0} \bar\chi^{m7}_{5} + 
\chi^{m7}_{\frac{1}{7}} \bar\chi^{m7}_{\frac{22}{7}} +  \chi^{m7}_{\frac{5}{7}}
\bar\chi^{m7}_{\frac{12}{7}} +  \chi^{m7}_{\frac{12}{7}}
\bar\chi^{m7}_{\frac{5}{7}} 
\nonumber\\
&\ \ \ \
+  \chi^{m7}_{\frac{22}{7}} \bar\chi^{m7}_{\frac{1}{7}} +  \chi^{m7}_{5}
\bar\chi^{m7}_{0} 
\nonumber \\ 
Z^{\eGau_{S_3}}_{a_2} &=  |\chi^{m7}_{\frac{4}{3}}|^2 + 
|\chi^{m7}_{\frac{10}{21}}|^2 
+  |\chi^{m7}_{\frac{1}{21}}|^2 
\nonumber \\ 
Z^{\eGau_{S_3}}_{b} &=  |\chi^{m7}_{\frac{4}{3}}|^2 + 
|\chi^{m7}_{\frac{10}{21}}|^2 +  |\chi^{m7}_{\frac{1}{21}}|^2 
\nonumber \\ 
Z^{\eGau_{S_3}}_{b_1} &=  \chi^{m7}_{\frac{4}{3}} \bar\chi^{m7}_{0} + 
\chi^{m7}_{\frac{4}{3}} \bar\chi^{m7}_{5} +  \chi^{m7}_{\frac{10}{21}}
\bar\chi^{m7}_{\frac{1}{7}} +  \chi^{m7}_{\frac{10}{21}}
\bar\chi^{m7}_{\frac{22}{7}} 
\nonumber\\
&\ \ \ \
+  \chi^{m7}_{\frac{1}{21}} \bar\chi^{m7}_{\frac{5}{7}} + 
\chi^{m7}_{\frac{1}{21}} \bar\chi^{m7}_{\frac{12}{7}} 
\nonumber \\ 
Z^{\eGau_{S_3}}_{b_2} &=  \chi^{m7}_{0} \bar\chi^{m7}_{\frac{4}{3}} + 
\chi^{m7}_{\frac{1}{7}} \bar\chi^{m7}_{\frac{10}{21}} +  \chi^{m7}_{\frac{5}{7}}
\bar\chi^{m7}_{\frac{1}{21}} +  \chi^{m7}_{\frac{12}{7}}
\bar\chi^{m7}_{\frac{1}{21}} 
\nonumber\\
&\ \ \ \
+  \chi^{m7}_{\frac{22}{7}} \bar\chi^{m7}_{\frac{10}{21}} +  \chi^{m7}_{5}
\bar\chi^{m7}_{\frac{4}{3}} 
\nonumber \\ 
Z^{\eGau_{S_3}}_{c} &=  |\chi^{m7}_{\frac{3}{8}}|^2 + 
|\chi^{m7}_{\frac{1}{56}}|^2 +  |\chi^{m7}_{\frac{5}{56}}|^2 + 
|\chi^{m7}_{\frac{33}{56}}|^2 
\nonumber\\
&\ \ \ \
+  |\chi^{m7}_{\frac{85}{56}}|^2 +  |\chi^{m7}_{\frac{23}{8}}|^2  
\nonumber \\ 
Z^{\eGau_{S_3}}_{c_1} &=  \chi^{m7}_{\frac{3}{8}} \bar\chi^{m7}_{\frac{23}{8}} +
\chi^{m7}_{\frac{1}{56}} \bar\chi^{m7}_{\frac{85}{56}} + 
\chi^{m7}_{\frac{5}{56}} \bar\chi^{m7}_{\frac{33}{56}} + 
\chi^{m7}_{\frac{33}{56}} \bar\chi^{m7}_{\frac{5}{56}} 
\nonumber\\
&\ \ \ \
+  \chi^{m7}_{\frac{85}{56}} \bar\chi^{m7}_{\frac{1}{56}} + 
\chi^{m7}_{\frac{23}{8}} \bar\chi^{m7}_{\frac{3}{8}} 
\end{align}
\endgroup
which is constructed from the $(7,6)$ minimal model.  The above CFT has two
relevant operators. It is a candidate CFT for one of the three tricritical
points $5$, $5'$, and $5''$, likely the tricritical point $5$.  We need find
more $\one$-condensed boundaries of $\eGau_{S_3}$ with two and only two $S_3$
symmetric relavent operators to describe the other two tricritical points.

We note that the critical points $5$, $5'$, and $5''$, having $\Z_3$ symmetry,
are also $\one$-condensed boundaries of topological order $\eGau_{\Z_3}$.
Indeed, we find the following multi-component partition function constructed
from the $(7,6)$ minimal model, realizing a $\one$-condensed boundary of
$\eGau_{\Z_3}$:
\begingroup
\allowdisplaybreaks
\begin{align}
\label{Z3m76}
Z^{\eGau_{\Z_3}}_{\bf 1} &=  |\chi^{m7}_0|^2 +  \chi^{m7}_0 \bar\chi^{m7}_5 + 
|\chi^{m7}_\frac{1}{7}|^2 +  \chi^{m7}_\frac{1}{7} \bar\chi^{m7}_\frac{22}{7} + 
|\chi^{m7}_\frac{5}{7}|^2 
\nonumber\\
&\ \ \ \
+  \chi^{m7}_\frac{5}{7} \bar\chi^{m7}_\frac{12}{7} +  \chi^{m7}_\frac{12}{7}
\bar\chi^{m7}_\frac{5}{7} +  |\chi^{m7}_\frac{12}{7}|^2
+  \chi^{m7}_\frac{22}{7} \bar\chi^{m7}_\frac{1}{7} 
\nonumber\\
&\ \ \ \
+  |\chi^{m7}_\frac{22}{7}|^2 +  \chi^{m7}_5 \bar\chi^{m7}_0 +  |\chi^{m7}_5|^2 
\nonumber \\ 
Z^{\eGau_{\Z_3}}_e &=  |\chi^{m7}_\frac{4}{3}|^2 +  |\chi^{m7}_\frac{10}{21}|^2
+  |\chi^{m7}_\frac{1}{21}|^2 
\nonumber \\ 
Z^{\eGau_{\Z_3}}_{e^2} &=  |\chi^{m7}_\frac{4}{3}|^2 + 
|\chi^{m7}_\frac{10}{21}|^2 +  |\chi^{m7}_\frac{1}{21}|^2 
\nonumber \\ 
Z^{\eGau_{\Z_3}}_m &=  |\chi^{m7}_\frac{4}{3}|^2 +  |\chi^{m7}_\frac{10}{21}|^2
+  |\chi^{m7}_\frac{1}{21}|^2 
\nonumber \\ 
Z^{\eGau_{\Z_3}}_{me} &=  \chi^{m7}_\frac{4}{3} \bar\chi^{m7}_0 + 
\chi^{m7}_\frac{4}{3} \bar\chi^{m7}_5 +  \chi^{m7}_\frac{10}{21}
\bar\chi^{m7}_\frac{1}{7} +  \chi^{m7}_\frac{10}{21} \bar\chi^{m7}_\frac{22}{7} 
\nonumber\\
&\ \ \ \
+  \chi^{m7}_\frac{1}{21} \bar\chi^{m7}_\frac{5}{7} +  \chi^{m7}_\frac{1}{21}
\bar\chi^{m7}_\frac{12}{7} 
\nonumber \\ 
Z^{\eGau_{\Z_3}}_{me^2} &=  \chi^{m7}_0 \bar\chi^{m7}_\frac{4}{3} + 
\chi^{m7}_\frac{1}{7} \bar\chi^{m7}_\frac{10}{21} +  \chi^{m7}_\frac{5}{7}
\bar\chi^{m7}_\frac{1}{21} +  \chi^{m7}_\frac{12}{7} \bar\chi^{m7}_\frac{1}{21} 
\nonumber\\
&\ \ \ \ 
+  \chi^{m7}_\frac{22}{7} \bar\chi^{m7}_\frac{10}{21} +  \chi^{m7}_5
\bar\chi^{m7}_\frac{4}{3} 
\nonumber \\ 
Z^{\eGau_{\Z_3}}_{m^2} &=  |\chi^{m7}_\frac{4}{3}|^2 + 
|\chi^{m7}_\frac{10}{21}|^2 +  |\chi^{m7}_\frac{1}{21}|^2 
\nonumber \\ 
Z^{\eGau_{\Z_3}}_{m^2e} &=  \chi^{m7}_0 \bar\chi^{m7}_\frac{4}{3} + 
\chi^{m7}_\frac{1}{7} \bar\chi^{m7}_\frac{10}{21} +  \chi^{m7}_\frac{5}{7}
\bar\chi^{m7}_\frac{1}{21} +  \chi^{m7}_\frac{12}{7} \bar\chi^{m7}_\frac{1}{21} 
\nonumber\\
&\ \ \ \
+  \chi^{m7}_\frac{22}{7} \bar\chi^{m7}_\frac{10}{21} +  \chi^{m7}_5
\bar\chi^{m7}_\frac{4}{3} 
\nonumber \\ 
Z^{\eGau_{\Z_3}}_{m^2e^2} &=  \chi^{m7}_\frac{4}{3} \bar\chi^{m7}_0 + 
\chi^{m7}_\frac{4}{3} \bar\chi^{m7}_5 +  \chi^{m7}_\frac{10}{21}
\bar\chi^{m7}_\frac{1}{7} +  \chi^{m7}_\frac{10}{21} \bar\chi^{m7}_\frac{22}{7} 
\nonumber\\
&\ \ \ \
+  \chi^{m7}_\frac{1}{21} \bar\chi^{m7}_\frac{5}{7} +  \chi^{m7}_\frac{1}{21}
\bar\chi^{m7}_\frac{12}{7}  
\end{align}
\endgroup

We also note that three tricritical points $5$, $5'$ and $5''$ can be viewed
as a $\one$-condensed boundary of topological order $\eGau_{\Z_2}$.  We do find
the following multi-component partition function constructed from the $(7,6)$
minimal model, realizing a $\one$-condensed boundary of $\eGau_{\Z_2}$:
\begingroup
\allowdisplaybreaks
\begin{align}
\label{Z2m76}
Z_{{\bf 1}}^{\eGau_{\Z_2}} &=  |\chi^{m7}_{0}|^2 +  |\chi^{m7}_{\frac{1}{7}}|^2
+  |\chi^{m7}_{\frac{5}{7}}|^2 +  |\chi^{m7}_{\frac{12}{7}}|^2 + 
|\chi^{m7}_{\frac{22}{7}}|^2 
\nonumber\\
&\ \ \ \
+  |\chi^{m7}_{5}|^2 +  |\chi^{m7}_{\frac{4}{3}}|^2 + 
|\chi^{m7}_{\frac{10}{21}}|^2 +  |\chi^{m7}_{\frac{1}{21}}|^2 
\nonumber \\ 
Z_{e}^{\eGau_{\Z_2}} &=  \chi^{m7}_{0} \bar\chi^{m7}_{5} + 
\chi^{m7}_{\frac{1}{7}} \bar\chi^{m7}_{\frac{22}{7}} +  \chi^{m7}_{\frac{5}{7}}
\bar\chi^{m7}_{\frac{12}{7}} +  \chi^{m7}_{\frac{12}{7}}
\bar\chi^{m7}_{\frac{5}{7}} 
\nonumber\\
&
+  \chi^{m7}_{\frac{22}{7}} \bar\chi^{m7}_{\frac{1}{7}} +  \chi^{m7}_{5}
\bar\chi^{m7}_{0} +  |\chi^{m7}_{\frac{4}{3}}|^2 + 
|\chi^{m7}_{\frac{10}{21}}|^2 +  |\chi^{m7}_{\frac{1}{21}}|^2 
\nonumber \\ 
Z_{m}^{\eGau_{\Z_2}} &=  |\chi^{m7}_{\frac{3}{8}}|^2 + 
|\chi^{m7}_{\frac{1}{56}}|^2 +  |\chi^{m7}_{\frac{5}{56}}|^2 + 
|\chi^{m7}_{\frac{33}{56}}|^2 +  |\chi^{m7}_{\frac{85}{56}}|^2 
\nonumber\\
&\ \ \ \
+  |\chi^{m7}_{\frac{23}{8}}|^2
\nonumber \\ 
Z_{f}^{\eGau_{\Z_2}} &=  \chi^{m7}_{\frac{3}{8}} \bar\chi^{m7}_{\frac{23}{8}} + 
\chi^{m7}_{\frac{1}{56}} \bar\chi^{m7}_{\frac{85}{56}} + 
\chi^{m7}_{\frac{5}{56}} \bar\chi^{m7}_{\frac{33}{56}} + 
\chi^{m7}_{\frac{33}{56}} \bar\chi^{m7}_{\frac{5}{56}} 
\nonumber\\
&\ \ \ \
+  \chi^{m7}_{\frac{85}{56}} \bar\chi^{m7}_{\frac{1}{56}} + 
\chi^{m7}_{\frac{23}{8}} \bar\chi^{m7}_{\frac{3}{8}}  
\end{align}
\endgroup
The above three multi-component partition functions are closely related.
In fact
\begin{align}
Z_\one^{\eGau_{S_3}} +
Z_{a_1}^{\eGau_{S_3}} +
2Z_{a_2}^{\eGau_{S_3}} 
&=
Z_\one^{\eGau_{\Z_3}} +
Z_{e}^{\eGau_{\Z_3}} +
Z_{e^2}^{\eGau_{\Z_3}} 
\nonumber\\
&=
Z_\one^{\eGau_{\Z_2}} +
Z_{e}^{\eGau_{\Z_2}}
\end{align}
This suggests that the three CFT's, \eqn{S3m76}, \eqn{Z3m76}, and \eqn{Z2m76},
are actually the same CFT.  This allows us to conclude that the CFT \eq{S3m76}
can be a candidate for one of three tricritical points $5$, $5'$, and $5''$ in
Fig.  \ref{S3phasetrans}.  Certainly, it is also possible that the three
tricritical points $5$, $5'$, and $5''$ are described by CFT's with $(c,\bar
c) \geq (1,1)$.

\section{1+1D anomalous $S_3$ symmetry}\label{anoS3}

In 1+1D, the anomalies for $S_3$ symmetry are classified by
$H^3(S_3;\RZ)=\Z_3\times\Z_2 \simeq \Z_6$.\cite{CGL1314}  We label those
anomalies by
$m\in \{0,1,2,3,4,5\}$.  The  \symmTO\ for an anomalous $S_3$ symmetry,
$S_3^{(m)}$, is given by a topological order $\eGau_{S_3}^{(m)}$ that is
described in the IR limit by the 2+1D Dijkgraaf-Witten gauge theory\cite{DW9093} with gauge
charges.  In this
section, we will use these \symmTOs\ to study the 1+1D gapped and gapless
states with anomalous $S_3$ symmetry.

\subsection{Anomalous $S_3^{(1)}$ symmetry}

\begin{table*}[tb]
\caption{ Possible gapped and gapless states for systems with anomalous
$S_3^{(1)}$ symmetry.  } \label{S3a1} 
\setlength\extrarowheight{4pt}
\setlength{\tabcolsep}{6pt}
\centering
\begin{tabular}{c|c|c}
\hline
condensable algebra $\cA$ & reduced \symmTO\ $\eM_{/\cA}$ & stable low energy
state\\
\hline
$\one\oplus a_1\oplus 2a_2$ & $(\eGau_{S_3}^{(1)})_{/\one\oplus a_1\oplus a_2} =
$ trivial & $S_3$-symmetry breaking
gapped $\Z_1$-state 
\\
\hline
$\one\oplus a_2$ & $(\eGau_{S_3}^{(1)})_{/\one\oplus a_2} = \eM_\text{DS}$ &
$K=\begin{pmatrix}
2 &0\\
0&-2\\
\end{pmatrix}$ chiral boson theory \eq{Kphi}\\
\hline
$\one\oplus a_1$ & $(\eGau_{S_3}^{(1)})_{/\one\oplus a_1} = \eM_{K_{04;3}}$ 
& $K=\begin{pmatrix}
0 &3\\
3&4\\
\end{pmatrix}$ chiral boson theory \eq{Kphi}\\
\hline
$\one$ & $(\eGau_{S_3}^{(1)})_{/\one} = \eGau_{S_3}^{(1)}$ 
&
$(c,\bar c) = (9,9)$ CFT 
$so(9)_2 \times
u(1)_2 \times \bar{u}(1)_2\times\bar{E}(8)_1$   \eq{S3a1one} 
\\
\hline
\end{tabular}
\end{table*}

The  $\eGau_{S_3}^{(1)}$ topological order has anyons given by
\begin{align}
\begin{matrix}
\text{anyons}: & \one &  a_1  & a_2 &  b  &  b_1  &  b_2  & c  & c_1    \\
d_a: &  1 & 1 & 2 & 2 & 2 & 2 & 3 & 3 \\
s_a: & 0 & 0 & 0 & \frac19 & \frac49 & \frac79 & \frac14 & \frac34 \\
\end{matrix}
\end{align}
The potential condensable algebras of $\eGau_{S_3}^{(1)}$ topological order are
given by
\begin{align}
& \one\oplus a_1\oplus 2 a_2, &
& \one\oplus a_2, 
&
& \one\oplus a_1, &
& \one .
\end{align}
The condensable algebra $\one\oplus a_1\oplus 2a_2$ is Lagrangian, and gives
rise to a gapped state that break the $S_3^{(1)}$ symmetry completely.  This is
the only gapped state allowed by the anomalous $S_3^{(1)}$ symmetry.

The potential condensable algebra $\one\oplus a_2$ is not Lagrangian.  If it is
a valid condensable algebra, its condensation will induce a 2+1D topological
order, that has a canonical domain wall with $\eGau_{S_3}^{(1)}$.  Indeed, we
find a canonical domain wall between $\eGau_{S_3}^{(1)}$ and $\eM_\text{DS}$.
Here $\eM_\text{DS}$ is the double-semion topological order, which has
excitations $\one,b,s_+,s_-$ with spins $s_a = 0,0,\frac14,\frac34$.  The
canonical domain wall is given by
\begin{align}
\left (A^{ai}_{\eGau_{S_3}^{(1)}\mid \eM_\text{DS}}\right )=
\begin{pmatrix}
1&  0&  0&  0 &\one\\
0&  1&  0&  0 & a_1\\
1&  1&  0&  0 & a_2\\
0&  0&  0&  0 & b\\
0&  0&  0&  0 & b_1\\
0&  0&  0&  0 & b_2\\
0&  0&  0&  1 & c\\
0&  0&  1&  0 & c_1\\
\one&  b&  s_+&  s_- & \\
\end{pmatrix}
\end{align}
From the first row and the first column of $A^{ai}_{\eGau_{S_3}^{(1)}\mid
\eM_\text{DS}}$, we can see that $\eM_\text{DS}$ is induced from
$\eGau_{S_3}^{(1)}$ via a condensation of $\one\oplus a_2$.  This indicates
that $\one\oplus a_2$ is a valid condensable algebra, and its condensation
induced topological order is 
\begin{align}
(\eGau_{S_3}^{(1)})_{/\one\oplus a_2} = \eM_\text{DS}.
\end{align}

The $\one\oplus a_2$-state is the canonical boundary of $\eM_\text{DS}$, which
breaks the $S_3$ symmetry down to $\Z_2$ symmetry.  Despite the symmetry
breaking, such a state still must be gapless.  To see this, we note that
$\one\oplus a_2$-state actually breaks the anomalous $S_3^{(1)}$ symmetry down
to
anomalous $\Z_2^{(1)}$ symmetry (as implied by the double-semion topological
order $\eM_\text{DS}$).  Since the unbroken $\Z_2$ symmetry is anomalous, the
$\one\oplus a_2$-state must be gapless since it does not break the anomalous
$\Z_2$ symmetry.  Such a gapless state is described by the following Lagrangian
\begin{align}
\label{Kphi}
\cL &= \frac{(K^{-1})_{IJ}}{4\pi}
\prt_x \phi_I \prt_t \phi_J - V_{IJ} \prt_x \phi_I \prt_x \phi_J ,
\nonumber\\
u_I & = \ee^{\ii  \phi_I}
\text{ generate all local operators},
\end{align}
with
\begin{align}
\label{KDS}
K = (K_{IJ}) = \begin{pmatrix}
2 &0 \\
0 & -2 \\
\end{pmatrix}
\end{align}

The condensable algebra $\one\oplus a_1$ can induce a 2+1D topological order
$\eM_{K_{04;3}}$, where $\eM_{K_{04;3}}$ is an Abelian topological order
described by the $K$-matrix \cite{BW9045,FK9169,WZ9290}
\begin{align}
\label{K043}
K_{04;3} = \begin{pmatrix}
0 & 3\\
3 & 4\\
\end{pmatrix}.
\end{align}
The nine anyons in $\eM_{K_{04;3}}$ have the following  spins $ 0, 0, 0,
\frac19, \frac19, \frac49, \frac49, \frac79, \frac79$.  Indeed, we find a
canonical domain wall between $\eGau_{S_3}^{(1)}$ and $\eM_{K_{04;3}}$ given
by
\begin{align}
\left (A^{ai}_{\eGau_{S_3}^{(1)}\mid \eM_{K_{04;3}}}\right )=
\begin{pmatrix}
1 &  0 &  0 &  0 &  0 &  0 &  0 &  0 &  0 &\one\\
1 &  0 &  0 &  0 &  0 &  0 &  0 &  0 &  0 & a_1\\
0 &  1 &  1 &  0 &  0 &  0 &  0 &  0 &  0 & a_2\\
0 &  0 &  0 &  1 &  1 &  0 &  0 &  0 &  0 & b\\
0 &  0 &  0 &  0 &  0 &  1 &  1 &  0 &  0 & b_1\\
0 &  0 &  0 &  0 &  0 &  0 &  0 &  1 &  1 & b_2\\
0 &  0 &  0 &  0 &  0 &  0 &  0 &  0 &  0 & c\\
0 &  0 &  0 &  0 &  0 &  0 &  0 &  0 &  0 & c_1\\
\end{pmatrix}
\end{align}
From the first row and the first column of $A^{ai}_{\eGau_{S_3}^{(1)}\mid
\eM_{K_{04;3}}}$, we can
see that $\eM_{K_{04;3}}$ is induced from $\eGau_{S_3}^{(1)}$ via a
condensation of $\one\oplus a_1$.  This indicates that $\one\oplus a_1$ is a
valid condensable algebra, and its condensation-induced topological order is
\begin{align}
(\eGau_{S_3}^{(1)})_{/\one\oplus a_1} = \eM_{K_{04;3}}.
\end{align}

The $\one\oplus a_1$-state is the canonical boundary of $\eM_{K_{04;3}}$, which
breaks the anomalous $S_3^{(1)}$ symmetry down to anomalous $\Z_3^{(1)}$
symmetry (as indicated by its \symmTO\ $\eM_{K_{04;3}}$).  The
$\one\oplus a_1$-state must be gapless since it does not break the anomalous
$\Z_3$ symmetry.  Such a gapless state is described by the Lagrangian \eq{Kphi}
with $K$ given by \eqn{K043}.  

The $\one$-state is the canonical boundary of $\eGau_{S_3}^{(1)}$, which has
the full \symmTO\ $\eGau_{S_3}^{(1)}$.  Such a state must be gapless since it
has a nontrivial reduced \symmTO.  The gapless state is described by the
following multi-component partition function:
\begin{widetext}
\begin{align}
\label{S3a1one}
Z_{\one}^{\eGau_{S_3}^{(1)}} &= \chi^{so(9)_2 \times u(1)_2 \times
\bar{u}(1)_2\times\bar{E}(8)_1}_{1,0; 1,0; 1,0} +  \chi^{so(9)_2 \times u(1)_2
\times \bar{u}(1)_2\times\bar{E}(8)_1}_{2,1; 2,\frac{1}{4}; 2,-\frac{1}{4}} 
\nonumber \\ 
Z_{a_1}^{\eGau_{S_3}^{(1)}} &= \chi^{so(9)_2 \times u(1)_2 \times
\bar{u}(1)_2\times\bar{E}(8)_1}_{1,0; 2,\frac{1}{4}; 2,-\frac{1}{4}} + 
\chi^{so(9)_2 \times u(1)_2 \times \bar{u}(1)_2\times\bar{E}(8)_1}_{2,1; 1,0;
1,0} 
\nonumber \\ 
Z_{a_2}^{\eGau_{S_3}^{(1)}} &= \chi^{so(9)_2 \times u(1)_2 \times
\bar{u}(1)_2\times\bar{E}(8)_1}_{6,1; 1,0; 1,0} +  \chi^{so(9)_2 \times u(1)_2
\times \bar{u}(1)_2\times\bar{E}(8)_1}_{6,1; 2,\frac{1}{4}; 2,-\frac{1}{4}} 
\nonumber \\ 
Z_{b}^{\eGau_{S_3}^{(1)}} &= \chi^{so(9)_2 \times u(1)_2 \times
\bar{u}(1)_2\times\bar{E}(8)_1}_{5,\frac{10}{9}; 1,0; 1,0} +  \chi^{so(9)_2
\times u(1)_2 \times \bar{u}(1)_2\times\bar{E}(8)_1}_{5,\frac{10}{9};
2,\frac{1}{4}; 2,-\frac{1}{4}} 
\nonumber \\ 
Z_{b_1}^{\eGau_{S_3}^{(1)}} &= \chi^{so(9)_2 \times u(1)_2 \times
\bar{u}(1)_2\times\bar{E}(8)_1}_{8,\frac{4}{9}; 1,0; 1,0} +  \chi^{so(9)_2
\times u(1)_2 \times \bar{u}(1)_2\times\bar{E}(8)_1}_{8,\frac{4}{9};
2,\frac{1}{4}; 2,-\frac{1}{4}} 
\nonumber \\ 
Z_{b_2}^{\eGau_{S_3}^{(1)}} &= \chi^{so(9)_2 \times u(1)_2 \times
\bar{u}(1)_2\times\bar{E}(8)_1}_{7,\frac{7}{9}; 1,0; 1,0} +  \chi^{so(9)_2
\times u(1)_2 \times \bar{u}(1)_2\times\bar{E}(8)_1}_{7,\frac{7}{9};
2,\frac{1}{4}; 2,-\frac{1}{4}} 
\nonumber \\ 
Z_{c}^{\eGau_{S_3}^{(1)}} &= \chi^{so(9)_2 \times u(1)_2 \times
\bar{u}(1)_2\times\bar{E}(8)_1}_{3,\frac{1}{2}; 1,0; 2,-\frac{1}{4}} + 
\chi^{so(9)_2 \times u(1)_2 \times \bar{u}(1)_2\times\bar{E}(8)_1}_{4,1;
2,\frac{1}{4}; 1,0} 
\nonumber \\ 
Z_{c_1}^{\eGau_{S_3}^{(1)}} &= \chi^{so(9)_2 \times u(1)_2 \times
\bar{u}(1)_2\times\bar{E}(8)_1}_{3,\frac{1}{2}; 2,\frac{1}{4}; 1,0} + 
\chi^{so(9)_2 \times u(1)_2 \times \bar{u}(1)_2\times\bar{E}(8)_1}_{4,1; 1,0;
2,-\frac{1}{4}}.
\end{align}
\end{widetext}
Here $\chi^{CFT_1\times CFT_2 \times \cdots}_{a_1,h_1;a_2,h_2;\cdots}$ is
product of conformal characters of $CFT_i$ for the primary fields labeled by
$a_i$ with scaling dimension $h_i$.
For example
\begin{align}
\label{cterm}
&\ \ \ \
\chi^{so(9)_2 \times u(1)_2 \times \bar{u}(1)_2\times\bar{E}(8)_1}_{2,1;
2,\frac{1}{4}; 2,-\frac{1}{4}} 
\nonumber\\
&= 
\chi^{so(9)_2 }_{2,1}(\tau) 
\chi^{u(1)_2 }_{ 2,\frac{1}{4} }(\tau) 
\chi^{\bar{u}(1)_2}_{ 2,-\frac{1}{4}} (\bar \tau)
\chi^{\bar{E}(8)_1}(\bar \tau),
\end{align}
where $\chi^{so(9)_2 }_{2,1}(\tau)$ is the conformal character of $so(9)_2$
CFT, for the second primary field with scaling dimension $h=1$; $\chi^{u(1)_2
}_{2,\frac{1}{4}}(\tau)$ is the conformal character of $u(1)_2$ CFT (the chiral
boson theory described by $K$-matrix $K=(2)$) , for the second primary field
with scaling dimension $h=\frac14$; $\chi^{\bar u(1)_2
}_{2,-\frac{1}{4}}(\bar\tau)$ is the conformal character of $\bar u(1)_2$ CFT
(the anti-chiral boson theory described by $K$-matrix $K=(-2)$) , for the
second primary field with scaling dimension $h=\frac14$; $\chi^{\bar E(8)_1 }$
is the conformal character of $\bar E(8)_1$ CFT (the complex conjugate of
$E(8)$ level-1 Kac-Moody algebra).  The $\bar E(8)_1$ CFT has only one primary
field (the identity), whose index is suppressed.

Eq. \eqref{S3a1one} describes a gapless state that does not break the anomalous
$S_3^{(1)}$ symmetry (or more precisely,  does not maximally condense and trivialize the \symmTO\
$\eGau_{S_3}^{(1)}$).  The gapless state  is described by a $so(9)_2 \times
u(1)_2 \times \bar{u}(1)_2\times\bar{E}(8)_1$ CFT with central charge $(c, \bar
c)=(9,9)$.  Such a CFT is chiral, where right-movers and left-movers have
different dynamics.  In particular, the right-movers are described by a $so(9)$
level-2 CFT and a $U(1)$ level-2 CFT (\ie $K$-matrix $K=(2)$.  The left-movers
are described by a $U(1)$ level-2 CFT (\ie $K$-matrix $K=(-2)$ and a $E(8)$
level-1 CFT.  Such a combined CFT corresponds to a stable gapless phase, since
there is no $S_3^{(1)}$ symmetric relevant perturbations.  We remark that the
primary field for the conformal character \eqref{cterm} in $
Z_{\one}^{\eGau_{S_3}^{(1)}}$ appears to be a  symmetric relavent operator
since its scaling dimension $h+\bar h = 1 +\frac14 + \frac14 = \frac 32 < 2$. 
But this
operator has $h-\bar h =  1 +\frac14 - \frac14 =1$, and hence describes a chiral
operator.  We recall that a chiral operator, such as $\psi_R \psi_R'$ that
couples two free
right-moving fermions, cannot open an energy gap even when they are formally
relevant.  In this paper, we regard them as irrelevant.  Thus, the anomalous
$S_3^{(1)}$ symmetry can give rise to a symmetry protected chiral gapless
phase.  The gapped and gapless phases for systems with anomalous $S_3^{(1)}$
symmetry is summarized in Table \ref{S3a1}.

\subsection{Anomalous $S_3^{(2)}$ symmetry}

\begin{table*}[tb]
\caption{ Possible gapped and gapless states for systems with anomalous
$S_3^{(2)}$ symmetry.  } \label{S3a2} 
\setlength\extrarowheight{4pt}
\setlength{\tabcolsep}{6pt}
\centering
\begin{tabular}{c|c|c}
\hline
condensable algebra $\cA$ & reduced \symmTO\ $\eM_{/\cA}$ & stable low energy
state\\
\hline
$\one\oplus a_1\oplus 2a_2$ 
& $(\eGau_{S_3}^{(2)})_{/\one\oplus a_1\oplus 2a_2} = $ trivial 
& $S_3$-symmetry breaking gapped $\Z_1$-state \\
\hline
$\one\oplus a_2\oplus c$ 
& $(\eGau_{S_3}^{(2)})_{/\one\oplus a_2\oplus c} = $ trivial 
& gapped $\Z_2$-state \\
\hline
$\one\oplus a_2$ 
& $(\eGau_{S_3}^{(2)})_{/\one\oplus a_2} = \eGau_{\Z_2}$ & 
$(c,\bar c)= \left( \frac12,\frac12 \right)$ Ising CFT
\\
\hline
$\one\oplus a_1$ 
& $(\eGau_{S_3}^{(2)})_{/\one\oplus a_1} = \eM_{-K_{04;3}}$ 
& $K=-\begin{pmatrix}
0 &3\\
3&4\\
\end{pmatrix}$ chiral boson theory \eq{Kphi}\\
\hline
$\one$ & $(\eGau_{S_3}^{(2)})_{/\one} = \eGau_{S_3}^{(2)}$ 
&
$(c,\bar c)= (8,8)$ CFT $E(8)_1 \times \overline{so}(9)_2$  
\\
\hline
\end{tabular}
\end{table*}

The  $\eGau_{S_3}^{(2)}$ topological order has anyons given by
\begin{align}
\label{anyonS3a2}
\begin{matrix}
\text{anyons}: & \one &  a_1  & a_2 &  b  &  b_1  &  b_2  & c  & c_1    \\
d_a: &  1 & 1 & 2 & 2 & 2 & 2 & 3 & 3 \\
s_a: & 0 & 0 & 0 & \frac29 & \frac59 & \frac89 & 0 & \frac12 \\
\end{matrix}
\end{align}
The potential condensable algebras of $\eGau_{S_3}^{(2)}$ topological order are
given by
\begin{align}
& \one\oplus a_1\oplus 2a_2, &
& \one\oplus a_2\oplus c, 
&
& \one\oplus a_2, &
& \one\oplus a_1, &
& \one .
\end{align}
The condensable algebra $\one\oplus a_1\oplus 2a_2$ is Lagrangian, and gives
rise
to a gapped state that break the $S_3^{(2)}$ symmetry completely.  The
condensable algebra $\one\oplus a_2\oplus c$ is also Lagrangian, and gives rise
to
a gapped state that break the $S_3^{(2)}$ symmetry down to anomaly-free $\Z_2$
symmetry.  These are the only two gapped states allowed by the anomalous
$S_3^{(2)}$ symmetry.

The condensable algebra $\one\oplus a_2$ is not Lagrangian.  We find a canonical
domain wall between $\eGau_{S_3}^{(2)}$ and $\eGau_{\Z_2}$:
\begin{align}
\left (A^{ai}_{\eGau_{S_3}^{(2)}\mid \eGau_{\Z_2} } \right )=
\begin{pmatrix}
1&  0&  0&  0 &\one\\
0&  1&  0&  0 & a_1\\
1&  1&  0&  0 & a_2\\
0&  0&  0&  0 & b\\
0&  0&  0&  0 & b_1\\
0&  0&  0&  0 & b_2\\
0&  0&  1&  0 & c\\
0&  0&  0&  1 & c_1\\
\one&  e&  m&  f & \\
\end{pmatrix}
\end{align}
which tells us that $\eGau_{\Z_2}$ is induced from $\eGau_{S_3}^{(2)}$ via a
condensation of $\one\oplus a_2$.  The $\one\oplus a_2$-state is the canonical
boundary of $\eM_{\Z_2}$, which breaks the anomalous $S_3^{(2)}$ symmetry down
to $\Z_2\vee \t \Z_2$ symmetry.  Such a state must be gapless and is described
by $(c,\bar c)=\left( \frac12,\frac12 \right)$ Ising CFT.\cite{JW190513279}

The condensable algebra $\one\oplus a_1$ is not Lagrangian.  We find a canonical
domain wall between $\eGau_{S_3}^{(2)}$ and $\eM_{-K_{04;3}}$: 
\begin{align}
\left (A^{ai}_{\eGau_{S_3}^{(2)}\mid \eM_{-K_{04;3}}}\right )=
\begin{pmatrix}
1 &  0 &  0 &  0 &  0 &  0 &  0 &  0 &  0 &\one\\
1 &  0 &  0 &  0 &  0 &  0 &  0 &  0 &  0 & a_1\\
0 &  1 &  1 &  0 &  0 &  0 &  0 &  0 &  0 & a_2\\
0 &  0 &  0 &  0 &  0 &  0 &  0 &  1 &  1 & b\\
0 &  0 &  0 &  0 &  0 &  1 &  1 &  0 &  0 & b_1\\
0 &  0 &  0 &  1 &  1 &  0 &  0 &  0 &  0 & b_2\\
0 &  0 &  0 &  0 &  0 &  0 &  0 &  0 &  0 & c\\
0 &  0 &  0 &  0 &  0 &  0 &  0 &  0 &  0 & c_1\\
\end{pmatrix}
\end{align}
which suggests that $\eM_{-K_{04;3}}$ is induced from $\eGau_{S_3}^{(2)}$ via a
condensation of $\one\oplus a_1$.  The $\one\oplus a_1$-state is the canonical
boundary of $\eM_{-K_{04;3}}$, which breaks the anomalous $S_3^{(2)}$ symmetry
down to anomalous $\Z_3^{(1)}$ symmetry.
The $\one\oplus a_1$-state must be
gapless since it does not break the anomalous $\Z_3$ symmetry.  Such a gapless
state is described by the Lagrangian \eq{Kphi} with $K$ given by the negative of
the $K$-matrix in \eqn{K043}.

The gapless $\one$-state is a canonical boundary of $\eGau_{S_3}^{(2)}$.  What
are the properties of such a gapless state?  It turns out that a canonical
boundary  of $\eGau_{S_3}^{(2)}$ is given by the $(c,\bar c)= (8,8)$ $E(8)_1
\times \overline{so}(9)_2$ CFT. In other words, the right movers are described
by $E(8)_1$ current algebra and the left movers are described by $so(9)_2$
current algebra.  The $E(8)_1$ current algebra has only one conformal character
which is modular invariant.  The $so(9)_2$ current algebra has $(c,\bar c)
=(0,8)$ and 8 conformal characters with the following quantum dimensions
($d_a$) and scaling dimensions ($\bar h_a$)
\begin{align}
\begin{matrix}
\text{characters}: & \one &  \bar a_1  &\bar  a_2  & \bar b & \bar b_1 &  \bar
b_2  &  \bar c &  \bar c_1      \\
d_a: &  1 & 1 & 2 & 2 & 2 & 2 & 3 & 3 \\
\bar h_a: & 0 & 1 & 1 & \frac79 & \frac{4}9 & \frac{10}9 & 0 & \frac12 \\
\end{matrix}
\end{align}
The above quantum dimensions $d_a$ and scaling dimensions ( $-\bar h_a$ mod 1) 
exactly match
those of anyons in $\eGau_{S_3}^{(2)}$ (see \eqn{anyonS3a2}).  Thus the 8
conformal characters of $so(9)_2$ transform according to the $S,T$-matrices of
$\eGau_{S_3}^{(2)}$.  We add the $E(8)_1$ to make $(c,\bar c) =(8,8)$.  This
matches the central charge of $\eGau_{S_3}^{(2)}$ that satisfies $c=\bar c$.
This is why the $E(8)_1 \times \overline{so}(9)_2$ CFT is a canonical boundary
of $\eGau_{S_3}^{(2)}$.  In particular, the multi-component partition function
for the canonical boundary is given by
\begin{align}
Z_{\one}^{\eGau_{S_3}^{(2)}} &= \chi^{{E(8)_1\times \overline{so}(9)_2}}_{1,0} 
\nonumber \\ 
Z_{a_1}^{\eGau_{S_3}^{(2)}} &= \chi^{{E(8)_1\times \overline{so}(9)_2}}_{2,-1} 
\nonumber \\ 
Z_{a_2}^{\eGau_{S_3}^{(2)}} &= \chi^{{E(8)_1\times \overline{so}(9)_2}}_{6,-1} 
\nonumber \\ 
Z_{b}^{\eGau_{S_3}^{(2)}} &= \chi^{{E(8)_1\times
\overline{so}(9)_2}}_{7,-\frac{7}{9}} 
\nonumber \\ 
Z_{b_1}^{\eGau_{S_3}^{(2)}} &= \chi^{{E(8)_1\times
\overline{so}(9)_2}}_{8,-\frac{4}{9}} 
\nonumber \\ 
Z_{b_2}^{\eGau_{S_3}^{(2)}} &= \chi^{{E(8)_1\times
\overline{so}(9)_2}}_{5,-\frac{10}{9}} 
\nonumber \\ 
Z_{c}^{\eGau_{S_3}^{(2)}} &= \chi^{{E(8)_1\times \overline{so}(9)_2}}_{4,-1} 
\nonumber \\ 
Z_{c_1}^{\eGau_{S_3}^{(2)}} &= \chi^{{E(8)_1\times
\overline{so}(9)_2}}_{3,-\frac{1}{2}} 
\end{align}

We would like to point out that the $E(8)_1 \times \overline{so}(9)_2$ CFT has
no $S_3^{(2)}$ symmetric relevant operators, since the $\one$-component of the
multi-component partition function is given by $Z_\one^{\eGau_{S_3}^{(2)}} = 
\chi^{{E(8)_1\times \overline{so}(9)_2}}_{1,0}=
\chi^{E(8)_1}(\tau) \bar \chi^{so(9)_2}_{1,0}(\bar \tau)$.  Apart from the
identity operator (the primary field with $(h,\bar h)=(0,0)$), other non-chiral
operators (the descendant fields of the current algebra) in this sector have
scaling dimensions at least $(h,\bar h)=(1,1)$.  The operators are at most
marginal.  Thus the gapless $\one$-state with the full \symmTO\
$\eGau_{S_3}^{(2)}$ is a stable gapless phase that has no unstable direction.
The gapped and gapless phases for systems with anomalous $S_3^{(2)}$ symmetry
is summarized in Table \ref{S3a2}.

\subsection{Anomalous $S_3^{(3)}$ symmetry}

\begin{table*}[tb]
\caption{ Possible gapped and gapless states for systems with anomalous
$S_3^{(3)}$ symmetry.  } \label{S3a3} 
\setlength\extrarowheight{4pt}
\setlength{\tabcolsep}{6pt}
\centering
\begin{tabular}{c|c|c}
\hline
condensable algebra $\cA$ & reduced \symmTO\ $\eM_{/\cA}$ & stable low energy
state\\
\hline
$\one\oplus a_1\oplus 2a_2$ 
& $(\eGau_{S_3}^{(3)})_{/\one\oplus a_1\oplus 2a_2} = $ trivial 
& $S_3$-symmetry breaking gapped $\Z_1$-state \\
\hline
$\one\oplus a_1\oplus 2b$ 
& $(\eGau_{S_3}^{(3)})_{/\one\oplus a_1\oplus 2b} = $ trivial 
& gapped $\Z_3$-state \\
\hline
$\one\oplus a_2$ 
& $(\eGau_{S_3}^{(3)})_{/\one\oplus a_2} = \eM_\text{DS}$ & 
$K=\begin{pmatrix}
2 &0\\
0&-2\\
\end{pmatrix}$ chiral boson theory \eq{Kphi}
\\
\hline
$\one\oplus b$ 
& $(\eGau_{S_3}^{(3)})_{/\one\oplus b} = \eM_\text{DS}$ & 
$K=\begin{pmatrix}
2 &0\\
0&-2\\
\end{pmatrix}$ chiral boson theory \eq{Kphi}
\\
\hline
$\one\oplus a_1$ 
& $(\eGau_{S_3}^{(3)})_{/\one\oplus a_1} = \eGau_{\Z_3}$ 
&  $(c,\bar c)= \left( \frac45,\frac45 \right)$ CFT \eq{Z3bdy} \\
\hline
$\one$ & $(\eGau_{S_3}^{(3)})_{/\one} = \eGau_{S_3}^{(3)}$ 
&
$(c, \bar c)=(\frac95,\frac95)$ CFT
$m6 \times u(1)_2
\times \bar m6 \times \bar u(1)_2 $ \eq{S3a3one}
\\
\hline
\end{tabular}
\end{table*}

The  $\eGau_{S_3}^{(3)}$ topological order has anyons given by
\begin{align}
\begin{matrix}
\text{anyons}: & \one &  a_1  & a_2 &  b  &  b_1  &  b_2  & c  & c_1    \\
d_a: &  1 & 1 & 2 & 2 & 2 & 2 & 3 & 3 \\
s_a: & 0 & 0 & 0 & 0 & \frac13 & \frac23 & \frac14 & \frac34 \\
\end{matrix}
\end{align}
The potential condensable algebras of $\eGau_{S_3}^{(3)}$ topological order are
given by
\begin{align}
& \one\oplus a_1\oplus 2a_2, &
& \one\oplus a_1\oplus 2b, 
\nonumber\\
& \one\oplus a_2, &
& \one\oplus b, &
& \one\oplus a_1, &
& \one .
\end{align}
The condensable algebra $\one\oplus a_1\oplus 2a_2$ is Lagrangian, and gives
rise to a gapped state that break the $S_3^{(3)}$ symmetry completely.  The
condensable algebra $\one\oplus a_1\oplus 2b$ is also Lagrangian, and gives
rise to a gapped state that break the $S_3^{(3)}$ symmetry down to anomaly-free
$\Z_3$ symmetry.  These are the only two gapped states allowed by the anomalous
$S_3^{(3)}$ symmetry.

The condensable algebra $\one\oplus a_2$ is not Lagrangian.  We find a
canonical domain wall between $\eGau_{S_3}^{(3)}$ and $\eM_\text{DS}$:
\begin{align}
\left (A^{ai}_{\eGau_{S_3}^{(3)}\mid \eM_\text{DS}}\right )=
\begin{pmatrix}
1&  0&  0&  0 &\one\\
0&  1&  0&  0 & a_1\\
1&  1&  0&  0 & a_2\\
0&  0&  0&  0 & b\\
0&  0&  0&  0 & b_1\\
0&  0&  0&  0 & b_2\\
0&  0&  1&  0 & c\\
0&  0&  0&  1 & c_1\\
\one&  b&  s_+&  s_- & \\
\end{pmatrix}
\end{align}
which tells us that $\eM_\text{DS}$ is induced from $\eGau_{S_3}^{(3)}$ via a
condensation of $\one\oplus a_2$.  The $\one\oplus a_2$-state is the canonical
boundary of $\eM_\text{DS}$, which breaks the anomalous $S_3^{(3)}$ symmetry
down to anomalous $\Z_2^{(1)}$ symmetry.  Such a state must be gapless and is
described by the Lagrangian \eq{Kphi} with $K$ given by \eqn{KDS}.

The $\one\oplus b$ condensation is similar to the $\one\oplus a_2$ condensation
discussed above, due to a $a_2\leftrightarrow b$ automorphism of
$\eGau_{S_3}^{(3)}$ topological order.  The $\one\oplus b$-state has the full
anomalous $S_3^{(3)}$ symmetry where the $a_2$ excitations are gapped (\ie the
$S_3$ charges, carrying the 2-dimensional representation, are gapped).  Such a
gapless state is described by the Lagrangian \eq{Kphi} with $K$ given by
\eqn{KDS}.  We note that $\one$-state also has the full anomalous $S_3^{(3)}$
symmetry. But in $\one$-state, the $a_2$ excitations are gapless.  So the
$\one$-state and $\one\oplus b$-state actually have different symmetry breaking
patterns, despite both states have the full  anomalous $S_3^{(3)}$ symmetry.

The condensable algebra $\one\oplus a_1$ is not Lagrangian.  We find a canonical
domain wall between $\eGau_{S_3}^{(3)}$ and $\eGau_{\Z_3}$: 
\begin{align}
\left  (A^{ai}_{\eGau_{S_3}^{(3)}\mid \eGau_{\Z_3} }\right )=
\begin{pmatrix}
1 &  0 &  0 &  0 &  0 &  0 &  0 &  0 &  0 &\one\\
1 &  0 &  0 &  0 &  0 &  0 &  0 &  0 &  0 & a_1\\
0 &  1 &  1 &  0 &  0 &  0 &  0 &  0 &  0 & a_2\\
0 &  0 &  0 &  1 &  1 &  0 &  0 &  0 &  0 & b\\
0 &  0 &  0 &  0 &  0 &  1 &  1 &  0 &  0 & b_1\\
0 &  0 &  0 &  0 &  0 &  0 &  0 &  1 &  1 & b_2\\
0 &  0 &  0 &  0 &  0 &  0 &  0 &  0 &  0 & c\\
0 &  0 &  0 &  0 &  0 &  0 &  0 &  0 &  0 & c_1\\
\end{pmatrix}
\end{align}
which suggests that $\eGau_{\Z_3}$ is induced from $\eGau_{S_3}^{(3)}$ via a
condensation of $\one\oplus a_1$.  The $\one\oplus a_1$-state is the canonical
boundary of $\eGau_{\Z_3}$, which breaks the anomalous $S_3^{(3)}$ symmetry
down to $\Z_3\vee \t \Z_3$ symmetry.  The $\one\oplus a_1$-state must be
gapless.  Such a gapless state is described by the $(c,\bar
c)=\left( \frac45,\frac45 \right)$ CFT constructed from (6,5) minimal model (see
\eqn{Z3bdy}).

The $\one$-state is the canonical boundary of $\eGau_{S_3}^{(3)}$, which has
the full \symmTO\ $\eGau_{S_3}^{(3)}$, which is gapless.  The gapless state is
described by the following multi-component partition function:
\begin{widetext}
\begingroup
\allowdisplaybreaks
\begin{align}
\label{S3a3one}
Z_{\one}^{\eGau_{S_3}^{(3)}} &= \chi^{m6 \times u(1)_2 \times \bar m6 \times
\bar u(1)_2}_{1,0; 1,0; 1,0; 1,0} +  \chi^{m6 \times u(1)_2 \times \bar m6
\times \bar u(1)_2}_{1,0; 2,\frac{1}{4}; 5,-3; 2,-\frac{1}{4}} +  \chi^{m6
\times u(1)_2 \times \bar m6 \times \bar u(1)_2}_{5,3; 1,0; 5,-3; 1,0} + 
\chi^{m6 \times u(1)_2 \times \bar m6 \times \bar u(1)_2}_{5,3; 2,\frac{1}{4};
1,0; 2,-\frac{1}{4}} 
\nonumber\\
& \ \ \ \
+  \chi^{m6 \times u(1)_2 \times \bar m6 \times \bar u(1)_2}_{6,\frac{2}{5};
1,0; 6,-\frac{2}{5}; 1,0} +  \chi^{m6 \times u(1)_2 \times \bar m6 \times \bar
u(1)_2}_{6,\frac{2}{5}; 2,\frac{1}{4}; 10,-\frac{7}{5}; 2,-\frac{1}{4}} + 
\chi^{m6 \times u(1)_2 \times \bar m6 \times \bar u(1)_2}_{10,\frac{7}{5}; 1,0;
10,-\frac{7}{5}; 1,0} +  \chi^{m6 \times u(1)_2 \times \bar m6 \times \bar
u(1)_2}_{10,\frac{7}{5}; 2,\frac{1}{4}; 6,-\frac{2}{5}; 2,-\frac{1}{4}} 
\nonumber \\ 
Z_{a_1}^{\eGau_{S_3}^{(3)}} &= \chi^{m6 \times u(1)_2 \times \bar m6 \times \bar
u(1)_2}_{1,0; 1,0; 5,-3; 1,0} +  \chi^{m6 \times u(1)_2 \times \bar m6 \times
\bar u(1)_2}_{1,0; 2,\frac{1}{4}; 1,0; 2,-\frac{1}{4}} +  \chi^{m6 \times u(1)_2
\times \bar m6 \times \bar u(1)_2}_{5,3; 1,0; 1,0; 1,0} +  \chi^{m6 \times
u(1)_2 \times \bar m6 \times \bar u(1)_2}_{5,3; 2,\frac{1}{4}; 5,-3;
2,-\frac{1}{4}}
\nonumber\\
& \ \ \ \
+  \chi^{m6 \times u(1)_2 \times \bar m6 \times \bar u(1)_2}_{6,\frac{2}{5};
1,0; 10,-\frac{7}{5}; 1,0} +  \chi^{m6 \times u(1)_2 \times \bar m6 \times \bar
u(1)_2}_{6,\frac{2}{5}; 2,\frac{1}{4}; 6,-\frac{2}{5}; 2,-\frac{1}{4}} + 
\chi^{m6 \times u(1)_2 \times \bar m6 \times \bar u(1)_2}_{10,\frac{7}{5}; 1,0;
6,-\frac{2}{5}; 1,0} +  \chi^{m6 \times u(1)_2 \times \bar m6 \times \bar
u(1)_2}_{10,\frac{7}{5}; 2,\frac{1}{4}; 10,-\frac{7}{5}; 2,-\frac{1}{4}} 
\nonumber \\ 
Z_{a_2}^{\eGau_{S_3}^{(3)}} &= \chi^{m6 \times u(1)_2 \times \bar m6 \times \bar
u(1)_2}_{3,\frac{2}{3}; 1,0; 3,-\frac{2}{3}; 1,0} +  \chi^{m6 \times u(1)_2
\times \bar m6 \times \bar u(1)_2}_{3,\frac{2}{3}; 2,\frac{1}{4};
3,-\frac{2}{3}; 2,-\frac{1}{4}} +  \chi^{m6 \times u(1)_2 \times \bar m6 \times
\bar u(1)_2}_{8,\frac{1}{15}; 1,0; 8,-\frac{1}{15}; 1,0} +  \chi^{m6 \times
u(1)_2 \times \bar m6 \times \bar u(1)_2}_{8,\frac{1}{15}; 2,\frac{1}{4};
8,-\frac{1}{15}; 2,-\frac{1}{4}} 
\nonumber \\ 
Z_{b}^{\eGau_{S_3}^{(3)}} &= \chi^{m6 \times u(1)_2 \times \bar m6 \times \bar
u(1)_2}_{3,\frac{2}{3}; 1,0; 3,-\frac{2}{3}; 1,0} +  \chi^{m6 \times u(1)_2
\times \bar m6 \times \bar u(1)_2}_{3,\frac{2}{3}; 2,\frac{1}{4};
3,-\frac{2}{3}; 2,-\frac{1}{4}} +  \chi^{m6 \times u(1)_2 \times \bar m6 \times
\bar u(1)_2}_{8,\frac{1}{15}; 1,0; 8,-\frac{1}{15}; 1,0} +  \chi^{m6 \times
u(1)_2 \times \bar m6 \times \bar u(1)_2}_{8,\frac{1}{15}; 2,\frac{1}{4};
8,-\frac{1}{15}; 2,-\frac{1}{4}} 
\nonumber \\ 
Z_{b_1}^{\eGau_{S_3}^{(3)}} &= \chi^{m6 \times u(1)_2 \times \bar m6 \times \bar
u(1)_2}_{1,0; 1,0; 3,-\frac{2}{3}; 1,0} +  \chi^{m6 \times u(1)_2 \times \bar m6
\times \bar u(1)_2}_{1,0; 2,\frac{1}{4}; 3,-\frac{2}{3}; 2,-\frac{1}{4}} + 
\chi^{m6 \times u(1)_2 \times \bar m6 \times \bar u(1)_2}_{5,3; 1,0;
3,-\frac{2}{3}; 1,0} +  \chi^{m6 \times u(1)_2 \times \bar m6 \times \bar
u(1)_2}_{5,3; 2,\frac{1}{4}; 3,-\frac{2}{3}; 2,-\frac{1}{4}}
\nonumber\\
& \ \ \ \
+  \chi^{m6 \times u(1)_2 \times \bar m6 \times \bar u(1)_2}_{6,\frac{2}{5};
1,0; 8,-\frac{1}{15}; 1,0} +  \chi^{m6 \times u(1)_2 \times \bar m6 \times \bar
u(1)_2}_{6,\frac{2}{5}; 2,\frac{1}{4}; 8,-\frac{1}{15}; 2,-\frac{1}{4}} + 
\chi^{m6 \times u(1)_2 \times \bar m6 \times \bar u(1)_2}_{10,\frac{7}{5}; 1,0;
8,-\frac{1}{15}; 1,0} +  \chi^{m6 \times u(1)_2 \times \bar m6 \times \bar
u(1)_2}_{10,\frac{7}{5}; 2,\frac{1}{4}; 8,-\frac{1}{15}; 2,-\frac{1}{4}} 
\nonumber \\ 
Z_{b_2}^{\eGau_{S_3}^{(3)}} &= \chi^{m6 \times u(1)_2 \times \bar m6 \times \bar
u(1)_2}_{3,\frac{2}{3}; 1,0; 1,0; 1,0} +  \chi^{m6 \times u(1)_2 \times \bar m6
\times \bar u(1)_2}_{3,\frac{2}{3}; 1,0; 5,-3; 1,0} +  \chi^{m6 \times u(1)_2
\times \bar m6 \times \bar u(1)_2}_{3,\frac{2}{3}; 2,\frac{1}{4}; 1,0;
2,-\frac{1}{4}} +  \chi^{m6 \times u(1)_2 \times \bar m6 \times \bar
u(1)_2}_{3,\frac{2}{3}; 2,\frac{1}{4}; 5,-3; 2,-\frac{1}{4}} 
\nonumber\\
& \ \ \ \
+  \chi^{m6 \times u(1)_2 \times \bar m6 \times \bar u(1)_2}_{8,\frac{1}{15};
1,0; 6,-\frac{2}{5}; 1,0} +  \chi^{m6 \times u(1)_2 \times \bar m6 \times \bar
u(1)_2}_{8,\frac{1}{15}; 1,0; 10,-\frac{7}{5}; 1,0} +  \chi^{m6 \times u(1)_2
\times \bar m6 \times \bar u(1)_2}_{8,\frac{1}{15}; 2,\frac{1}{4};
6,-\frac{2}{5}; 2,-\frac{1}{4}} +  \chi^{m6 \times u(1)_2 \times \bar m6 \times
\bar u(1)_2}_{8,\frac{1}{15}; 2,\frac{1}{4}; 10,-\frac{7}{5}; 2,-\frac{1}{4}} 
\nonumber \\ 
Z_{c}^{\eGau_{S_3}^{(3)}} &= \chi^{m6 \times u(1)_2 \times \bar m6 \times \bar
u(1)_2}_{2,\frac{1}{8}; 1,0; 4,-\frac{13}{8}; 2,-\frac{1}{4}} +  \chi^{m6 \times
u(1)_2 \times \bar m6 \times \bar u(1)_2}_{2,\frac{1}{8}; 2,\frac{1}{4};
2,-\frac{1}{8}; 1,0} +  \chi^{m6 \times u(1)_2 \times \bar m6 \times \bar
u(1)_2}_{4,\frac{13}{8}; 1,0; 2,-\frac{1}{8}; 2,-\frac{1}{4}} +  \chi^{m6 \times
u(1)_2 \times \bar m6 \times \bar u(1)_2}_{4,\frac{13}{8}; 2,\frac{1}{4};
4,-\frac{13}{8}; 1,0} 
\nonumber\\
& \ \ \ \
+  \chi^{m6 \times u(1)_2 \times \bar m6 \times \bar u(1)_2}_{7,\frac{1}{40};
1,0; 9,-\frac{21}{40}; 2,-\frac{1}{4}} +  \chi^{m6 \times u(1)_2 \times \bar m6
\times \bar u(1)_2}_{7,\frac{1}{40}; 2,\frac{1}{4}; 7,-\frac{1}{40}; 1,0} + 
\chi^{m6 \times u(1)_2 \times \bar m6 \times \bar u(1)_2}_{9,\frac{21}{40}; 1,0;
7,-\frac{1}{40}; 2,-\frac{1}{4}} +  \chi^{m6 \times u(1)_2 \times \bar m6 \times
\bar u(1)_2}_{9,\frac{21}{40}; 2,\frac{1}{4}; 9,-\frac{21}{40}; 1,0} 
\nonumber \\ 
Z_{c_1}^{\eGau_{S_3}^{(3)}} &= \chi^{m6 \times u(1)_2 \times \bar m6 \times \bar
u(1)_2}_{2,\frac{1}{8}; 1,0; 2,-\frac{1}{8}; 2,-\frac{1}{4}} +  \chi^{m6 \times
u(1)_2 \times \bar m6 \times \bar u(1)_2}_{2,\frac{1}{8}; 2,\frac{1}{4};
4,-\frac{13}{8}; 1,0} +  \chi^{m6 \times u(1)_2 \times \bar m6 \times \bar
u(1)_2}_{4,\frac{13}{8}; 1,0; 4,-\frac{13}{8}; 2,-\frac{1}{4}} +  \chi^{m6
\times u(1)_2 \times \bar m6 \times \bar u(1)_2}_{4,\frac{13}{8}; 2,\frac{1}{4};
2,-\frac{1}{8}; 1,0} 
\nonumber\\
& \ \ \ \
+  \chi^{m6 \times u(1)_2 \times \bar m6 \times \bar u(1)_2}_{7,\frac{1}{40};
1,0; 7,-\frac{1}{40}; 2,-\frac{1}{4}} +  \chi^{m6 \times u(1)_2 \times \bar m6
\times \bar u(1)_2}_{7,\frac{1}{40}; 2,\frac{1}{4}; 9,-\frac{21}{40}; 1,0} + 
\chi^{m6 \times u(1)_2 \times \bar m6 \times \bar u(1)_2}_{9,\frac{21}{40}; 1,0;
9,-\frac{21}{40}; 2,-\frac{1}{4}} +  \chi^{m6 \times u(1)_2 \times \bar m6
\times \bar u(1)_2}_{9,\frac{21}{40}; 2,\frac{1}{4}; 7,-\frac{1}{40}; 1,0} 
\end{align}
\endgroup
\end{widetext}
Eq. \eqref{S3a3one} describes a gapless state that does not break the anomalous
$S_3^{(3)}$ symmetry.  The gapless state  is described by a $m6 \times u(1)_2
\times \bar m6 \times \bar u(1)_2 $ CFT with central charge $(c, \bar
c)=(\frac95,\frac95)$.  Such a CFT is non-chiral, where right-movers and
left-movers have the dynamics.  In particular, the right-movers (and
left-movers) are described by a $(6,5)$ minimal model CFT (denoted as $m6$)
and a $U(1)$ level-2 CFT (\ie $K$-matrix $K=(2)$).  Such a CFT corresponds a
gapless state with one $S_3^{(3)}$ symmetric relevant operator.  Thus, the CFT
may describe a stable continuous phase transition.  The gapped and gapless
phases for systems with anomalous $S_3^{(1)}$ symmetry is summarized in Table
\ref{S3a3}.

\section{Summary}

It is well known that symmetry and anomaly can constrain the low energy
properties of quantum systems.  However, even given a symmetry and/or an
anomaly, there still can be a lot of allowed possible low energy properties,
which are hard to organize and hard to understand.  In this paper, we used
Symm/TO correspondence proposed in
\Rf{JW190513279,JW191213492,KZ200514178,CW220303596}, to view symmetry and
anomaly from a new point of view, and also to place them in a more generalized
framework. This allows us to organize the low energy properties according to
the condensation patterns and their reduced \symmTO.  These patterns of
condensations can describe, in a unified way, symmetry breaking phases, symmetry
enriched topological phases, symmetry protected topological phases, and gapless
critical points connecting these phases.  These patterns of \symmTO\
reductions, and the associated gapped/gapless phases, are classified by the
condensable algebras $\cA$ in the \symmTO\ $\eM$.  

In order to similarly study phases and symmetry in $n$-dimensional space for $n
> 1$, the theory of condensable algebra needs to be further developed.  In some
sense, a condensable algebra should correspond to an $n$-dimensional domain
wall in a topological order in $(n+1)$-dimensional space, which describes a
symmetry for a quantum system in $n$ spatial dimensions.  These  domain walls
are necessarily descendant excitations (\ie formed by the condensation of
$(n-1)$-dimensional, $(n-2)$-dimensional, \etc. excitations).  Under such a
generalization of condensable algebra, one must also include topological orders
in $n$-dimensional space without any symmetry. This is because condensation of
trivial excitations in the \symmTO\ (\ie topological order in
$(n+1)$-dimensional space) can give rise to topological order in
$n$-dimensional space.  Condensation of nontrivial excitations, on the other
hand, can give rise to symmetry enriched topological order in $n$-dimensional
space. In this way, an appropriately  generalized analogue of condensable
algebra should be able to describe symmetry-enriched topologically ordered
gapped phases of quantum systems in $ n >1$ spatial dimensions.

For gapless states, the possible low energy properties with a reduced \symmTO\
$\eM_{/\cA}$ are the same as the possible low energy properties of the
$\one$-condensed boundary of $\eM_{/\cA}$. In the language of bulk topological
order, this refers to the $\one$-condensed boundary of the topological order
induced from $\eM$ via the condensation of $\cA$, which we denote by
$\eM_{/\cA}$.  We find that possible low energy properties, such as scaling
dimensions, are determined by the reduced \symmTO\ $\eM_{/\cA}$, and can be
computed using an algebraic number theoretical method. 

Different condensable algebras $\cA$'s of $\eM$ can give rise to the same
reduced \symmTO\ $\eM_{/\cA}$, which implies that different patterns of
condensation associated to a \symmTO\ can give rise to the same set of low
energy properties. This allows us to show that some seemingly different
continuous quantum phase transitions are described by the same critical theory.
It appears that Symm/TO correspondence is a powerful way to use \symmTO\ (also
referred to as \catsymm\ before) to study, or even to classify, gapless quantum
states and the associated quantum field theories (up to local low-energy
equivalence). In higher than 1+1D, similar techniques are lacking, partly due
to a lack of systematic understanding of gapped boundaries of 3+1D and higher
topological orders. This constitutes one major direction for future research.

~

We thank Liang Kong, Ethan Lake, Ho Tat Lam, Ryan Lanzetta, Richard Ng,
Salvatore Pace, Eric Rowell, Nathan Seiberg, Rokas Veitas, and Hao Zheng for
many enlightening discussions, and Nathanan Tantivasadakarn for a critical
reading of the manuscript. This work is partially supported by NSF DMR-2022428
and by the Simons Collaboration on Ultra-Quantum Matter, which is a grant from
the Simons Foundation (651446, XGW). 

\appendix

\section{Some remarks on the term \catsymm}
\label{name}

\begin{figure}[t]
\begin{center}
\includegraphics[scale=0.6]{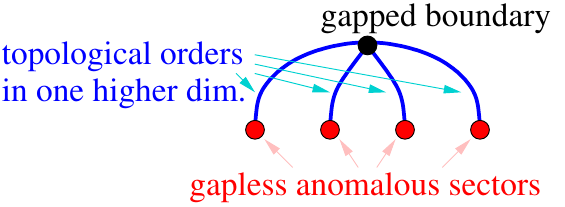}
\end{center}
\caption{ 
Symm/TO correspondence where (emergent) symmetry is viewed as anomaly
(Fig. 1 in \Rf{JW191209391}).
} \label{asect} 
\end{figure}

To address some comments from referee about the term \catsymm, we make some
remarks here.  The term \catsymm\ was introduced in 2019 \cite{JW191213492},
which is a way to describe a symmetry by viewing it as a (non-invertible)
gravitational anomaly \cite{JW190513279}, or by including both symmetry charges
and symmetry defects \emph{at an equal footing}.  We stress that \emph{at equal
footing} is the key here.  If we only include symmetry charges, and use the
fusion ring (\ie conservation law) of symmetry charges to describe the
symmetry, it will lead to a group theory (or fusion category) description of
symmetry.\footnote{We do not need to include the ``braiding'' properties of
symmetry charges, since their are always trivial for anomaly-free and anomalous
symmetries.} If we only use symmetry transformations (or symmetry defects) to
describe the symmetry, it will also lead to a group theory (or fusion category
\cite{PZh0011021,CSh0107001,FSh0204148,FSh0607247,DR11070495,CY180204445,TW191202817,I210315588,Q200509072})
description of symmetry.  On the other hand, if we include both symmetry
charges and symmetry defects at an equal footing, and use the fusion rings (\ie
conservation laws) of symmetry charges and symmetry defects to describe the
symmetry, we find that we also need to include the ``braiding'' properties of
symmetry charges and symmetry defects.  Thus, we need to use non-invertible
gappable-boundary topological order in one higher dimension (called \symmTO),
or more precisely, ``non-degenerate braided fusion $n$-category in trivial Witt
class'' to describe such a structure, if the system is in $n$-dimensional
space.  This way, the \catsymm\ point of view leads to Symm/TO correspondence
\cite{JW190513279,JW191213492} (see Fig.  \ref{asect}).  

In an earlier work \Rf{KZ170501087}, \catsymm\ appeared in 1+1D CFT as the
ambient category of enriched fusion category of all the topological defect
lines and is, at the same time, the category of modules over a chiral or
non-chiral symmetry (\ie a VOA or a full field algebra).  Topological field
theory (TFT) in one higher dimension was also used  in \Rf{FT180600008} to
discuss a duality relation in 1+1D Ising model. 

Later,  ``categorical symmetry'' was also used to refer to ``non-invertible
symmetry'' (which was called ``fusion category symmetry'', a term first
introduced also in 2019 \cite{TW191202817}).  ``Simons Collaboration on Global
Categorical Symmetries'' founded in 2021 used the term ``categorical
symmetry'' with ``non-invertible'' meaning.

In 2021, motivated by \Rf{FreedSymm}, \Rf{AS211202092} introduced ``symmetry
TFT'', which is closely related to  ``\symmTO''.  A possible difference is
that, for example, in 2+1D, the $\Z_2\times \Z_2$-DW theory and the
$\Z_4$-gauge theory are usually regarded as different field theories.  Thus
they may be viewed as different TFT's, but they correspond to the same \symmTO.
In other words, symmetry TFT may carry extra information about the field theory
representations, which is not needed here.

\begin{figure}[t]
\begin{center}
\includegraphics[height=1.0in]{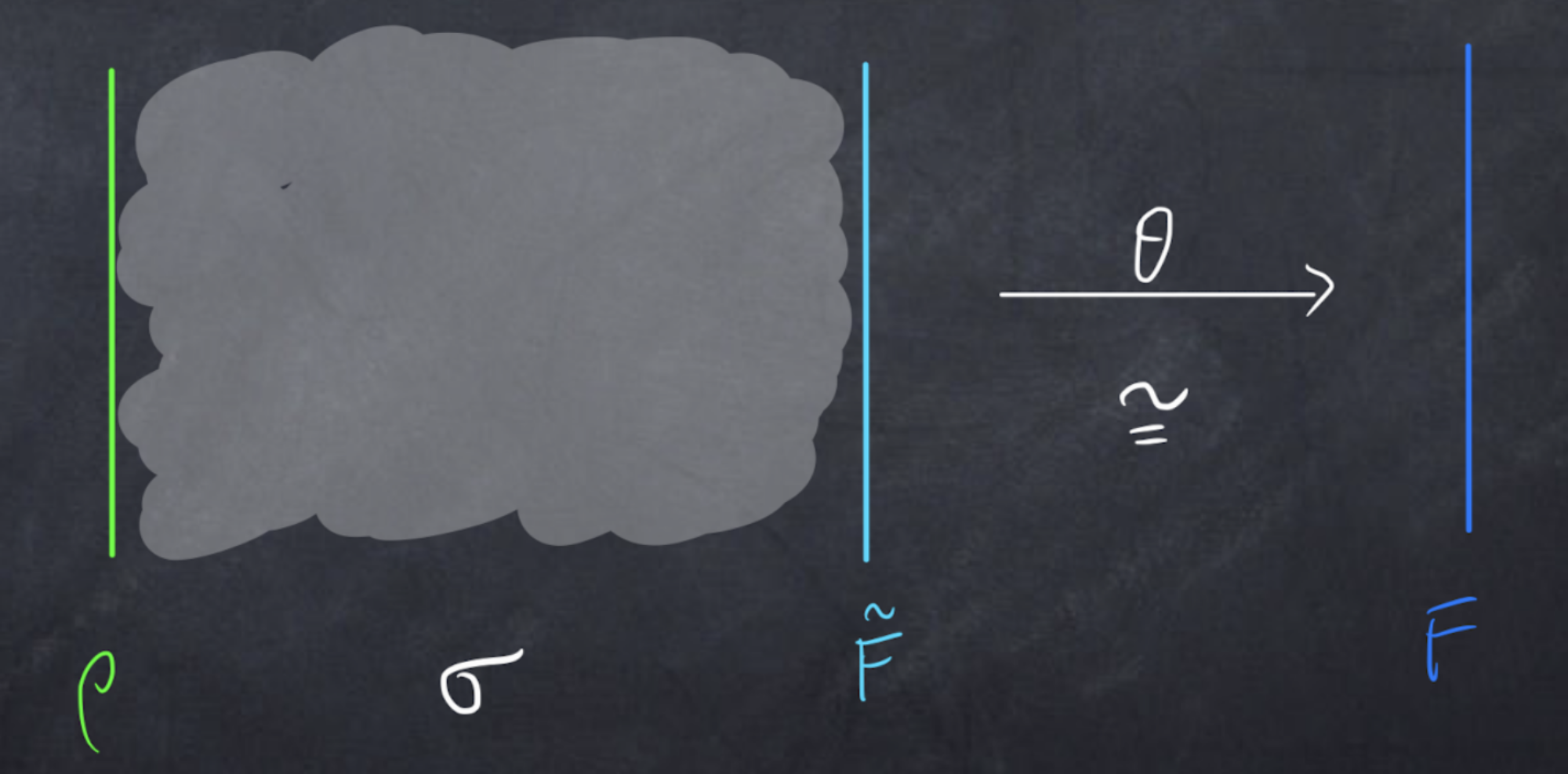}
\end{center}
\caption{ A pair $(\rho,\si)$ describes a ``topological symmetry'' in an
anomaly-free field theory $F$ (Fig. 1 in \Rf{FT220907471}).  } \label{sandwich} 
\end{figure}

In 2022, ``topological symmetry'' was introduced \cite{FT220907471}.
``Topological symmetry'' corresponds to a pair $(\rho,\si)$, where $\si$ is the
\symmTO\ discussed above,  and $\rho$  a
gapped boundary of the \symmTO: $\text{bulk}(\rho)=\si$ (see Fig.
\ref{sandwich}).  The pair $(\rho,\si)$ describes a (generalized) symmetry in a
quantum field theory $F$ (using the notations in \Rf{KZ200514178}): $F \cong
\rho \boxtimes_{\si} \t F$ where $\t F$ is a boundary of $\si$, \ie
$\text{bulk}(\t F)=\si$.  

\begin{figure}[t]
\begin{center}
\includegraphics[scale=0.6]{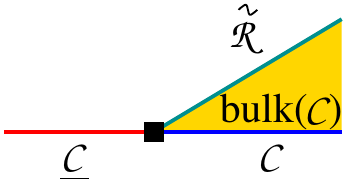}
\end{center}
\caption{ In \Rf{KZ200514178}, Fig. \ref{sandwich} 
was drawn as the above, where
$F = \underline{\cC}$, $\t F = \cC$, $\rho = \t \cR$, and $\si =
\text{bulk}(\cC)$.  (See Fig. 24 and Fig. 29 in \Rf{KZ200514178}).  }
\label{CCmoph} 
\end{figure}

In 2020, \Rf{KZ200514178} also used a similar pair $(\t \cR,\text{bulk}(\cC))$
(see Fig. \ref{CCmoph}) to describe an anomaly-free algebraic higher symmetry
(\ie non-invertible higher symmetry), where $\text{bulk}(\cC)$ is a
non-degenerate braided fusion higher category in trivial Witt class
(\ie a \symmTO\ corresponding
to $\si$ in the above) and $\t \cR$ is a local
fusion higher category that satisfy $\eZ(\t \cR) = \text{bulk}(\cC)$ (\ie a
gapped boundary of the \symmTO\, corresponding to $\rho$ in the above).
\Rf{KZ200514178} used the pair $(\t \cR,\text{bulk}(\cC))$ to classify symmetry
protected topological orders and symmetry enriched topological orders with the
anomaly-free algebraic higher symmetry.  Here $\cC$ corresponds to $\t F$ in
the above.  In \Rf{KZ200514178}, $\t \cR$ in the pair is assumed to be a local
fusion higher category.  Since, $\text{bulk}(\cC)= \text{bulk}(\t \cR)$,
\Rf{KZ200514178} usually used $\t \cR$, or its dual $\cR$, to describe the
symmetry, which is referred to as algebraic higher symmetry (that is
anomaly-free since  $\cR$ and $\t \cR$ is assumed to be local fusion higher
categories).

\begin{figure}[t]
\begin{center}
\includegraphics[scale=0.6]{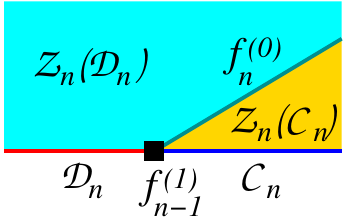}
\end{center}
\caption{ A morphism between $n+1$D (gapped or gapless) quantum field theories
$\cC_n$ and $\cD_n$ with (non-invertible) gravitational anomalies. See (4.3)
of \Rf{KZ150201690}.  } \label{CDmorph} 
\end{figure}

It turns out that the pair $(\rho,\si)$ also appeared in a 2015 work
\cite{KZ150201690,KZ170200673}, but was interpreted differently as a morphism
from $\t F$ to $F$, which leads an equivalence $F \stackrel{\th}{\cong} \rho
\boxtimes_\si \t F$ (see Fig. \ref{sandwich}). Such an equivalence corresponds
to a symmetry described by $(\rho,\si)$, as pointed out in
\Rf{KZ200514178,FT220907471}.

More specifically, in \Rf{KZ150201690,KZ170200673},  morphism between
$n+1$D (gapped or gapless) quantum field theories $\cC_n$ and $\cD_n$ with
(non-perturbative) gravitational anomalies \cite{KW1458} are studied.  The
partition function of a $n+1$D anomalous field theory $\cD_n$ on spacetime
$M^{n+1}$ is define only after we view  $M^{n+1}$ as a boundary of $N^{n+2}$
(like Wess-Zumino-Witten theory \cite{WZ7195,W8322}) and is denoted as
\begin{align}
Z(\cD_n; M^{n+1},N^{n+2}),\ \ \ \ M^{n+1} = \prt N^{n+2}.
\end{align}
The morphism is a \emph{topological domain wall} $(f_{n-1}^{(1)}, f_{n}^{(0)})$
between anomalous field theories (see Fig. \ref{CDmorph}, (4.3) of
\Rf{KZ150201690}), \emph{where $f_{n-1}^{(1)}$ is invertible}. 
Since $f_{n-1}^{(1)}$ is invertible, the morphism $(f_{n-1}^{(1)},
f_{n}^{(0)})$ (\ie the presence of topological domain wall) give rises to an
equivalence relation (see (4.3) in \Rf{KZ150201690}, which is called a
decomposition in \Rf{CW221214432}):
\begin{align}
\label{fCfD}
\cD_n \stackrel{f_{n-1}^{(1)}}{\cong}  f_{n}^{(0)} \boxtimes_{\eZ_n(\cC_n)}
\cC_n
\end{align}
where $\cD_n$ and $f_{n}^{(0)} \boxtimes_{\eZ_n(\cC_n)} \cC_n$ have the same
partition function  \cite{CW221214432}
\begin{align}
\label{ZZCD}
Z(\cD_n; M^{n+1},N^{n+2}) =
Z(f_{n}^{(0)} \boxtimes_{\eZ_n(\cC_n)} \cC_n; M^{n+1},N^{n+2}),
\end{align}
which is the defining property of the equivalence relation or the decomposition
\eqref{fCfD} \cite{CW221214432}.  Since the domain wall $(f_{n-1}^{(1)},
f_{n}^{(0)})$ is topological and $f_{n-1}^{(1)}$ is invertible, the above
implies that the two anomalous field theories $\cC_n$ and $\cD_n$ have the same
\emph{local low energy properties} defined in Footnote \ref{locallow}.  Some
explicit examples of such local low energy equivalence were discussed in
\Rf{CW221214432,PW230105261}.

The equivalence relation or the decomposition, \eqref{fCfD} and \eqref{ZZCD},
reveals the symmetry described by the pair $(f_{n}^{(0)}, \eZ_n(\cC_n))$
\cite{CW221214432,PW230105261}, in the anomalous field theories $\cD_n$ and
$\cC_n$. Thus, morphism between between (anomalous) quantum field theories
defined in \Rf{KZ150201690} corresponds to symmetry.

\begin{figure}[t]
\begin{center}
\includegraphics[scale=0.6]{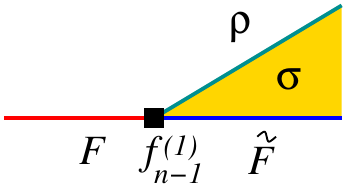}
\end{center}
\caption{ When $\cD_n \equiv  F$ is anomaly-free, Fig. \ref{CDmorph} becomes
the above, which is another way to represent Fig. \ref{sandwich}.  }
\label{FFmorph} 
\end{figure}

In the special case when $\cD_n$ is an anomaly-free field theory (denoted as
$F$), its center (\ie the bulk) $\eZ_n(\cD_n) = \text{bulk}(\cD_n)$ is trivial
and Fig.  \ref{CDmorph} becomes Fig. \ref{FFmorph}, if we rename $\cC_n$ as $\t
F$, $\eZ_n(\cC_n)$ as $\si$, and $f_n^{(0)}$ as $\rho$.  Due to the equivalence
relation \eqref{fCfD} and \eqref{ZZCD}, the pair $(\rho,\si)$ can be viewed as
a symmetry of an anomaly-free field theory $F$ as pointed out in
\Rf{KZ200514178,FT220907471}.  In other words, the equivalence relation or the
decomposition, \eqref{fCfD} and \eqref{ZZCD}, reveals the symmetry $(\rho,\si)$
in the anomaly-free field theory $F$ and in the anomalous field theory $\t F$.
\Rf{CW221214432} used the equivalence relation or decomposition, \eqref{fCfD}
and \eqref{ZZCD}, to identify maximal \catsymm\ in an anomaly-free
field theory $F$.  Note that the anomalous field theory $\t F$ corresponds to
anomaly-free field theory $F$ restricted in the symmetric sub-Hilbert space
$\cV_\text{symmetric}$ \cite{JW191213492}.

The \catsymm\ of the anomaly-free field theory $F$ only corresponds to the
$\si$ in the pair.  We see a clear distinction between \catsymm\ and symmetry:
a \catsymm\ is a holo-equivalence class of symmetries, as indicated by
\eqref{eqsymms}. (Two symmetries, $(\rho,\si)$ and $(\rho',\si')$, are
holo-equivalent if $\si \cong \si'$ \cite{KZ200514178}.)

We have been using \catsymm\ to just mean ``including both symmetry charges and
symmetry defects \emph{at an equal footing}, and including their `braiding'
properties''. This leads to Symm/TO correspondence.  However, ``categorical
symmetry'' has since been used to mean different things.
This causes some confusions.

Another source of confusion comes from the fact that \catsymm\ has several
equivalent descriptions, that emphasis different aspects of Symm/TO
correspondence: A \catsymm\ can be, equivalently, described by 
\begin{enumerate}

\item
a non-invertible gravitational anomaly\cite{JW190513279} (see Fig.
\ref{asect}).  Here we view symmetry by restricting to symmetric sub-Hilbert
space.  The symmetric sub-Hilbert space does not have a tensor product
decomposition $\cV_{symm} \neq \bigotimes_i \cV_i$, where $\cV_i$'s are vector
spaces on lattice sites $i$.  This implies a non-invertible gravitational
anomaly, and thus a symmetry can be described by a non-invertible gravitational
anomaly.

\item 
a symmetry + dual symmetry + braiding \cite{BT170402330,JW191213492}.  Conservation (\ie
the fusion ring) of symmetry charges corresponds to symmetry.  Conservation
(\ie the fusion ring) of symmetry defects corresponds to dual-symmetry.  Here
we treat  symmetry charges and symmetry defects at an equal footing. The fusion
ring  of symmetry charges/defects corresponds to ``symmetry'' in \catsymm.  The
braiding properties of symmetry charges/defects corresponds to ``categorical''
in \catsymm.  In fact, the term \catsymm\ is a parallel generalization of the
term ``anomalous symmetry''.  The fusion ring of the symmetry charges
correspond to ``symmetry'' in ``anomalous symmetry'' and the braiding
properties of symmetry defects corresponds to ``anomalous'' in  ``anomalous
symmetry'' \cite{LG1209,W181202517}.

\item
a gappable-boundary topological order in one higher dimension (symmetry
TO)\cite{JW191213492,KZ200514178}.
This is because gravitational anomaly = topological order in one higher
dimension \cite{KW1458}.

\item a part of topological skeleton introduced in \Rf{KZ201102859}.

\item a non-degenerate braided fusion higher category in trivial Witt class
\cite{TW191202817,KZ200514178}.  This is because topological order is described
by non-degenerate braided fusion higher category.  We use a short name
``\nBFcat'' to refer to ``non-degenerate braided fusion higher category in
trivial Witt class''.  Thus \nBFcat, replacing group and higher group, is used
to describe (generalized) symmetry.  This leads to  a unified frame work to
classify spontaneous symmetry breaking order, topological order, symmetry
protect topological order, symmetry enriched topological order, \etc\ in any
dimension \cite{KZ200514178}.

\item
an equivalence class of algebras of commutant patch operators
(also called transparent patch operators)
\cite{KZ220105726,CW220303596}.  This is a non-holographic point of view that
does not go to one higher dimension, and leads to the notion of \psymm (see
\eqref{eqsymms}).  Here a symmetry is defined via the algebra of local
symmetric operators.  Using an algebra formed by commutant patch operators
(that are constructed from local symmetric operators and define the \psymm),
we can compute a non-degenerate braided fusion category in trivial Witt class 
that describes a
\catsymm.

\end{enumerate}

Let us use some simple examples to illustrate the notion of \catsymm.  In 1+1D
Ising model with $\Z_2$ symmetry. The $\Z_2$-symmetry charge is denoted by $e$
and $\Z_2$-symmetry defect is denoted by $m$.  The $\Z_2$-symmetry is described
by the fusion ring $e\otimes e = \one$ (the conservation law).  The
$\Z_2$-symmetry is also described by transformation law $U^2 =\id$.  On the
other hand, the \catsymm\ of the Ising model is described by the fusion ring of
$\Z_2$-symmetry charges $e\otimes e = \one$ and the fusion ring of
$\Z_2$-symmetry defects $m\otimes m = \one$, as well as a non-trivial
``braiding'' property between $e$ and $m$.  In other words, \catsymm\ treats
the symmetry charges and symmetry transformations (or symmetry defects) at
equal footing.  Such a treatment leads to \nBFcat\ description of symmetry
(instead of group description of symmetry).

In an 1+1D model with $\Z_4$ symmetry. The $\Z_4$-symmetry charges are denoted
by $e^k$, $k=1,2,3$, and $\Z_4$-symmetry defects are denoted by $m^k$,
$k=1,2,3$.  The $\Z_4$-symmetry is described by the fusion ring $ e\otimes e
\otimes e \otimes e = \one$ (the conservation law).  The $\Z_4$-symmetry is
also described by transformation law $U^4 =\id$, or $ m\otimes m \otimes m
\otimes m = \one$.  In contrast, the \catsymm\ of the $\Z_4$ model is described
by the fusion ring of $\Z_4$-symmetry charges $e\otimes e \otimes e \otimes e =
\one$ and the fusion ring of $\Z_4$-symmetry defects $ m\otimes m \otimes m
\otimes m = \one$, as well as a non-trivial ``braiding'' property between $e$
and $m$.

However, if we call $e^2=e_1$ the charge and $m = m_1$ the defect of first
$\Z_2$-symmetry, and $m^2=e_2$ the charge and $e = m_2$ the defect of second
$\Z_2'$-symmetry, then the same \catsymm\ will describe $\Z_2\times
\Z_2'$-symmetry with a mixed anomaly.\cite{CW220303596} This example
demonstrates a difference between the usual symmetry point of view and
\catsymm\ point of view.  The \catsymm\ point of view allows us to see certain
relations more easily.


Since many people use ``categorical symmetry'' to mean non-invertible symmetry,
in this paper, we will use an equivalent notion \emph{\symmTO} to refer to
\catsymm, hopping to avoid confusions.  In fact,  \symmTO\ (\ie \catsymm) is
the ``Drinfeld'' center of global symmetry or fusion category symmetry.

\section{Structure of phase diagram} 
\label{phasestructure}

Condensable algebras $\cA$ have many relations, such as
algebra-subalgebra relation, overlap relations, \etc.  These relations can
constrain the phase diagram of condensation patterns $\cA$ in $\eM$
symmetric systems.  To describe such a  phase diagram, let $\cX_{\eM}$ be the
space of all $\eM$-systems (which is called moduli space), that have
liquid ground states
\cite{ZW1490,SM1403}.  $\cX_{\eM}$ is parametrized by the coupling constants in
the Hamiltonians with the symmetry.  

The moduli space $\cX_{\eM}$ can be divided in to many regions, each described
by a different condensation $\cA$, which will be called $\cA$-phase. The state
in the $\cA$-phase will be called $\cA$-state.  Here $\cA$-phase can be gapped.
$\cA$-phase can also be gapless, in which case, the gapless $\cA$-state has no
symmetric relevant operators.

The boundary between two  regions of condensations $\cA_1$ and $\cA_2$
describes the phase transition between $\cA_1$ and $\cA_2$.  If the phase
transition is first order, at the boundary, the system has degenerate ground
states: one described by $\cA_1$-condensation and the other by
$\cA_2$-condensation.  If the phase transition is continuous, at the boundary,
the system has a condensation described by $\cA_{c}$.  Starting in the
$\cA_1$-phase, as we approach the $\cA_{c}$-boundary, certain condensation
becomes weaker and weaker. At the boundary, we reach a smaller condensation
$\cA_{c} \subset \cA_1$.  Similarly, we have $\cA_{c} \subset \cA_2$.  Thus
\frmbox{A stable continuous transition between $\cA_1$-phase and $\cA_2$-phase
is described by a critical point with $\cA_{c}$ condensation that satisfies
\begin{align}
\cA_{c} \subset \cA_1,\ \ \ \ \cA_{c} \subset \cA_2.
\end{align}
The gapless
$\cA_{c}$-state has only one symmetric relevant operator.  } Here $\cA_1$ and
$\cA_2$ can be the same, in which case, the gapless $\cA_{c}$-state describes
a continuous transition between the same phase.

\begin{figure}[t]
\begin{center}
\includegraphics[scale=0.6]{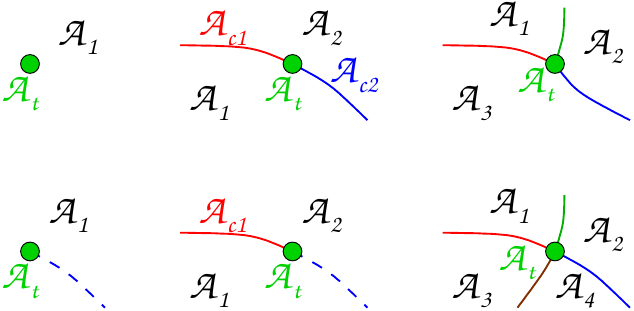}
\end{center}
\caption{Some possible structures of the phases near a tricritical point
$\cA_t$ (the dot).  The solid curves are continuous transitions (described by
$\cA_{c1}$, $\cA_{c2}$, \etc.).  The dashed curves are first-order transitions.
} \label{tricritical} 
\end{figure}

To summarize, $\cA$-state's with no symmetric relevant operator form the stable
phases. Gapless $\cA_c$-state's with one and only one symmetric relevant
operator describe stable continuous transitions between stable phases.
Similarly, gapless $\cA_t$-state's with two and only two symmetric relevant
operators are tricritical points, a kind of multicritical points.  Some
structures of the phases near a tricritical point are described in Fig.
\ref{tricritical}.

\begin{figure}[t]
\begin{center}
\includegraphics[scale=0.65]{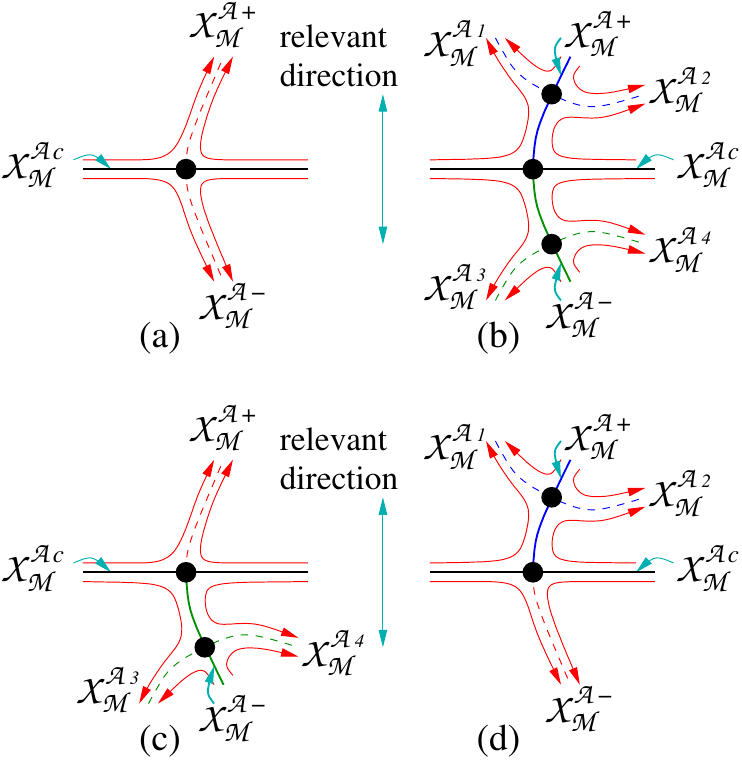}
\end{center}
\caption{The curves with arrow represent the RG flow. The dots represent the
fixed points of the RG flow.  The plane is a subspace of the total moduli space
$\cX_{\eM}$ (the space formed by $\eM$-systems).  Let
$\cX_{\eM}^{\cA}$ be the subspace formed by $\cA$-states.  The horizontal line
is a subspace of $\cX_{\eM}^{\cA_c}$.
In (a), the sub-spaces $\cX_{\eM}^{\cA_{+}}$ and $\cX_{\eM}^{\cA_{-}}$ are upper
and lower half planes.  In (b), the sub-spaces $\cX_{\eM}^{\cA_{+}}$ and
$\cX_{\eM}^{\cA_{-}}$ are two marked curves. (c) and (d) are combinations of
(a) and (b).  } \label{phasetrans} \end{figure}

Let us consider a stable continuous transition, whose critical point is
described by a gapless $\cA_c$-state that has one symmetric relevant operator.
The possible renormalization-group (RG) flows are presented schematically in
Fig.  \ref{phasetrans}.  But what are the resulting states after a long RG
flow?  To address this question, let us introduce a notion of allowed competing
pair for the condensable algebra $\cA_c$, which is a pair of excitations $a_+$
and $a_-$ with nontrivial mutual statistics, but they both have trivial mutual
statistics with respect to $\cA_c$.\footnote{More precisely, an allowed
competing pair $(a_+,a_-)$ for a condensable algebra $\cA_c$ has the following
defining properties: (1) $a_+$ can be added to $\cA_c$ to generate a larger
condensable algebra and so does $a_-$.  (2) $a_+$ and $a_-$ cannot be added
together to generate a larger condensable algebra.  } If $a_+$ condenses, $a_-$
will be confined and uncondense. If $a_-$ condenses, $a_+$ will be confined and
uncondense.  Thus we can imagine that there is some parameter $\veps$ in the
Hamiltonian that controls whether $a_+$ condenses or $a_-$ condenses.  For
example, we could have a situation that $\veps >0$ causes $a_+$ to condense and
$\veps < 0$ causes $a_-$ to condense.  $a_+$ and $a_-$ cannot both condense due
to their nontrivial mutual statistics, but $a_+$ and $a_-$ can be both
uncondensed. Let us assume this can happen only if we fine tune $\veps$, so as
to set $\veps=0$ (otherwise, we would have had a stable gapless phase).  

If the gapless $\cA_c$-state only allows one competing pair, then the two
different condensations of the one competing pair should correspond to the
relevant direction. However if the gapless $\cA_c$-state allows several
competing pairs, then the only relevant direction should correspond to one of
these competing pairs.  With these considerations, we propose that \emph{the
switching between two different condensations of a competing pair is the basic
mechanism for continuous phase transition.  The resulting two condensable
algebras $\cA_+$ and $\cA_-$ from the two condensations must contain the
condensing particle and must contain $\cA_c$ as a sub algebra.} 

To be more concrete, let us assume the competing pair $(a_+,a_-)$ corresponds
to the relevant direction.  After the condensation of $a_+,$ or $a_-$, the
condensable algebra $\cA_c$ will change to $\cA_+ = \cA_c\oplus a_+\oplus
+\cdots$ or $\cA_- = \cA_c\oplus a_- \oplus +\cdots$, where $\cdots$ represent
any additional excitations that condense together with the $a_+$ or $a_-$
condensations.

Now, we need to consider several cases separately.  If $\cA_+$ and $\cA_-$ are
Lagrangian, then the switching between two different condensations of the
competing pair $(a_+,a_-)$ will cause a stable continuous phase transition
between $\cA_+$-state and $\cA_-$-state. We will have a local phase diagram as
in Fig. \ref{phasetrans}(a), where we have assumed that the parameter $\veps$
mentioned above has an overlap with the relevant direction of the RG flow.

If $\cA_+$ and $\cA_-$ are both non-Lagrangian, then the switching between two
different condensations of the competing pair $(a_+,a_-)$ will cause a
continuous phase transition between the gapless $\cA_+$-state and the gapless
$\cA_-$-state.  Let us further assume that both $\cA_+$-state and the
$\cA_-$-state have one relevant operator (if neither has a relevant operator,
the local phase diagram will be given by Fig. \ref{phasetrans}(a) as in the
previous case).  In this case, the continuous transition will be
multicritical. The local phase diagram  will be controlled by the relevant
operator and the dangerously irrelevant operator, a mechanism discussed in
\Rf{BLS191012856}.  The unstable $\cA_+$-state can become $\cA_1$-state or
$\cA_2$-state.  The unstable $\cA_-$-state can become $\cA_3$-state or
$\cA_4$-state.  Thus we find the phase diagram shown schematically in Fig.
\ref{phasetrans}(b).

From the phase diagram Fig. \ref{phasetrans}(b), we see that there are stable
continuous transitions $\cA_1 \leftrightarrow \cA_3$ and $\cA_2 \leftrightarrow
\cA_4$. Whether we get a $\cA_1 \leftrightarrow \cA_3$ transition or a $\cA_2
\leftrightarrow \cA_4$ transition is controlled by dangerously irrelevant
operators.  From the phase diagram, we also see a direct continuous transition
$\cA_1 \leftrightarrow \cA_4$, and a direct continuous transition $\cA_2
\leftrightarrow \cA_3$.  These two transitions are not stable and are
controlled by a multicritical point.  The critical points for all the four
transitions $\cA_1 \leftrightarrow \cA_3$, $\cA_2 \leftrightarrow \cA_4$ $\cA_1
\leftrightarrow \cA_4$, and  $\cA_2 \leftrightarrow \cA_3$, are described by
the same critical theory with condensation pattern $\cA_c$ and with only
one relevant operator.  How can a critical theory with  only one relevant
operator sometimes describe stable continuous transition, and other times
describe multicritical continuous transition?  This is because sometimes
tuning dangerously irrelevant operators can also cause phase transitions.  Thus
a critical theory with  only one relevant operator can sometimes describe a
multicritical point.

If $\cA_+$ is Lagrangian but $\cA_-$ is non-Lagrangian, the local phase diagram
will be a combination of the above two cases, and is given by Fig.
\ref{phasetrans}(c).  Similarly, If $\cA_+$ is non-Lagrangian but $\cA_-$ is
Lagrangian, we get a phase diagram   Fig. \ref{phasetrans}(d).

From the above discussion, we see that the properties of a continuous phase
transition are not only determined by the number of relevant operators of the
critical point, as we usually expect, they are also determined by the
condensation pattern $\cA_c$ of the critical point.  In particular, the
number of condensations needed to change $\cA_c$ into Lagrangian condensable
algebra will strongly influence the critical properties.  Compared to Landau
symmetry breaking theory, the holographic theory replaces the group-subgroup
relation  by the relations of condensable algebras.  The new theory applies to
beyond-Landau continuous phase transitions, as well as non-invertible
symmetries (\ie algebraic higher symmetries), as we discuss in the
main text.

\section{An algebraic number theoretical method to calculate 
condensable algebras and  gapped/gapless boundaries/domain-walls}

\label{MMA}

We have see that the \symmTO\ $\eM$, its condensable algebras $\cA$, and the
induced topological orders $\eM_{/\cA}$ are very important in understanding the
patterns of possible condensations and the allowed gaplessness by the
\symmTO\ $\eM$.  For 1+1D symmetry, there are some simple relations among
$\cA$, $\eM$, and  $\eM_{/\cA}$, where $\eM$, and  $\eM_{/\cA}$ are viewed 2+1D
as topological orders.  

Let us use $a,b,c$ to label the anyons in $\eM$.  As a 2+1D topological order,
$\eM$ is characterized by modular matrices $\t S_\eM = (\t S_\eM^{ab})$ and $\t
T_\eM = (\t T_\eM^{ab})$, whose indices are labeled by the anyons.  $\t S_\eM,
\t T_\eM$ are unitary matrices that generate a representation of $SL(2,\Z_n)$,
where $n$ is the smallest integer that satisfy $\t T_\eM^n =\id$. We call $n$
as the order of $\t T_\eM$ and denote it as $n=\ord(\t T_\eM)$.  $\t T_\eM$ is
a diagonal matrix and $\t S_\eM$ is a symmetric matrix.

From $\t S_\eM$ and $\t T_\eM$, we define
normalized $S^\text{cat},T^\text{cat}$-matrices and unitary $S,T$-matrices
\begin{align}
S_\eM^\text{cat} &= \t S_\eM/\t S_\eM^{\one\one}, &
T_\eM^\text{cat} &= \t T_\eM/\t T_\eM^{\one\one}.
\nonumber\\
S_\eM &= S_\eM^\text{cat}/D_{\eM}, &
T_\eM &= T_\eM^\text{cat}.
\end{align}
Let $d_a$ be the quantum dimension of anyon $a$, which is given by
$d_a=(S_\eM^\text{cat})^{a\one}$.  Let $s_a$ be the topological spin of anyon
$a$, which is given by $\ee^{\ii 2\pi s_a} = (T_\eM^\text{cat})^{aa}$.  The
total dimension of $\eM$ is defined as $D^2_{\eM} \equiv \sum_{a\in \eM}
d_a^2$.  Also let $d_{\cA}$ be the quantum dimension of the condensable algebra
$\cA$, \ie if 
\begin{align}
\cA=\bigoplus_{a\in \eM} A^a a
\end{align}
then $d_\cA =\sum_a A^a d_a$.  We also have a particle to anti-particle
conjugation $a \to \bar a$.
Similarly, we use $i,j,k$ to label the anyons in $\eM_{/\cA}$.  Following the
above, we can define $\t S_{\eM_{/\cA}} = (\t S_{\eM_{/\cA}} ^{ij})$, $\t
T_{\eM_{/\cA}} = (\t T_{\eM_{/\cA}} ^{ij})$, $S_{\eM_{/\cA}}^\text{cat}$,
$T_{\eM_{/\cA}}^\text{cat}$, $S_{\eM_{/\cA}}$, $T_{\eM_{/\cA}}$, as well as
$d_i$, $s_i$, and $D^2_{\eM_{/\cA}}$.  Then we have the following properties
\begin{itemize}

\item The distinct $s_i$'s form a sub set of $\{s_a \mid a \in \eM\}$.

\item $(S_\eM^\text{cat})^{ab}$, $(T_\eM^\text{cat})^{aa}$, $D^2_{\eM}$, $d_a$,
and $d_\cA$ are cyclotomic integers, whose conductors divide
$\ord(T_\eM^\text{cat})$.  $D_{\eM}$ is a real cyclotomic integer whose
conductors divide $\ord(\t T_\eM)$ (assuming $\t S_\eM^{\one\one}$ is real).

\item $(S_{\eM_{/\cA}}^\text{cat})^{ij}$, $(T_{\eM_{/\cA}}^\text{cat})^{ii}$,
$D^2_{\eM_{/\cA}}$ and $d_i$ are cyclotomic integers, whose conductors divide
$\ord(T_{\eM_{/\cA}}^\text{cat})$.  $D_{\eM_{/\cA}}$ is a real cyclotomic
integer whose conductors divide $\ord(\t T_{\eM_{/\cA}})$ (assuming $\t
S_{\eM_{/\cA}}^{\one\one}$ is real).

\item $D_{\eM} = D_{\eM_{/\cA}} d_\cA$.

\item $A^a$ in $\cA$ are non-negative integers, $A^a=A^{\bar a}$, and
$A^\one =1$. 

\item For $a \in \cA$ (\ie for $A^a \neq 0$), the corresponding  $s_a  = 0$ mod
1.  \ie the anyons in $\cA$ are all bosonic. 

\item if $a,b \in \cA$, then at least one of the fusion products in $a\otimes b$
must be contained $\cA$, \ie $ \exists  c\in \cA $ such that
$a\otimes b = c\oplus \cdots$.

\end{itemize}

Now, let us assume $\cA$ to be Lagrangian, then the
$\cA$-condensed boundary of $\eM$ is gapped. 
Let us use $x$ to label the (simple) excitations on the gapped boundary.
If we bring a bulk excitation $a$
to such a boundary, it will become a (composite) boundary excitation $X$ 
\begin{align}
X = \oplus M^a_x x, \ \ \ M^a_x \in \N.
\end{align}
Then $A^a$ is given by $A^a =M^a_\one$.  In other words, $A^a\neq 0$ means that
$a$ condenses on the boundary (\ie the bulk $a$ can become the null excitation
$\one$ on the boundary).  $M^a_x$ satisfies
\begin{align}
\label{NMMK}
\sum_c N^{ab}_{\eM,c} M^c_x = \sum_{y,z} M^a_y M^b_{z} K^{yz}_x 
\end{align}
where $N^{ab}_{\eM,c}$ describes the fusion ring of the bulk excitations in
$\eM$ and $K^{xy}_z$ describes the fusion ring of the boundary excitations.  By
rewriting $\sum_y = \sum_{y=\one} +\sum_{y\neq \one}$, we find
\begin{align}
\sum_c N^{ab}_{\eM,c} M^c_x & \geq A^a M^b_{x}  
\end{align}
Let $\bar A^a \equiv \sum_{x\neq \one} M^a_x$ and do $\sum_{x\neq \one}$ to the
above, we obtain (noticing $\bar A^\one=0$)
\begin{align}
\sum_{c\neq \one} N^{ab}_{\eM,c} \bar A^c & \geq A^a \bar A^b  
\end{align}
Taking $x=\one$, \eqn{NMMK} reduces to
\begin{align}
\label{AAAMM}
\sum_c N^{ab}_{\eM,c} A^c 
&=
A^a A^b + \sum_{x\neq \one}  M^a_x M^b_{\bar x}
\end{align}
Since $M^a_x\geq 0$, we obtain an additional condition
on $A^a$
\begin{align}
\label{NAAA}
\sum_c N^{ab}_{\eM,c} A^c & \geq A^a A^b.
\end{align}

We can try to obtain a stronger condition, by showing $\sum_{x\neq \one}  M^a_x
M^b_{\bar x}$ is equal or larger than a positive integer.  Summing over $x$,
\eqn{NMMK} implies
\begin{align}
\sum_c N^{ab}_{\eM,c} (A^c+\bar A^c) 
&\geq
A^a A^b 
+ A^a \bar A^b 
+ A^b \bar A^a 
.
\end{align}
Combining the above two equations, we find
\begin{align}
\label{AANA}
\sum_{x\neq \one}  M^a_x M^b_{\bar x}
&\geq
A^a \bar A^b 
+ A^b \bar A^a 
-\sum_{c\neq \one} N^{ab}_{\eM,c} \bar A^c
.
\end{align}
Taking $b=\bar a$ in
\eqn{AAAMM}, we find
\begin{align}
\label{AaAa}
\sum_c N^{a\bar a}_{\eM,c} A^c 
&\geq
(A^a)^2  + \bar A^a.
\end{align}
From the conservation of quantum dimensions, we have
\begin{align}
d_a = \sum_x M^a_x d_x 
= A^a + \sum_{x\neq \one}  M^a_x d_x,
\end{align}
which implies
\begin{align}
\label{AdA}
\del(d_a) \leq \bar A^a \leq d_a-A^a.
\end{align}
where $\del(d_a)$ is defined as
\begin{align}
\del(d) = \begin{cases}
0 \text{ if } d \in \N,\\
1 \text{ if } d \not\in \N.\\
\end{cases}
\end{align}
Let us define $ \bar A^a_\text{max}$ to be the largest integer that is less
than both $d_a-A^a$  and $\sum_c N^{a\bar a}_{\eM,c} A^c - (A^a)^2$.  Note that
both terms must be larger than $\del(d_a)$ and $ \bar A^a_\text{max}$ is equal
or larger than $\del(d_a)$:
\begin{align}
\label{Amaxdel}
\bar A^a_\text{max} \geq \del(d_a).
\end{align}
Substituting $ \bar A^a \leq \bar A^a_\text{max} $ into \eqn{AANA}, we obtain
\begin{align}
&\ \ \ \
\sum_{x\neq \one}  M^a_x M^b_{\bar x}
\geq
\text{max}(0,
A^a \bar A^b 
+ A^b \bar A^a 
-\sum_{c\neq \one} N^{ab}_{\eM,c} \bar A^c_\text{max} )
\nonumber\\
&\geq
\text{max}(0,
A^a \del(d_b) 
+ A^b \del(d_a) 
-\sum_{c\neq \one} N^{ab}_{\eM,c}  \bar A^c_\text{max}
).
\end{align}

For Lagrangian $\cA$, the condensation of $\cA$ give rises to a gapped boundary.
\Rf{JW190513279} gave a physical picture of the multi-component
$\tau$-independent partition function $Z_a$ of the corresponding gapped
boundary.  From such a physical picture, we find that $A^a = Z_a$ and satisfies
\eqn{STZ}.  Summarizing the above discussions, we see that,
\begin{align}
\label{Acond1}
\v A &= S_\eM\v A, & \v A &= T_\eM\v A,
\nonumber\\
A^a & \leq  d_a -\del(d_a),  &
A^a 
A^{\bar a} 
&\leq
\sum_c N^{a\bar a}_{\eM,c} A^c 
-\del(d_a) , 
\nonumber\\
A^a A^b &\leq \sum_c N^{ab}_{\eM,c} A^c 
,
\end{align}
where $\v A =(A^\one,A^a,\cdots)^\top$.
The last condition can be improved to
\begin{align}
A^a A^b &\leq \sum_c N^{ab}_{\eM,c} A^c 
\\
&\ \ \ \
- \text{max}(0,
A^a \del(d_b) 
+ A^b \del(d_a) 
-\sum_{c\neq \one} N^{ab}_{\eM,c}  \bar A^c_\text{max}
)
\nonumber 
\end{align}

Now, let us assume $\cA$ not to be Lagrangian.  In this case $\eM_{/\cA}$ is
nontrivial.  Let us consider the domain wall between $\eM$ and $\eM_{/\cA}$.
Such a domain wall can be viewed as a boundary of $\eM \boxtimes \overline
\eM_{/\cA}$ topological order form by stacking $\eM$ and the spatial reflection
of $\eM_{/\cA}$.  Since the domain wall, and hence the boundary, is gapped,
there must be a Lagrangian condensable algebra $\cA_{\eM \boxtimes \overline
\eM_{/\cA}}$ in $\eM \boxtimes \overline \eM_{/\cA}$, whose condensation gives
rise to the boundary.  Let 
\begin{align}
\cA_{\eM \boxtimes \overline \eM_{/\cA}} =
\bigoplus_{a\in \eM,\ i \in \eM_{/\cA}} A^{ai}\ a\otimes i
,
\end{align}
then the matrix $ A = (A^{ai})$ satisfies
\begin{align}
\label{STA}
& S_{\eM} A = A  S_{\eM_{/\cA}}, \   T_{\eM} A = A  T_{\eM_{/\cA}},
\ 
A^{ai} \leq d_ad_i -\del(d_ad_i),
\nonumber \\
&A^{ai} A^{bj}  \leq \sum_{c,k} 
N^{ab}_{\eM,c} 
N^{ij}_{\eM_{/\cA},k} 
A^{ck} - \del_{a,\bar b}\del_{i,\bar j} \del(d_a d_i) ,
\end{align}
where \eqn{AdA} and \eqn{AaAa} are used.
The above conditions only require the domain wall between $\eM$ and
$\eM_{/\cA}$ to be gapped.  However, since $\eM$ and $\eM_{/\cA}$ are related
by a condensation of $\cA$, there is a special domain wall (called the
canonical domain wall) such that all the excitation in $\eM_{/\cA}$ can pass
through the  domain wall to go into $\eM$ without leaving any nontrivial
excitations on the wall.  For the  canonical domain wall, the corresponding
$A^{ai}$ must satisfy the following condition: 
\begin{align}
\label{Aainzero}
\text{For any $i$, there exists an $a$
such that }\ A^{ai} \neq 0 .
\end{align}

The canonical domain wall can be viewed as $\cA_{\eM\to \eM_{/\cA}}$-condensed
boundary
of $\eM$ with 
\begin{align}
\cA_{\eM\to \eM_{/\cA}} = \bigoplus A^{a\one} a .
\end{align}
We note that anyon $a$ in $\eM$ condenses on the canonical domain wall between
$\eM$ and $\eM_{/\cA}$, if and only if $A^{a\one}\neq 0$.  This implies that
\begin{align}
\label{AMA1}
\cA_{\eM\to \eM_{/\cA}}  = \cA ,\ \ \ \ A^a = A^{a\one}.
\end{align}
The  domain wall can also be viewed as $\cA_{\eM_{/\cA}\to\eM}$-condensed
boundary of $\eM_{/\cA}$ with 
\begin{align}
\cA_{\eM_{/\cA}\to\eM} = \bigoplus A^{\one i} i .
\end{align}
Since $\eM_{/\cA}$ comes from a condensation of $\eM$, the canonical domain
wall must be an $\one$-condensed boundary of $\eM_{/\cA}$, \ie
\begin{align}
\label{AMA2}
\cA_{\eM_{/\cA}\to\eM} = \one,\ \ \
A^{\one i} = \del_{\one,i}.
\end{align}

We can obtain more conditions on $A^a$. From \eqn{STA}, we find
\begin{align}
\frac{D_{\eM_{/\cA}}}{D_\eM}\sum_{b\in \eM} (S_{\eM}^\text{cat})^{ab} A^{bi} 
= \sum_{j\in \eM_{/\cA} }
A^{aj} (S_{\eM_{/\cA}}^\text{cat})^{ji} , 
\end{align}
which implies
\begin{align}
\frac{\sum_{b\in \eM} 
(S_{\eM}^\text{cat})^{ab} A^{bi} }{\sum_{b\in \eM} d_b A^b}
&= 
\text{cyclotomic integer}
\nonumber\\
&\ \ \ \
\text{for all }   
a\in \eM ,\ \  
i\in \eM_{/\cA}  
\end{align} 
In particular,
$\cA  = \bigoplus_a A^{a} a$ must satisfies
\begin{align}
\frac{\sum_{b\in \eM} (S_{\eM}^\text{cat})^{ab} A^b}{\sum_{b\in \eM} d_b A^b}
= 
\text{cyclotomic integer for all } &  a\in \eM  
\end{align} 
From \eqn{STA}, we also obtain
\begin{align}
A^{a} &\leq d_a -\del(d_a), 
\nonumber\\
A^a A^b &\leq \sum_c N^{ab}_{\eM,c} A^c - \del_{a,\bar b} \del(d_a).
\end{align}

These conditions can help us to find possible condensable algebras $\cA =
\bigoplus_a A^{a} a$, which we call different condensation patterns of the system.  These conditions can also help us to find possible
condensation-induced topological orders $\eM_{/\cA}$, which determine the low
energy properties of the gapless $\cA$-state.  When
combined with conformal character of CFT's (see \eqn{STAgapless}), these
conditions allow us to obtain gapless (and gapped) boundaries of topological
order $\eM$ and $\eM_{/\cA}$.  This represents an algebraic number
theoretical way to calculate properties of critical points.

\section{SPT order as automorphism of \hcatsymm}
\label{SPTauto}

In a 1+1D $\Z_2\times \Z_2'$-symmetric system, its \hcatsymm\ is described
\symmTO\ $\eGau_{\Z_2\times \Z_2'}$ (\ie a 2+1D $\Z_2\times \Z_2'$-gauge
theory).  Let us elaborate on one of the boundary of $\eGau_{\Z_2\times
\Z_2'}$, the  $ \one\oplus e_2m_1\oplus e_1m_2\oplus f_1f_2 $-condensed
boundary, to make contact with \Rf{KZ200308898,KZ200514178}, where a
classification of SPT order for finite symmetries, higher symmetries, and
algebraic higher symmetries \footnote{also known as non-invertible symmetries}
was given in terms of certain automorphisms of the corresponding \symmTO. The
argument is based on the fact that the boundary of a bulk topological order can
be changed by stacking with a domain wall of the bulk TO. It is assumed that
all the changes of a gapped boundary phase to other gapped boundary phases can
be obtained this way. The gapped boundaries that correspond to trivial and
nontrivial SPT orders form a group, \ie they all have inverses.  This motivates
the association of SPT orders with certain invertible domain walls in the bulk
topological order which correspond to automorphisms of the
TO.\cite{KZ200308898,KZ200514178} The $ \Z_2 \times \Z_2' $-SPT state, the
$\one\oplus e_2m_1\oplus e_1m_2\oplus f_1f_2$-state, furnishes a simple example
of this result.

Let us spell this out in more detail. According to
\Rf{KZ200308898,KZ200514178}, all anomaly-free generalized symmetries are
described and classified by local fusion higher category $\cR$ formed by the
symmetry charges.  For 1+1D $ \Z_2 \times \Z_2' $ symmetry, $\cR$ is a fusion
1-category consisting of the anyons $(\one,e_1,e_2,e_1e_2)$ as objects.  The
dual symmetry $\t \Z_2 \times \t \Z_2' $ is described by the dual fusion
1-category, $\t \cR$, similarly formed by $(\one,m_1,m_2,m_1m_2)$.  The
\symmTO\ of $\cR$ and $\t \cR$ are the same and is given by their Drinfeld
center
\begin{align}
\eGau_{\Z_2\times \Z_2'}  = \eZ(\cR) = \eZ(\t \cR).
\end{align}
Since \emph{center is bulk} \cite{KW1458,KZ150201690,KZ170200673}, the above
expression means that the 2+1D TO $\eGau_{\Z_2\times \Z_2'}$ has two gapped
boundaries with excitations described by $\cR$ and $\t\cR$.  The $\cR$-boundary
is induced by condensing $\cA_{\cR}=\one\oplus m_1\oplus m_2\oplus m_1m_2$.
The $\t\cR$-boundary is induced by condensing $\cA_{\t\cR}=\one\oplus e_1\oplus
e_2\oplus e_1e_2$. The $\cA_{\cR}$-condensed boundary is the $\Z_2\times
\Z_2'$-symmetric state with trivial SPT order.  The gapped state with
nontrivial SPT order is given by condensation of $\al(\cA_{\cR})$ on the
boundary, where $\al$ is an automorphism of the bulk topological order
$\eGau_{\Z_2\times \Z_2'}$ that satisfies $\al(\cA_{\t\cR})=\cA_{\t\cR}$. In
other words, the automorphism acts trivially on the so-called \emph{electric}
Lagrangian condensable algebra.  This requirement for the automorphism $ \al $
is required so that it does not alter the description of the symmetry on the
boundary system.  For our example $\eGau_{\Z_2\times \Z_2'}$, one of the
automorphisms is $(e_1\leftrightarrow e_2, m_1\leftrightarrow m_2)$, which
exchanges $\Z_2$ and $\Z_2'$ and changes $\one\oplus e_1\oplus e_2\oplus
e_1e_2$ to $\one\oplus e_2\oplus e_1\oplus e_1e_2$.  However, such an
automorphism changes the symmetry since, for instance, $\Z_2$ symmetry may
correspond to spin rotation while $\Z_2'$ to charge conjugation. So we need to
exclude such automorphisms.\footnote{Note that, here, we view $\one\oplus
e_1\oplus e_2\oplus e_1e_2$ and $\one\oplus e_2\oplus e_1\oplus e_1e_2$ as
unequal condensable algebras: $\one\oplus e_1\oplus e_2\oplus e_1e_2\neq
\one\oplus e_2\oplus e_1\oplus e_1e_2$.} By observation, we find another
nontrivial automorphism of $ \eGau_{\Z_2\times \Z_2'} $:
\begin{align*}
\al(e_1) = e_1,\ 
\al(e_2) = e_2,\  
\al(m_1) = e_2m_1,\ 
\al(m_2) = e_1m_2,
\end{align*}
which maps $\cA_{\t \cR}$ to $\cA_{\t \cR}$ and maps $\cA_\cR$ to
\begin{align}
&\ \ \ \ \al(\one\oplus m_1\oplus m_2\oplus m_1m_2) 
\nonumber\\
&=
\al(\one)\oplus \al(m_1)\oplus \al(m_2)\oplus \al(m_1)\al(m_2) 
\nonumber\\
&=
\one \oplus e_2m_1 \oplus e_1m_2 \oplus f_1f_2.
\end{align} 
Thus the gapped $\one \oplus e_2m_1 \oplus
e_1m_2 \oplus f_1f_2$-state is a nontrivial $\Z_2\times\Z_2'$-SPT state.

\allowdisplaybreaks

\section{Gapless boundaries of $\eGau_{S_3}$ with central charge $(c,\bar c)
\leq (\frac56, \frac56)$}

\label{S3boundaries}

In this section, we list all the multi-component boundary partition functions
with central charge $(c,\bar c) \leq (\frac56, \frac56)$ for the 2+1D
topological order $\eGau_{S_3}$. These gapless boundaries are described by
CFT's constructed from minimal models. Here $\chi^{m4}_{h}$ are conformal
characters with conformal dimension $h$, for $(4,3)$ minimal model.
$\chi^{m5}_{h}$ are conformal characters for $(5,4)$ minimal model, \etc.

We can determine the condensable algebra $\cA$ that produces the boundary by
examine the appearances of $|\chi^{m\#}_0|^2$ term.  From partition function
$Z_\one$, we can also determine the number of relevant operators.

$\one\oplus a_2$-condensed boundary with 1 relevant operator:
\begin{align}
Z_\one &=  |\chi^{m4}_{0}|^2 +  |\chi^{m4}_{\frac{1}{2}}|^2 
\nonumber \\ 
Z_{a_1} &=  |\chi^{m4}_{\frac{1}{16}}|^2 
\nonumber \\ 
Z_{a_2} &=  |\chi^{m4}_{0}|^2 +  |\chi^{m4}_{\frac{1}{16}}|^2 + 
|\chi^{m4}_{\frac{1}{2}}|^2 
\nonumber \\ 
Z_{b} &= 0
\nonumber \\ 
Z_{b_1} &= 0
\nonumber \\ 
Z_{b_2} &= 0
\nonumber \\ 
Z_{c} &=  |\chi^{m4}_{\frac{1}{16}}|^2 
\nonumber \\ 
Z_{c_1} &=  \chi^{m4}_{0} \bar\chi^{m4}_{\frac{1}{2}} +  \chi^{m4}_{\frac{1}{2}}
\bar\chi^{m4}_{0} 
\end{align}

\begin{align}
Z_\one &=  |\chi^{m4}_{0}|^2 +  |\chi^{m4}_{\frac{1}{16}}|^2 + 
|\chi^{m4}_{\frac{1}{2}}|^2 
\nonumber \\ 
Z_{a_1} &=  |\chi^{m4}_{0}|^2 +  |\chi^{m4}_{\frac{1}{16}}|^2 + 
|\chi^{m4}_{\frac{1}{2}}|^2 
\nonumber \\ 
Z_{a_2} &=  2 |\chi^{m4}_{0}|^2 +  2 |\chi^{m4}_{\frac{1}{16}}|^2 +  2
|\chi^{m4}_{\frac{1}{2}}|^2 
\nonumber \\ 
Z_{b} &= 0
\nonumber \\ 
Z_{b_1} &= 0
\nonumber \\ 
Z_{b_2} &= 0
\nonumber \\ 
Z_{c} &= 0
\nonumber \\ 
Z_{c_1} &= 0
\end{align}

$\one\oplus b$-condensed boundary with 1 relevant operator:
\begin{align}
Z_\one &=  |\chi^{m4}_{0}|^2 +  |\chi^{m4}_{\frac{1}{2}}|^2 
\nonumber \\ 
Z_{a_1} &=  |\chi^{m4}_{\frac{1}{16}}|^2 
\nonumber \\ 
Z_{a_2} &= 0
\nonumber \\ 
Z_{b} &=  |\chi^{m4}_{0}|^2 +  |\chi^{m4}_{\frac{1}{16}}|^2 + 
|\chi^{m4}_{\frac{1}{2}}|^2 
\nonumber \\ 
Z_{b_1} &= 0
\nonumber \\ 
Z_{b_2} &= 0
\nonumber \\ 
Z_{c} &=  |\chi^{m4}_{\frac{1}{16}}|^2 
\nonumber \\ 
Z_{c_1} &=  \chi^{m4}_{0} \bar\chi^{m4}_{\frac{1}{2}} +  \chi^{m4}_{\frac{1}{2}}
\bar\chi^{m4}_{0} 
\end{align}

\begin{align}
Z_\one &=  |\chi^{m4}_{0}|^2 +  |\chi^{m4}_{\frac{1}{16}}|^2 + 
|\chi^{m4}_{\frac{1}{2}}|^2 
\nonumber \\ 
Z_{a_1} &=  |\chi^{m4}_{0}|^2 +  |\chi^{m4}_{\frac{1}{16}}|^2 + 
|\chi^{m4}_{\frac{1}{2}}|^2 
\nonumber \\ 
Z_{a_2} &= 0
\nonumber \\ 
Z_{b} &=  2 |\chi^{m4}_{0}|^2 +  2 |\chi^{m4}_{\frac{1}{16}}|^2 +  2
|\chi^{m4}_{\frac{1}{2}}|^2 
\nonumber \\ 
Z_{b_1} &= 0
\nonumber \\ 
Z_{b_2} &= 0
\nonumber \\ 
Z_{c} &= 0
\nonumber \\ 
Z_{c_1} &= 0
\end{align}

\begin{align}
Z_\one &=  |\chi^{m4}_{0}|^2 +  |\chi^{m4}_{\frac{1}{16}}|^2 + 
|\chi^{m4}_{\frac{1}{2}}|^2 
\nonumber \\ 
Z_{a_1} &= 0
\nonumber \\ 
Z_{a_2} &=  |\chi^{m4}_{0}|^2 +  |\chi^{m4}_{\frac{1}{16}}|^2 + 
|\chi^{m4}_{\frac{1}{2}}|^2 
\nonumber \\ 
Z_{b} &= 0
\nonumber \\ 
Z_{b_1} &= 0
\nonumber \\ 
Z_{b_2} &= 0
\nonumber \\ 
Z_{c} &=  |\chi^{m4}_{0}|^2 +  |\chi^{m4}_{\frac{1}{16}}|^2 + 
|\chi^{m4}_{\frac{1}{2}}|^2 
\nonumber \\ 
Z_{c_1} &= 0
\end{align}

\begin{align}
Z_\one &=  |\chi^{m4}_{0}|^2 +  |\chi^{m4}_{\frac{1}{16}}|^2 + 
|\chi^{m4}_{\frac{1}{2}}|^2 
\nonumber \\ 
Z_{a_1} &= 0
\nonumber \\ 
Z_{a_2} &= 0
\nonumber \\ 
Z_{b} &=  |\chi^{m4}_{0}|^2 +  |\chi^{m4}_{\frac{1}{16}}|^2 + 
|\chi^{m4}_{\frac{1}{2}}|^2 
\nonumber \\ 
Z_{b_1} &= 0
\nonumber \\ 
Z_{b_2} &= 0
\nonumber \\ 
Z_{c} &=  |\chi^{m4}_{0}|^2 +  |\chi^{m4}_{\frac{1}{16}}|^2 + 
|\chi^{m4}_{\frac{1}{2}}|^2 
\nonumber \\ 
Z_{c_1} &= 0
\end{align}

$\one\oplus a_2$-condensed boundary with 2 relevant operators:
\begin{align}
Z_\one &=  |\chi^{m5}_{0}|^2 +  |\chi^{m5}_{\frac{1}{10}}|^2 + 
|\chi^{m5}_{\frac{3}{5}}|^2 +  |\chi^{m5}_{\frac{3}{2}}|^2 
\nonumber \\ 
Z_{a_1} &=  |\chi^{m5}_{\frac{7}{16}}|^2 +  |\chi^{m5}_{\frac{3}{80}}|^2 
\nonumber \\ 
Z_{a_2} &=  |\chi^{m5}_{0}|^2 +  |\chi^{m5}_{\frac{1}{10}}|^2 + 
|\chi^{m5}_{\frac{3}{5}}|^2 +  |\chi^{m5}_{\frac{3}{2}}|^2 + 
|\chi^{m5}_{\frac{7}{16}}|^2
\nonumber\\
&\ \ \ \ 
+  |\chi^{m5}_{\frac{3}{80}}|^2 
\nonumber \\ 
Z_{b} &= 0
\nonumber \\ 
Z_{b_1} &= 0
\nonumber \\ 
Z_{b_2} &= 0
\nonumber \\ 
Z_{c} &=  |\chi^{m5}_{\frac{7}{16}}|^2 +  |\chi^{m5}_{\frac{3}{80}}|^2 
\nonumber \\ 
Z_{c_1} &=  \chi^{m5}_{0} \bar\chi^{m5}_{\frac{3}{2}} + 
\chi^{m5}_{\frac{1}{10}} \bar\chi^{m5}_{\frac{3}{5}} +  \chi^{m5}_{\frac{3}{5}}
\bar\chi^{m5}_{\frac{1}{10}} +  \chi^{m5}_{\frac{3}{2}} \bar\chi^{m5}_{0} 
\end{align}

\begin{align}
Z_\one &=  |\chi^{m5}_{0}|^2 +  |\chi^{m5}_{\frac{1}{10}}|^2 + 
|\chi^{m5}_{\frac{3}{5}}|^2 +  |\chi^{m5}_{\frac{3}{2}}|^2 + 
|\chi^{m5}_{\frac{7}{16}}|^2 
\nonumber\\
&\ \ \ \
+  |\chi^{m5}_{\frac{3}{80}}|^2 
\nonumber \\ 
Z_{a_1} &=  |\chi^{m5}_{0}|^2 +  |\chi^{m5}_{\frac{1}{10}}|^2 + 
|\chi^{m5}_{\frac{3}{5}}|^2 +  |\chi^{m5}_{\frac{3}{2}}|^2 + 
|\chi^{m5}_{\frac{7}{16}}|^2 
\nonumber\\
&\ \ \ \
+  |\chi^{m5}_{\frac{3}{80}}|^2 
\nonumber \\ 
Z_{a_2} &=  2 |\chi^{m5}_{0}|^2 +  2 |\chi^{m5}_{\frac{1}{10}}|^2 +  2
|\chi^{m5}_{\frac{3}{5}}|^2 +  2 |\chi^{m5}_{\frac{3}{2}}|^2 +  2
|\chi^{m5}_{\frac{7}{16}}|^2 
\nonumber\\
&\ \ \ \
+  2 |\chi^{m5}_{\frac{3}{80}}|^2 
\nonumber \\ 
Z_{b} &= 0
\nonumber \\ 
Z_{b_1} &= 0
\nonumber \\ 
Z_{b_2} &= 0
\nonumber \\ 
Z_{c} &= 0
\nonumber \\ 
Z_{c_1} &= 0
\end{align}

$\one\oplus b$-condensed boundary with 2 relevant operators:
\begin{align}
Z_\one &=  |\chi^{m5}_{0}|^2 +  |\chi^{m5}_{\frac{1}{10}}|^2 + 
|\chi^{m5}_{\frac{3}{5}}|^2 +  |\chi^{m5}_{\frac{3}{2}}|^2 
\nonumber \\ 
Z_{a_1} &=  |\chi^{m5}_{\frac{7}{16}}|^2 +  |\chi^{m5}_{\frac{3}{80}}|^2 
\nonumber \\ 
Z_{a_2} &= 0
\nonumber \\ 
Z_{b} &=  |\chi^{m5}_{0}|^2 +  |\chi^{m5}_{\frac{1}{10}}|^2 + 
|\chi^{m5}_{\frac{3}{5}}|^2 +  |\chi^{m5}_{\frac{3}{2}}|^2 + 
|\chi^{m5}_{\frac{7}{16}}|^2 
\nonumber\\
&\ \ \ \
+  |\chi^{m5}_{\frac{3}{80}}|^2 
\nonumber \\ 
Z_{b_1} &= 0
\nonumber \\ 
Z_{b_2} &= 0
\nonumber \\ 
Z_{c} &=  |\chi^{m5}_{\frac{7}{16}}|^2 +  |\chi^{m5}_{\frac{3}{80}}|^2 
\nonumber \\ 
Z_{c_1} &=  \chi^{m5}_{0} \bar\chi^{m5}_{\frac{3}{2}} + 
\chi^{m5}_{\frac{1}{10}} \bar\chi^{m5}_{\frac{3}{5}} +  \chi^{m5}_{\frac{3}{5}}
\bar\chi^{m5}_{\frac{1}{10}} +  \chi^{m5}_{\frac{3}{2}} \bar\chi^{m5}_{0} 
\end{align}

\begin{align}
Z_\one &=  |\chi^{m5}_{0}|^2 +  |\chi^{m5}_{\frac{1}{10}}|^2 + 
|\chi^{m5}_{\frac{3}{5}}|^2 +  |\chi^{m5}_{\frac{3}{2}}|^2 + 
|\chi^{m5}_{\frac{7}{16}}|^2 
\nonumber\\
&\ \ \ \
+  |\chi^{m5}_{\frac{3}{80}}|^2 
\nonumber \\ 
Z_{a_1} &=  |\chi^{m5}_{0}|^2 +  |\chi^{m5}_{\frac{1}{10}}|^2 + 
|\chi^{m5}_{\frac{3}{5}}|^2 +  |\chi^{m5}_{\frac{3}{2}}|^2 + 
|\chi^{m5}_{\frac{7}{16}}|^2 
\nonumber\\
&\ \ \ \
+  |\chi^{m5}_{\frac{3}{80}}|^2 
\nonumber \\ 
Z_{a_2} &= 0
\nonumber \\ 
Z_{b} &=  2 |\chi^{m5}_{0}|^2 +  2 |\chi^{m5}_{\frac{1}{10}}|^2 +  2
|\chi^{m5}_{\frac{3}{5}}|^2 +  2 |\chi^{m5}_{\frac{3}{2}}|^2 +  2
|\chi^{m5}_{\frac{7}{16}}|^2 
\nonumber\\
&\ \ \ \
+  2 |\chi^{m5}_{\frac{3}{80}}|^2 
\nonumber \\ 
Z_{b_1} &= 0
\nonumber \\ 
Z_{b_2} &= 0
\nonumber \\ 
Z_{c} &= 0
\nonumber \\ 
Z_{c_1} &= 0
\end{align}

\begin{align}
Z_\one &=  |\chi^{m5}_{0}|^2 +  |\chi^{m5}_{\frac{1}{10}}|^2 + 
|\chi^{m5}_{\frac{3}{5}}|^2 +  |\chi^{m5}_{\frac{3}{2}}|^2 + 
|\chi^{m5}_{\frac{7}{16}}|^2 
\nonumber\\
&\ \ \ \
+  |\chi^{m5}_{\frac{3}{80}}|^2 
\nonumber \\ 
Z_{a_1} &= 0
\nonumber \\ 
Z_{a_2} &=  |\chi^{m5}_{0}|^2 +  |\chi^{m5}_{\frac{1}{10}}|^2 + 
|\chi^{m5}_{\frac{3}{5}}|^2 +  |\chi^{m5}_{\frac{3}{2}}|^2 + 
|\chi^{m5}_{\frac{7}{16}}|^2 
\nonumber\\
&\ \ \ \
+  |\chi^{m5}_{\frac{3}{80}}|^2 
\nonumber \\ 
Z_{b} &= 0
\nonumber \\ 
Z_{b_1} &= 0
\nonumber \\ 
Z_{b_2} &= 0
\nonumber \\ 
Z_{c} &=  |\chi^{m5}_{0}|^2 +  |\chi^{m5}_{\frac{1}{10}}|^2 + 
|\chi^{m5}_{\frac{3}{5}}|^2 +  |\chi^{m5}_{\frac{3}{2}}|^2 + 
|\chi^{m5}_{\frac{7}{16}}|^2 
\nonumber\\
&\ \ \ \
+  |\chi^{m5}_{\frac{3}{80}}|^2 
\nonumber \\ 
Z_{c_1} &= 0
\end{align}

\begin{align}
Z_\one &=  |\chi^{m5}_{0}|^2 +  |\chi^{m5}_{\frac{1}{10}}|^2 + 
|\chi^{m5}_{\frac{3}{5}}|^2 +  |\chi^{m5}_{\frac{3}{2}}|^2 + 
|\chi^{m5}_{\frac{7}{16}}|^2 
\nonumber\\
&\ \ \ \
+  |\chi^{m5}_{\frac{3}{80}}|^2 
\nonumber \\ 
Z_{a_1} &= 0
\nonumber \\ 
Z_{a_2} &= 0
\nonumber \\ 
Z_{b} &=  |\chi^{m5}_{0}|^2 +  |\chi^{m5}_{\frac{1}{10}}|^2 + 
|\chi^{m5}_{\frac{3}{5}}|^2 +  |\chi^{m5}_{\frac{3}{2}}|^2 + 
|\chi^{m5}_{\frac{7}{16}}|^2 
\nonumber\\
&\ \ \ \
+  |\chi^{m5}_{\frac{3}{80}}|^2 
\nonumber \\ 
Z_{b_1} &= 0
\nonumber \\ 
Z_{b_2} &= 0
\nonumber \\ 
Z_{c} &=  |\chi^{m5}_{0}|^2 +  |\chi^{m5}_{\frac{1}{10}}|^2 + 
|\chi^{m5}_{\frac{3}{5}}|^2 +  |\chi^{m5}_{\frac{3}{2}}|^2 + 
|\chi^{m5}_{\frac{7}{16}}|^2 
\nonumber\\
&\ \ \ \
+  |\chi^{m5}_{\frac{3}{80}}|^2 
\nonumber \\ 
Z_{c_1} &= 0
\end{align}

$\one\oplus a_2$-condensed boundary with 3 relevant operators:
\begin{align}
Z_\one &=  |\chi^{m6}_{0}|^2 +  |\chi^{m6}_{\frac{2}{3}}|^2 + 
|\chi^{m6}_{3}|^2 +  |\chi^{m6}_{\frac{2}{5}}|^2 +  |\chi^{m6}_{\frac{1}{15}}|^2
\nonumber\\
&\ \ \ \
+  |\chi^{m6}_{\frac{7}{5}}|^2 
\nonumber \\ 
Z_{a_1} &=  |\chi^{m6}_{\frac{1}{8}}|^2 +  |\chi^{m6}_{\frac{13}{8}}|^2 + 
|\chi^{m6}_{\frac{1}{40}}|^2 +  |\chi^{m6}_{\frac{21}{40}}|^2 
\nonumber \\ 
Z_{a_2} &=  |\chi^{m6}_{0}|^2 +  |\chi^{m6}_{\frac{1}{8}}|^2 + 
|\chi^{m6}_{\frac{2}{3}}|^2 +  |\chi^{m6}_{\frac{13}{8}}|^2 +  |\chi^{m6}_{3}|^2
\nonumber\\
&\ \ \ \
+  |\chi^{m6}_{\frac{2}{5}}|^2 +  |\chi^{m6}_{\frac{1}{40}}|^2 + 
|\chi^{m6}_{\frac{1}{15}}|^2 +  |\chi^{m6}_{\frac{21}{40}}|^2 + 
|\chi^{m6}_{\frac{7}{5}}|^2 
\nonumber \\ 
Z_{b} &= 0
\nonumber \\ 
Z_{b_1} &= 0
\nonumber \\ 
Z_{b_2} &= 0
\nonumber \\ 
Z_{c} &=  \chi^{m6}_{0} \bar\chi^{m6}_{3} +  |\chi^{m6}_{\frac{2}{3}}|^2 + 
\chi^{m6}_{3} \bar\chi^{m6}_{0} +  \chi^{m6}_{\frac{2}{5}}
\bar\chi^{m6}_{\frac{7}{5}} 
\nonumber\\
&\ \ \ \
+  |\chi^{m6}_{\frac{1}{15}}|^2 +  \chi^{m6}_{\frac{7}{5}}
\bar\chi^{m6}_{\frac{2}{5}} 
\nonumber \\ 
Z_{c_1} &=  \chi^{m6}_{\frac{1}{8}} \bar\chi^{m6}_{\frac{13}{8}} + 
\chi^{m6}_{\frac{13}{8}} \bar\chi^{m6}_{\frac{1}{8}} +  \chi^{m6}_{\frac{1}{40}}
\bar\chi^{m6}_{\frac{21}{40}} +  \chi^{m6}_{\frac{21}{40}}
\bar\chi^{m6}_{\frac{1}{40}} 
\end{align}

$\one\oplus b$-condensed boundary with 3 relevant operators:
\begin{align}
Z_\one &=  |\chi^{m6}_{0}|^2 +  |\chi^{m6}_{\frac{2}{3}}|^2 + 
|\chi^{m6}_{3}|^2 +  |\chi^{m6}_{\frac{2}{5}}|^2 +  |\chi^{m6}_{\frac{1}{15}}|^2
\nonumber\\
&\ \ \ \
+  |\chi^{m6}_{\frac{7}{5}}|^2 
\nonumber \\ 
Z_{a_1} &=  |\chi^{m6}_{\frac{1}{8}}|^2 +  |\chi^{m6}_{\frac{13}{8}}|^2 + 
|\chi^{m6}_{\frac{1}{40}}|^2 +  |\chi^{m6}_{\frac{21}{40}}|^2 
\nonumber \\ 
Z_{a_2} &= 0
\nonumber \\ 
Z_{b} &=  |\chi^{m6}_{0}|^2 +  |\chi^{m6}_{\frac{1}{8}}|^2 + 
|\chi^{m6}_{\frac{2}{3}}|^2 +  |\chi^{m6}_{\frac{13}{8}}|^2 +  |\chi^{m6}_{3}|^2
\nonumber\\
&\ \ \ \
+  |\chi^{m6}_{\frac{2}{5}}|^2 +  |\chi^{m6}_{\frac{1}{40}}|^2 + 
|\chi^{m6}_{\frac{1}{15}}|^2 +  |\chi^{m6}_{\frac{21}{40}}|^2 + 
|\chi^{m6}_{\frac{7}{5}}|^2 
\nonumber \\ 
Z_{b_1} &= 0
\nonumber \\ 
Z_{b_2} &= 0
\nonumber \\ 
Z_{c} &=  \chi^{m6}_{0} \bar\chi^{m6}_{3} +  |\chi^{m6}_{\frac{2}{3}}|^2 + 
\chi^{m6}_{3} \bar\chi^{m6}_{0} +  \chi^{m6}_{\frac{2}{5}}
\bar\chi^{m6}_{\frac{7}{5}} 
\nonumber\\
&\ \ \ \
+  |\chi^{m6}_{\frac{1}{15}}|^2 +  \chi^{m6}_{\frac{7}{5}}
\bar\chi^{m6}_{\frac{2}{5}} 
\nonumber \\ 
Z_{c_1} &=  \chi^{m6}_{\frac{1}{8}} \bar\chi^{m6}_{\frac{13}{8}} + 
\chi^{m6}_{\frac{13}{8}} \bar\chi^{m6}_{\frac{1}{8}} +  \chi^{m6}_{\frac{1}{40}}
\bar\chi^{m6}_{\frac{21}{40}} +  \chi^{m6}_{\frac{21}{40}}
\bar\chi^{m6}_{\frac{1}{40}} 
\end{align}

\begin{align}
Z_\one &=  |\chi^{m6}_{0}|^2 +  \chi^{m6}_{0} \bar\chi^{m6}_{3} +  2
|\chi^{m6}_{\frac{2}{3}}|^2 +  \chi^{m6}_{3} \bar\chi^{m6}_{0} + 
|\chi^{m6}_{3}|^2 
\nonumber\\
&\ \ \ \
+  |\chi^{m6}_{\frac{2}{5}}|^2 +  \chi^{m6}_{\frac{2}{5}}
\bar\chi^{m6}_{\frac{7}{5}} +  2 |\chi^{m6}_{\frac{1}{15}}|^2 + 
\chi^{m6}_{\frac{7}{5}} \bar\chi^{m6}_{\frac{2}{5}} + 
|\chi^{m6}_{\frac{7}{5}}|^2 
\nonumber \\ 
Z_{a_1} &=  |\chi^{m6}_{0}|^2 +  \chi^{m6}_{0} \bar\chi^{m6}_{3} +  2
|\chi^{m6}_{\frac{2}{3}}|^2 +  \chi^{m6}_{3} \bar\chi^{m6}_{0} + 
|\chi^{m6}_{3}|^2 
\nonumber\\
&\ \ \ \
+  |\chi^{m6}_{\frac{2}{5}}|^2 +  \chi^{m6}_{\frac{2}{5}}
\bar\chi^{m6}_{\frac{7}{5}} +  2 |\chi^{m6}_{\frac{1}{15}}|^2 + 
\chi^{m6}_{\frac{7}{5}} \bar\chi^{m6}_{\frac{2}{5}} + 
|\chi^{m6}_{\frac{7}{5}}|^2 
\nonumber \\ 
Z_{a_2} &= 0
\nonumber \\ 
Z_{b} &=  2 |\chi^{m6}_{0}|^2 +  2 \chi^{m6}_{0} \bar\chi^{m6}_{3} +  4
|\chi^{m6}_{\frac{2}{3}}|^2 +  2 \chi^{m6}_{3} \bar\chi^{m6}_{0} +  2
|\chi^{m6}_{3}|^2 
\nonumber\\
&
+  2 |\chi^{m6}_{\frac{2}{5}}|^2 +  2 \chi^{m6}_{\frac{2}{5}}
\bar\chi^{m6}_{\frac{7}{5}} +  4 |\chi^{m6}_{\frac{1}{15}}|^2 +  2
\chi^{m6}_{\frac{7}{5}} \bar\chi^{m6}_{\frac{2}{5}} +  2
|\chi^{m6}_{\frac{7}{5}}|^2 
\nonumber \\ 
Z_{b_1} &= 0
\nonumber \\ 
Z_{b_2} &= 0
\nonumber \\ 
Z_{c} &= 0
\nonumber \\ 
Z_{c_1} &= 0
\end{align}

$\one\oplus a_1$-condensed boundary with 1 relevant operator:
\begin{align}
Z_\one &=  |\chi^{m6}_{0}|^2 +  \chi^{m6}_{0} \bar\chi^{m6}_{3} + 
\chi^{m6}_{3} \bar\chi^{m6}_{0} +  |\chi^{m6}_{3}|^2 + 
|\chi^{m6}_{\frac{2}{5}}|^2 
\nonumber\\
&\ \ \ \
+  \chi^{m6}_{\frac{2}{5}} \bar\chi^{m6}_{\frac{7}{5}} + 
\chi^{m6}_{\frac{7}{5}} \bar\chi^{m6}_{\frac{2}{5}} + 
|\chi^{m6}_{\frac{7}{5}}|^2 
\nonumber \\ 
Z_{a_1} &=  |\chi^{m6}_{0}|^2 +  \chi^{m6}_{0} \bar\chi^{m6}_{3} + 
\chi^{m6}_{3} \bar\chi^{m6}_{0} +  |\chi^{m6}_{3}|^2 + 
|\chi^{m6}_{\frac{2}{5}}|^2 
\nonumber\\
&\ \ \ \
+  \chi^{m6}_{\frac{2}{5}} \bar\chi^{m6}_{\frac{7}{5}} + 
\chi^{m6}_{\frac{7}{5}} \bar\chi^{m6}_{\frac{2}{5}} + 
|\chi^{m6}_{\frac{7}{5}}|^2 
\nonumber \\ 
Z_{a_2} &=  2 |\chi^{m6}_{\frac{2}{3}}|^2 +  2 |\chi^{m6}_{\frac{1}{15}}|^2 
\nonumber \\ 
Z_{b} &=  2 |\chi^{m6}_{\frac{2}{3}}|^2 +  2 |\chi^{m6}_{\frac{1}{15}}|^2 
\nonumber \\ 
Z_{b_1} &=  2 \chi^{m6}_{0} \bar\chi^{m6}_{\frac{2}{3}} +  2 \chi^{m6}_{3}
\bar\chi^{m6}_{\frac{2}{3}} +  2 \chi^{m6}_{\frac{2}{5}}
\bar\chi^{m6}_{\frac{1}{15}} +  2 \chi^{m6}_{\frac{7}{5}}
\bar\chi^{m6}_{\frac{1}{15}} 
\nonumber \\ 
Z_{b_2} &=  2 \chi^{m6}_{\frac{2}{3}} \bar\chi^{m6}_{0} +  2
\chi^{m6}_{\frac{2}{3}} \bar\chi^{m6}_{3} +  2 \chi^{m6}_{\frac{1}{15}}
\bar\chi^{m6}_{\frac{2}{5}} +  2 \chi^{m6}_{\frac{1}{15}}
\bar\chi^{m6}_{\frac{7}{5}} 
\nonumber \\ 
Z_{c} &= 0
\nonumber \\ 
Z_{c_1} &= 0
\end{align}

\begin{align}
Z_\one &=  |\chi^{m6}_{0}|^2 +  \chi^{m6}_{0} \bar\chi^{m6}_{3} +  2
|\chi^{m6}_{\frac{2}{3}}|^2 +  \chi^{m6}_{3} \bar\chi^{m6}_{0} + 
|\chi^{m6}_{3}|^2 
\nonumber\\
&\ \ \ \
+  |\chi^{m6}_{\frac{2}{5}}|^2 +  \chi^{m6}_{\frac{2}{5}}
\bar\chi^{m6}_{\frac{7}{5}} +  2 |\chi^{m6}_{\frac{1}{15}}|^2 + 
\chi^{m6}_{\frac{7}{5}} \bar\chi^{m6}_{\frac{2}{5}} + 
|\chi^{m6}_{\frac{7}{5}}|^2 
\nonumber \\ 
Z_{a_1} &= 0
\nonumber \\ 
Z_{a_2} &=  |\chi^{m6}_{0}|^2 +  \chi^{m6}_{0} \bar\chi^{m6}_{3} +  2
|\chi^{m6}_{\frac{2}{3}}|^2 +  \chi^{m6}_{3} \bar\chi^{m6}_{0} + 
|\chi^{m6}_{3}|^2 
\nonumber\\
&\ \ \ \
+  |\chi^{m6}_{\frac{2}{5}}|^2 +  \chi^{m6}_{\frac{2}{5}}
\bar\chi^{m6}_{\frac{7}{5}} +  2 |\chi^{m6}_{\frac{1}{15}}|^2 + 
\chi^{m6}_{\frac{7}{5}} \bar\chi^{m6}_{\frac{2}{5}} + 
|\chi^{m6}_{\frac{7}{5}}|^2 
\nonumber \\ 
Z_{b} &= 0
\nonumber \\ 
Z_{b_1} &= 0
\nonumber \\ 
Z_{b_2} &= 0
\nonumber \\ 
Z_{c} &=  |\chi^{m6}_{0}|^2 +  \chi^{m6}_{0} \bar\chi^{m6}_{3} +  2
|\chi^{m6}_{\frac{2}{3}}|^2 +  \chi^{m6}_{3} \bar\chi^{m6}_{0} + 
|\chi^{m6}_{3}|^2 
\nonumber\\
&\ \ \ \
+  |\chi^{m6}_{\frac{2}{5}}|^2 +  \chi^{m6}_{\frac{2}{5}}
\bar\chi^{m6}_{\frac{7}{5}} +  2 |\chi^{m6}_{\frac{1}{15}}|^2 + 
\chi^{m6}_{\frac{7}{5}} \bar\chi^{m6}_{\frac{2}{5}} + 
|\chi^{m6}_{\frac{7}{5}}|^2 
\nonumber \\ 
Z_{c_1} &= 0
\end{align}

\begin{align}
Z_\one &=  |\chi^{m6}_{0}|^2 +  \chi^{m6}_{0} \bar\chi^{m6}_{3} +  2
|\chi^{m6}_{\frac{2}{3}}|^2 +  \chi^{m6}_{3} \bar\chi^{m6}_{0} + 
|\chi^{m6}_{3}|^2 
\nonumber\\
&\ \ \ \
+  |\chi^{m6}_{\frac{2}{5}}|^2 +  \chi^{m6}_{\frac{2}{5}}
\bar\chi^{m6}_{\frac{7}{5}} +  2 |\chi^{m6}_{\frac{1}{15}}|^2 + 
\chi^{m6}_{\frac{7}{5}} \bar\chi^{m6}_{\frac{2}{5}} + 
|\chi^{m6}_{\frac{7}{5}}|^2 
\nonumber \\ 
Z_{a_1} &= 0
\nonumber \\ 
Z_{a_2} &= 0
\nonumber \\ 
Z_{b} &=  |\chi^{m6}_{0}|^2 +  \chi^{m6}_{0} \bar\chi^{m6}_{3} +  2
|\chi^{m6}_{\frac{2}{3}}|^2 +  \chi^{m6}_{3} \bar\chi^{m6}_{0} + 
|\chi^{m6}_{3}|^2 
\nonumber\\
&\ \ \ \
+  |\chi^{m6}_{\frac{2}{5}}|^2 +  \chi^{m6}_{\frac{2}{5}}
\bar\chi^{m6}_{\frac{7}{5}} +  2 |\chi^{m6}_{\frac{1}{15}}|^2 + 
\chi^{m6}_{\frac{7}{5}} \bar\chi^{m6}_{\frac{2}{5}} + 
|\chi^{m6}_{\frac{7}{5}}|^2 
\nonumber \\ 
Z_{b_1} &= 0
\nonumber \\ 
Z_{b_2} &= 0
\nonumber \\ 
Z_{c} &=  |\chi^{m6}_{0}|^2 +  \chi^{m6}_{0} \bar\chi^{m6}_{3} +  2
|\chi^{m6}_{\frac{2}{3}}|^2 +  \chi^{m6}_{3} \bar\chi^{m6}_{0} + 
|\chi^{m6}_{3}|^2 
\nonumber\\
&\ \ \ \
+  |\chi^{m6}_{\frac{2}{5}}|^2 +  \chi^{m6}_{\frac{2}{5}}
\bar\chi^{m6}_{\frac{7}{5}} +  2 |\chi^{m6}_{\frac{1}{15}}|^2 + 
\chi^{m6}_{\frac{7}{5}} \bar\chi^{m6}_{\frac{2}{5}} + 
|\chi^{m6}_{\frac{7}{5}}|^2 
\nonumber \\ 
Z_{c_1} &= 0
\end{align}

$\one$-condensed boundary with 1 relevant operator:
\begin{align}
Z_\one &=  |\chi^{m6}_{0}|^2 +  |\chi^{m6}_{3}|^2 + 
|\chi^{m6}_{\frac{2}{5}}|^2 +  |\chi^{m6}_{\frac{7}{5}}|^2 
\nonumber \\ 
Z_{a_1} &=  \chi^{m6}_{0} \bar\chi^{m6}_{3} +  \chi^{m6}_{3} \bar\chi^{m6}_{0} +
\chi^{m6}_{\frac{2}{5}} \bar\chi^{m6}_{\frac{7}{5}} +  \chi^{m6}_{\frac{7}{5}}
\bar\chi^{m6}_{\frac{2}{5}} 
\nonumber \\ 
Z_{a_2} &=  |\chi^{m6}_{\frac{2}{3}}|^2 +  |\chi^{m6}_{\frac{1}{15}}|^2 
\nonumber \\ 
Z_{b} &=  |\chi^{m6}_{\frac{2}{3}}|^2 +  |\chi^{m6}_{\frac{1}{15}}|^2 
\nonumber \\ 
Z_{b_1} &=  \chi^{m6}_{0} \bar\chi^{m6}_{\frac{2}{3}} +  \chi^{m6}_{3}
\bar\chi^{m6}_{\frac{2}{3}} +  \chi^{m6}_{\frac{2}{5}}
\bar\chi^{m6}_{\frac{1}{15}} +  \chi^{m6}_{\frac{7}{5}}
\bar\chi^{m6}_{\frac{1}{15}} 
\nonumber \\ 
Z_{b_2} &=  \chi^{m6}_{\frac{2}{3}} \bar\chi^{m6}_{0} +  \chi^{m6}_{\frac{2}{3}}
\bar\chi^{m6}_{3} +  \chi^{m6}_{\frac{1}{15}} \bar\chi^{m6}_{\frac{2}{5}} + 
\chi^{m6}_{\frac{1}{15}} \bar\chi^{m6}_{\frac{7}{5}} 
\nonumber \\ 
Z_{c} &=  |\chi^{m6}_{\frac{1}{8}}|^2 +  |\chi^{m6}_{\frac{13}{8}}|^2 + 
|\chi^{m6}_{\frac{1}{40}}|^2 +  |\chi^{m6}_{\frac{21}{40}}|^2 
\nonumber \\ 
Z_{c_1} &=  \chi^{m6}_{\frac{1}{8}} \bar\chi^{m6}_{\frac{13}{8}} + 
\chi^{m6}_{\frac{13}{8}} \bar\chi^{m6}_{\frac{1}{8}} +  \chi^{m6}_{\frac{1}{40}}
\bar\chi^{m6}_{\frac{21}{40}} +  \chi^{m6}_{\frac{21}{40}}
\bar\chi^{m6}_{\frac{1}{40}} 
\end{align}

\begin{align}
Z_\one &=  |\chi^{m6}_{0}|^2 +  |\chi^{m6}_{\frac{1}{8}}|^2 + 
|\chi^{m6}_{\frac{2}{3}}|^2 +  |\chi^{m6}_{\frac{13}{8}}|^2 +  |\chi^{m6}_{3}|^2
\nonumber\\
&\ \ \ \
+  |\chi^{m6}_{\frac{2}{5}}|^2 +  |\chi^{m6}_{\frac{1}{40}}|^2 + 
|\chi^{m6}_{\frac{1}{15}}|^2 +  |\chi^{m6}_{\frac{21}{40}}|^2 + 
|\chi^{m6}_{\frac{7}{5}}|^2 
\nonumber \\ 
Z_{a_1} &= 0
\nonumber \\ 
Z_{a_2} &=  |\chi^{m6}_{0}|^2 +  |\chi^{m6}_{\frac{1}{8}}|^2 + 
|\chi^{m6}_{\frac{2}{3}}|^2 +  |\chi^{m6}_{\frac{13}{8}}|^2 +  |\chi^{m6}_{3}|^2
\nonumber\\
&\ \ \ \
+  |\chi^{m6}_{\frac{2}{5}}|^2 +  |\chi^{m6}_{\frac{1}{40}}|^2 + 
|\chi^{m6}_{\frac{1}{15}}|^2 +  |\chi^{m6}_{\frac{21}{40}}|^2 + 
|\chi^{m6}_{\frac{7}{5}}|^2 
\nonumber \\ 
Z_{b} &= 0
\nonumber \\ 
Z_{b_1} &= 0
\nonumber \\ 
Z_{b_2} &= 0
\nonumber \\ 
Z_{c} &=  |\chi^{m6}_{0}|^2 +  |\chi^{m6}_{\frac{1}{8}}|^2 + 
|\chi^{m6}_{\frac{2}{3}}|^2 +  |\chi^{m6}_{\frac{13}{8}}|^2 +  |\chi^{m6}_{3}|^2
\nonumber\\
&\ \ \ \
+  |\chi^{m6}_{\frac{2}{5}}|^2 +  |\chi^{m6}_{\frac{1}{40}}|^2 + 
|\chi^{m6}_{\frac{1}{15}}|^2 +  |\chi^{m6}_{\frac{21}{40}}|^2 + 
|\chi^{m6}_{\frac{7}{5}}|^2 
\nonumber \\ 
Z_{c_1} &= 0
\end{align}

$\one\oplus a_2$-condensed boundary with 3 relevant operators:
\begin{align}
Z_\one &=  |\chi^{m6}_{0}|^2 +  |\chi^{m6}_{\frac{2}{3}}|^2 + 
|\chi^{m6}_{3}|^2 +  |\chi^{m6}_{\frac{2}{5}}|^2 +  |\chi^{m6}_{\frac{1}{15}}|^2
\nonumber\\
&\ \ \ \
+  |\chi^{m6}_{\frac{7}{5}}|^2 
\nonumber \\ 
Z_{a_1} &=  \chi^{m6}_{0} \bar\chi^{m6}_{3} +  |\chi^{m6}_{\frac{2}{3}}|^2 + 
\chi^{m6}_{3} \bar\chi^{m6}_{0} +  \chi^{m6}_{\frac{2}{5}}
\bar\chi^{m6}_{\frac{7}{5}} +  |\chi^{m6}_{\frac{1}{15}}|^2 
\nonumber\\
&\ \ \ \
+  \chi^{m6}_{\frac{7}{5}} \bar\chi^{m6}_{\frac{2}{5}} 
\nonumber \\ 
Z_{a_2} &=  |\chi^{m6}_{0}|^2 +  \chi^{m6}_{0} \bar\chi^{m6}_{3} +  2
|\chi^{m6}_{\frac{2}{3}}|^2 +  \chi^{m6}_{3} \bar\chi^{m6}_{0} + 
|\chi^{m6}_{3}|^2 
\nonumber\\
&\ \ \ \
+  |\chi^{m6}_{\frac{2}{5}}|^2 +  \chi^{m6}_{\frac{2}{5}}
\bar\chi^{m6}_{\frac{7}{5}} +  2 |\chi^{m6}_{\frac{1}{15}}|^2 + 
\chi^{m6}_{\frac{7}{5}} \bar\chi^{m6}_{\frac{2}{5}} + 
|\chi^{m6}_{\frac{7}{5}}|^2 
\nonumber \\ 
Z_{b} &= 0
\nonumber \\ 
Z_{b_1} &= 0
\nonumber \\ 
Z_{b_2} &= 0
\nonumber \\ 
Z_{c} &=  |\chi^{m6}_{\frac{1}{8}}|^2 +  |\chi^{m6}_{\frac{13}{8}}|^2 + 
|\chi^{m6}_{\frac{1}{40}}|^2 +  |\chi^{m6}_{\frac{21}{40}}|^2 
\nonumber \\ 
Z_{c_1} &=  \chi^{m6}_{\frac{1}{8}} \bar\chi^{m6}_{\frac{13}{8}} + 
\chi^{m6}_{\frac{13}{8}} \bar\chi^{m6}_{\frac{1}{8}} +  \chi^{m6}_{\frac{1}{40}}
\bar\chi^{m6}_{\frac{21}{40}} +  \chi^{m6}_{\frac{21}{40}}
\bar\chi^{m6}_{\frac{1}{40}} 
\end{align}

\begin{align}
Z_\one &=  |\chi^{m6}_{0}|^2 +  |\chi^{m6}_{\frac{1}{8}}|^2 + 
|\chi^{m6}_{\frac{2}{3}}|^2 +  |\chi^{m6}_{\frac{13}{8}}|^2 +  |\chi^{m6}_{3}|^2
\nonumber\\
&\ \ \ \
+  |\chi^{m6}_{\frac{2}{5}}|^2 +  |\chi^{m6}_{\frac{1}{40}}|^2 + 
|\chi^{m6}_{\frac{1}{15}}|^2 +  |\chi^{m6}_{\frac{21}{40}}|^2 + 
|\chi^{m6}_{\frac{7}{5}}|^2 
\nonumber \\ 
Z_{a_1} &= 0
\nonumber \\ 
Z_{a_2} &= 0
\nonumber \\ 
Z_{b} &=  |\chi^{m6}_{0}|^2 +  |\chi^{m6}_{\frac{1}{8}}|^2 + 
|\chi^{m6}_{\frac{2}{3}}|^2 +  |\chi^{m6}_{\frac{13}{8}}|^2 +  |\chi^{m6}_{3}|^2
\nonumber\\
&\ \ \ \
+  |\chi^{m6}_{\frac{2}{5}}|^2 +  |\chi^{m6}_{\frac{1}{40}}|^2 + 
|\chi^{m6}_{\frac{1}{15}}|^2 +  |\chi^{m6}_{\frac{21}{40}}|^2 + 
|\chi^{m6}_{\frac{7}{5}}|^2 
\nonumber \\ 
Z_{b_1} &= 0
\nonumber \\ 
Z_{b_2} &= 0
\nonumber \\ 
Z_{c} &=  |\chi^{m6}_{0}|^2 +  |\chi^{m6}_{\frac{1}{8}}|^2 + 
|\chi^{m6}_{\frac{2}{3}}|^2 +  |\chi^{m6}_{\frac{13}{8}}|^2 +  |\chi^{m6}_{3}|^2
\nonumber\\
&\ \ \ \
+  |\chi^{m6}_{\frac{2}{5}}|^2 +  |\chi^{m6}_{\frac{1}{40}}|^2 + 
|\chi^{m6}_{\frac{1}{15}}|^2 +  |\chi^{m6}_{\frac{21}{40}}|^2 + 
|\chi^{m6}_{\frac{7}{5}}|^2 
\nonumber \\ 
Z_{c_1} &= 0
\end{align}

$\one\oplus b$-condensed boundary with 3 relevant operators:
\begin{align}
Z_\one &=  |\chi^{m6}_{0}|^2 +  |\chi^{m6}_{\frac{2}{3}}|^2 + 
|\chi^{m6}_{3}|^2 +  |\chi^{m6}_{\frac{2}{5}}|^2 +  |\chi^{m6}_{\frac{1}{15}}|^2
+  |\chi^{m6}_{\frac{7}{5}}|^2 
\nonumber \\ 
Z_{a_1} &=  \chi^{m6}_{0} \bar\chi^{m6}_{3} +  |\chi^{m6}_{\frac{2}{3}}|^2 + 
\chi^{m6}_{3} \bar\chi^{m6}_{0} +  \chi^{m6}_{\frac{2}{5}}
\bar\chi^{m6}_{\frac{7}{5}} +  |\chi^{m6}_{\frac{1}{15}}|^2 
\nonumber\\
&\ \ \ \
+  \chi^{m6}_{\frac{7}{5}} \bar\chi^{m6}_{\frac{2}{5}} 
\nonumber \\ 
Z_{a_2} &= 0
\nonumber \\ 
Z_{b} &=  |\chi^{m6}_{0}|^2 +  \chi^{m6}_{0} \bar\chi^{m6}_{3} +  2
|\chi^{m6}_{\frac{2}{3}}|^2 +  \chi^{m6}_{3} \bar\chi^{m6}_{0} + 
|\chi^{m6}_{3}|^2 
\nonumber\\
&\ \ \ \
+  |\chi^{m6}_{\frac{2}{5}}|^2 +  \chi^{m6}_{\frac{2}{5}}
\bar\chi^{m6}_{\frac{7}{5}} +  2 |\chi^{m6}_{\frac{1}{15}}|^2 + 
\chi^{m6}_{\frac{7}{5}} \bar\chi^{m6}_{\frac{2}{5}} + 
|\chi^{m6}_{\frac{7}{5}}|^2 
\nonumber \\ 
Z_{b_1} &= 0
\nonumber \\ 
Z_{b_2} &= 0
\nonumber \\ 
Z_{c} &=  |\chi^{m6}_{\frac{1}{8}}|^2 +  |\chi^{m6}_{\frac{13}{8}}|^2 + 
|\chi^{m6}_{\frac{1}{40}}|^2 +  |\chi^{m6}_{\frac{21}{40}}|^2 
\nonumber \\ 
Z_{c_1} &=  \chi^{m6}_{\frac{1}{8}} \bar\chi^{m6}_{\frac{13}{8}} + 
\chi^{m6}_{\frac{13}{8}} \bar\chi^{m6}_{\frac{1}{8}} +  \chi^{m6}_{\frac{1}{40}}
\bar\chi^{m6}_{\frac{21}{40}} +  \chi^{m6}_{\frac{21}{40}}
\bar\chi^{m6}_{\frac{1}{40}} 
\end{align}

\begin{align}
Z_\one &=  |\chi^{m6}_{0}|^2 +  |\chi^{m6}_{\frac{1}{8}}|^2 + 
|\chi^{m6}_{\frac{2}{3}}|^2 +  |\chi^{m6}_{\frac{13}{8}}|^2 +  |\chi^{m6}_{3}|^2
\nonumber\\
&\ \ \ \
+  |\chi^{m6}_{\frac{2}{5}}|^2 +  |\chi^{m6}_{\frac{1}{40}}|^2 + 
|\chi^{m6}_{\frac{1}{15}}|^2 +  |\chi^{m6}_{\frac{21}{40}}|^2 + 
|\chi^{m6}_{\frac{7}{5}}|^2 
\nonumber \\ 
Z_{a_1} &=  |\chi^{m6}_{0}|^2 +  |\chi^{m6}_{\frac{1}{8}}|^2 + 
|\chi^{m6}_{\frac{2}{3}}|^2 +  |\chi^{m6}_{\frac{13}{8}}|^2 +  |\chi^{m6}_{3}|^2
\nonumber\\
&\ \ \ \
+  |\chi^{m6}_{\frac{2}{5}}|^2 +  |\chi^{m6}_{\frac{1}{40}}|^2 + 
|\chi^{m6}_{\frac{1}{15}}|^2 +  |\chi^{m6}_{\frac{21}{40}}|^2 + 
|\chi^{m6}_{\frac{7}{5}}|^2 
\nonumber \\ 
Z_{a_2} &=  2 |\chi^{m6}_{0}|^2 +  2 |\chi^{m6}_{\frac{1}{8}}|^2 +  2
|\chi^{m6}_{\frac{2}{3}}|^2 +  2 |\chi^{m6}_{\frac{13}{8}}|^2 +  2
|\chi^{m6}_{3}|^2 
\nonumber\\
&
+  2 |\chi^{m6}_{\frac{2}{5}}|^2 +  2 |\chi^{m6}_{\frac{1}{40}}|^2 +  2
|\chi^{m6}_{\frac{1}{15}}|^2 +  2 |\chi^{m6}_{\frac{21}{40}}|^2 +  2
|\chi^{m6}_{\frac{7}{5}}|^2 
\nonumber \\ 
Z_{b} &= 0
\nonumber \\ 
Z_{b_1} &= 0
\nonumber \\ 
Z_{b_2} &= 0
\nonumber \\ 
Z_{c} &= 0
\nonumber \\ 
Z_{c_1} &= 0
\end{align}

\begin{align}
Z_\one &=  |\chi^{m6}_{0}|^2 +  |\chi^{m6}_{\frac{1}{8}}|^2 + 
|\chi^{m6}_{\frac{2}{3}}|^2 +  |\chi^{m6}_{\frac{13}{8}}|^2 +  |\chi^{m6}_{3}|^2
\nonumber\\
&\ \ \ \
+  |\chi^{m6}_{\frac{2}{5}}|^2 +  |\chi^{m6}_{\frac{1}{40}}|^2 + 
|\chi^{m6}_{\frac{1}{15}}|^2 +  |\chi^{m6}_{\frac{21}{40}}|^2 + 
|\chi^{m6}_{\frac{7}{5}}|^2 
\nonumber \\ 
Z_{a_1} &=  |\chi^{m6}_{0}|^2 +  |\chi^{m6}_{\frac{1}{8}}|^2 + 
|\chi^{m6}_{\frac{2}{3}}|^2 +  |\chi^{m6}_{\frac{13}{8}}|^2 +  |\chi^{m6}_{3}|^2
\nonumber\\
&\ \ \ \
+  |\chi^{m6}_{\frac{2}{5}}|^2 +  |\chi^{m6}_{\frac{1}{40}}|^2 + 
|\chi^{m6}_{\frac{1}{15}}|^2 +  |\chi^{m6}_{\frac{21}{40}}|^2 + 
|\chi^{m6}_{\frac{7}{5}}|^2 
\nonumber \\ 
Z_{a_2} &= 0
\nonumber \\ 
Z_{b} &=  2 |\chi^{m6}_{0}|^2 +  2 |\chi^{m6}_{\frac{1}{8}}|^2 +  2
|\chi^{m6}_{\frac{2}{3}}|^2 +  2 |\chi^{m6}_{\frac{13}{8}}|^2 +  2
|\chi^{m6}_{3}|^2 
\nonumber\\
&
+  2 |\chi^{m6}_{\frac{2}{5}}|^2 +  2 |\chi^{m6}_{\frac{1}{40}}|^2 +  2
|\chi^{m6}_{\frac{1}{15}}|^2 +  2 |\chi^{m6}_{\frac{21}{40}}|^2 +  2
|\chi^{m6}_{\frac{7}{5}}|^2 
\nonumber \\ 
Z_{b_1} &= 0
\nonumber \\ 
Z_{b_2} &= 0
\nonumber \\ 
Z_{c} &= 0
\nonumber \\ 
Z_{c_1} &= 0
\end{align}

$\one\oplus a_2$-condensed boundary with 4 relevant operators:
\begin{align}
Z_\one &=  |\chi^{m7}_{0}|^2 +  |\chi^{m7}_{\frac{1}{7}}|^2 + 
|\chi^{m7}_{\frac{5}{7}}|^2 +  |\chi^{m7}_{\frac{12}{7}}|^2 + 
|\chi^{m7}_{\frac{22}{7}}|^2 
\nonumber\\
&\ \ \ \
+  |\chi^{m7}_{5}|^2 +  |\chi^{m7}_{\frac{4}{3}}|^2 + 
|\chi^{m7}_{\frac{10}{21}}|^2 +  |\chi^{m7}_{\frac{1}{21}}|^2 
\nonumber \\ 
Z_{a_1} &=  |\chi^{m7}_{\frac{3}{8}}|^2 +  |\chi^{m7}_{\frac{1}{56}}|^2 + 
|\chi^{m7}_{\frac{5}{56}}|^2 +  |\chi^{m7}_{\frac{33}{56}}|^2 + 
|\chi^{m7}_{\frac{85}{56}}|^2 
\nonumber\\
&\ \ \ \
+  |\chi^{m7}_{\frac{23}{8}}|^2 
\nonumber \\ 
Z_{a_2} &=  |\chi^{m7}_{0}|^2 +  |\chi^{m7}_{\frac{1}{7}}|^2 + 
|\chi^{m7}_{\frac{5}{7}}|^2 +  |\chi^{m7}_{\frac{12}{7}}|^2 + 
|\chi^{m7}_{\frac{22}{7}}|^2 
\nonumber\\
&\ \ \ \
+  |\chi^{m7}_{5}|^2 +  |\chi^{m7}_{\frac{3}{8}}|^2 + 
|\chi^{m7}_{\frac{1}{56}}|^2 +  |\chi^{m7}_{\frac{5}{56}}|^2 + 
|\chi^{m7}_{\frac{33}{56}}|^2 
\nonumber\\
&\ \ \ \
+  |\chi^{m7}_{\frac{85}{56}}|^2 +  |\chi^{m7}_{\frac{23}{8}}|^2 + 
|\chi^{m7}_{\frac{4}{3}}|^2 +  |\chi^{m7}_{\frac{10}{21}}|^2 + 
|\chi^{m7}_{\frac{1}{21}}|^2 
\nonumber \\ 
Z_{b} &= 0
\nonumber \\ 
Z_{b_1} &= 0
\nonumber \\ 
Z_{b_2} &= 0
\nonumber \\ 
Z_{c} &=  \chi^{m7}_{0} \bar\chi^{m7}_{5} +  \chi^{m7}_{\frac{1}{7}}
\bar\chi^{m7}_{\frac{22}{7}} +  \chi^{m7}_{\frac{5}{7}}
\bar\chi^{m7}_{\frac{12}{7}} +  \chi^{m7}_{\frac{12}{7}}
\bar\chi^{m7}_{\frac{5}{7}} 
\nonumber\\
&\ \ \ \
+  \chi^{m7}_{\frac{22}{7}} \bar\chi^{m7}_{\frac{1}{7}} +  \chi^{m7}_{5}
\bar\chi^{m7}_{0} +  |\chi^{m7}_{\frac{4}{3}}|^2 + 
|\chi^{m7}_{\frac{10}{21}}|^2 +  |\chi^{m7}_{\frac{1}{21}}|^2 
\nonumber \\ 
Z_{c_1} &=  \chi^{m7}_{\frac{3}{8}} \bar\chi^{m7}_{\frac{23}{8}} + 
\chi^{m7}_{\frac{1}{56}} \bar\chi^{m7}_{\frac{85}{56}} + 
\chi^{m7}_{\frac{5}{56}} \bar\chi^{m7}_{\frac{33}{56}} + 
\chi^{m7}_{\frac{33}{56}} \bar\chi^{m7}_{\frac{5}{56}} 
\nonumber\\
&\ \ \ \
+  \chi^{m7}_{\frac{85}{56}} \bar\chi^{m7}_{\frac{1}{56}} + 
\chi^{m7}_{\frac{23}{8}} \bar\chi^{m7}_{\frac{3}{8}} 
\end{align}

$\one\oplus b$-condensed boundary with 4 relevant operators:
\begin{align}
Z_\one &=  |\chi^{m7}_{0}|^2 +  |\chi^{m7}_{\frac{1}{7}}|^2 + 
|\chi^{m7}_{\frac{5}{7}}|^2 +  |\chi^{m7}_{\frac{12}{7}}|^2 + 
|\chi^{m7}_{\frac{22}{7}}|^2 
\nonumber\\
&\ \ \ \
+  |\chi^{m7}_{5}|^2 +  |\chi^{m7}_{\frac{4}{3}}|^2 + 
|\chi^{m7}_{\frac{10}{21}}|^2 +  |\chi^{m7}_{\frac{1}{21}}|^2 
\nonumber \\ 
Z_{a_1} &=  |\chi^{m7}_{\frac{3}{8}}|^2 +  |\chi^{m7}_{\frac{1}{56}}|^2 + 
|\chi^{m7}_{\frac{5}{56}}|^2 +  |\chi^{m7}_{\frac{33}{56}}|^2 + 
|\chi^{m7}_{\frac{85}{56}}|^2 
\nonumber\\
&\ \ \ \
+  |\chi^{m7}_{\frac{23}{8}}|^2 
\nonumber \\ 
Z_{a_2} &= 0
\nonumber \\ 
Z_{b} &=  |\chi^{m7}_{0}|^2 +  |\chi^{m7}_{\frac{1}{7}}|^2 + 
|\chi^{m7}_{\frac{5}{7}}|^2 +  |\chi^{m7}_{\frac{12}{7}}|^2 + 
|\chi^{m7}_{\frac{22}{7}}|^2 
\nonumber\\
&\ \ \ \
+  |\chi^{m7}_{5}|^2 +  |\chi^{m7}_{\frac{3}{8}}|^2 + 
|\chi^{m7}_{\frac{1}{56}}|^2 +  |\chi^{m7}_{\frac{5}{56}}|^2 + 
|\chi^{m7}_{\frac{33}{56}}|^2 
\nonumber\\
&\ \ \ \
+  |\chi^{m7}_{\frac{85}{56}}|^2 +  |\chi^{m7}_{\frac{23}{8}}|^2 + 
|\chi^{m7}_{\frac{4}{3}}|^2 +  |\chi^{m7}_{\frac{10}{21}}|^2 + 
|\chi^{m7}_{\frac{1}{21}}|^2 
\nonumber \\ 
Z_{b_1} &= 0
\nonumber \\ 
Z_{b_2} &= 0
\nonumber \\ 
Z_{c} &=  \chi^{m7}_{0} \bar\chi^{m7}_{5} +  \chi^{m7}_{\frac{1}{7}}
\bar\chi^{m7}_{\frac{22}{7}} +  \chi^{m7}_{\frac{5}{7}}
\bar\chi^{m7}_{\frac{12}{7}} +  \chi^{m7}_{\frac{12}{7}}
\bar\chi^{m7}_{\frac{5}{7}} 
\nonumber\\
&\ \ \ \
+  \chi^{m7}_{\frac{22}{7}} \bar\chi^{m7}_{\frac{1}{7}} +  \chi^{m7}_{5}
\bar\chi^{m7}_{0} +  |\chi^{m7}_{\frac{4}{3}}|^2 + 
|\chi^{m7}_{\frac{10}{21}}|^2 +  |\chi^{m7}_{\frac{1}{21}}|^2 
\nonumber \\ 
Z_{c_1} &=  \chi^{m7}_{\frac{3}{8}} \bar\chi^{m7}_{\frac{23}{8}} + 
\chi^{m7}_{\frac{1}{56}} \bar\chi^{m7}_{\frac{85}{56}} + 
\chi^{m7}_{\frac{5}{56}} \bar\chi^{m7}_{\frac{33}{56}} + 
\chi^{m7}_{\frac{33}{56}} \bar\chi^{m7}_{\frac{5}{56}} 
\nonumber\\
&\ \ \ \
+  \chi^{m7}_{\frac{85}{56}} \bar\chi^{m7}_{\frac{1}{56}} + 
\chi^{m7}_{\frac{23}{8}} \bar\chi^{m7}_{\frac{3}{8}} 
\end{align}

\begin{align}
Z_\one &=  |\chi^{m7}_{0}|^2 +  \chi^{m7}_{0} \bar\chi^{m7}_{5} + 
|\chi^{m7}_{\frac{1}{7}}|^2 +  \chi^{m7}_{\frac{1}{7}}
\bar\chi^{m7}_{\frac{22}{7}} +  |\chi^{m7}_{\frac{5}{7}}|^2 
\nonumber\\
&\ \ \ \
+  \chi^{m7}_{\frac{5}{7}} \bar\chi^{m7}_{\frac{12}{7}} + 
\chi^{m7}_{\frac{12}{7}} \bar\chi^{m7}_{\frac{5}{7}} + 
|\chi^{m7}_{\frac{12}{7}}|^2 +  \chi^{m7}_{\frac{22}{7}}
\bar\chi^{m7}_{\frac{1}{7}} 
\nonumber\\
&\ \ \ \
+  |\chi^{m7}_{\frac{22}{7}}|^2 +  \chi^{m7}_{5} \bar\chi^{m7}_{0} + 
|\chi^{m7}_{5}|^2 +  2 |\chi^{m7}_{\frac{4}{3}}|^2 
\nonumber\\
&\ \ \ \
+  2 |\chi^{m7}_{\frac{10}{21}}|^2 +  2 |\chi^{m7}_{\frac{1}{21}}|^2 
\nonumber \\ 
Z_{a_1} &=  |\chi^{m7}_{0}|^2 +  \chi^{m7}_{0} \bar\chi^{m7}_{5} + 
|\chi^{m7}_{\frac{1}{7}}|^2 +  \chi^{m7}_{\frac{1}{7}}
\bar\chi^{m7}_{\frac{22}{7}} +  |\chi^{m7}_{\frac{5}{7}}|^2 
\nonumber\\
&\ \ \ \
+  \chi^{m7}_{\frac{5}{7}} \bar\chi^{m7}_{\frac{12}{7}} + 
\chi^{m7}_{\frac{12}{7}} \bar\chi^{m7}_{\frac{5}{7}} + 
|\chi^{m7}_{\frac{12}{7}}|^2 +  \chi^{m7}_{\frac{22}{7}}
\bar\chi^{m7}_{\frac{1}{7}} 
\nonumber\\
&\ \ \ \
+  |\chi^{m7}_{\frac{22}{7}}|^2 +  \chi^{m7}_{5} \bar\chi^{m7}_{0} + 
|\chi^{m7}_{5}|^2 +  2 |\chi^{m7}_{\frac{4}{3}}|^2 
\nonumber\\
&\ \ \ \
+  2 |\chi^{m7}_{\frac{10}{21}}|^2 +  2 |\chi^{m7}_{\frac{1}{21}}|^2 
\nonumber \\ 
Z_{a_2} &= 0
\nonumber \\ 
Z_{b} &=  2 |\chi^{m7}_{0}|^2 +  2 \chi^{m7}_{0} \bar\chi^{m7}_{5} +  2
|\chi^{m7}_{\frac{1}{7}}|^2 +  2 \chi^{m7}_{\frac{1}{7}}
\bar\chi^{m7}_{\frac{22}{7}} +  2 |\chi^{m7}_{\frac{5}{7}}|^2 
\nonumber\\
&\ \ \ \
+  2 \chi^{m7}_{\frac{5}{7}} \bar\chi^{m7}_{\frac{12}{7}} +  2
\chi^{m7}_{\frac{12}{7}} \bar\chi^{m7}_{\frac{5}{7}} +  2
|\chi^{m7}_{\frac{12}{7}}|^2 +  2 \chi^{m7}_{\frac{22}{7}}
\bar\chi^{m7}_{\frac{1}{7}} 
\nonumber\\
&\ \ \ \
+  2 |\chi^{m7}_{\frac{22}{7}}|^2 +  2 \chi^{m7}_{5} \bar\chi^{m7}_{0} +  2
|\chi^{m7}_{5}|^2 +  4 |\chi^{m7}_{\frac{4}{3}}|^2 
\nonumber\\
&\ \ \ \
+  4 |\chi^{m7}_{\frac{10}{21}}|^2 +  4 |\chi^{m7}_{\frac{1}{21}}|^2 
\nonumber \\ 
Z_{b_1} &= 0
\nonumber \\ 
Z_{b_2} &= 0
\nonumber \\ 
Z_{c} &= 0
\nonumber \\ 
Z_{c_1} &= 0
\end{align}

$\one\oplus a_1$-condensed boundary with 2 relevant operators:
\begin{align}
Z_\one &=  |\chi^{m7}_{0}|^2 +  \chi^{m7}_{0} \bar\chi^{m7}_{5} + 
|\chi^{m7}_{\frac{1}{7}}|^2 +  \chi^{m7}_{\frac{1}{7}}
\bar\chi^{m7}_{\frac{22}{7}} +  |\chi^{m7}_{\frac{5}{7}}|^2 
\nonumber\\
&\ \ \ \
+  \chi^{m7}_{\frac{5}{7}} \bar\chi^{m7}_{\frac{12}{7}} + 
\chi^{m7}_{\frac{12}{7}} \bar\chi^{m7}_{\frac{5}{7}} + 
|\chi^{m7}_{\frac{12}{7}}|^2 +  \chi^{m7}_{\frac{22}{7}}
\bar\chi^{m7}_{\frac{1}{7}} 
\nonumber\\
&\ \ \ \
+  |\chi^{m7}_{\frac{22}{7}}|^2 +  \chi^{m7}_{5} \bar\chi^{m7}_{0} + 
|\chi^{m7}_{5}|^2 
\nonumber \\ 
Z_{a_1} &=  |\chi^{m7}_{0}|^2 +  \chi^{m7}_{0} \bar\chi^{m7}_{5} + 
|\chi^{m7}_{\frac{1}{7}}|^2 +  \chi^{m7}_{\frac{1}{7}}
\bar\chi^{m7}_{\frac{22}{7}} +  |\chi^{m7}_{\frac{5}{7}}|^2 
\nonumber\\
&\ \ \ \
+  \chi^{m7}_{\frac{5}{7}} \bar\chi^{m7}_{\frac{12}{7}} + 
\chi^{m7}_{\frac{12}{7}} \bar\chi^{m7}_{\frac{5}{7}} + 
|\chi^{m7}_{\frac{12}{7}}|^2 +  \chi^{m7}_{\frac{22}{7}}
\bar\chi^{m7}_{\frac{1}{7}} 
\nonumber\\
&\ \ \ \
+  |\chi^{m7}_{\frac{22}{7}}|^2 +  \chi^{m7}_{5} \bar\chi^{m7}_{0} + 
|\chi^{m7}_{5}|^2 
\nonumber \\ 
Z_{a_2} &=  2 |\chi^{m7}_{\frac{4}{3}}|^2 +  2 |\chi^{m7}_{\frac{10}{21}}|^2 + 
2 |\chi^{m7}_{\frac{1}{21}}|^2 
\nonumber \\ 
Z_{b} &=  2 |\chi^{m7}_{\frac{4}{3}}|^2 +  2 |\chi^{m7}_{\frac{10}{21}}|^2 +  2
|\chi^{m7}_{\frac{1}{21}}|^2 
\nonumber \\ 
Z_{b_1} &=  2 \chi^{m7}_{\frac{4}{3}} \bar\chi^{m7}_{0} +  2
\chi^{m7}_{\frac{4}{3}} \bar\chi^{m7}_{5} +  2 \chi^{m7}_{\frac{10}{21}}
\bar\chi^{m7}_{\frac{1}{7}} +  2 \chi^{m7}_{\frac{10}{21}}
\bar\chi^{m7}_{\frac{22}{7}} 
\nonumber\\
&\ \ \ \
+  2 \chi^{m7}_{\frac{1}{21}} \bar\chi^{m7}_{\frac{5}{7}} +  2
\chi^{m7}_{\frac{1}{21}} \bar\chi^{m7}_{\frac{12}{7}} 
\nonumber \\ 
Z_{b_2} &=  2 \chi^{m7}_{0} \bar\chi^{m7}_{\frac{4}{3}} +  2
\chi^{m7}_{\frac{1}{7}} \bar\chi^{m7}_{\frac{10}{21}} +  2
\chi^{m7}_{\frac{5}{7}} \bar\chi^{m7}_{\frac{1}{21}} +  2
\chi^{m7}_{\frac{12}{7}} \bar\chi^{m7}_{\frac{1}{21}} 
\nonumber\\
&\ \ \ \
+  2 \chi^{m7}_{\frac{22}{7}} \bar\chi^{m7}_{\frac{10}{21}} +  2 \chi^{m7}_{5}
\bar\chi^{m7}_{\frac{4}{3}} 
\nonumber \\ 
Z_{c} &= 0
\nonumber \\ 
Z_{c_1} &= 0
\end{align}

$\one$-condensed boundary with 2 relevant operators:
\begin{align}
Z_\one &=  |\chi^{m7}_{0}|^2 +  |\chi^{m7}_{\frac{1}{7}}|^2 + 
|\chi^{m7}_{\frac{5}{7}}|^2 +  |\chi^{m7}_{\frac{12}{7}}|^2 + 
|\chi^{m7}_{\frac{22}{7}}|^2 
\nonumber\\
&\ \ \ \
+  |\chi^{m7}_{5}|^2 
\nonumber \\ 
Z_{a_1} &=  \chi^{m7}_{0} \bar\chi^{m7}_{5} +  \chi^{m7}_{\frac{1}{7}}
\bar\chi^{m7}_{\frac{22}{7}} +  \chi^{m7}_{\frac{5}{7}}
\bar\chi^{m7}_{\frac{12}{7}} +  \chi^{m7}_{\frac{12}{7}}
\bar\chi^{m7}_{\frac{5}{7}} 
\nonumber\\
&\ \ \ \
+  \chi^{m7}_{\frac{22}{7}} \bar\chi^{m7}_{\frac{1}{7}} +  \chi^{m7}_{5}
\bar\chi^{m7}_{0} 
\nonumber \\ 
Z_{a_2} &=  |\chi^{m7}_{\frac{4}{3}}|^2 +  |\chi^{m7}_{\frac{10}{21}}|^2 + 
|\chi^{m7}_{\frac{1}{21}}|^2 
\nonumber \\ 
Z_{b} &=  |\chi^{m7}_{\frac{4}{3}}|^2 +  |\chi^{m7}_{\frac{10}{21}}|^2 + 
|\chi^{m7}_{\frac{1}{21}}|^2 
\nonumber \\ 
Z_{b_1} &=  \chi^{m7}_{\frac{4}{3}} \bar\chi^{m7}_{0} +  \chi^{m7}_{\frac{4}{3}}
\bar\chi^{m7}_{5} +  \chi^{m7}_{\frac{10}{21}} \bar\chi^{m7}_{\frac{1}{7}} + 
\chi^{m7}_{\frac{10}{21}} \bar\chi^{m7}_{\frac{22}{7}} 
\nonumber\\
&\ \ \ \
+  \chi^{m7}_{\frac{1}{21}} \bar\chi^{m7}_{\frac{5}{7}} + 
\chi^{m7}_{\frac{1}{21}} \bar\chi^{m7}_{\frac{12}{7}} 
\nonumber \\ 
Z_{b_2} &=  \chi^{m7}_{0} \bar\chi^{m7}_{\frac{4}{3}} +  \chi^{m7}_{\frac{1}{7}}
\bar\chi^{m7}_{\frac{10}{21}} +  \chi^{m7}_{\frac{5}{7}}
\bar\chi^{m7}_{\frac{1}{21}} +  \chi^{m7}_{\frac{12}{7}}
\bar\chi^{m7}_{\frac{1}{21}} 
\nonumber\\
&\ \ \ \
+  \chi^{m7}_{\frac{22}{7}} \bar\chi^{m7}_{\frac{10}{21}} +  \chi^{m7}_{5}
\bar\chi^{m7}_{\frac{4}{3}} 
\nonumber \\ 
Z_{c} &=  |\chi^{m7}_{\frac{3}{8}}|^2 +  |\chi^{m7}_{\frac{1}{56}}|^2 + 
|\chi^{m7}_{\frac{5}{56}}|^2 +  |\chi^{m7}_{\frac{33}{56}}|^2 + 
|\chi^{m7}_{\frac{85}{56}}|^2 
\nonumber\\
&\ \ \ \
+  |\chi^{m7}_{\frac{23}{8}}|^2 
\nonumber \\ 
Z_{c_1} &=  \chi^{m7}_{\frac{3}{8}} \bar\chi^{m7}_{\frac{23}{8}} + 
\chi^{m7}_{\frac{1}{56}} \bar\chi^{m7}_{\frac{85}{56}} + 
\chi^{m7}_{\frac{5}{56}} \bar\chi^{m7}_{\frac{33}{56}} + 
\chi^{m7}_{\frac{33}{56}} \bar\chi^{m7}_{\frac{5}{56}} 
\nonumber\\
&\ \ \ \
+  \chi^{m7}_{\frac{85}{56}} \bar\chi^{m7}_{\frac{1}{56}} + 
\chi^{m7}_{\frac{23}{8}} \bar\chi^{m7}_{\frac{3}{8}} 
\end{align}

\begin{align}
Z_\one &=  |\chi^{m7}_{0}|^2 +  |\chi^{m7}_{\frac{1}{7}}|^2 + 
|\chi^{m7}_{\frac{5}{7}}|^2 +  |\chi^{m7}_{\frac{12}{7}}|^2 + 
|\chi^{m7}_{\frac{22}{7}}|^2 
\nonumber\\
&\ \ \ \ 
+  |\chi^{m7}_{5}|^2 +  |\chi^{m7}_{\frac{3}{8}}|^2 + 
|\chi^{m7}_{\frac{1}{56}}|^2 +  |\chi^{m7}_{\frac{5}{56}}|^2 + 
|\chi^{m7}_{\frac{33}{56}}|^2 
\nonumber\\
&\ \ \ \ 
+  |\chi^{m7}_{\frac{85}{56}}|^2 +  |\chi^{m7}_{\frac{23}{8}}|^2 + 
|\chi^{m7}_{\frac{4}{3}}|^2 +  |\chi^{m7}_{\frac{10}{21}}|^2 + 
|\chi^{m7}_{\frac{1}{21}}|^2 
\nonumber \\ 
Z_{a_1} &= 0
\nonumber \\ 
Z_{a_2} &=  |\chi^{m7}_{0}|^2 +  |\chi^{m7}_{\frac{1}{7}}|^2 + 
|\chi^{m7}_{\frac{5}{7}}|^2 +  |\chi^{m7}_{\frac{12}{7}}|^2 + 
|\chi^{m7}_{\frac{22}{7}}|^2 
\nonumber\\
&\ \ \ \ 
+  |\chi^{m7}_{5}|^2 +  |\chi^{m7}_{\frac{3}{8}}|^2 + 
|\chi^{m7}_{\frac{1}{56}}|^2 +  |\chi^{m7}_{\frac{5}{56}}|^2 + 
|\chi^{m7}_{\frac{33}{56}}|^2 
\nonumber\\
&\ \ \ \ 
+  |\chi^{m7}_{\frac{85}{56}}|^2 +  |\chi^{m7}_{\frac{23}{8}}|^2 + 
|\chi^{m7}_{\frac{4}{3}}|^2 +  |\chi^{m7}_{\frac{10}{21}}|^2 + 
|\chi^{m7}_{\frac{1}{21}}|^2 
\nonumber \\ 
Z_{b} &= 0
\nonumber \\ 
Z_{b_1} &= 0
\nonumber \\ 
Z_{b_2} &= 0
\nonumber \\ 
Z_{c} &=  |\chi^{m7}_{0}|^2 +  |\chi^{m7}_{\frac{1}{7}}|^2 + 
|\chi^{m7}_{\frac{5}{7}}|^2 +  |\chi^{m7}_{\frac{12}{7}}|^2 + 
|\chi^{m7}_{\frac{22}{7}}|^2 
\nonumber\\
&\ \ \ \ 
+  |\chi^{m7}_{5}|^2 +  |\chi^{m7}_{\frac{3}{8}}|^2 + 
|\chi^{m7}_{\frac{1}{56}}|^2 +  |\chi^{m7}_{\frac{5}{56}}|^2 + 
|\chi^{m7}_{\frac{33}{56}}|^2 
\nonumber\\
&\ \ \ \ 
+  |\chi^{m7}_{\frac{85}{56}}|^2 +  |\chi^{m7}_{\frac{23}{8}}|^2 + 
|\chi^{m7}_{\frac{4}{3}}|^2 +  |\chi^{m7}_{\frac{10}{21}}|^2 + 
|\chi^{m7}_{\frac{1}{21}}|^2 
\nonumber \\ 
Z_{c_1} &= 0
\end{align}

$\one \oplus a_2$-condensed boundary with 4 relevant operators:
\begin{align}
Z_\one &=  |\chi^{m7}_{0}|^2 +  |\chi^{m7}_{\frac{1}{7}}|^2 + 
|\chi^{m7}_{\frac{5}{7}}|^2 +  |\chi^{m7}_{\frac{12}{7}}|^2 + 
|\chi^{m7}_{\frac{22}{7}}|^2 
\nonumber\\
&\ \ \ \ 
+  |\chi^{m7}_{5}|^2 +  |\chi^{m7}_{\frac{4}{3}}|^2 + 
|\chi^{m7}_{\frac{10}{21}}|^2 +  |\chi^{m7}_{\frac{1}{21}}|^2 
\nonumber \\ 
Z_{a_1} &=  \chi^{m7}_{0} \bar\chi^{m7}_{5} +  \chi^{m7}_{\frac{1}{7}}
\bar\chi^{m7}_{\frac{22}{7}} +  \chi^{m7}_{\frac{5}{7}}
\bar\chi^{m7}_{\frac{12}{7}} +  \chi^{m7}_{\frac{12}{7}}
\bar\chi^{m7}_{\frac{5}{7}} +  \chi^{m7}_{\frac{22}{7}}
\bar\chi^{m7}_{\frac{1}{7}} 
\nonumber\\
&\ \ \ \ 
+  \chi^{m7}_{5} \bar\chi^{m7}_{0} +  |\chi^{m7}_{\frac{4}{3}}|^2 + 
|\chi^{m7}_{\frac{10}{21}}|^2 +  |\chi^{m7}_{\frac{1}{21}}|^2 
\nonumber \\ 
Z_{a_2} &=  |\chi^{m7}_{0}|^2 +  \chi^{m7}_{0} \bar\chi^{m7}_{5} + 
|\chi^{m7}_{\frac{1}{7}}|^2 +  \chi^{m7}_{\frac{1}{7}}
\bar\chi^{m7}_{\frac{22}{7}} +  |\chi^{m7}_{\frac{5}{7}}|^2 
\nonumber\\
&\ \ \ \ 
+  \chi^{m7}_{\frac{5}{7}} \bar\chi^{m7}_{\frac{12}{7}} + 
\chi^{m7}_{\frac{12}{7}} \bar\chi^{m7}_{\frac{5}{7}} + 
|\chi^{m7}_{\frac{12}{7}}|^2 +  \chi^{m7}_{\frac{22}{7}}
\bar\chi^{m7}_{\frac{1}{7}} 
\nonumber\\
&\ \ \ \ 
+  |\chi^{m7}_{\frac{22}{7}}|^2 +  \chi^{m7}_{5} \bar\chi^{m7}_{0} + 
|\chi^{m7}_{5}|^2 +  2 |\chi^{m7}_{\frac{4}{3}}|^2 +  2
|\chi^{m7}_{\frac{10}{21}}|^2 
\nonumber\\
&\ \ \ \ 
+  2 |\chi^{m7}_{\frac{1}{21}}|^2 
\nonumber \\ 
Z_{b} &= 0
\nonumber \\ 
Z_{b_1} &= 0
\nonumber \\ 
Z_{b_2} &= 0
\nonumber \\ 
Z_{c} &=  |\chi^{m7}_{\frac{3}{8}}|^2 +  |\chi^{m7}_{\frac{1}{56}}|^2 + 
|\chi^{m7}_{\frac{5}{56}}|^2 +  |\chi^{m7}_{\frac{33}{56}}|^2 + 
|\chi^{m7}_{\frac{85}{56}}|^2 +  |\chi^{m7}_{\frac{23}{8}}|^2 
\nonumber \\ 
Z_{c_1} &=  \chi^{m7}_{\frac{3}{8}} \bar\chi^{m7}_{\frac{23}{8}} + 
\chi^{m7}_{\frac{1}{56}} \bar\chi^{m7}_{\frac{85}{56}} + 
\chi^{m7}_{\frac{5}{56}} \bar\chi^{m7}_{\frac{33}{56}} + 
\chi^{m7}_{\frac{33}{56}} \bar\chi^{m7}_{\frac{5}{56}} 
\nonumber\\
&\ \ \ \ 
+  \chi^{m7}_{\frac{85}{56}} \bar\chi^{m7}_{\frac{1}{56}} + 
\chi^{m7}_{\frac{23}{8}} \bar\chi^{m7}_{\frac{3}{8}} 
\end{align}

\begin{align}
Z_\one &=  |\chi^{m7}_{0}|^2 +  |\chi^{m7}_{\frac{1}{7}}|^2 + 
|\chi^{m7}_{\frac{5}{7}}|^2 +  |\chi^{m7}_{\frac{12}{7}}|^2 + 
|\chi^{m7}_{\frac{22}{7}}|^2 
\nonumber\\
&\ \ \ \ 
+  |\chi^{m7}_{5}|^2 +  |\chi^{m7}_{\frac{3}{8}}|^2 + 
|\chi^{m7}_{\frac{1}{56}}|^2 +  |\chi^{m7}_{\frac{5}{56}}|^2 + 
|\chi^{m7}_{\frac{33}{56}}|^2 
\nonumber\\
&\ \ \ \ 
+  |\chi^{m7}_{\frac{85}{56}}|^2 +  |\chi^{m7}_{\frac{23}{8}}|^2 + 
|\chi^{m7}_{\frac{4}{3}}|^2 +  |\chi^{m7}_{\frac{10}{21}}|^2 + 
|\chi^{m7}_{\frac{1}{21}}|^2 
\nonumber \\ 
Z_{a_1} &= 0
\nonumber \\ 
Z_{a_2} &= 0
\nonumber \\ 
Z_{b} &=  |\chi^{m7}_{0}|^2 +  |\chi^{m7}_{\frac{1}{7}}|^2 + 
|\chi^{m7}_{\frac{5}{7}}|^2 +  |\chi^{m7}_{\frac{12}{7}}|^2 + 
|\chi^{m7}_{\frac{22}{7}}|^2 
\nonumber\\
&\ \ \ \ 
+  |\chi^{m7}_{5}|^2 +  |\chi^{m7}_{\frac{3}{8}}|^2 + 
|\chi^{m7}_{\frac{1}{56}}|^2 +  |\chi^{m7}_{\frac{5}{56}}|^2 + 
|\chi^{m7}_{\frac{33}{56}}|^2 
\nonumber\\
&\ \ \ \ 
+  |\chi^{m7}_{\frac{85}{56}}|^2 +  |\chi^{m7}_{\frac{23}{8}}|^2 + 
|\chi^{m7}_{\frac{4}{3}}|^2 +  |\chi^{m7}_{\frac{10}{21}}|^2 + 
|\chi^{m7}_{\frac{1}{21}}|^2 
\nonumber \\ 
Z_{b_1} &= 0
\nonumber \\ 
Z_{b_2} &= 0
\nonumber \\ 
Z_{c} &=  |\chi^{m7}_{0}|^2 +  |\chi^{m7}_{\frac{1}{7}}|^2 + 
|\chi^{m7}_{\frac{5}{7}}|^2 +  |\chi^{m7}_{\frac{12}{7}}|^2 + 
|\chi^{m7}_{\frac{22}{7}}|^2 
\nonumber\\
&\ \ \ \ 
+  |\chi^{m7}_{5}|^2 +  |\chi^{m7}_{\frac{3}{8}}|^2 + 
|\chi^{m7}_{\frac{1}{56}}|^2 +  |\chi^{m7}_{\frac{5}{56}}|^2 + 
|\chi^{m7}_{\frac{33}{56}}|^2 
\nonumber\\
&\ \ \ \ 
+  |\chi^{m7}_{\frac{85}{56}}|^2 +  |\chi^{m7}_{\frac{23}{8}}|^2 + 
|\chi^{m7}_{\frac{4}{3}}|^2 +  |\chi^{m7}_{\frac{10}{21}}|^2 + 
|\chi^{m7}_{\frac{1}{21}}|^2 
\nonumber \\ 
Z_{c_1} &= 0
\end{align}

$\one \oplus b$-condensed boundary with 4 relevant operators:
\begin{align}
Z_\one &=  |\chi^{m7}_{0}|^2 +  |\chi^{m7}_{\frac{1}{7}}|^2 + 
|\chi^{m7}_{\frac{5}{7}}|^2 +  |\chi^{m7}_{\frac{12}{7}}|^2 + 
|\chi^{m7}_{\frac{22}{7}}|^2 
\nonumber\\
&\ \ \ \ 
+  |\chi^{m7}_{5}|^2 +  |\chi^{m7}_{\frac{4}{3}}|^2 + 
|\chi^{m7}_{\frac{10}{21}}|^2 +  |\chi^{m7}_{\frac{1}{21}}|^2 
\nonumber \\ 
Z_{a_1} &=  \chi^{m7}_{0} \bar\chi^{m7}_{5} +  \chi^{m7}_{\frac{1}{7}}
\bar\chi^{m7}_{\frac{22}{7}} +  \chi^{m7}_{\frac{5}{7}}
\bar\chi^{m7}_{\frac{12}{7}} +  \chi^{m7}_{\frac{12}{7}}
\bar\chi^{m7}_{\frac{5}{7}} 
\nonumber\\
&\ \ \ \ 
+  \chi^{m7}_{\frac{22}{7}} \bar\chi^{m7}_{\frac{1}{7}} +  \chi^{m7}_{5}
\bar\chi^{m7}_{0} +  |\chi^{m7}_{\frac{4}{3}}|^2 + 
|\chi^{m7}_{\frac{10}{21}}|^2 +  |\chi^{m7}_{\frac{1}{21}}|^2 
\nonumber \\ 
Z_{a_2} &= 0
\nonumber \\ 
Z_{b} &=  |\chi^{m7}_{0}|^2 +  \chi^{m7}_{0} \bar\chi^{m7}_{5} + 
|\chi^{m7}_{\frac{1}{7}}|^2 +  \chi^{m7}_{\frac{1}{7}}
\bar\chi^{m7}_{\frac{22}{7}} +  |\chi^{m7}_{\frac{5}{7}}|^2 
\nonumber\\
&\ \ \ \ 
+  \chi^{m7}_{\frac{5}{7}} \bar\chi^{m7}_{\frac{12}{7}} + 
\chi^{m7}_{\frac{12}{7}} \bar\chi^{m7}_{\frac{5}{7}} + 
|\chi^{m7}_{\frac{12}{7}}|^2 +  \chi^{m7}_{\frac{22}{7}}
\bar\chi^{m7}_{\frac{1}{7}} +  |\chi^{m7}_{\frac{22}{7}}|^2 
\nonumber\\
&\ \ \ \ 
+  \chi^{m7}_{5} \bar\chi^{m7}_{0} +  |\chi^{m7}_{5}|^2 +  2
|\chi^{m7}_{\frac{4}{3}}|^2 +  2 |\chi^{m7}_{\frac{10}{21}}|^2 +  2
|\chi^{m7}_{\frac{1}{21}}|^2 
\nonumber \\ 
Z_{b_1} &= 0
\nonumber \\ 
Z_{b_2} &= 0
\nonumber \\ 
Z_{c} &=  |\chi^{m7}_{\frac{3}{8}}|^2 +  |\chi^{m7}_{\frac{1}{56}}|^2 + 
|\chi^{m7}_{\frac{5}{56}}|^2 +  |\chi^{m7}_{\frac{33}{56}}|^2 + 
|\chi^{m7}_{\frac{85}{56}}|^2 +  |\chi^{m7}_{\frac{23}{8}}|^2 
\nonumber \\ 
Z_{c_1} &=  \chi^{m7}_{\frac{3}{8}} \bar\chi^{m7}_{\frac{23}{8}} + 
\chi^{m7}_{\frac{1}{56}} \bar\chi^{m7}_{\frac{85}{56}} + 
\chi^{m7}_{\frac{5}{56}} \bar\chi^{m7}_{\frac{33}{56}} + 
\chi^{m7}_{\frac{33}{56}} \bar\chi^{m7}_{\frac{5}{56}} + 
\chi^{m7}_{\frac{85}{56}} \bar\chi^{m7}_{\frac{1}{56}} 
\nonumber\\
&\ \ \ \ 
+  \chi^{m7}_{\frac{23}{8}} \bar\chi^{m7}_{\frac{3}{8}} 
\end{align}

\begin{align}
Z_\one &=  |\chi^{m7}_{0}|^2 +  \chi^{m7}_{0} \bar\chi^{m7}_{5} + 
|\chi^{m7}_{\frac{1}{7}}|^2 +  \chi^{m7}_{\frac{1}{7}}
\bar\chi^{m7}_{\frac{22}{7}} +  |\chi^{m7}_{\frac{5}{7}}|^2 
\nonumber\\
&\ \ \ \ 
+  \chi^{m7}_{\frac{5}{7}} \bar\chi^{m7}_{\frac{12}{7}} + 
\chi^{m7}_{\frac{12}{7}} \bar\chi^{m7}_{\frac{5}{7}} + 
|\chi^{m7}_{\frac{12}{7}}|^2 +  \chi^{m7}_{\frac{22}{7}}
\bar\chi^{m7}_{\frac{1}{7}} 
\nonumber\\
&\ \ \ \ 
+  |\chi^{m7}_{\frac{22}{7}}|^2 +  \chi^{m7}_{5} \bar\chi^{m7}_{0} + 
|\chi^{m7}_{5}|^2 +  2 |\chi^{m7}_{\frac{4}{3}}|^2 +  2
|\chi^{m7}_{\frac{10}{21}}|^2 
\nonumber\\
&\ \ \ \ 
+  2 |\chi^{m7}_{\frac{1}{21}}|^2 
\nonumber \\ 
Z_{a_1} &= 0
\nonumber \\ 
Z_{a_2} &=  |\chi^{m7}_{0}|^2 +  \chi^{m7}_{0} \bar\chi^{m7}_{5} + 
|\chi^{m7}_{\frac{1}{7}}|^2 +  \chi^{m7}_{\frac{1}{7}}
\bar\chi^{m7}_{\frac{22}{7}} +  |\chi^{m7}_{\frac{5}{7}}|^2 
\nonumber\\
&\ \ \ \ 
+  \chi^{m7}_{\frac{5}{7}} \bar\chi^{m7}_{\frac{12}{7}} + 
\chi^{m7}_{\frac{12}{7}} \bar\chi^{m7}_{\frac{5}{7}} + 
|\chi^{m7}_{\frac{12}{7}}|^2 +  \chi^{m7}_{\frac{22}{7}}
\bar\chi^{m7}_{\frac{1}{7}} 
\nonumber\\
&\ \ \ \ 
+  |\chi^{m7}_{\frac{22}{7}}|^2 +  \chi^{m7}_{5} \bar\chi^{m7}_{0} + 
|\chi^{m7}_{5}|^2 +  2 |\chi^{m7}_{\frac{4}{3}}|^2 +  2
|\chi^{m7}_{\frac{10}{21}}|^2 
\nonumber\\
&\ \ \ \ 
+  2 |\chi^{m7}_{\frac{1}{21}}|^2 
\nonumber \\ 
Z_{b} &= 0
\nonumber \\ 
Z_{b_1} &= 0
\nonumber \\ 
Z_{b_2} &= 0
\nonumber \\ 
Z_{c} &=  |\chi^{m7}_{0}|^2 +  \chi^{m7}_{0} \bar\chi^{m7}_{5} + 
|\chi^{m7}_{\frac{1}{7}}|^2 +  \chi^{m7}_{\frac{1}{7}}
\bar\chi^{m7}_{\frac{22}{7}} +  |\chi^{m7}_{\frac{5}{7}}|^2 
\nonumber\\
&\ \ \ \ 
+  \chi^{m7}_{\frac{5}{7}} \bar\chi^{m7}_{\frac{12}{7}} + 
\chi^{m7}_{\frac{12}{7}} \bar\chi^{m7}_{\frac{5}{7}} + 
|\chi^{m7}_{\frac{12}{7}}|^2 +  \chi^{m7}_{\frac{22}{7}}
\bar\chi^{m7}_{\frac{1}{7}} 
\nonumber\\
&\ \ \ \ 
+  |\chi^{m7}_{\frac{22}{7}}|^2 +  \chi^{m7}_{5} \bar\chi^{m7}_{0} + 
|\chi^{m7}_{5}|^2 +  2 |\chi^{m7}_{\frac{4}{3}}|^2 +  2
|\chi^{m7}_{\frac{10}{21}}|^2 
\nonumber\\
&\ \ \ \ 
+  2 |\chi^{m7}_{\frac{1}{21}}|^2 
\nonumber \\ 
Z_{c_1} &= 0
\end{align}

\begin{align}
Z_\one &=  |\chi^{m7}_{0}|^2 +  \chi^{m7}_{0} \bar\chi^{m7}_{5} + 
|\chi^{m7}_{\frac{1}{7}}|^2 +  \chi^{m7}_{\frac{1}{7}}
\bar\chi^{m7}_{\frac{22}{7}} +  |\chi^{m7}_{\frac{5}{7}}|^2 
\nonumber\\
&\ \ \ \ 
+  \chi^{m7}_{\frac{5}{7}} \bar\chi^{m7}_{\frac{12}{7}} + 
\chi^{m7}_{\frac{12}{7}} \bar\chi^{m7}_{\frac{5}{7}} + 
|\chi^{m7}_{\frac{12}{7}}|^2 +  \chi^{m7}_{\frac{22}{7}}
\bar\chi^{m7}_{\frac{1}{7}} 
\nonumber\\
&\ \ \ \ 
+  |\chi^{m7}_{\frac{22}{7}}|^2 +  \chi^{m7}_{5} \bar\chi^{m7}_{0} + 
|\chi^{m7}_{5}|^2 +  2 |\chi^{m7}_{\frac{4}{3}}|^2 +  2
|\chi^{m7}_{\frac{10}{21}}|^2 
\nonumber\\
&\ \ \ \ 
+  2 |\chi^{m7}_{\frac{1}{21}}|^2 
\nonumber \\ 
Z_{a_1} &= 0
\nonumber \\ 
Z_{a_2} &= 0
\nonumber \\ 
Z_{b} &=  |\chi^{m7}_{0}|^2 +  \chi^{m7}_{0} \bar\chi^{m7}_{5} + 
|\chi^{m7}_{\frac{1}{7}}|^2 +  \chi^{m7}_{\frac{1}{7}}
\bar\chi^{m7}_{\frac{22}{7}} +  |\chi^{m7}_{\frac{5}{7}}|^2 
\nonumber\\
&\ \ \ \ 
+  \chi^{m7}_{\frac{5}{7}} \bar\chi^{m7}_{\frac{12}{7}} + 
\chi^{m7}_{\frac{12}{7}} \bar\chi^{m7}_{\frac{5}{7}} + 
|\chi^{m7}_{\frac{12}{7}}|^2 +  \chi^{m7}_{\frac{22}{7}}
\bar\chi^{m7}_{\frac{1}{7}} 
\nonumber\\
&\ \ \ \ 
+  |\chi^{m7}_{\frac{22}{7}}|^2 +  \chi^{m7}_{5} \bar\chi^{m7}_{0} + 
|\chi^{m7}_{5}|^2 +  2 |\chi^{m7}_{\frac{4}{3}}|^2 +  2
|\chi^{m7}_{\frac{10}{21}}|^2 
\nonumber\\
&\ \ \ \ 
+  2 |\chi^{m7}_{\frac{1}{21}}|^2 
\nonumber \\ 
Z_{b_1} &= 0
\nonumber \\ 
Z_{b_2} &= 0
\nonumber \\ 
Z_{c} &=  |\chi^{m7}_{0}|^2 +  \chi^{m7}_{0} \bar\chi^{m7}_{5} + 
|\chi^{m7}_{\frac{1}{7}}|^2 +  \chi^{m7}_{\frac{1}{7}}
\bar\chi^{m7}_{\frac{22}{7}} +  |\chi^{m7}_{\frac{5}{7}}|^2 
\nonumber\\
&\ \ \ \ 
+  \chi^{m7}_{\frac{5}{7}} \bar\chi^{m7}_{\frac{12}{7}} + 
\chi^{m7}_{\frac{12}{7}} \bar\chi^{m7}_{\frac{5}{7}} + 
|\chi^{m7}_{\frac{12}{7}}|^2 +  \chi^{m7}_{\frac{22}{7}}
\bar\chi^{m7}_{\frac{1}{7}} 
\nonumber\\
&\ \ \ \ 
+  |\chi^{m7}_{\frac{22}{7}}|^2 +  \chi^{m7}_{5} \bar\chi^{m7}_{0} + 
|\chi^{m7}_{5}|^2 +  2 |\chi^{m7}_{\frac{4}{3}}|^2 +  2
|\chi^{m7}_{\frac{10}{21}}|^2 
\nonumber\\
&\ \ \ \ 
+  2 |\chi^{m7}_{\frac{1}{21}}|^2 
\nonumber \\ 
Z_{c_1} &= 0
\end{align}

\begin{align}
Z_\one &=  |\chi^{m7}_{0}|^2 +  |\chi^{m7}_{\frac{1}{7}}|^2 + 
|\chi^{m7}_{\frac{5}{7}}|^2 +  |\chi^{m7}_{\frac{12}{7}}|^2 + 
|\chi^{m7}_{\frac{22}{7}}|^2 
\nonumber\\
&\ \ \ \ 
+  |\chi^{m7}_{5}|^2 +  |\chi^{m7}_{\frac{3}{8}}|^2 + 
|\chi^{m7}_{\frac{1}{56}}|^2 +  |\chi^{m7}_{\frac{5}{56}}|^2 + 
|\chi^{m7}_{\frac{33}{56}}|^2 
\nonumber\\
&\ \ \ \ 
+  |\chi^{m7}_{\frac{85}{56}}|^2 +  |\chi^{m7}_{\frac{23}{8}}|^2 + 
|\chi^{m7}_{\frac{4}{3}}|^2 +  |\chi^{m7}_{\frac{10}{21}}|^2 + 
|\chi^{m7}_{\frac{1}{21}}|^2 
\nonumber \\ 
Z_{a_1} &=  |\chi^{m7}_{0}|^2 +  |\chi^{m7}_{\frac{1}{7}}|^2 + 
|\chi^{m7}_{\frac{5}{7}}|^2 +  |\chi^{m7}_{\frac{12}{7}}|^2 + 
|\chi^{m7}_{\frac{22}{7}}|^2 
\nonumber\\
&\ \ \ \ 
+  |\chi^{m7}_{5}|^2 +  |\chi^{m7}_{\frac{3}{8}}|^2 + 
|\chi^{m7}_{\frac{1}{56}}|^2 +  |\chi^{m7}_{\frac{5}{56}}|^2 + 
|\chi^{m7}_{\frac{33}{56}}|^2 
\nonumber\\
&\ \ \ \ 
+  |\chi^{m7}_{\frac{85}{56}}|^2 +  |\chi^{m7}_{\frac{23}{8}}|^2 + 
|\chi^{m7}_{\frac{4}{3}}|^2 +  |\chi^{m7}_{\frac{10}{21}}|^2 + 
|\chi^{m7}_{\frac{1}{21}}|^2 
\nonumber \\ 
Z_{a_2} &=  2 |\chi^{m7}_{0}|^2 +  2 |\chi^{m7}_{\frac{1}{7}}|^2 +  2
|\chi^{m7}_{\frac{5}{7}}|^2 +  2 |\chi^{m7}_{\frac{12}{7}}|^2 +  2
|\chi^{m7}_{\frac{22}{7}}|^2 
\nonumber\\
&\ \ \ \ 
+  2 |\chi^{m7}_{5}|^2 +  2 |\chi^{m7}_{\frac{3}{8}}|^2 +  2
|\chi^{m7}_{\frac{1}{56}}|^2 +  2 |\chi^{m7}_{\frac{5}{56}}|^2 +  2
|\chi^{m7}_{\frac{33}{56}}|^2 
\nonumber\\
&\ \ \ \ 
+  2 |\chi^{m7}_{\frac{85}{56}}|^2 +  2 |\chi^{m7}_{\frac{23}{8}}|^2 +  2
|\chi^{m7}_{\frac{4}{3}}|^2 +  2 |\chi^{m7}_{\frac{10}{21}}|^2 +  2
|\chi^{m7}_{\frac{1}{21}}|^2 
\nonumber \\ 
Z_{b} &= 0
\nonumber \\ 
Z_{b_1} &= 0
\nonumber \\ 
Z_{b_2} &= 0
\nonumber \\ 
Z_{c} &= 0
\nonumber \\ 
Z_{c_1} &= 0
\end{align}

\begin{align}
Z_\one &=  |\chi^{m7}_{0}|^2 +  |\chi^{m7}_{\frac{1}{7}}|^2 + 
|\chi^{m7}_{\frac{5}{7}}|^2 +  |\chi^{m7}_{\frac{12}{7}}|^2 + 
|\chi^{m7}_{\frac{22}{7}}|^2 
\nonumber\\
&\ \ \ \ 
+  |\chi^{m7}_{5}|^2 +  |\chi^{m7}_{\frac{3}{8}}|^2 + 
|\chi^{m7}_{\frac{1}{56}}|^2 +  |\chi^{m7}_{\frac{5}{56}}|^2 + 
|\chi^{m7}_{\frac{33}{56}}|^2 
\nonumber\\
&\ \ \ \ 
+  |\chi^{m7}_{\frac{85}{56}}|^2 +  |\chi^{m7}_{\frac{23}{8}}|^2 + 
|\chi^{m7}_{\frac{4}{3}}|^2 +  |\chi^{m7}_{\frac{10}{21}}|^2 + 
|\chi^{m7}_{\frac{1}{21}}|^2 
\nonumber \\ 
Z_{a_1} &=  |\chi^{m7}_{0}|^2 +  |\chi^{m7}_{\frac{1}{7}}|^2 + 
|\chi^{m7}_{\frac{5}{7}}|^2 +  |\chi^{m7}_{\frac{12}{7}}|^2 + 
|\chi^{m7}_{\frac{22}{7}}|^2 
\nonumber\\
&\ \ \ \ 
+  |\chi^{m7}_{5}|^2 +  |\chi^{m7}_{\frac{3}{8}}|^2 + 
|\chi^{m7}_{\frac{1}{56}}|^2 +  |\chi^{m7}_{\frac{5}{56}}|^2 + 
|\chi^{m7}_{\frac{33}{56}}|^2 
\nonumber\\
&\ \ \ \ 
+  |\chi^{m7}_{\frac{85}{56}}|^2 +  |\chi^{m7}_{\frac{23}{8}}|^2 + 
|\chi^{m7}_{\frac{4}{3}}|^2 +  |\chi^{m7}_{\frac{10}{21}}|^2 + 
|\chi^{m7}_{\frac{1}{21}}|^2 
\nonumber \\ 
Z_{a_2} &= 0
\nonumber \\ 
Z_{b} &=  2 |\chi^{m7}_{0}|^2 +  2 |\chi^{m7}_{\frac{1}{7}}|^2 +  2
|\chi^{m7}_{\frac{5}{7}}|^2 +  2 |\chi^{m7}_{\frac{12}{7}}|^2 +  2
|\chi^{m7}_{\frac{22}{7}}|^2 
\nonumber\\
&\ \ \ \ 
+  2 |\chi^{m7}_{5}|^2 +  2 |\chi^{m7}_{\frac{3}{8}}|^2 +  2
|\chi^{m7}_{\frac{1}{56}}|^2 +  2 |\chi^{m7}_{\frac{5}{56}}|^2 +  2
|\chi^{m7}_{\frac{33}{56}}|^2 
\nonumber\\
&\ \ \ \ 
+  2 |\chi^{m7}_{\frac{85}{56}}|^2 +  2 |\chi^{m7}_{\frac{23}{8}}|^2 +  2
|\chi^{m7}_{\frac{4}{3}}|^2 +  2 |\chi^{m7}_{\frac{10}{21}}|^2 +  2
|\chi^{m7}_{\frac{1}{21}}|^2 
\nonumber \\ 
Z_{b_1} &= 0
\nonumber \\ 
Z_{b_2} &= 0
\nonumber \\ 
Z_{c} &= 0
\nonumber \\ 
Z_{c_1} &= 0
\end{align}

\section{ 1+1D non-invertible symmetry $\t S_3$ -- dual symmetry of $S_3$ }

In \Rf{JW191213492,KZ200514178}, a 1+1D model with a non-invertible symmetry,
denoted as $\t S_3$, is constructed.  The model has degrees of freedom on the
links $ij$, which are labeled by the $S_3$ group elements $g_{ij}\in S_3$.  The
$\t S_3$ symmetry transformation are generated by
\begin{align}
W_R = \Tr\Big(\prod_i R(g_{i,i+1})\Big)
\end{align}
for all irreducible representations $R$ of $S_3$, \ie $R=\one,a_1,a_2$.
Using
\begin{align}
W_{R} W_{R'}
=\Tr\Big(\prod_i R(g_{i,i+1})\otimes R'(g_{i,i+1}) \Big)
\end{align}
we find that
the symmetry transformations satisfy the following algebra
\begin{align}
W_\one W_\one &= W_\one, \ \ \
W_\one W_{a_1} = W_{a_1}, \ \ \ 
W_\one W_{a_2} = W_{a_2},
\nonumber\\
W_{a_1} W_\one &= W_{a_1}, \ \ \
W_{a_1} W_{a_1} = W_\one, \ \ \ 
W_{a_1} W_{a_2} = W_{a_2},
\nonumber\\
W_{a_2} W_\one &= W_{a_2}, \ \ \
W_{a_2} W_{a_1} = W_{a_2},  
\nonumber\\
W_{a_2} W_{a_2} &= W_\one+W_{a_1}+W_{a_2}.
\end{align}
For example, $R=a_2$ is a 2-dimensional irreducible representation of $S_3$.
$a_2\otimes a_2$ is a 4-dimensional reducible representation of $S_3$, which is
a direction sum of an 1-dimensional trivial representation $\one$, an
1-dimensional nontrivial representation $a_1$, and a 2-dimensional irreducible
representation $a_2$: $a_2\otimes a_2 = \one\oplus a_1\oplus a_2$.  This leads
to the last expression in the above.

The algebra for the symmetry transformations is not a group algebra like
$W_RW_{R'}=W_{R''}$.  The composition of two $a_2$-symmetry transformations
$W_{a_2} W_{a_2} = W_\one+W_{a_1}+W_{a_2}$ makes the $\t S_3$ symmetry
non-invertible.  Such kind of symmetry was referred to as algebraic symmetry,
or fusion category symmetry, \etc.

\Rf{JW191213492,KZ200514178} showed that $\t S_3$ and $S_3$ symmetries are
equivalent symmetries, \ie they have the same \hcatsymm.  \Rf{CW220303596} shows
that the symmetries with the same \hcatsymm\ have isomorphic algebras of local
symmetric operators, which is the meaning of equivalence. A \hcatsymm\ is
nothing but an isomorphic class of algebras of local symmetric operators.

From the holographic point of view, both 1+1D $S_3$ and $\t S_3$ symmetry are
described by the same 2+1D topological order $\eGau_{S_3}$.  In other words,
\texttt{systems} with $S_3$-symmetry are \emph{exactly locally reproduced} by
\texttt{boundaries} of $\eGau_{S_3}$ topological orders, in the sense that the
local symmetric operators for a \texttt{system} have identical correlations
with the local symmetric operators for the corresponding \texttt{boundary}.
Similarly, \texttt{systems} with $\t S_3$-symmetry are also \emph{exactly
locally reproduced} by \texttt{boundaries} of $\eGau_{S_3}$ topological orders.

The charges of $S_3$-symmetry correspond to $a_1$ and $a_2$ anyons in
$\eGau_{S_3}$, while the $S_3$ symmetry transformations correspond to
string operators that produce $b$ and $c$ anyons in $\eGau_{S_3}$ (at the string
ends).  Similarly, the charges of $\t S_3$-symmetry correspond to $b$ and $c$
anyons in $\eGau_{S_3}$, while the $\t S_3$ symmetry transformations
correspond to string operators that produce $a_1$ and $a_2$ anyons in
$\eGau_{S_3}$.

Certainly, we can also divide the anyons in $\eGau_{S_3}$ differently. Call some
of them charges of a symmetry and others as the transformation of the symmetry.
This way, we get a different symmetry or an anomalous symmetry, or even a
symmetry beyond anomaly.

This example demonstrates that the notions of symmetry and anomaly are not
essential notions that reflect the physical properties of a quantum system.
They are notions that depend on our point of views to look at the system.
Different angles to look at the same system will leads to different points of
view.  In contrast, \hcatsymm\ reflects the essence of (anomalous) symmetries.
It more directly reflects the physical properties of a quantum system.

\bibliography{../../bib/all,../../bib/allnew,../../bib/publst,../../bib/publstnew,../../bib/ACrefs.bib}

\end{document}